\documentclass{article}

\usepackage{graphicx} 
\usepackage[a4paper, total={6in, 9in}]{geometry}
\usepackage{amsmath, amssymb, amsthm, physics}
\usepackage{mathtools, shuffle}
\usepackage{float}
\usepackage{tcolorbox}
\usepackage{makecell}
\usepackage{hyperref}
\hypersetup{hidelinks}
\usepackage{tikz,tikz-3dplot}

\numberwithin{equation}{section}

\makeatletter
 \tikzoption{canvas is plane}[]{\@setOxy#1}
 \def\@setOxy O(#1,#2,#3)x(#4,#5,#6)y(#7,#8,#9)%
   {\def\tikz@plane@origin{\pgfpointxyz{#1}{#2}{#3}}%
    \def\tikz@plane@x{\pgfpointxyz{#4}{#5}{#6}}%
    \def\tikz@plane@y{\pgfpointxyz{#7}{#8}{#9}}%
    \tikz@canvas@is@plane
   }
\makeatother  

\usepackage[compat=1.1.0]{tikz-feynman}
\usetikzlibrary{shapes.geometric, arrows, patterns, decorations.markings}
\usetikzlibrary{calc}
\tikzstyle{top} = [rectangle, rounded corners, minimum width=3cm, minimum height=1cm, text centered, text width=3cm, draw=black, fill=red!30]
\allowdisplaybreaks

\newcommand{\kperiod}{
\begin{minipage}{1.7cm}
\begin{tikzpicture}
      \draw [thin] (-0.5,0.8) -- (0.5,1.2);
      \draw [thin] (-0.5,0.8) -- (0.5,0.4);
      \draw [thin] (-0.5,0.8) -- (0.5,-0.4);
      \draw [thin] (-0.5,0.8) -- (0.5,-1.2);
      \draw [thin] (-0.5,0) -- (0.5,1.2);
      \draw [thin] (-0.5,0) -- (0.5,0.4);
      \draw [thin] (-0.5,0) -- (0.5,-0.4);
      \draw [thin] (-0.5,0) -- (0.5,-1.2);
      \draw [thin] (-0.5,-0.8) -- (0.5,1.2);
      \draw [thin] (-0.5,-0.8) -- (0.5,0.4);
      \draw [thin] (-0.5,-0.8) -- (0.5,-0.4);
      \draw [thin] (-0.5,-0.8) -- (0.5,-1.2);
      \draw [thin] (0.5,-1.2)  -- (1, -1.2);
      \draw [thin] (0.5,-0.4)  -- (1, -0.4);
      \draw [thin] (0.5,1.2)  -- (1, 1.2);
      \draw [thin] (0.5,0.4)  -- (1, 0.4);
      \filldraw (-0.5, 0.8) circle (1pt);
      \filldraw (-0.5, 0) circle (1pt);
      \filldraw (-0.5, -0.8) circle (1pt);
      \filldraw (0.5, 1.2) circle (1pt);
      \filldraw (0.5, 0.4) circle (1pt);
      \filldraw (0.5, -0.4) circle (1pt);
      \filldraw (0.5, -1.2) circle (1pt);
\end{tikzpicture}
\end{minipage}}

\newcommand{\kperiodlem}{
\begin{minipage}{1.5cm}
\begin{tikzpicture}
      \draw [thin] (-0.5,0.8) -- (0.5,1.2);
      \draw [thin] (-0.5,0.8) -- (0.5,0.4);
      \draw [thin] (-0.5,0.8) -- (0.5,-0.4);
      \draw [thin] (-0.5,0.8) -- (0.5,-1.2);
      \draw [thin] (-0.5,0) -- (0.5,1.2);
      \draw [thin] (-0.5,0) -- (0.5,0.4);
      \draw [thin] (-0.5,0) -- (0.5,-0.4);
      \draw [thin] (-0.5,0) -- (0.5,-1.2);
      \draw [thin] (-0.5,-0.8) -- (0.5,1.2);
      \draw [thin] (-0.5,-0.8) -- (0.5,0.4);
      \draw [thin] (-0.5,-0.8) -- (0.5,-0.4);
      \draw [thin] (-0.5,-0.8) -- (0.5,-1.2);
      \draw (-0.7,0.8) node{${\scriptstyle 0}$};
      \draw (-0.7,0) node{${\scriptstyle 1}$};
      \draw (-0.7,-0.8) node{${\scriptstyle z}$};
      \filldraw (-0.5, 0.8) circle (1pt);
      \filldraw (-0.5, 0) circle (1pt);
      \filldraw (-0.5, -0.8) circle (1pt);
      \filldraw (0.5, 1.2) circle (1pt);
      \filldraw (0.5, 0.4) circle (1pt);
      \filldraw (0.5, -0.4) circle (1pt);
      \filldraw (0.5, -1.2) circle (1pt);
\end{tikzpicture}
\end{minipage}}

\newcommand{\thefourwexternal}{
\begin{minipage}{1cm}
\begin{tikzpicture}[scale=0.5]
    \def\L{0.5}

    \draw[thin] (1,0) -- (1+\L, 0.866);
    \draw[thin] (1,0) -- (1+\L,-0.866);

    \draw[thin, bend left=-45]  (1,0) to (-1,0);
    \draw[thin, bend right=-45] (1,0) to (-1,0);

    \draw[draw=black, fill=black, thin, solid] (1,0) circle (2pt);
    \draw[draw=black, fill=black, thin, solid] (-1,0) circle (2pt);
    \draw[thin] (-1,0) -- (-1-\L, 0.866);
    \draw[thin] (-1,0) -- (-1-\L,-0.866);
\end{tikzpicture}
\end{minipage}
}

\newcommand{\TriangleWithOuterCurves}{
\begin{minipage}{1cm}
\begin{tikzpicture}[scale=0.5]
    \coordinate (A) at (0,1.2);
    \coordinate (B) at (-1.1,-0.8);
    \coordinate (C) at (1.1,-0.8);

    \draw[thin] (A) -- (B) -- (C) -- (A);

    \draw[thin, bend left=25] (A) to (B);
    \draw[thin, bend left=25] (B) to (C);
    \draw[thin, bend left=25] (C) to (A);

    \draw[draw=black, fill=black, thin, solid] (A) circle (2pt);
    \draw[draw=black, fill=black, thin, solid] (B) circle (2pt);
    \draw[draw=black, fill=black, thin, solid] (C) circle (2pt);

\end{tikzpicture}
\end{minipage}
}

\newcommand{\thetree}{
\begin{minipage}{2.5cm}
\begin{tikzpicture}
      \draw [thin] (-1,0) -- (0,0);
      \draw [thin] (0,0) -- (0.5,0.866);
      \draw [thin] (0,0) -- (0.5,-0.866);
      \draw (0.7,0.866) node{${\scriptstyle 1}$};
      \draw (-1.2,0) node{${\scriptstyle 0}$};
      \draw (0.7,-0.866) node{${\scriptstyle z}$};
      \filldraw (0, 0) circle (1pt);
      \filldraw (0.5, 0.866) circle (1pt);
      \filldraw (-1, 0) circle (1pt);
      \filldraw (0.5, -0.866) circle (1pt);
\end{tikzpicture}
\end{minipage}}

\newcommand{\thetwo}{
\begin{minipage}{2.5cm}
\begin{tikzpicture}
      \draw [thin] (-1,0) -- (0,0);
      \draw [thin] (0,0) -- (0.5,0.866);
      \draw (0.7,0.866) node{${\scriptstyle 1}$};
      \draw (-1.2,0) node{${\scriptstyle 0}$};
      \draw (0.2,0) node{${\scriptstyle z}$};
      \draw (0.7,-0.866) node{};
      \filldraw (0, 0) circle (1pt);
      \filldraw (0.5, 0.866) circle (1pt);
      \filldraw (-1, 0) circle (1pt);
\end{tikzpicture}
\end{minipage}}

\newcommand{\blobline}{
\begin{minipage}{2.2cm}
\begin{tikzpicture}
      \fill[black!20] (-0.5,0) circle (0.5);
      \draw [black, thin] (0,0) -- (0.7,0);
      \draw (-0.5,0) circle (0.5);
      \draw (-0.5,-0.7) node{${\scriptstyle 1}$};
      \draw (-0.5,0.7) node{${\scriptstyle 0}$};
      \draw (0.9,0) node{${\scriptstyle z}$};
      \filldraw (-0.5, -0.5) circle (1pt);
      \filldraw (-0.5, 0.5) circle (1pt);
      \filldraw (0, 0) circle (1pt);
      \filldraw (0.7, 0) circle (1pt);
\end{tikzpicture}
\end{minipage}}

\newcommand{\blob}{
\begin{minipage}{1.5cm}
\begin{tikzpicture}
      \fill[black!20] (-0.5,0) circle (0.5);
      \draw (-0.5,0) circle (0.5);
      \draw (-0.5,-0.7) node{${\scriptstyle 1}$};
      \draw (-0.5,0.7) node{${\scriptstyle 0}$};
      \draw (0.2,0) node{${\scriptstyle z}$};
      \filldraw (-0.5, -0.5) circle (1pt);
      \filldraw (-0.5, 0.5) circle (1pt);
      \filldraw (0, 0) circle (1pt);
\end{tikzpicture}
\end{minipage}}

\newcommand{\momentumthreepoint}{
\begin{minipage}{2cm}
\begin{tikzpicture}[rotate=-90]
    \def\circleradius{0.5cm}
    \def\linelength{1cm}
    \def\arcradius{0.35cm}
    \def\shrink{0.4cm} 
    \def\offset{-0.3cm} 

    \coordinate (Afull) at (120:{\circleradius+\linelength});
    \coordinate (Bfull) at (240:{\circleradius+\linelength});
    \coordinate (Cfull) at (0:{\circleradius+\linelength});
    
    \coordinate (A) at ($ (120:{\circleradius+\linelength-\shrink}) + (30:\offset) $);
    \coordinate (B) at ($ (240:{\circleradius+\linelength-\shrink}) + (330:\offset)$);
    \coordinate (C) at ($ (0:{\circleradius+\linelength-\shrink}) + (270:\offset) $);
    \coordinate (AS) at ($ (120:{\circleradius+\linelength-\shrink}) - (30:\offset) $);

    \draw[thin] (0,0) circle (\circleradius);
    
    \foreach \angle in {0,120,240} {
        \draw[thin] 
            (\angle:\circleradius) -- (\angle:{\circleradius + \linelength});
        \filldraw (\angle:\circleradius) circle (1pt);
        
    }
    
    \draw[thin, ->]
        (-90:\arcradius) arc (-90:90:\arcradius)
        node[pos=1.0, left] {$\scriptstyle q$};
    
    \draw[thin, ->, bend right=30] (B) to (A);

    \draw[thin, ->, bend right=30] (AS) to (C);

    \node at (C) [right = 0.15] {$\scriptstyle p_1$};
    \node at (A) [left = 0.15] {$\scriptstyle p_2$};
\end{tikzpicture}
\end{minipage}}

\newcommand{\threepointdual}{
\begin{minipage}{2cm}
\begin{tikzpicture}[rotate=-90]
    \def\circleradius{0.5cm}
    \def\linelength{1cm}
    \def\arcradius{0.35cm}
    \def\shrink{0.4cm} 
    \def\offset{-0.3cm} 
    \def\duallenght{1.5cm}

    \coordinate (Afull) at (120:{\circleradius+\linelength});
    \coordinate (Bfull) at (240:{\circleradius+\linelength});
    \coordinate (Cfull) at (0:{\circleradius+\linelength});
    
    \coordinate (A) at ($ (120:{\circleradius+\linelength-\shrink}) + (30:\offset) $);
    \coordinate (B) at ($ (240:{\circleradius+\linelength-\shrink}) + (330:\offset)$);
    \coordinate (C) at ($ (0:{\circleradius+\linelength-\shrink}) + (270:\offset) $);
    \coordinate (AS) at ($ (120:{\circleradius+\linelength-\shrink}) - (30:\offset) $);

    \draw[thin] (0,0) circle (\circleradius);
    
    \foreach \angle in {0,120,240} {
        \draw[thin] 
            (\angle:\circleradius) -- (\angle:{\circleradius + \linelength});
        \filldraw (\angle:\circleradius) circle (1pt);
    }
    
    \draw[thin, red, dashed] (0,0) -- (-\duallenght,0);
    \draw[thin, red, dashed] (0,0) -- (60:{\duallenght});
    \draw[thin, red, dashed] (0,0) -- (-60:{\duallenght});
\end{tikzpicture}
\end{minipage}}

\newcommand{\thetreeResc}{
\begin{minipage}{2.5cm}
\begin{tikzpicture}
      \draw [thin] (-1,0) -- (0,0);
      \draw [thin] (0,0) -- (0.5,0.866);
      \draw [thin] (0,0) -- (0.5,-0.866);
      \draw (0.7,0.866) node{${\scriptstyle z_1}$};
      \draw (-1.2,0) node{${\scriptstyle 0}$};
      \draw (0.7,-0.866) node{${\scriptstyle z_2}$};
      \filldraw (0, 0) circle (1pt);
      \filldraw (0.5, 0.866) circle (1pt);
      \filldraw (-1, 0) circle (1pt);
      \filldraw (0.5, -0.866) circle (1pt);
\end{tikzpicture}
\end{minipage}}

\newcommand{\thetreeX}{
\begin{minipage}{2.5cm}
\begin{tikzpicture}
      \draw [thin] (-1,0) -- (0,0);
      \draw [thin] (0,0) -- (0.5,0.866);
      \draw [thin] (0,0) -- (0.5,-0.866);
      \draw (0.7,0.866) node{${\scriptstyle 1}$};
      \draw (-1.2,0) node{${\scriptstyle 0}$};
      \draw (0.7,-0.866) node{${\scriptstyle z}$};
      \node at (0,0) [right = 0.1] {$\scriptstyle x$};
      \filldraw (0, 0) circle (1pt);
      \filldraw (0.5, 0.866) circle (1pt);
      \filldraw (-1, 0) circle (1pt);
      \filldraw (0.5, -0.866) circle (1pt);
\end{tikzpicture}
\end{minipage}}

\newcommand{\periodX}{
\begin{minipage}{2.3cm}
\begin{tikzpicture}
       Parameters
    \def\circleradius{0.6cm}
    \def\dotradius{1pt}
    \draw[thin] (0,0) circle (\circleradius);

    \draw[draw=black, fill=black, thin, solid] (0, \circleradius) circle (\dotradius);
    \draw[draw=black, fill=black, thin, solid] (0,-\circleradius) circle (\dotradius);
    \draw[draw=black, fill=black, thin, solid] (\circleradius, 0) circle (\dotradius);
    \draw[draw=black, fill=black, thin, solid] (-\circleradius, 0) circle (\dotradius);
        
    \draw [thin] (0, \circleradius) -- (0,-\circleradius);
    \node at (0, \circleradius) [above = 0.05cm] {$\scriptstyle x_1$};
    \node at (0, -\circleradius) [below = 0.05cm] {$\scriptstyle x_2$};
    \node at (\circleradius, 0) [right = 0.05cm] {$\scriptstyle z_1$};
    \node at (-\circleradius, 0) [left = 0.05cm] {$\scriptstyle 0$};
\end{tikzpicture}
\end{minipage}}

\newcommand{\periodZ}{
\begin{minipage}{2.3cm}
\begin{tikzpicture}
       Parameters
    \def\circleradius{0.6cm}
    \def\dotradius{1pt}
    \draw[thin] (0,0) circle (\circleradius);

    \draw[draw=black, fill=black, thin, solid] (0, \circleradius) circle (\dotradius);
    \draw[draw=black, fill=black, thin, solid] (0,-\circleradius) circle (\dotradius);
    \draw[draw=black, fill=black, thin, solid] (\circleradius, 0) circle (\dotradius);
    \draw[draw=black, fill=black, thin, solid] (-\circleradius, 0) circle (\dotradius);
        
    \draw [thin] (0, \circleradius) -- (0,-\circleradius);
    \node at (0, \circleradius) [above = 0.05cm] {$\scriptstyle z_2$};
    \node at (0, -\circleradius) [below = 0.05cm] {$\scriptstyle x$};
    \node at (\circleradius, 0) [right = 0.05cm] {$\scriptstyle z_1$};
    \node at (-\circleradius, 0) [left = 0.05cm] {$\scriptstyle 0$};
\end{tikzpicture}
\end{minipage}}

\newcommand{\periodA}{
\begin{minipage}{2.5cm}
\begin{tikzpicture}
    \def\circleradius{0.6cm}
    \def\dotradius{0.05cm}
    \def\ext{0.4cm}
    \draw[thin] (0,0) circle (\circleradius);
        
    \draw [thin] (0, \circleradius) -- (0,-\circleradius);
    \draw [thin] (\circleradius, 0) -- (\circleradius + \ext, 0);
    \draw [thin] (-\circleradius, 0) -- (-\circleradius - \ext, 0);
    \filldraw (0, \circleradius) circle (1pt);
    \filldraw (0, -\circleradius) circle (1pt);
    \filldraw (\circleradius, 0) circle (1pt);
    \filldraw (-\circleradius, 0) circle (1pt);
\end{tikzpicture}
\end{minipage}}

\newcommand{\periodB}{
\begin{minipage}{2.5cm}
\begin{tikzpicture}
    \def\circleradius{0.6cm}
    \def\dotradius{0.05cm}
    \def\ext{0.4cm}
    \def\inter{0.1cm}
    \draw[thin] (0,0) circle (\circleradius);
        
    \draw [thin] (0, \circleradius) -- (0,-\circleradius);
    \draw [thin] (\circleradius, 0) -- (\circleradius + \ext, 0);
    \draw [thin] (-\circleradius, 0) -- (-\circleradius - \ext, 0);
    \draw [thin] (0, \circleradius) -- (0, \circleradius + \ext);
    \draw [thin] (0, -\circleradius) -- (0, -\circleradius - \ext);
    \draw [thin] (-\circleradius, 0) -- (-\inter, 0);
    \draw [thin] (\circleradius, 0) -- (\inter, 0);
    \draw [thin] (\circleradius, 0) -- (-\circleradius, 0);
    \draw[draw=black, fill=black, thin, solid] (0, \circleradius) circle (1pt);
    \draw[draw=black, fill=black, thin, solid] (0,-\circleradius) circle (1pt);
    \draw[draw=black, fill=black, thin, solid] (\circleradius, 0) circle (1pt);
    \draw[draw=black, fill=black, thin, solid] (-\circleradius, 0) circle (1pt);
\end{tikzpicture}
\end{minipage}}

\newcommand{\periodBwithoutexternal}{
\begin{minipage}{1.3cm}
\begin{tikzpicture}[scale=0.7]
    \def\circleradius{0.6cm}
    \def\inter{0.1cm}

    \draw[thin] (0,0) circle (\circleradius);

    \draw [thin] (0, \circleradius) -- (0,-\circleradius);
    \draw [thin] (\circleradius, 0) -- (-\circleradius, 0);
    \draw[draw=black, fill=black, thin, solid] (0, \circleradius) circle (1.42pt);
    \draw[draw=black, fill=black, thin, solid] (0,-\circleradius) circle (1.42pt);
    \draw[draw=black, fill=black, thin, solid] (\circleradius, 0) circle (1.42pt);
    \draw[draw=black, fill=black, thin, solid] (-\circleradius, 0) circle (1.42pt);

\end{tikzpicture}
\end{minipage}}

\newcommand{\periodBconnectedtoInfinity}{
\begin{minipage}{1.5cm}
\begin{tikzpicture}[scale=0.7]
    \def\R{0.6cm}
    \def\inter{0.1cm}
    \def\infradius{2.0cm}

    \coordinate (INF) at (45:\infradius);

    \coordinate (C) at (0,0);
    \coordinate (N) at (0,\R);
    \coordinate (S) at (0,-\R);
    \coordinate (E) at (\R,0);
    \coordinate (W) at (-\R,0);

    \draw[thin] (C) circle (\R);

    \draw[thin] (N) -- (S);
    \draw[thin] (E) -- (W);

    \draw[draw=black, fill=black, thin, solid] (N) circle (1.42pt);
    \draw[draw=black, fill=black, thin, solid] (E) circle (1.42pt);
    \draw[draw=black, fill=black, thin, solid] (S) circle (1.42pt);
    \draw[draw=black, fill=black, thin, solid] (W) circle (1.42pt);

    \draw[thin] (N) -- (INF);

    \draw[thin] (E) -- (INF);

    \draw[thin, bend right=60] (S) to (INF);

    \draw[thin, bend left=60] (W) to (INF);

    \draw[draw=black, fill=black, thin, solid] (INF) circle (1.2pt);
    \node[above right] at (INF) {$\infty$};
\end{tikzpicture}
\end{minipage}}

\newcommand{\kfive}{
\begin{minipage}{1.5cm}
\begin{tikzpicture}[scale=0.7, x=3ex, y=3ex,
  baseline={([yshift=-5ex]current bounding box.center)}]

  \coordinate (v1) at (90:2.0);
  \coordinate (v2) at (18:2.0);
  \coordinate (v3) at (-54:2.0);
  \coordinate (v4) at (-126:2.0);
  \coordinate (v5) at (162:2.0);

  \foreach \v in {v1,v2,v3,v4,v5}{
    \draw[draw=black, fill=black, thin, solid] (\v) circle (1.42pt);
  }
  
  \draw[draw=black, thin, solid] (v1)--(v2);
  \draw[draw=black, thin, solid] (v1)--(v3);
  \draw[draw=black, thin, solid] (v1)--(v4);
  \draw[draw=black, thin, solid] (v1)--(v5);

  \draw[draw=black, thin, solid] (v2)--(v3);
  \draw[draw=black, thin, solid] (v2)--(v4);
  \draw[draw=black, thin, solid] (v2)--(v5);

  \draw[draw=black, thin, solid] (v3)--(v4);
  \draw[draw=black, thin, solid] (v3)--(v5);

  \draw[draw=black, thin, solid] (v4)--(v5);

\end{tikzpicture}
\end{minipage}
}

\newcommand{\periodC}{
\begin{minipage}{1.8cm}
\begin{tikzpicture}
       Parameters
    \def\circleradius{0.6cm}
    \def\dotradius{0.05cm}
    \draw[thin] (0,0) circle (\circleradius);

    \draw[thin] (0,0) -- (90:\circleradius);
    \draw[thin] (0,0) -- (-30:\circleradius);
    \draw[thin] (0,0) -- (-150:\circleradius);
    \filldraw (0,0) circle (1pt);
    \filldraw (90:\circleradius) circle (1pt);
    \filldraw (-30:\circleradius) circle (1pt);
    \filldraw (-150:\circleradius) circle (1pt);
\end{tikzpicture}
\end{minipage}}

\newcommand{\externalA}{
\begin{minipage}{1.5cm}
\begin{tikzpicture}[x=1.5ex,y=1.5ex,baseline={([yshift=-.6ex]current bounding box.center)}]
    \coordinate (v0);
    \coordinate [above right=1.41 and 2 of v0] (v1);
    \coordinate [below right=1.41 and 2 of v0] (v2);
    \node [left =.1 of v0] (v0n) { $z$};
    \coordinate [right =1 of v0] (v0u);
    \node [right=.07 of v1] (v1n) { $1$};
    \coordinate [below left=.71 and .71 of v1] (v1u);
    \node [right=.07 of v2] (v2n) { $0$};
    \coordinate [above left=.71 and .71 of v2] (v2u);
    \draw[fill=black!20] (v0) .. controls (v0u) and (v1u) .. (v1) .. controls (v1u) and (v2u) .. (v2) .. controls (v2u) and (v0u) .. (v0);
    \filldraw (v0) circle (1pt);
    \filldraw (v1) circle (1pt);
    \filldraw (v2) circle (1pt);
\end{tikzpicture}
\end{minipage}
}

\newcommand{\externalB}{
\begin{minipage}{1.5cm}
\begin{tikzpicture}[x=1.5ex,y=1.5ex,baseline={([yshift=-.6ex]current bounding box.center)}]
    \coordinate (v0);
    \coordinate [above right=1.41 and 2 of v0] (v1);
    \coordinate [below right=1.41 and 2 of v0] (v2);
    \node [left =.1 of v0] (v0n) { $z$};
    \coordinate [right =1 of v0] (v0u);
    \node [right=.07 of v1] (v1n) { $1$};
    \coordinate [below left=.71 and .71 of v1] (v1u);
    \node [right=.07 of v2] (v2n) { $0$};
    \coordinate [above left=.71 and .71 of v2] (v2u);
    \draw[fill=black!20] (v0) .. controls (v0u) and (v1u) .. (v1) .. controls (v1u) and (v2u) .. (v2) .. controls (v2u) and (v0u) .. (v0);
    \draw (v1) to [bend left=90] (v2);
    \filldraw (v0) circle (1pt);
    \filldraw (v1) circle (1pt);
    \filldraw (v2) circle (1pt);
\end{tikzpicture}
\end{minipage}
}

\newcommand{\externalC}{
\begin{minipage}{1.5cm}
\begin{tikzpicture}[x=1.5ex,y=1.5ex,baseline={([yshift=-.6ex]current bounding box.center)}]
    \coordinate (v0);
    \coordinate [above right=1.41 and 2 of v0] (v1);
    \coordinate [below right=1.41 and 2 of v0] (v2);
    \node [left =.1 of v0] (v0n) { $z$};
    \coordinate [right =1 of v0] (v0u);
    \node [right=.07 of v1] (v1n) { $1$};
    \coordinate [below left=.71 and .71 of v1] (v1u);
    \node [right=.07 of v2] (v2n) { $0$};
    \coordinate [above left=.71 and .71 of v2] (v2u);
    \draw[fill=black!20] (v0) .. controls (v0u) and (v1u) .. (v1) .. controls (v1u) and (v2u) .. (v2) .. controls (v2u) and (v0u) .. (v0);
    \draw (v2) to [bend left=90] (v0);
    \filldraw (v0) circle (1pt);
    \filldraw (v1) circle (1pt);
    \filldraw (v2) circle (1pt);
\end{tikzpicture}
\end{minipage}
}

\newcommand{\externalD}{
\begin{minipage}{1.5cm}
\begin{tikzpicture}[x=1.5ex,y=1.5ex,baseline={([yshift=-.6ex]current bounding box.center)}]
    \coordinate (v0);
    \coordinate [above right=1.41 and 2 of v0] (v1);
    \coordinate [below right=1.41 and 2 of v0] (v2);
    \node [left =.1 of v0] (v0n) { $z$};
    \coordinate [right =1 of v0] (v0u);
    \node [right=.07 of v1] (v1n) { $1$};
    \coordinate [below left=.71 and .71 of v1] (v1u);
    \node [right=.07 of v2] (v2n) { $0$};
    \coordinate [above left=.71 and .71 of v2] (v2u);
    \draw[fill=black!20] (v0) .. controls (v0u) and (v1u) .. (v1) .. controls (v1u) and (v2u) .. (v2) .. controls (v2u) and (v0u) .. (v0);
    \draw (v0) to [bend left=90] (v1);
    \filldraw (v0) circle (1pt);
    \filldraw (v1) circle (1pt);
    \filldraw (v2) circle (1pt);
\end{tikzpicture}
\end{minipage}
}

\newcommand{\productA}{
\begin{minipage}{2cm}
\begin{tikzpicture}[x=2.5ex, y=2.5ex, baseline={([yshift=-10.0ex]current bounding box.center)}]
  \coordinate (v0);
  \coordinate[below=2 of v0] (v1);
  \coordinate[below=2 of v1] (v2);
  \filldraw (v0) circle (1pt);
  \filldraw (v1) circle (1pt);
  \filldraw (v2) circle (1pt);
  \filldraw[fill=black!20] (v0) to [bend left=90] (v1) to [bend left=90] (v2) to [bend right=90, looseness=1.5] (v0);
  \filldraw[fill=black!20] (v0) to [bend right=90] (v1) to [bend right=90] (v2) to [bend left=90, looseness=1.5] (v0);

  \node [above=.07 of v0] (v0n) {$0$};
  \node [above=.07 of v1] (v1n) {$1$};
  \node [above=.07 of v2] (v2n) {$z$};

\end{tikzpicture}

\end{minipage}
}

\newcommand{\productB}{
\begin{minipage}{1.15cm}
\begin{tikzpicture}[x=2.5ex, y=2.5ex, baseline={([yshift=-10.0ex]current bounding box.center)}]
  \coordinate (v0);
  \coordinate[below=2 of v0] (v1);
  \coordinate[below=2 of v1] (v2);

  \filldraw (v0) circle (1pt);
  \filldraw (v1) circle (1pt);
  \filldraw (v2) circle (1pt);
  \filldraw[fill=black!20] (v0) to [bend right=90] (v1) to [bend right=90] (v2) to [bend left=90, looseness=1.5] (v0);

  \node [above=.07 of v0] (v0n) {$0$};
  \node [above=.07 of v1] (v1n) {$1$};
  \node [above=.07 of v2] (v2n) {$z$};

\end{tikzpicture}

\end{minipage}
}

\newcommand{\productC}{
\begin{minipage}{1cm}
\begin{tikzpicture}[x=2.5ex, y=2.5ex, baseline={([yshift=-10.0ex]current bounding box.center)}]
  \coordinate (v0);
  \coordinate[below=2 of v0] (v1);
  \coordinate[below=2 of v1] (v2);

  \filldraw (v0) circle (1pt);
  \filldraw (v1) circle (1pt);
  \filldraw (v2) circle (1pt);
  \filldraw[fill=black!20] (v0) to [bend left=90] (v1) to [bend left=90] (v2) to [bend right=90, looseness=1.5] (v0);

  \node [above=.07 of v0] (v0n) {$0$};
  \node [above=.07 of v1] (v1n) {$1$};
  \node [above=.07 of v2] (v2n) {$z$};

\end{tikzpicture}

\end{minipage}
}

\newcommand{\YukawaG}{
\begin{minipage}{2.8cm}
\begin{tikzpicture}[x=2.5ex, y=2.5ex,
  baseline={([yshift=-4.0ex]current bounding box.center)}, >=latex]

  \def\bumpx{0.3}
  \def\bumpy{0.3}

  \coordinate (v0);
  \coordinate[below=2 of v0] (v1);
  \coordinate[below=2 of v1] (v2);

  \filldraw (v0) circle (1pt);
  \filldraw (v1) circle (1pt);
  \filldraw (v2) circle (1pt);

  \coordinate (v1U) at ($(v1)+(0,\bumpy)$);
  \coordinate (v1D) at ($(v1)+(0,-\bumpy)$);
  \coordinate (v1B) at ($(v1)+(\bumpx,0)$); 

  \filldraw[fill=black!20]
    (v0)
      to[bend right=90] (v1U)
      .. controls ($(v1U)+(0.35,0)$) and ($(v1B)+(0,0.35)$) .. (v1B)
      .. controls ($(v1B)+(0,-0.35)$) and ($(v1D)+(0.35,0)$) .. (v1D)
      to[bend right=90] (v2)
      to[bend left=90, looseness=1.5] (v0);

  \coordinate (w0) at ($(v0)+(3,0)$);
  \coordinate (w1) at ($(v1)+(3,0)$);
  \coordinate (w2) at ($(v2)+(3,0)$);

  \filldraw (w0) circle (1pt);
  \filldraw (w1) circle (1pt);
  \filldraw (w2) circle (1pt);

  \coordinate (w1U) at ($(w1)+(0,\bumpy)$);
  \coordinate (w1D) at ($(w1)+(0,-\bumpy)$);
  \coordinate (w1B) at ($(w1)-(\bumpx,0)$); 

  \path (v1U) -- (w1D) coordinate[midway] (c);
  \filldraw (c) circle (1pt);

  \filldraw[fill=black!20]
    (w0)
      to[bend left=90] (w1U)
      .. controls ($(w1U)+(-0.35,0)$) and ($(w1B)+(0,0.35)$) .. (w1B)
      .. controls ($(w1B)+(0,-0.35)$) and ($(w1D)+(-0.35,0)$) .. (w1D)
      to[bend left=90] (w2)
      to[bend right=90, looseness=1.5] (w0);

  \draw[thin, postaction={decorate, decoration={markings, mark=at position 0.6 with {\arrow{>}}}}] (w0) -- (v0);
  \draw[thin, postaction={decorate, decoration={markings, mark=at position 0.6 with {\arrow{>}}}}] (v2) -- (w2);

  \draw[thin] (v1U) -- (w1D);
  \draw[thin] (v1D) -- (w1U);

\end{tikzpicture}
\end{minipage}
}

\newcommand{\YukawaGone}{
\begin{minipage}{2.2cm}
\begin{tikzpicture}[x=2.5ex, y=2.5ex,
  baseline={([yshift=-4.0ex]current bounding box.center)}, >=latex]

  \def\bumpx{0.9}
  \def\bumpy{0.55}

  \coordinate (v0);
  \coordinate[below=2 of v0] (v1);
  \coordinate[below=2 of v1] (v2);

  \filldraw (v0) circle (1pt);
  \filldraw (v1) circle (1pt);
  \filldraw (v2) circle (1pt);

  \coordinate (v1U) at ($(v1)+(0,\bumpy)$);
  \coordinate (v1D) at ($(v1)+(0,-\bumpy)$);
  \coordinate (v1B) at ($(v1)+(\bumpx,0)$);

  \filldraw[fill=black!20]
    (v0)
      to[bend right=90] (v1U)
      .. controls ($(v1U)+(0.35,0)$) and ($(v1B)+(0,0.35)$) .. (v1B)
      .. controls ($(v1B)+(0,-0.35)$) and ($(v1D)+(0.35,0)$) .. (v1D)
      to[bend right=90] (v2)
      to[bend left=90, looseness=1.5] (v0);

  \coordinate (w0) at ($(v0)+(3.2,0)$);
  \coordinate (w2) at ($(v2)+(3.2,0)$);

  \filldraw (w0) circle (1pt);
  \filldraw (w2) circle (1pt);

  \draw[thin, postaction={decorate, decoration={markings, mark=at position 0.6 with {\arrow{>}}}}] (w2) -- (w0);

  \draw[thin, postaction={decorate, decoration={markings, mark=at position 0.6 with {\arrow{>}}}}] (w0) -- (v0);
  \draw[thin, postaction={decorate, decoration={markings, mark=at position 0.6 with {\arrow{>}}}}] (v2) -- (w2);

  \coordinate (C) at ($(v1)+(1.8,0)$);
  \filldraw (C) circle (1pt);

  \draw[thin] (v1U) -- (C);
  \draw[thin] (v1D) -- (C);

  \draw[thin] (C) -- (w0);
  \draw[thin] (C) -- (w2);

\end{tikzpicture}
\end{minipage}
}

\newcommand{\YukawaGtwo}{
\begin{minipage}{2.2cm}
\begin{tikzpicture}[x=2.5ex, y=2.5ex,
  baseline={([yshift=-4.0ex]current bounding box.center)}, >=latex]

  \def\bumpx{0.9}
  \def\bumpy{0.55}

  \coordinate (l0) at (0,0);
  \coordinate (l1) at (0,-2); 
  \coordinate (l2) at (0,-4);

  \filldraw (l0) circle (1pt);
  \filldraw (l2) circle (1pt);

  \draw[thin, postaction={decorate, decoration={markings, mark=at position 0.6 with {\arrow{>}}}}] (l2) -- (l0);

  \coordinate (r0) at (3.2,0);
  \coordinate (r1) at (3.2,-2);
  \coordinate (r2) at (3.2,-4);

  \filldraw (r0) circle (1pt);
  \filldraw (r1) circle (1pt);
  \filldraw (r2) circle (1pt);

  \coordinate (r1U) at ($(r1)+(0,\bumpy)$);
  \coordinate (r1D) at ($(r1)+(0,-\bumpy)$);
  \coordinate (r1B) at ($(r1)-(\bumpx,0)$);

  \filldraw[fill=black!20]
    (r0)
      to[bend left=90] (r1U)
      .. controls ($(r1U)+(-0.35,0)$) and ($(r1B)+(0,0.35)$) .. (r1B)
      .. controls ($(r1B)+(0,-0.35)$) and ($(r1D)+(-0.35,0)$) .. (r1D)
      to[bend left=90] (r2)
      to[bend right=90, looseness=1.5] (r0);

  \draw[thin, postaction={decorate, decoration={markings, mark=at position 0.6 with {\arrow{>}}}}] (l0) -- (r0);
  \draw[thin, postaction={decorate, decoration={markings, mark=at position 0.6 with {\arrow{>}}}}] (r2) -- (l2);

  \coordinate (C) at (1.4,-2);
  \filldraw (C) circle (1pt);

  \draw[thin] (r1U) -- (C);
  \draw[thin] (r1D) -- (C);

  \draw[thin] (C) -- (l0);
  \draw[thin] (C) -- (l2);

\end{tikzpicture}
\end{minipage}
}

\newcommand{\productK}{
\begin{minipage}{2.5cm}
\begin{tikzpicture}[x=3ex, y = 3ex, baseline={([yshift=-8.0ex]current bounding box.center)}]
  \draw[draw=black, fill=black, thin, solid] (-1.00,1.00) circle (1pt);
  \draw[draw=black, fill=black, thin, solid] (1.00,1.00) circle (1pt);
  \draw[draw=black, fill=black, thin, solid] (3.00,1.00) circle (1pt);
  \draw[draw=black, fill=black, thin, solid] (2.00,-1.00) circle (1pt);
  \draw[draw=black, fill=black, thin, solid] (0.00,-1.00) circle (1pt);
  \draw[draw=black, thin, solid] (-1.00,1.00) -- (0.00,-1.00);
  \draw[draw=black, thin, solid] (0.00,-1.00) -- (1.00,1.00);
  \draw[draw=black, thin, solid] (1.00,1.00) -- (2.00,-1.00);
  \draw[draw=black, thin, solid] (2.00,-1.00) -- (3.00,1.00);
  \draw[draw=black, thin, solid] (3.00,1.00) -- (0.00,-1.00);
  \draw[draw=black, thin, solid] (-1.00,1.00) -- (2.00,-1.00);
  \node[black, anchor=south west] at (-1.50,1.05) {$0$};
  \node[black, anchor=south west] at (0.67,1.05) {$1$};
  \node[black, anchor=south west] at (2.67,1.05) {$z$};
\end{tikzpicture}
\end{minipage}
}

\newcommand{\completionthreestar}{
\begin{minipage}{2.5cm}
\begin{tikzpicture}
    \coordinate (v) ;
    \def \rad {1}
    \coordinate[label=above:$1$] (v1) at ([shift=(90:\rad)]v);
    \coordinate[label=left:$0$] (v2) at ([shift=(180:\rad)]v);
    \coordinate[label=below:$\infty$] (v3) at ([shift=(270:\rad)]v);
    \coordinate[label=right:$z$] (v4) at ([shift=(0:\rad)]v);

    \draw (v) -- (v1);
    \draw (v) -- (v2);
    \draw (v) -- (v3);
    \draw (v) -- (v4);

    \draw (v1) -- node[inner sep=0pt,above left] {\tiny{$-1$}} (v2);
    \draw (v3) -- node[inner sep=0pt,below right] {\tiny{$-1$}} (v4);

    \filldraw (v) circle (1pt);
    \filldraw (v1) circle (1pt);
    \filldraw (v2) circle (1pt);
    \filldraw (v3) circle (1pt);
    \filldraw (v4) circle (1pt);

\end{tikzpicture}
\end{minipage}
}

\newcommand{\completionKfivetwo}{
\begin{minipage}{2.5cm}
\begin{tikzpicture}[x=3ex, y = 3ex, baseline={([yshift=-8.0ex]current bounding box.center)}]
  \coordinate[label=above:$0$] (v1) at (-1.00, 1.00);
  \coordinate[label=above right:$1$] (v2) at (1.00, 1.00);
  \coordinate[label=above:$z$] (v3) at (3.00, 1.00);
  \coordinate[label=above:$\infty$] (v4) at (1.00, 3.00);
  \coordinate (v5) at (0.00, -1.00);
  \coordinate (v6) at (2.00, -1.00);
  \draw[draw=black, fill=black, thin, solid] (v1) circle (1pt);
  \draw[draw=black, fill=black, thin, solid] (v2) circle (1pt);
  \draw[draw=black, fill=black, thin, solid] (v3) circle (1pt);
  \draw[draw=black, fill=black, thin, solid] (v4) circle (1pt);
  \draw[draw=black, fill=black, thin, solid] (v5) circle (1pt);
  \draw[draw=black, fill=black, thin, solid] (v6) circle (1pt);
  \draw[draw=black, thin, solid] (v1) -- (v5);
  \draw[draw=black, thin, solid] (v1) -- (v6);
  \draw[draw=black, thin, solid] (v2) -- (v5);
  \draw[draw=black, thin, solid] (v2) -- (v6);
  \draw[draw=black, thin, solid] (v3) -- (v5);
  \draw[draw=black, thin, solid] (v3) -- (v6);
  \draw[draw=black, thin, solid] (v5) -- (v6);
  \draw[draw=black, thin, solid] (v1) -- node[inner sep=0pt,above] {\tiny{$-3$}} (v2);
  \draw[draw=black, thin, solid] (v1) -- (v4);
  \draw[draw=black, thin, solid] (v2) -- (v4);
  \draw[draw=black, thin, solid] (v3) -- node[inner sep=0pt,above right] {\tiny{$-2$}} (v4);

\end{tikzpicture}
\end{minipage}
}

\newcommand{\completionKfivetwoinverted}{
\begin{minipage}{2.5cm}
\begin{tikzpicture}[
    x=3ex,
    y=3ex,
    yscale=-1,
    baseline={([yshift=-8.0ex]current bounding box.center)}
]
  \coordinate (v1) at (-1.00, 1.00);
  \coordinate (v2) at (1.00, 1.00);
  \coordinate (v3) at (3.00, 1.00);
  \coordinate (v4) at (1.00, 3.00);
  \coordinate (v5) at (0.00, -1.00);
  \coordinate (v6) at (2.00, -1.00);

  \draw[draw=black, fill=black, thin, solid] (v1) circle (1pt);
  \draw[draw=black, fill=black, thin, solid] (v2) circle (1pt);
  \draw[draw=black, fill=black, thin, solid] (v3) circle (1pt);
  \draw[draw=black, fill=black, thin, solid] (v4) circle (1pt);
  \draw[draw=black, fill=black, thin, solid] (v5) circle (1pt);
  \draw[draw=black, fill=black, thin, solid] (v6) circle (1pt);

  \draw[draw=black, thin, solid] (v1) -- (v5);
  \draw[draw=black, thin, solid] (v1) -- (v6);
  \draw[draw=black, thin, solid] (v2) -- (v5);
  \draw[draw=black, thin, solid] (v2) -- (v6);
  \draw[draw=black, thin, solid] (v3) -- (v5);
  \draw[draw=black, thin, solid] (v3) -- (v6);
  \draw[draw=black, thin, solid] (v5) -- (v6);
  \draw[draw=black, thin, solid] (v1) -- (v4);
  \draw[draw=black, thin, solid] (v2) -- (v4);
  \draw[draw=black, thin, solid] (v3) -- (v4);

\end{tikzpicture}
\end{minipage}
}

\newcommand{\completionKseven}{
\begin{minipage}{2.5cm}
\begin{tikzpicture}[x=3ex,y=3ex,
  baseline={([yshift=-7.5ex]current bounding box.center)}]

  \def\h{1.6}

  \coordinate (C) at (-2.0, 0.0);
  \coordinate (D) at ( 2.0, 0.0);
  \coordinate (G) at ( 0.0, 0.0);

  \coordinate (A) at (-1.0, 2.0);
  \coordinate (B) at ( 1.0, 2.0);

  \coordinate (E) at (-0.9,-2.0);
  \coordinate (F) at ( 0.9,-2.0);

  \draw[thin] (A)--(B);
  \draw[thin] (B)--(D);
  \draw[thin] (D)--(F);
  \draw[thin] (C)--(F);
  \draw[thin] (F)--(E);
  \draw[thin] (E)--(C);
  \draw[thin] (E)--(D);
  \draw[thin] (C)--(A);

  \draw[thin] (A)--(D);
  \draw[thin] (B)--(C);

  \draw[thin] (A)--(G);
  \draw[thin] (B)--(G);
  \draw[thin] (G)--(E);
  \draw[thin] (G)--(F);

  \foreach \v in {A,B,C,D,E,F,G}{
    \draw[draw=black, fill=black, thin, solid] (\v) circle (1pt);
  }
\end{tikzpicture}
\end{minipage}
}

\newcommand{\kperiodfour}{
\begin{minipage}{1.8cm}
\begin{tikzpicture}
    \def\xL{-0.55}
    \def\xR{ 0.55}

    \coordinate (L1) at (\xL,  1.2);
    \coordinate (L2) at (\xL,  0.4);
    \coordinate (L3) at (\xL, -0.4);
    \coordinate (L4) at (\xL, -1.2);

    \coordinate (R1) at (\xR,  1.2);
    \coordinate (R2) at (\xR,  0.4);
    \coordinate (R3) at (\xR, -0.4);
    \coordinate (R4) at (\xR, -1.2);

    \foreach \L in {L1,L2,L3,L4}{
      \foreach \R in {R1,R2,R3,R4}{
        \draw[thin] (\L) -- (\R);
      }
    }

    \foreach \v in {L1,L2,L3,L4,R1,R2,R3,R4}{
      \draw[draw=black, fill=black, thin, solid] (\v) circle (1pt);
    }
\end{tikzpicture}
\end{minipage}}

\newcommand{\TriangleWithExternals}{
\begin{minipage}{2.2cm}
\begin{tikzpicture}[scale=0.5]
  \coordinate (A) at (0,1.0);
  \coordinate (B) at (-1.0,-0.7);
  \coordinate (C) at (1.0,-0.7);

  \draw[thin] (A) -- (B) -- (C) -- (A);

  \draw[thin] (A) -- ++(0,0.5);        
  \draw[thin] (B) -- ++(-0.45,0);   
  \draw[thin] (C) -- ++(0.45,0);    

  \draw[draw=black, fill=black, thin, solid] (A) circle (2pt);
  \draw[draw=black, fill=black, thin, solid] (B) circle (2pt);
  \draw[draw=black, fill=black, thin, solid] (C) circle (2pt);

\end{tikzpicture}
\end{minipage}}

\newcommand{\YukawaTriangle}{
\begin{minipage}{2cm}
\begin{tikzpicture}[scale=0.5, >=latex]
  \coordinate (A) at (0,1.0);
  \coordinate (B) at (-1.0,-0.7);
  \coordinate (C) at (1.0,-0.7);

  \draw[draw=black, fill=black, thin, solid] (A) circle (2pt);
  \draw[draw=black, fill=black, thin, solid] (B) circle (2pt);
  \draw[draw=black, fill=black, thin, solid] (C) circle (2pt);
  
  \draw[thin, postaction={decorate, decoration={markings, mark=at position 0.6 with {\arrow{>}}}}] (B) -- (A); 
  \draw[thin, postaction={decorate, decoration={markings, mark=at position 0.6 with {\arrow{>}}}}] (A) -- (C);
  \draw[thin] (C) -- (B);

 \draw[thin] (A) -- ++(0,0.5);        
  \draw[thin, postaction={decorate, decoration={markings, mark=at position 0.9 with {\arrow{>}}}}] (B)++(-0.45,0) -- (B);   
  \draw[thin, postaction={decorate, decoration={markings, mark=at position 0.9 with {\arrow{>}}}}] (C) -- ++(0.45,0);    
\end{tikzpicture}
\end{minipage}}

\newcommand{\Trianglecompletion}{
\begin{minipage}{2cm}
\begin{tikzpicture}[scale=0.4]

  \coordinate (A) at (0, 1.2);       
  \coordinate (B) at (-1.1,-0.8);     
  \coordinate (C) at ( 1.1,-0.8);

  \coordinate (INF) at (90:2.4);

  \draw[draw=black, fill=black, thin, solid] (A) circle (2.5pt);
  \draw[draw=black, fill=black, thin, solid] (B) circle (2.5pt);
  \draw[draw=black, fill=black, thin, solid] (C) circle (2.5pt);
  \draw[draw=black, fill=black, thin, solid] (INF) circle (2.5pt);

  \draw[thin] (A) -- (B) -- (C) -- cycle;

  \draw[thin] (A) -- (INF);

  \draw[thin, bend left=25] (B) to (INF);
  \draw[thin, bend right=25] (C) to (INF);

  \node[above] at (INF) {$\infty$};

\end{tikzpicture}
\end{minipage}}

\newcommand{\BoxWithExternals}{
\begin{minipage}{2.4cm}
\begin{tikzpicture}[scale=0.5, >=latex]

  \coordinate (A) at (-1, 1);   
  \coordinate (B) at ( 1, 1);   
  \coordinate (C) at ( 1,-1);   
  \coordinate (D) at (-1,-1);

  \draw[thin, postaction={decorate, decoration={markings, mark=at position 0.6 with {\arrow{>}}}}] (A) -- (B); 
  \draw[thin, postaction={decorate, decoration={markings, mark=at position 0.6 with {\arrow{>}}}}] (B) -- (C); 
  \draw[thin, postaction={decorate, decoration={markings, mark=at position 0.6 with {\arrow{>}}}}] (C) -- (D); 
  \draw[thin, postaction={decorate, decoration={markings, mark=at position 0.6 with {\arrow{>}}}}] (D) -- (A);

 \draw[thin] (A) -- ++(-0.5, 0.5);   
  \draw[thin] (B) -- ++( 0.5, 0.5);   
  \draw[thin] (C) -- ++( 0.5,-0.5);   
  \draw[thin] (D) -- ++(-0.5,-0.5);

  \draw[draw=black, fill=black, thin, solid] (A) circle (2pt);
  \draw[draw=black, fill=black, thin, solid] (B) circle (2pt);
  \draw[draw=black, fill=black, thin, solid] (C) circle (2pt);
  \draw[draw=black, fill=black, thin, solid] (D) circle (2pt);

\end{tikzpicture}
\end{minipage}}

\newcommand{\BoxWithDiagonals}{
\begin{minipage}{2cm}
\begin{tikzpicture}[scale=0.5, >=latex]

  \coordinate (A) at (-1,  1);   
  \coordinate (B) at ( 1,  1);   
  \coordinate (C) at ( 1, -1);   
  \coordinate (D) at (-1, -1);   

  \draw[draw=black, fill=black, thin, solid] (A) circle (2pt);
  \draw[draw=black, fill=black, thin, solid] (B) circle (2pt);
  \draw[draw=black, fill=black, thin, solid] (C) circle (2pt);
  \draw[draw=black, fill=black, thin, solid] (D) circle (2pt);

  \draw[thin, postaction={decorate, decoration={markings, mark=at position 0.6 with {\arrow{>}}}}] (A) -- (B); 
  \draw[thin, postaction={decorate, decoration={markings, mark=at position 0.6 with {\arrow{>}}}}] (B) -- (C); 
  \draw[thin, postaction={decorate, decoration={markings, mark=at position 0.6 with {\arrow{>}}}}] (C) -- (D); 
  \draw[thin, postaction={decorate, decoration={markings, mark=at position 0.6 with {\arrow{>}}}}] (D) -- (A); 

  \draw[thin] (A) -- (C);
  \draw[thin] (B) -- (D);

\end{tikzpicture}
\end{minipage}}

\newcommand{\BoxWithDiagonalsVertex}{
\begin{minipage}{2cm}
\begin{tikzpicture}[scale=0.5, >=latex]

  \coordinate (A) at (-1,  1);   
  \coordinate (B) at ( 1,  1);   
  \coordinate (C) at ( 1, -1);   
  \coordinate (D) at (-1, -1);   

  \draw[thin, postaction={decorate, decoration={markings, mark=at position 0.6 with {\arrow{>}}}}] (A) -- (B); 
  \draw[thin, postaction={decorate, decoration={markings, mark=at position 0.6 with {\arrow{>}}}}] (B) -- (C); 
  \draw[thin, postaction={decorate, decoration={markings, mark=at position 0.6 with {\arrow{>}}}}] (C) -- (D); 
  \draw[thin, postaction={decorate, decoration={markings, mark=at position 0.6 with {\arrow{>}}}}] (D) -- (A); 

  \draw[thin] (A) -- (C);
  \draw[thin] (B) -- (D);
  \draw[draw=black, fill=black, thin, solid] (A) circle (2pt);
  \draw[draw=black, fill=black, thin, solid] (B) circle (2pt);
  \draw[draw=black, fill=black, thin, solid] (C) circle (2pt);
  \draw[draw=black, fill=black, thin, solid] (D) circle (2pt);
  \draw[draw=black, fill=black, thin, solid] (0,0) circle (2pt);

\end{tikzpicture}
\end{minipage}}

\newcommand{\Yukawavertex}{
\begin{minipage}{2.4cm}
\begin{tikzpicture}[scale=0.6, >=latex]

  \coordinate (L)   at (-1.2, 0.0);   
  \coordinate (C)   at ( 0.0, 0.0);   
  \coordinate (UL)  at ( 0.0, 1.2);   

  \coordinate (BR)  at ( 1.4, 0.0);   
  \coordinate (TR)  at ( 1.4, 1.2);   

  \coordinate (UR)  at ( 1.4, 1.2);   

  \coordinate (R)   at ( 2.4, 1.2);   

  \draw[thin, postaction={decorate, decoration={markings, mark=at position 0.6 with {\arrow{>}}}}] (L) -- (C); 

  \draw[thin] (UL) -- (BR);

  \draw[thin] (C)  -- (UR);

  \draw[thin, postaction={decorate, decoration={markings, mark=at position 0.6 with {\arrow{>}}}}] (C) -- (BR); 
  \draw[thin, postaction={decorate, decoration={markings, mark=at position 0.6 with {\arrow{>}}}}] (BR) -- (TR); 
  \draw[thin, postaction={decorate, decoration={markings, mark=at position 0.6 with {\arrow{>}}}}] (TR) -- (R); 

  \draw[draw=black, fill=black, thin, solid] (TR) circle (1.67pt);
  \draw[draw=black, fill=black, thin, solid] (BR) circle (1.67pt);
  \draw[draw=black, fill=black, thin, solid] (C) circle (1.67pt);
  \draw[draw=black, fill=black, thin, solid] (0.7, 0.6) circle (1.67pt);

\end{tikzpicture}
\end{minipage}}

\newcommand{\threeloopYukawa}{
\begin{minipage}{2.5cm}
\begin{tikzpicture}[scale=0.7, >=latex]

  \coordinate (L)  at (-2.6, 0.0);
  \coordinate (V)  at (-1.8, 0.0);
  \coordinate (A)  at (-1.2, 0.8);   
  \coordinate (B)  at (-1.2,-0.8);   
  
  \coordinate (D)  at ( 0.0, 0.8);   
  \coordinate (E)  at ( 0.0,-0.8);   

  \draw[draw=black, fill=black, thin, solid] (A) circle (1.42pt);
  \draw[draw=black, fill=black, thin, solid] (B) circle (1.42pt);
  \draw[draw=black, fill=black, thin, solid] (D) circle (1.42pt);
  \draw[draw=black, fill=black, thin, solid] (E) circle (1.42pt);
  \draw[draw=black, fill=black, thin, solid] (V) circle (1.42pt);
  
  \coordinate (Rt) at ( 1.2, 0.8);
  \coordinate (Rb) at ( 1.2,-0.8);

  \draw[thin] (L) -- (V);
  \draw[thin, postaction={decorate, decoration={markings, mark=at position 0.6 with {\arrow{>}}}}] (A) -- (V); 
  \draw[thin, postaction={decorate, decoration={markings, mark=at position 0.6 with {\arrow{>}}}}] (V) -- (B); 
  \draw[thin, postaction={decorate, decoration={markings, mark=at position 0.6 with {\arrow{>}}}}] (D) -- (A); 
  \draw[thin, postaction={decorate, decoration={markings, mark=at position 0.6 with {\arrow{>}}}}] (B) -- (E); 
  \draw[thin, postaction={decorate, decoration={markings, mark=at position 0.6 with {\arrow{>}}}}] (Rt) -- (D); 
  \draw[thin, postaction={decorate, decoration={markings, mark=at position 0.6 with {\arrow{>}}}}] (E) -- (Rb); 

  \draw[thin] (A) -- (E);
  \draw[thin] (B) -- (D);

\end{tikzpicture}
\end{minipage}}

\newcommand{\CompletionthreeLoopYukawa}{
\begin{minipage}{1.8cm}
\begin{tikzpicture}[scale=0.65, >=latex]

  \coordinate (v1) at (0, 1.2);    
  \coordinate (v2) at (1.0, 0.6);  
  \coordinate (v3) at (1.0,-0.6);  
  \coordinate (v4) at (0,-1.2);    
  \coordinate (v5) at (-1.0,-0.6); 
  \coordinate (v6) at (-1.0, 0.6); 
  \coordinate (mL) at ($(v6)!0.5!(v5)$);
  \coordinate (mR) at ($(v2)!0.5!(v3)$);
  \draw[thin, postaction={decorate, decoration={markings, mark=at position 0.6 with {\arrow{>}}}}] (v1) -- (v2); 
  \draw[thin, postaction={decorate, decoration={markings, mark=at position 0.6 with {\arrow{>}}}}] (v2) -- (v3); 
  \draw[thin, postaction={decorate, decoration={markings, mark=at position 0.6 with {\arrow{>}}}}] (v3) -- (v4); 
  \draw[thin, postaction={decorate, decoration={markings, mark=at position 0.6 with {\arrow{>}}}}] (v4) -- (v5); 
  \draw[thin, postaction={decorate, decoration={markings, mark=at position 0.6 with {\arrow{>}}}}] (v5) -- (v6); 
  \draw[thin, postaction={decorate, decoration={markings, mark=at position 0.6 with {\arrow{>}}}}] (v6) -- (v1); 

  \draw[draw=black, fill=black, thin, solid] (v1) circle (1.54pt);
  \draw[draw=black, fill=black, thin, solid] (v2) circle (1.54pt);
  \draw[draw=black, fill=black, thin, solid] (v3) circle (1.54pt);
  \draw[draw=black, fill=black, thin, solid] (v4) circle (1.54pt);
  \draw[draw=black, fill=black, thin, solid] (v5) circle (1.54pt);
  \draw[draw=black, fill=black, thin, solid] (v6) circle (1.54pt);

  \draw[thin] (v6) -- (v3);
  \draw[thin] (v2) -- (v5);
  \draw[thin] (v1) -- (v4);

\end{tikzpicture}
\end{minipage}}

\newcommand{\CompletionthreeLoopYukawatwo}{
\begin{minipage}{2.4cm}
\begin{tikzpicture}[scale=0.55, >=latex]

  \coordinate (v1) at ( 0.8,  1.6);
  \coordinate (v2) at ( 1.6,  0.8);
  \coordinate (v3) at ( 1.6, -0.8);
  \coordinate (v4) at ( 0.8, -1.6);
  \coordinate (v5) at (-0.8, -1.6);
  \coordinate (v6) at (-1.6, -0.8);
  \coordinate (v7) at (-1.6,  0.8);
  \coordinate (v8) at (-0.8,  1.6);

  \draw[draw=black, fill=black, thin, solid] (v1) circle (1.81pt);
  \draw[draw=black, fill=black, thin, solid] (v2) circle (1.81pt);
  \draw[draw=black, fill=black, thin, solid] (v3) circle (1.81pt);
  \draw[draw=black, fill=black, thin, solid] (v4) circle (1.81pt);
  \draw[draw=black, fill=black, thin, solid] (v5) circle (1.81pt);
  \draw[draw=black, fill=black, thin, solid] (v6) circle (1.81pt);
  \draw[draw=black, fill=black, thin, solid] (v7) circle (1.81pt);
  \draw[draw=black, fill=black, thin, solid] (v8) circle (1.81pt);

  \foreach \a/\b in {v1/v2,v2/v3,v3/v4,v4/v5,v5/v6,v6/v7,v7/v8,v8/v1}{
    \draw[thin, postaction={decorate, decoration={markings, mark=at position 0.6 with {\arrow{>}}}}] (\a) -- (\b); 
  }

  \coordinate (sL) at ($(v8)!0.0!(v1)$);
  \coordinate (sR) at ($(v1)!0.0!(v2)$);

  \draw[thin] (sL) -- (v5);
  \draw[thin] (sR) -- (v4);

  \draw[thin] (v2) -- (v6);
  \draw[thin] (v7) -- (v3);

\end{tikzpicture}
\end{minipage}}

\newcommand{\Kfivetwo}{
\begin{minipage}{2.5cm}
\begin{tikzpicture}[x=3ex, y = 3ex, baseline={([yshift=-8.0ex]current bounding box.center)}]
  \coordinate[label=above:$0$] (v1) at (-1.00, 1.00);
  \coordinate[label=above:$1$] (v2) at (1.00, 1.00);
  \coordinate[label=above:$z$] (v3) at (3.00, 1.00);
  \coordinate (v5) at (0.00, -1.00);
  \coordinate (v6) at (2.00, -1.00);
  \draw[draw=black, fill=black, thin, solid] (v1) circle (1pt);
  \draw[draw=black, fill=black, thin, solid] (v2) circle (1pt);
  \draw[draw=black, fill=black, thin, solid] (v3) circle (1pt);
  \draw[draw=black, fill=black, thin, solid] (v5) circle (1pt);
  \draw[draw=black, fill=black, thin, solid] (v6) circle (1pt);
  \draw[draw=black, thin, solid] (v1) -- (v5);
  \draw[draw=black, thin, solid] (v1) -- (v6);
  \draw[draw=black, thin, solid] (v2) -- (v5);
  \draw[draw=black, thin, solid] (v2) -- (v6);
  \draw[draw=black, thin, solid] (v3) -- (v5);
  \draw[draw=black, thin, solid] (v3) -- (v6);
  \draw[draw=black, thin, solid] (v5) -- (v6);
\end{tikzpicture}
\end{minipage}
}

\newcommand{\completionKfivetwoPermuted}{
\begin{minipage}{2.5cm}
\begin{tikzpicture}[x=3ex, y = 3ex, baseline={([yshift=-8.0ex]current bounding box.center)}]
  \coordinate[label=above:$1$] (v1) at (-1.00, 1.00);
  \coordinate[label=above right:$0$] (v2) at (1.00, 1.00);
  \coordinate[label=above:$\infty$] (v3) at (3.00, 1.00);
  \coordinate[label=above:$z$] (v4) at (1.00, 3.00);
  \coordinate (v5) at (0.00, -1.00);
  \coordinate (v6) at (2.00, -1.00);
  \draw[draw=black, fill=black, thin, solid] (v1) circle (1pt);
  \draw[draw=black, fill=black, thin, solid] (v2) circle (1pt);
  \draw[draw=black, fill=black, thin, solid] (v3) circle (1pt);
  \draw[draw=black, fill=black, thin, solid] (v4) circle (1pt);
  \draw[draw=black, fill=black, thin, solid] (v5) circle (1pt);
  \draw[draw=black, fill=black, thin, solid] (v6) circle (1pt);
  \draw[draw=black, thin, solid] (v1) -- (v5);
  \draw[draw=black, thin, solid] (v1) -- (v6);
  \draw[draw=black, thin, solid] (v2) -- (v5);
  \draw[draw=black, thin, solid] (v2) -- (v6);
  \draw[draw=black, thin, solid] (v3) -- (v5);
  \draw[draw=black, thin, solid] (v3) -- (v6);
  \draw[draw=black, thin, solid] (v5) -- (v6);
  \draw[draw=black, thin, solid] (v1) -- node[inner sep=0pt,above] {\tiny{$-3$}} (v2);
  \draw[draw=black, thin, solid] (v1) -- (v4);
  \draw[draw=black, thin, solid] (v2) -- (v4);
  \draw[draw=black, thin, solid] (v3) -- node[inner sep=0pt,above right] {\tiny{$-2$}} (v4);

\end{tikzpicture}
\end{minipage}
}

\newcommand{\KfivetwoPermutedDecompleted}{
\begin{minipage}{2cm}
\begin{tikzpicture}[x=3ex, y = 3ex, baseline={([yshift=-8.0ex]current bounding box.center)}]
  \coordinate[label=above:$1$] (v1) at (-1.00, 1.00);
  \coordinate[label=above right:$0$] (v2) at (1.00, 1.00);
  \coordinate[label=above:$z$] (v4) at (1.00, 3.00);
  \coordinate (v5) at (0.00, -1.00);
  \coordinate (v6) at (2.00, -1.00);
  \draw[draw=black, fill=black, thin, solid] (v1) circle (1pt);
  \draw[draw=black, fill=black, thin, solid] (v2) circle (1pt);
  \draw[draw=black, fill=black, thin, solid] (v4) circle (1pt);
  \draw[draw=black, fill=black, thin, solid] (v5) circle (1pt);
  \draw[draw=black, fill=black, thin, solid] (v6) circle (1pt);
  \draw[draw=black, thin, solid] (v1) -- (v5);
  \draw[draw=black, thin, solid] (v1) -- (v6);
  \draw[draw=black, thin, solid] (v2) -- (v5);
  \draw[draw=black, thin, solid] (v2) -- (v6);
  \draw[draw=black, thin, solid] (v5) -- (v6);
  \draw[draw=black, thin, solid] (v1) -- (v4);
  \draw[draw=black, thin, solid] (v2) -- (v4);

\end{tikzpicture}
\end{minipage}
}

\newcommand{\KfivetwoPermutedDecompletedExternalRemoved}{
\begin{minipage}{2cm}
\begin{tikzpicture}[x=3ex, y = 3ex, baseline={([yshift=-8.0ex]current bounding box.center)}]
  \coordinate[label=above:$1$] (v1) at (-1.00, 1.00);
  \coordinate[label=above:$0$] (v2) at (1.00, 1.00);
  \coordinate[label=above:$z$] (v4) at (1.00, 3.00);
  \coordinate (v5) at (0.00, -1.00);
  \coordinate (v6) at (2.00, -1.00);
  \draw[draw=black, fill=black, thin, solid] (v1) circle (1pt);
  \draw[draw=black, fill=black, thin, solid] (v2) circle (1pt);
  \draw[draw=black, fill=black, thin, solid] (v4) circle (1pt);
  \draw[draw=black, fill=black, thin, solid] (v5) circle (1pt);
  \draw[draw=black, fill=black, thin, solid] (v6) circle (1pt);
  \draw[draw=black, thin, solid] (v1) -- (v5);
  \draw[draw=black, thin, solid] (v1) -- (v6);
  \draw[draw=black, thin, solid] (v2) -- (v5);
  \draw[draw=black, thin, solid] (v2) -- (v6);
  \draw[draw=black, thin, solid] (v5) -- (v6);

\end{tikzpicture}
\end{minipage}
}

\newcommand{\TwoPointDiv}{
\begin{minipage}{2cm}
\begin{tikzpicture}[x=3ex, y = 3ex, baseline={(current bounding box.center)}]
  \coordinate (v1) at (0.00, 0.00);
  \coordinate (v2) at (4.00, 0.00);
  \draw[draw=black, fill=black, thin, solid] (v1) circle (1pt);
  \draw[draw=black, fill=black, thin, solid] (v2) circle (1pt);
  \draw[draw=black, thin, solid, bend left=30] (v1) to (v2);
  \draw[draw=black, thin, solid, bend left=30] (v2) to (v1);
\end{tikzpicture}
\end{minipage}
}

\newcommand{\WheelDiv}{
\begin{minipage}{2cm}
\begin{tikzpicture}[x=3ex, y = 3ex, baseline={([yshift=-7ex]current bounding box.center)}]
  \coordinate (v1) at (0.00, 0.00);
  \coordinate (v2) at (-2.00, 0.00);
  \coordinate (v3) at (1.00, 1.73);
  \coordinate (v4) at (1.00, -1.73);
  \draw[draw=black, fill=black, thin, solid] (v1) circle (1pt);
  \draw[draw=black, fill=black, thin, solid] (v2) circle (1pt);
  \draw[draw=black, fill=black, thin, solid] (v3) circle (1pt);
  \draw[draw=black, fill=black, thin, solid] (v4) circle (1pt);
  \draw[draw=black, thin, solid] (v1) -- (v2);
  \draw[draw=black, thin, solid] (v1) -- (v3);
  \draw[draw=black, thin, solid] (v1) -- (v4);
  \draw[draw=black, thin, solid] (v2) -- (v3);
  \draw[draw=black, thin, solid] (v2) -- (v4);
  \draw[draw=black, thin, solid] (v3) -- (v4);
\end{tikzpicture}
\end{minipage}
}

\newcommand{\WheelDivrotated}{
\begin{minipage}{3cm}
\begin{tikzpicture}[
  x=3ex, y=3ex,
  rotate=270,
  baseline={([yshift=-7ex]current bounding box.center)}
]
  \coordinate (v1) at (0.00, 0.00);
  \coordinate (v2) at (-2.00, 0.00);
  \coordinate (v3) at (1.00, 1.73);
  \coordinate (v4) at (1.00, -1.73);

  \fill (v1) circle (0.08);
  \fill (v2) circle (0.08);
  \fill (v3) circle (0.08);
  \fill (v4) circle (0.08);
  \draw[draw=black, fill=black, thin, solid] (v1) circle (1pt);
  \draw[draw=black, fill=black, thin, solid] (v2) circle (1pt);
  \draw[draw=black, fill=black, thin, solid] (v3) circle (1pt);
  \draw[draw=black, fill=black, thin, solid] (v4) circle (1pt);

  \draw[thin] (v1) -- (v2);
  \draw[thin] (v1) -- (v3);
  \draw[thin] (v1) -- (v4);
  \draw[thin] (v2) -- (v3);
  \draw[thin] (v2) -- (v4);
  \draw[thin] (v3) -- (v4);
\end{tikzpicture}
\end{minipage}
}

\newcommand{\WheelDivrotatedwdots}{
\begin{minipage}{4cm}
\begin{tikzpicture}[
  x=4.5ex, y=4.5ex,
  baseline={([yshift=-7ex]current bounding box.center)}
]
  \coordinate (v1) at (0.00, 0.00);
  \coordinate (v6) at (0.60, 0.00);
  \coordinate (v7) at (-0.35, 0.55);
  \coordinate (v8) at (-0.35, -0.55);
  \coordinate (v2) at (-2.00, 0.00);
  \coordinate (v3) at (1.00, 1.73);
  \coordinate (v5) at (2.40, 0.00);
  \coordinate (v4) at (1.00, -1.73);

  \draw[draw=black, fill=black, thin, solid] (v1) circle (1pt);
  \draw[draw=black, fill=black, thin, solid] (v2) circle (1pt);
  \draw[draw=black, fill=black, thin, solid] (v3) circle (1pt);
  \draw[draw=black, fill=black, thin, solid] (v4) circle (1pt);

  \draw[draw=red, fill=red, thin, solid] (v5) circle (1pt);
  \draw[draw=red, fill=red, thin, solid] (v6) circle (1pt);
  \draw[draw=red, fill=red, thin, solid] (v7) circle (1pt);
  \draw[draw=red, fill=red, thin, solid] (v8) circle (1pt);

  \draw[thin] (v1) -- (v2);
  \draw[thin] (v1) -- (v3);
  \draw[thin] (v1) -- (v4);
  \draw[thin] (v2) -- (v3);
  \draw[thin] (v2) -- (v4);
  \draw[thin] (v3) -- (v4);

  \draw[thin, red, dashed] (v7) -- (v8);
  \draw[thin, red, dashed] (v7) -- (v6);
  \draw[thin, red, dashed] (v8) -- (v6);
  \draw[thin, red, dashed] (v6) -- (v5);

  \draw[thin, red, dashed] (v7) .. controls (-1.90, 2.20) and (2.40, 3.20) .. (v5);

  \draw[thin, red, dashed] (v8) .. controls (-1.90, -2.20) and (2.40, -3.20) .. (v5);
\end{tikzpicture}
\end{minipage}
}

\newcommand{\WheelDivrotatedwselfdual}{
\begin{minipage}{3.5cm}
\begin{tikzpicture}[x=4.5ex,y=4.5ex, baseline=(current bounding box.center)]
  \coordinate (v6) at (0.60, 0.00);
  \coordinate (v7) at (-0.35, 0.55);
  \coordinate (v8) at (-0.35, -0.55);
  \coordinate (v5) at (2.40, 0.00);

  \draw[draw=blue, fill=blue, thin, solid] (v5) circle (1pt);
  \draw[draw=blue, fill=blue, thin, solid] (v6) circle (1pt);
  \draw[draw=blue, fill=blue, thin, solid] (v7) circle (1pt);
  \draw[draw=blue, fill=blue, thin, solid] (v8) circle (1pt);

  \draw[thin, black, solid] (v7) -- (v8);
  \draw[thin, black, solid] (v7) -- (v6);
  \draw[thin, black, solid] (v8) -- (v6);
  \draw[thin, black, solid] (v6) -- (v5);
  \draw[thin, black, solid] (v7) .. controls (-1.90, 2.20) and (2.40, 3.20) .. (v5);
  \draw[thin, black, solid] (v8) .. controls (-1.90, -2.20) and (2.40, -3.20) .. (v5);
\end{tikzpicture}
\end{minipage}
}

\newcommand{\SelfDualKFour}{
\begin{tikzpicture}[x=4.5ex,y=4.5ex,rotate=270,baseline=(current bounding box.center)]
  \coordinate (v1) at (0.00, 1.73);
  \coordinate (v2) at (-1.50, -0.87);
  \coordinate (v3) at (1.50, -0.87);
  \coordinate (v4) at (0.00, 0.00);

  \draw[thin] (v1) -- (v2);
  \draw[thin] (v2) -- (v3);
  \draw[thin] (v3) -- (v1);
  \draw[thin] (v4) -- (v1);
  \draw[thin] (v4) -- (v2);
  \draw[thin] (v4) -- (v3);

  \draw[draw=blue, fill=blue, thin, solid] (v1) circle (1pt);
  \draw[draw=blue, fill=blue, thin, solid] (v2) circle (1pt);
  \draw[draw=blue, fill=blue, thin, solid] (v3) circle (1pt);
  \draw[draw=blue, fill=blue, thin, solid] (v4) circle (1pt);
\end{tikzpicture}
}

\newcommand{\Triangle}{
\begin{minipage}{2.5cm}
\begin{tikzpicture}[scale=0.7, x=3ex, y = 3ex, baseline={([yshift=-10ex]current bounding box.center)}]
  \coordinate (v1) at (0.00, 0.00);
  \coordinate (v2) at (0.00, -2.00);
  \coordinate (v3) at (1.73, 1.00);
  \coordinate (v4) at (-1.73, 1.00);
  \coordinate[label=left:$0$] (v5) at (-3.46, -2.00);
  \coordinate[label=right:$z$] (v6) at (3.46, -2.00);
  \coordinate[label=above:$1$] (v7) at (0.00, 4.00);
  \draw[draw=black, fill=black, thin, solid] (v1) circle (1.42pt);
  \draw[draw=black, fill=black, thin, solid] (v2) circle (1.42pt);
  \draw[draw=black, fill=black, thin, solid] (v3) circle (1.42pt);
  \draw[draw=black, fill=black, thin, solid] (v4) circle (1.42pt);
  \draw[draw=black, fill=black, thin, solid] (v5) circle (1.42pt);
  \draw[draw=black, fill=black, thin, solid] (v6) circle (1.42pt);
  \draw[draw=black, fill=black, thin, solid] (v7) circle (1.42pt);
  \draw[draw=black, thin, solid] (v1) -- (v2);
  \draw[draw=black, thin, solid] (v1) -- (v3);
  \draw[draw=black, thin, solid] (v1) -- (v4);
  \draw[draw=black, thin, solid] (v2) -- (v5);
  \draw[draw=black, thin, solid] (v2) -- (v6);
  \draw[draw=black, thin, solid] (v3) -- (v6);
  \draw[draw=black, thin, solid] (v3) -- (v7);
  \draw[draw=black, thin, solid] (v4) -- (v5);
  \draw[draw=black, thin, solid] (v4) -- (v7);
\end{tikzpicture}
\end{minipage}
}

\newcommand{\TriangleWithOneZero}{
\begin{minipage}{2.5cm}
\begin{tikzpicture}[scale=0.7, x=3ex, y = 3ex, baseline={([yshift=-10ex]current bounding box.center)}]
  \coordinate (v1) at (0.00, 0.00);
  \coordinate (v2) at (0.00, -2.00);
  \coordinate (v3) at (1.73, 1.00);
  \coordinate (v4) at (-1.73, 1.00);
  \coordinate[label=left:$0$] (v5) at (-3.46, -2.00);
  \coordinate[label=right:$z$] (v6) at (3.46, -2.00);
  \coordinate[label=above:$1$] (v7) at (0.00, 4.00);
  \draw[draw=black, fill=black, thin, solid] (v1) circle (1.42pt);
  \draw[draw=black, fill=black, thin, solid] (v2) circle (1.42pt);
  \draw[draw=black, fill=black, thin, solid] (v3) circle (1.42pt);
  \draw[draw=black, fill=black, thin, solid] (v4) circle (1.42pt);
  \draw[draw=black, fill=black, thin, solid] (v5) circle (1.42pt);
  \draw[draw=black, fill=black, thin, solid] (v6) circle (1.42pt);
  \draw[draw=black, fill=black, thin, solid] (v7) circle (1.42pt);
  \draw[draw=black, thin, solid] (v1) -- (v2);
  \draw[draw=black, thin, solid] (v1) -- (v3);
  \draw[draw=black, thin, solid] (v1) -- (v4);
  \draw[draw=black, thin, solid] (v2) -- (v5);
  \draw[draw=black, thin, solid] (v2) -- (v6);
  \draw[draw=black, thin, solid] (v3) -- (v6);
  \draw[draw=black, thin, solid] (v3) -- (v7);
  \draw[draw=black, thin, solid] (v4) -- (v5);
  \draw[draw=black, thin, solid] (v4) -- (v7);
  \draw[draw=black, thin, solid, bend left=30] (v5) to (v7);
\end{tikzpicture}
\end{minipage}
}

\newcommand{\TriangleDual}{
\begin{minipage}{2.5cm}
\begin{tikzpicture}[scale=0.7, x=3ex, y = 3ex, baseline={([yshift=-5ex]current bounding box.center)}]
  \coordinate (v1) at (0.00, 2.00);
  \coordinate (v2) at (1.73, -1.00);
  \coordinate (v3) at (-1.73, -1.00);
  \coordinate[label=below:$1$] (v4) at (0.00, -4.00);
  \coordinate (v5) at (-3.46, 2.00);
  \coordinate[label=above:$0$] (v6) at (3.46, 2.00);
  \coordinate[label=above:$z$] (v7) at (-6.05, 3.50);
  \draw[draw=black, fill=black, thin, solid] (v1) circle (1.42pt);
  \draw[draw=black, fill=black, thin, solid] (v2) circle (1.42pt);
  \draw[draw=black, fill=black, thin, solid] (v3) circle (1.42pt);
  \draw[draw=black, fill=black, thin, solid] (v4) circle (1.42pt);
  \draw[draw=black, fill=black, thin, solid] (v5) circle (1.42pt);
  \draw[draw=black, fill=black, thin, solid] (v6) circle (1.42pt);
  \draw[draw=black, fill=black, thin, solid] (v7) circle (1.42pt);
  \draw[draw=black, thin, solid] (v1) -- (v2);
  \draw[draw=black, thin, solid] (v2) -- (v3);
  \draw[draw=black, thin, solid] (v3) -- (v1);
  \draw[draw=black, thin, solid] (v2) -- (v4);
  \draw[draw=black, thin, solid] (v3) -- (v4);
  \draw[draw=black, thin, solid] (v1) -- (v5);
  \draw[draw=black, thin, solid] (v3) -- (v5);
  \draw[draw=black, thin, solid] (v1) -- (v6);
  \draw[draw=black, thin, solid] (v2) -- (v6);
  \draw[draw=black, thin, solid] (v5) -- (v7);
\end{tikzpicture}
\end{minipage}
}

\newcommand{\TriangleDualWOEdge}{
\begin{minipage}{2.5cm}
\begin{tikzpicture}[scale=0.7, x=3ex, y = 3ex, baseline={([yshift=-5ex]current bounding box.center)}]
  \coordinate (v1) at (0.00, 2.00);
  \coordinate (v2) at (1.73, -1.00);
  \coordinate (v3) at (-1.73, -1.00);
  \coordinate[label=below:$1$] (v4) at (0.00, -4.00);
  \coordinate[label=above:$z$] (v5) at (-3.46, 2.00);
  \coordinate[label=above:$0$] (v6) at (3.46, 2.00);
  \draw[draw=black, fill=black, thin, solid] (v1) circle (1.42pt);
  \draw[draw=black, fill=black, thin, solid] (v2) circle (1.42pt);
  \draw[draw=black, fill=black, thin, solid] (v3) circle (1.42pt);
  \draw[draw=black, fill=black, thin, solid] (v4) circle (1.42pt);
  \draw[draw=black, fill=black, thin, solid] (v5) circle (1.42pt);
  \draw[draw=black, fill=black, thin, solid] (v6) circle (1.42pt);
  \draw[draw=black, thin, solid] (v1) -- (v2);
  \draw[draw=black, thin, solid] (v2) -- (v3);
  \draw[draw=black, thin, solid] (v3) -- (v1);
  \draw[draw=black, thin, solid] (v2) -- (v4);
  \draw[draw=black, thin, solid] (v3) -- (v4);
  \draw[draw=black, thin, solid] (v1) -- (v5);
  \draw[draw=black, thin, solid] (v3) -- (v5);
  \draw[draw=black, thin, solid] (v1) -- (v6);
  \draw[draw=black, thin, solid] (v2) -- (v6);
\end{tikzpicture}
\end{minipage}
}

\newcommand{\TriangleDualCompleted}{
\begin{minipage}{2.5cm}
\begin{tikzpicture}[scale=0.7, x=3ex, y = 3ex, baseline={([yshift=-5ex]current bounding box.center)}]
  \coordinate (v1) at (0.00, 2.00);
  \coordinate (v2) at (1.73, -1.00);
  \coordinate (v3) at (-1.73, -1.00);
  \coordinate[label=below:$1$] (v4) at (0.00, -4.00);
  \coordinate[label=below left:$z$] (v5) at (-3.46, 2.00);
  \coordinate[label=above:$0$] (v6) at (3.46, 2.00);
  \coordinate[label=above:$\infty$] (v7) at (-6.05, 3.50);
  \draw[draw=black, fill=black, thin, solid] (v1) circle (1.42pt);
  \draw[draw=black, fill=black, thin, solid] (v2) circle (1.42pt);
  \draw[draw=black, fill=black, thin, solid] (v3) circle (1.42pt);
  \draw[draw=black, fill=black, thin, solid] (v4) circle (1.42pt);
  \draw[draw=black, fill=black, thin, solid] (v5) circle (1.42pt);
  \draw[draw=black, fill=black, thin, solid] (v6) circle (1.42pt);
  \draw[draw=black, fill=black, thin, solid] (v7) circle (1.42pt);
  \draw[draw=black, thin, solid] (v1) -- (v2);
  \draw[draw=black, thin, solid] (v2) -- (v3);
  \draw[draw=black, thin, solid] (v3) -- (v1);
  \draw[draw=black, thin, solid] (v2) -- (v4);
  \draw[draw=black, thin, solid] (v3) -- (v4);
  \draw[draw=black, thin, solid] (v1) -- (v5);
  \draw[draw=black, thin, solid] (v3) -- (v5);
  \draw[draw=black, thin, solid] (v1) -- (v6);
  \draw[draw=black, thin, solid] (v2) -- (v6);
  \draw[draw=black, thin, solid] (v5) -- node[inner sep=0pt,above right] {\tiny{$-2$}} (v7);
  \draw[draw=black, thin, solid, bend left=30] (v4) to (v7);
  \draw[draw=black, thin, solid, bend right=30] (v6) to (v7);
  \draw[draw=black, thin, solid, bend left=30] (v6) to node[inner sep=0pt,right] {\tiny{$-3$}} (v4);
\end{tikzpicture}
\end{minipage}
}

\newcommand{\TriangleDualCompletedPermuted}{
\begin{minipage}{2.5cm}
\begin{tikzpicture}[scale=0.7, x=3ex, y = 3ex, baseline={([yshift=-5ex]current bounding box.center)}]
  \coordinate (v1) at (0.00, 2.00);
  \coordinate (v2) at (1.73, -1.00);
  \coordinate (v3) at (-1.73, -1.00);
  \coordinate[label=below:$0$] (v4) at (0.00, -4.00);
  \coordinate[label=below left:$\infty$] (v5) at (-3.46, 2.00);
  \coordinate[label=above:$1$] (v6) at (3.46, 2.00);
  \coordinate[label=above:$z$] (v7) at (-6.05, 3.50);
  \draw[draw=black, fill=black, thin, solid] (v1) circle (1.42pt);
  \draw[draw=black, fill=black, thin, solid] (v2) circle (1.42pt);
  \draw[draw=black, fill=black, thin, solid] (v3) circle (1.42pt);
  \draw[draw=black, fill=black, thin, solid] (v4) circle (1.42pt);
  \draw[draw=black, fill=black, thin, solid] (v5) circle (1.42pt);
  \draw[draw=black, fill=black, thin, solid] (v6) circle (1.42pt);
  \draw[draw=black, fill=black, thin, solid] (v7) circle (1.42pt);
  \draw[draw=black, thin, solid] (v1) -- (v2);
  \draw[draw=black, thin, solid] (v2) -- (v3);
  \draw[draw=black, thin, solid] (v3) -- (v1);
  \draw[draw=black, thin, solid] (v2) -- (v4);
  \draw[draw=black, thin, solid] (v3) -- (v4);
  \draw[draw=black, thin, solid] (v1) -- (v5);
  \draw[draw=black, thin, solid] (v3) -- (v5);
  \draw[draw=black, thin, solid] (v1) -- (v6);
  \draw[draw=black, thin, solid] (v2) -- (v6);
  \draw[draw=black, thin, solid] (v5) -- node[inner sep=0pt,above right] {\tiny{$-2$}} (v7);
  \draw[draw=black, thin, solid] (v1) -- (v3);
  \draw[draw=black, thin, solid, bend left=30] (v4) to (v7);
  \draw[draw=black, thin, solid, bend right=30] (v6) to (v7);
  \draw[draw=black, thin, solid, bend left=30] (v6) to node[inner sep=0pt,right] {\tiny{$-3$}} (v4);
\end{tikzpicture}
\end{minipage}
}

\newcommand{\TriangleDualCompletedPermutedDecompleted}{
\begin{minipage}{2.5cm}
\begin{tikzpicture}[scale=0.7, x=3ex, y = 3ex, baseline={([yshift=-5ex]current bounding box.center)}]
  \coordinate (v1) at (0.00, 2.00);
  \coordinate (v2) at (1.73, -1.00);
  \coordinate (v3) at (-1.73, -1.00);
  \coordinate[label=below:$0$] (v4) at (0.00, -4.00);
  \coordinate[label=above:$1$] (v6) at (3.46, 2.00);
  \coordinate[label=above:$z$] (v7) at (-3.46, 2.00);
  \draw[draw=black, fill=black, thin, solid] (v1) circle (1.42pt);
  \draw[draw=black, fill=black, thin, solid] (v2) circle (1.42pt);
  \draw[draw=black, fill=black, thin, solid] (v3) circle (1.42pt);
  \draw[draw=black, fill=black, thin, solid] (v4) circle (1.42pt);
  \draw[draw=black, fill=black, thin, solid] (v6) circle (1.42pt);
  \draw[draw=black, fill=black, thin, solid] (v7) circle (1.42pt);
  \draw[draw=black, thin, solid] (v1) -- (v2);
  \draw[draw=black, thin, solid] (v2) -- (v3);
  \draw[draw=black, thin, solid] (v3) -- (v1);
  \draw[draw=black, thin, solid] (v2) -- (v4);
  \draw[draw=black, thin, solid] (v3) -- (v4);
  \draw[draw=black, thin, solid] (v1) -- (v6);
  \draw[draw=black, thin, solid] (v2) -- (v6);
  \draw[draw=black, thin, solid, bend left=30] (v4) to (v7);
  \draw[draw=black, thin, solid, bend right=30] (v6) to (v7);
\end{tikzpicture}
\end{minipage}
}

\newcommand{\TriangleDualCompletedPermutedDecompletedAttached}{
\begin{minipage}{2.5cm}
\begin{tikzpicture}[scale=0.7, x=3ex, y = 3ex, baseline={([yshift=-5ex]current bounding box.center)}]
  \coordinate (v1) at (0.00, 2.00);
  \coordinate (v2) at (1.73, -1.00);
  \coordinate (v3) at (-1.73, -1.00);
  \coordinate[label=below:$0$] (v4) at (0.00, -4.00);
  \coordinate (v5) at (-3.46, 2.00);
  \coordinate[label=above:$1$] (v6) at (3.46, 2.00);
  \coordinate[label=above:$z$] (v7) at (-6.05, 3.50);
  \draw[draw=black, fill=black, thin, solid] (v1) circle (1.42pt);
  \draw[draw=black, fill=black, thin, solid] (v2) circle (1.42pt);
  \draw[draw=black, fill=black, thin, solid] (v3) circle (1.42pt);
  \draw[draw=black, fill=black, thin, solid] (v4) circle (1.42pt);
  \draw[draw=black, fill=black, thin, solid] (v5) circle (1.42pt);
  \draw[draw=black, fill=black, thin, solid] (v6) circle (1.42pt);
  \draw[draw=black, fill=black, thin, solid] (v7) circle (1.42pt);
  \draw[draw=black, thin, solid] (v1) -- (v2);
  \draw[draw=black, thin, solid] (v2) -- (v3);
  \draw[draw=black, thin, solid] (v3) -- (v1);
  \draw[draw=black, thin, solid] (v2) -- (v4);
  \draw[draw=black, thin, solid] (v3) -- (v4);
  \draw[draw=black, thin, solid] (v1) -- (v6);
  \draw[draw=black, thin, solid] (v2) -- (v6);
  \draw[draw=black, thin, solid] (v5) -- (v7);
  \draw[draw=black, thin, solid, bend left=30] (v4) to (v5);
  \draw[draw=black, thin, solid, bend right=30] (v6) to (v5);
\end{tikzpicture}
\end{minipage}
}

\newcommand{\TriangleRightPart}{
\begin{minipage}{2.5cm}
\begin{tikzpicture}[scale=0.7, x=3ex, y = 3ex, baseline={([yshift=-5ex]current bounding box.center)}]
  \coordinate (v1) at (0.00, 2.00);
  \coordinate (v2) at (1.73, -1.00);
  \coordinate (v3) at (-1.73, -1.00);
  \coordinate[label=below:$0$] (v4) at (0.00, -4.00);
  \coordinate[label=above:$1$] (v6) at (3.46, 2.00);
  \draw[draw=black, fill=black, thin, solid] (v1) circle (1.42pt);
  \draw[draw=black, fill=black, thin, solid] (v2) circle (1.42pt);
  \draw[draw=black, fill=black, thin, solid] (v3) circle (1.42pt);
  \draw[draw=black, fill=black, thin, solid] (v4) circle (1.42pt);
  \draw[draw=black, fill=black, thin, solid] (v6) circle (1.42pt);
  \draw[draw=black, thin, solid] (v1) -- (v2);
  \draw[draw=black, thin, solid] (v2) -- (v3);
  \draw[draw=black, thin, solid] (v3) -- (v1);
  \draw[draw=black, thin, solid] (v2) -- (v4);
  \draw[draw=black, thin, solid] (v3) -- (v4);
  \draw[draw=black, thin, solid] (v1) -- (v6);
  \draw[draw=black, thin, solid] (v2) -- (v6);
\end{tikzpicture}
\end{minipage}
}

\newcommand{\TreeSlim}{
\begin{minipage}{1.6cm}
\begin{tikzpicture}[scale=0.7, x=3ex, y = 3ex, baseline={([yshift=-5ex]current bounding box.center)}]
  \coordinate (v1) at (0.00, 0.00);
  \coordinate[label=below:$0$] (v2) at (0.00, -2.00);
  \coordinate[label=above:$1$] (v3) at (1.73, 1.00);
  \coordinate[label=above:$z$] (v4) at (-1.73, 1.00);
  \draw[draw=black, fill=black, thin, solid] (v1) circle (1.42pt);
  \draw[draw=black, fill=black, thin, solid] (v2) circle (1.42pt);
  \draw[draw=black, fill=black, thin, solid] (v3) circle (1.42pt);
  \draw[draw=black, fill=black, thin, solid] (v4) circle (1.42pt);
  \draw[draw=black, thin, solid] (v1) -- (v2);
  \draw[draw=black, thin, solid] (v1) -- (v3);
  \draw[draw=black, thin, solid] (v1) -- (v4);
\end{tikzpicture}
\end{minipage}}

\newcommand{\HardExampleOne}{
\begin{minipage}{2cm}
\begin{tikzpicture}[x=3ex, y = 3ex, baseline={([yshift=-5ex]current bounding box.center)}]
  \coordinate (v1) at (0.00, 0.00);
  \coordinate (v2) at (-2.00, 0.00);
  \coordinate (v3) at (2.00, 0.00);
  \coordinate (v4) at (-2.00, 2.00);
  \coordinate (v5) at (-2.00, -2.00);
  \coordinate (v6) at (2.00, 2.00);
  \coordinate (v7) at (2.00, -2.00);
  \draw[draw=black, fill=black, thin, solid] (v1) circle (1pt);
  \draw[draw=black, fill=black, thin, solid] (v2) circle (1pt);
  \draw[draw=black, fill=black, thin, solid] (v3) circle (1pt);
  \draw[draw=black, fill=black, thin, solid] ([xshift=-0.8,yshift=-0.8]v4) rectangle ++(0.2, 0.2);
  \draw[draw=black, fill=black, thin, solid] ([xshift=-0.8,yshift=-0.8]v5) rectangle ++(0.2, 0.2);
  \draw[draw=black, fill=black, thin, solid] ([xshift=-0.8,yshift=-0.8]v6) rectangle ++(0.2, 0.2);
  \draw[draw=black, fill=black, thin, solid] ([xshift=-0.8,yshift=-0.8]v7) rectangle ++(0.2, 0.2);
  \draw[draw=black, thin, solid] (v1) -- (v2);
  \draw[draw=black, thin, solid] (v1) -- (v3);
  \draw[draw=black, thin, solid] (v1) -- (v4);
  \draw[draw=black, thin, solid] (v2) -- (v4);
  \draw[draw=black, thin, solid] (v1) -- (v6);
  \draw[draw=black, thin, solid] (v3) -- (v6);
  \draw[draw=black, thin, solid] (v2) -- (v4);
  \draw[draw=black, thin, solid] (v2) -- (v5);
  \draw[draw=black, thin, solid] (v3) -- (v5);
  \draw[draw=black, thin, solid] (v2) -- (v7);
  \draw[draw=black, thin, solid] (v3) -- (v7);
\end{tikzpicture}
\end{minipage}}

\newcommand{\HardExampleTwo}{
\begin{minipage}{2cm}
\begin{tikzpicture}[x=3ex, y = 3ex, baseline={([yshift=-5ex]current bounding box.center)}]
  \coordinate (v1) at (0.00, 1.33);
  \coordinate (v2) at (-0.90, 0.00);
  \coordinate (v3) at (0.90, 0.00);
  \coordinate (v4) at (-2.00, 2.66);
  \coordinate (v5) at (-2.00, -1.34);
  \coordinate (v6) at (2.00, 2.66);
  \coordinate (v7) at (2.00, -1.34);
  \draw[draw=black, fill=black, thin, solid] (v1) circle (1pt);
  \draw[draw=black, fill=black, thin, solid] (v2) circle (1pt);
  \draw[draw=black, fill=black, thin, solid] (v3) circle (1pt);
  \draw[draw=black, fill=black, thin, solid] ([xshift=-0.8,yshift=-0.8]v4) rectangle ++(0.2, 0.2);
  \draw[draw=black, fill=black, thin, solid] ([xshift=-0.8,yshift=-0.8]v5) rectangle ++(0.2, 0.2);
  \draw[draw=black, fill=black, thin, solid] ([xshift=-0.8,yshift=-0.8]v6) rectangle ++(0.2, 0.2);
  \draw[draw=black, fill=black, thin, solid] ([xshift=-0.8,yshift=-0.8]v7) rectangle ++(0.2, 0.2);
  \draw[draw=black, thin, solid] (v1) -- (v2);
  \draw[draw=black, thin, solid] (v1) -- (v3);
  \draw[draw=black, thin, solid] (v2) -- (v3);
  \draw[draw=black, thin, solid] (v1) -- (v4);
  \draw[draw=black, thin, solid] (v2) -- node[inner sep=0pt,below left] {\tiny{$-1$}} (v4);
  \draw[draw=black, thin, solid] (v1) -- (v6);
  \draw[draw=black, thin, solid] (v2) -- (v6);
  \draw[draw=black, thin, solid] (v2) -- (v4);
  \draw[draw=black, thin, solid] (v2) -- (v5);
  \draw[draw=black, thin, solid] (v3) -- (v5);
  \draw[draw=black, thin, solid] (v2) -- (v7);
  \draw[draw=black, thin, solid] (v3) -- (v7);
\end{tikzpicture}
\end{minipage}}

\newcommand{\HardExampleThree}{
\begin{minipage}{2cm}
\begin{tikzpicture}[x=3ex, y = 3ex, baseline={([yshift=-5ex]current bounding box.center)}]
  \coordinate (v1) at (0.75, 0.75);
  \coordinate (v2) at (-0.75, 0.75);
  \coordinate (v3) at (0.75, -0.75);
  \coordinate (v4) at (-0.75, -0.75);
  \coordinate (v5) at (2.00, 2.00);
  \coordinate (v6) at (2.00, -2.00);
  \coordinate (v7) at (-2.00, 2.00);
  \coordinate (v8) at (-2.00, -2.00);
  \draw[draw=black, fill=black, thin, solid] (v1) circle (1pt);
  \draw[draw=black, fill=black, thin, solid] (v2) circle (1pt);
  \draw[draw=black, fill=black, thin, solid] (v3) circle (1pt);
  \draw[draw=black, fill=black, thin, solid] (v4) circle (1pt);
  \draw[draw=black, fill=black, thin, solid] ([xshift=-0.8,yshift=-0.8]v5) rectangle ++(0.2, 0.2);
  \draw[draw=black, fill=black, thin, solid] ([xshift=-0.8,yshift=-0.8]v6) rectangle ++(0.2, 0.2);
  \draw[draw=black, fill=black, thin, solid] ([xshift=-0.8,yshift=-0.8]v7) rectangle ++(0.2, 0.2);
  \draw[draw=black, fill=black, thin, solid] ([xshift=-0.8,yshift=-0.8]v8) rectangle ++(0.2, 0.2);
  \draw[draw=black, thin, solid] (v1) -- (v2);
  \draw[draw=black, thin, solid] (v2) -- (v4);
  \draw[draw=black, thin, solid] (v3) -- (v4);
  \draw[draw=black, thin, solid] (v3) -- (v1);
  \draw[draw=black, thin, solid] (v5) -- (v2);
  \draw[draw=black, thin, solid] (v5) -- (v3);
  \draw[draw=black, thin, solid] (v8) -- (v2);
  \draw[draw=black, thin, solid] (v8) -- (v3);
  \draw[draw=black, thin, solid] (v7) -- (v1);
  \draw[draw=black, thin, solid] (v7) -- (v4);
  \draw[draw=black, thin, solid] (v6) -- (v1);
  \draw[draw=black, thin, solid] (v6) -- (v4);
\end{tikzpicture}
\end{minipage}}

\newcommand{\bubblethreestar}{
\begin{minipage}{2.5cm}
\begin{tikzpicture}
    \coordinate (v);
    \def \rad {1}
    \coordinate[label=above:$1$] (v1) at ([shift=(60:\rad)]v);
    \coordinate[label=left:$0$] (v2) at ([shift=(180:\rad)]v);
    \coordinate[label=below:$z$] (v3) at ([shift=(300:\rad)]v);

    \draw[draw=black, fill=black, thin, solid] (v) circle (1pt);
    \draw[draw=black, fill=black, thin, solid] (v1) circle (1pt);
    \draw[draw=black, fill=black, thin, solid] (v2) circle (1pt);
    \draw[draw=black, fill=black, thin, solid] (v3) circle (1pt);
    \draw[draw=black, thin, solid, bend left=30] (v) to (v2);
    \draw[draw=black, thin, solid, bend right=30] (v) to (v2);
    \draw[draw=black, thin, solid] (v) -- (v1);
    \draw[draw=black, thin, solid] (v) -- (v3);

\end{tikzpicture}
\end{minipage}
}

\newcommand{\bubblethreestarremoved}{
\begin{minipage}{2.5cm}
\begin{tikzpicture}[baseline={([yshift=-4ex]current bounding box.center)}]
    \coordinate[label=above:$z$] (v);
    \def \rad {1}
    \coordinate[label=above:$1$] (v1) at ([shift=(0:\rad)]v);
    \coordinate[label=above:$0$] (v2) at ([shift=(180:\rad)]v);

    \draw[draw=black, fill=black, thin, solid] (v) circle (1pt);
    \draw[draw=black, fill=black, thin, solid] (v1) circle (1pt);
    \draw[draw=black, fill=black, thin, solid] (v2) circle (1pt);
    \draw[draw=black, thin, solid, bend left=30] (v) to (v2);
    \draw[draw=black, thin, solid, bend right=30] (v) to (v2);
    \draw[draw=black, thin, solid] (v) -- (v1);

\end{tikzpicture}
\end{minipage}
}

\newcommand{\bubblethreestarinduced}{
\begin{minipage}{1.5cm}
\begin{tikzpicture}[baseline={([yshift=-4ex]current bounding box.center)}]
    \coordinate[label=above:$z$] (v);
    \def \rad {1}
    \coordinate[label=above:$0$] (v2) at ([shift=(180:\rad)]v);

    \draw[draw=black, fill=black, thin, solid] (v) circle (1pt);
    \draw[draw=black, fill=black, thin, solid] (v2) circle (1pt);
    \draw[draw=black, thin, solid, bend left=30] (v) to (v2);
    \draw[draw=black, thin, solid, bend right=30] (v) to (v2);

\end{tikzpicture}
\end{minipage}
}

\newcommand{\bubblethreestarfactor}{
\begin{minipage}{1.5cm}
\begin{tikzpicture}[baseline={([yshift=-4ex]current bounding box.center)}]
    \coordinate[label=above:$1$] (v);
    \def \rad {1}
    \coordinate[label=above:$0$] (v2) at ([shift=(180:\rad)]v);

    \draw[draw=black, fill=black, thin, solid] (v2) circle (1pt);
    \draw[draw=black, fill=black, thin, solid] (v) circle (1pt);
    \draw[draw=black, thin, solid] (v) -- (v2);

\end{tikzpicture}
\end{minipage}
}

\newcommand{\bubblethreestarfactoredattached}{
\begin{minipage}{2.5cm}
\begin{tikzpicture}[baseline={([yshift=-4ex]current bounding box.center)}]
    \coordinate (v);
    \def \rad {1}
    \coordinate[label=above:$z$] (v1) at ([shift=(0:\rad)]v);
    \coordinate[label=above:$0$] (v2) at ([shift=(180:\rad)]v);

    \draw[draw=black, fill=black, thin, solid] (v) circle (1pt);
    \draw[draw=black, fill=black, thin, solid] (v1) circle (1pt);
    \draw[draw=black, fill=black, thin, solid] (v2) circle (1pt);
    \draw[draw=black, thin, solid, bend left=30] (v) to (v2);
    \draw[draw=black, thin, solid, bend right=30] (v) to (v2);
    \draw[draw=black, thin, solid] (v) -- (v1);

\end{tikzpicture}
\end{minipage}
}

\newcommand{\IRthreestar}{
\begin{minipage}{2.2cm}
\begin{tikzpicture}
    \coordinate (v);
    \def \rad {1}
    \coordinate[label=above:$1$] (v1) at ([shift=(60:\rad)]v);
    \coordinate[label=left:$0$] (v2) at ([shift=(180:\rad)]v);
    \coordinate[label=below:$z$] (v3) at ([shift=(300:\rad)]v);

    \draw[draw=black, fill=black, thin, solid] (v) circle (1pt);
    \draw[draw=black, fill=black, thin, solid] (v1) circle (1pt);
    \draw[draw=black, fill=black, thin, solid] (v2) circle (1pt);
    \draw[draw=black, fill=black, thin, solid] (v3) circle (1pt);
    \draw[draw=black, thin, solid] (v) -- node[inner sep=0pt,above] {\tiny $\frac{a\ep}{\lambda}$} (v2);
    \draw[draw=black, thin, solid] (v) -- (v1);
    \draw[draw=black, thin, solid] (v) -- (v3);

\end{tikzpicture}
\end{minipage}
}

\newcommand{\IRthreestarfactor}{
\begin{minipage}{1.5cm}
\begin{tikzpicture}[baseline={([yshift=-4ex]current bounding box.center)}]
    \coordinate[label=above:$z$] (v);
    \def \rad {1}
    \coordinate[label=above:$0$] (v2) at ([shift=(180:\rad)]v);

    \draw[draw=black, fill=black, thin, solid] (v2) circle (1pt);
    \draw[draw=black, fill=black, thin, solid] (v) circle (1pt);
    \draw[draw=black, thin, solid] (v) -- node[inner sep=0pt,above] {\tiny $\frac{\lambda + a\ep}{\lambda}$} (v2);

\end{tikzpicture}
\end{minipage}
}

\newcommand{\IRthreestarfactoredattached}{
\begin{minipage}{2.5cm}
\begin{tikzpicture}[baseline={([yshift=-4ex]current bounding box.center)}]
    \coordinate (v);
    \def \rad {1}
    \coordinate[label=above:$z$] (v1) at ([shift=(0:\rad)]v);
    \coordinate[label=above:$0$] (v2) at ([shift=(180:\rad)]v);

    \draw[draw=black, fill=black, thin, solid] (v) circle (1pt);
    \draw[draw=black, fill=black, thin, solid] (v1) circle (1pt);
    \draw[draw=black, fill=black, thin, solid] (v2) circle (1pt);
    \draw[draw=black, thin, solid] (v) -- (v1);
    \draw[draw=black, thin, solid] (v) -- node[inner sep=0pt,above] {\tiny $\frac{\lambda + a\ep}{\lambda}$} (v2);

\end{tikzpicture}
\end{minipage}
}

\newcommand{\GFTwistLeft}{
\begin{minipage}{4.5cm}
\begin{tikzpicture}[scale = 0.7, baseline={([yshift=-1.3ex]current bounding box.center)}]
    \coordinate[label=above:$a$] (va) at (0,3);
    \coordinate[label=above:$b$] (vb) at (0,2);
    \coordinate (vm) at (0, 2.15);
    \coordinate[label=above:$c$] (vc) at (0,1);
    \coordinate[label=above:$d$] (vd) at (0,0);
    \filldraw (va) circle (1.42pt);
    \filldraw (vb) circle (1.42pt);
    \filldraw (vc) circle (1.42pt);
    \filldraw (vd) circle (1.42pt);
    \draw[fill=black!20] (va) arc (90:270:3 and 1.5) arc (270:90:.5) arc (270:90:.5) arc (270:90:.5);
    \draw[fill=black!20] (va) arc (90:-90:3 and 1.5) arc (-90:90:.5) arc (-90:90:.5) arc (-90:90:.5);
    \node (G1) at (-1.5,1.5) {$\overline{G}_1$};
    \node (G2) at (+1.5,1.5) {$\overline{G}_2$};
    \node (G) at (-2,0) {$\overline{G}$};
    \coordinate[label=above right:$0$] (v0)  at ([shift={(vm)}]60:3 and 1.5);
    \coordinate[label=above right:$1$] (v1)  at ([shift={(vm)}]30:3 and 1.5);
    \coordinate[label=below right:$z$] (vz)  at ([shift={(vm)}]-30:3 and 1.5);
    \coordinate[label=below right:$\infty$] (voo) at ([shift={(vm)}]-60:3 and 1.5);
    \filldraw (v0) circle (1.42pt);
    \filldraw (v1) circle (1.42pt);
    \filldraw (vz) circle (1.42pt);
    \filldraw (voo) circle (1.42pt);
\end{tikzpicture}
\end{minipage}
}

\newcommand{\GFTwistRight}{
\begin{minipage}{4.5cm}
\begin{tikzpicture}[scale = 0.7, baseline={([yshift=-1.3ex]current bounding box.center)}]
    \coordinate[label=above:$a$] (va) at (0,3);
    \coordinate[label=above:$b$] (vb) at (0,2);
    \coordinate (vm) at (0, 2.15);
    \coordinate[label=below:$c$] (vc) at (0,1);
    \coordinate[label=above:$d$] (vd) at (0,0);
    \draw[fill=black!20] (vb) .. controls (-1,3) and (-3,3) .. (-3,1.5) .. controls (-3,0) and (-1,0) .. (vc) .. controls (-1,.5) .. (-1,.75) .. controls (-1,1) .. (vd) .. controls (-1,1.5) .. (va) .. controls (-1,2) .. (-1,2.25) .. controls (-1,2.5) .. (vb);
    \draw[fill=black!20] (-1,.75) .. controls (-1,1) .. (vd) .. controls (-1,1.5) .. (va) .. controls (-1,2) .. (-1,2.25);
    \draw[fill=black!20] (va) arc (90:-90:3 and 1.5) arc (-90:90:.5) arc (-90:90:.5) arc (-90:90:.5);
    \node (G1) at (-1.75,1.5) {$\overline{G}_1$};
    \node (G2) at (+2,1.5) {$\overline{G}_2$};
    \node (G) at (-2,0) {$\overline{G}'$};
    \coordinate[label=above right:$0$] (v0)  at ([shift={(vm)}]60:3 and 1.5);
    \coordinate[label=above right:$1$] (v1)  at ([shift={(vm)}]30:3 and 1.5);
    \coordinate[label=below right:$z$] (vz)  at ([shift={(vm)}]-30:3 and 1.5);
    \coordinate[label=below right:$\infty$] (voo) at ([shift={(vm)}]-60:3 and 1.5);
    \filldraw (v0) circle (1.42pt);
    \filldraw (v1) circle (1.42pt);
    \filldraw (vz) circle (1.42pt);
    \filldraw (voo) circle (1.42pt);
    \filldraw (va) circle (1.42pt);
    \filldraw (vb) circle (1.42pt);
    \filldraw (vc) circle (1.42pt);
    \filldraw (vd) circle (1.42pt);
\end{tikzpicture}
\end{minipage}
}

\newcommand{\EDExample}{
\begin{minipage}{2.5cm}
\begin{tikzpicture}[scale=0.7, x=3ex, y = 3ex, baseline={([yshift=-5ex]current bounding box.center)}]
  \coordinate[label=above:$x_1$] (v1) at (0.00, 2.00);
  \coordinate[label=below:$x_3$] (v2) at (1.73, -1.00);
  \coordinate[label=below left:$x_2$](v3) at (-1.73, -1.00);
  \coordinate[label=below:$0$] (v4) at (0.00, -4.00);
  \coordinate[label=above:$z$] (v5) at (-3.46, 2.00);
  \coordinate[label=above:$1$] (v6) at (3.46, 2.00);
  \draw[draw=black, fill=black, thin, solid] (v1) circle (1.42pt);
  \draw[draw=black, fill=black, thin, solid] (v2) circle (1.42pt);
  \draw[draw=black, fill=black, thin, solid] (v3) circle (1.42pt);
  \draw[draw=black, fill=black, thin, solid] (v4) circle (1.42pt);
  \draw[draw=black, fill=black, thin, solid] (v5) circle (1.42pt);
  \draw[draw=black, fill=black, thin, solid] (v6) circle (1.42pt);
  \draw[draw=black, thin, solid] (v1) -- (v2);
  \draw[draw=black, thin, solid] (v2) -- (v3);
  \draw[draw=black, thin, solid] (v3) -- (v1);
  \draw[draw=black, thin, solid] (v1) -- (v5);
  \draw[draw=black, thin, solid] (v3) -- (v5);
  \draw[draw=black, thin, solid] (v1) -- (v6);
  \draw[draw=black, thin, solid] (v2) -- (v6);
  \draw[draw=black, thin, solid] (v4) -- node[inner sep=0pt,right, pos=0.25] {\tiny $-\frac{1}{\lambda}$ }(v1);
  \draw[draw=black, thin, solid] (v3) -- (v4);
\end{tikzpicture}
\end{minipage}
}

\newcommand{\EDExampleTwo}{
\begin{minipage}{2.5cm}
\begin{tikzpicture}[scale=0.7, x=3ex, y = 3ex, baseline={([yshift=-5ex]current bounding box.center)}]
  \coordinate (v1) at (0.00, 2.00);
  \coordinate (v2) at (1.73, -1.00);
  \coordinate (v3) at (-1.73, -1.00);
  \coordinate[label=below:$0$] (v4) at (0.00, -4.00);
  \coordinate[label=above:$z$] (v5) at (-3.46, 2.00);
  \coordinate[label=above:$1$] (v6) at (3.46, 2.00);
  \draw[draw=black, fill=black, thin, solid] (v1) circle (1.42pt);
  \draw[draw=black, fill=black, thin, solid] (v2) circle (1.42pt);
  \draw[draw=black, fill=black, thin, solid] (v3) circle (1.42pt);
  \draw[draw=black, fill=black, thin, solid] (v4) circle (1.42pt);
  \draw[draw=black, fill=black, thin, solid] (v5) circle (1.42pt);
  \draw[draw=black, fill=black, thin, solid] (v6) circle (1.42pt);
  \draw[draw=black, thin, solid] (v1) -- (v2);
  \draw[draw=black, thin, solid] (v2) -- (v3);
  \draw[draw=black, thin, solid] (v3) -- (v1);
  \draw[draw=black, thin, solid] (v1) -- (v5);
  \draw[draw=black, thin, solid] (v3) -- (v5);
  \draw[draw=black, thin, solid] (v1) -- (v6);
  \draw[draw=black, thin, solid] (v2) -- (v6);
  \draw[draw=black, thin, solid] (v3) -- (v4);
\end{tikzpicture}
\end{minipage}
}

\newcommand{\EDExampleThree}{
\begin{minipage}{2.5cm}
\begin{tikzpicture}[scale=0.7, x=3ex, y = 3ex, baseline={([yshift=-5ex]current bounding box.center)}]
  \coordinate (v1) at (0.00, 2.00);
  \coordinate (v2) at (1.73, -1.00);
  \coordinate (v3) at (-1.73, -1.00);
  \coordinate[label=below:$0$] (v4) at (0.00, -4.00);
  \coordinate[label=above:$z$] (v5) at (-3.46, 2.00);
  \coordinate[label=above:$1$] (v6) at (3.46, 2.00);
  \draw[draw=black, fill=black, thin, solid] (v1) circle (1.42pt);
  \draw[draw=black, fill=black, thin, solid] (v2) circle (1.42pt);
  \draw[draw=black, fill=black, thin, solid] (v3) circle (1.42pt);
  \draw[draw=black, fill=black, thin, solid] (v4) circle (1.42pt);
  \draw[draw=black, fill=black, thin, solid] (v5) circle (1.42pt);
  \draw[draw=black, fill=black, thin, solid] (v6) circle (1.42pt);
  \draw[draw=black, thin, solid] (v1) -- (v2);
  \draw[draw=black, thin, solid] (v2) -- (v3);
  \draw[draw=black, thin, solid] (v3) -- (v1);
  \draw[draw=black, thin, solid] (v3) -- (v5);
  \draw[draw=black, thin, solid] (v1) -- (v6);
  \draw[draw=black, thin, solid] (v2) -- (v6);
  \draw[draw=black, thin, solid] (v5) -- node[inner sep=0pt,above] {\tiny $1\!-\!\frac{1}{\lambda}$ }(v1);
  \draw[draw=black, thin, solid] (v3) -- (v4);
\end{tikzpicture}
\end{minipage}
}

\newcommand{\EDExampleFour}{
\begin{minipage}{2.5cm}
\begin{tikzpicture}[scale=0.7, x=3ex, y = 3ex, baseline={([yshift=-5ex]current bounding box.center)}]
  \coordinate (v1) at (0.00, 2.00);
  \coordinate (v2) at (1.73, -1.00);
  \coordinate (v3) at (-1.73, -1.00);
  \coordinate[label=below:$0$] (v4) at (0.00, -4.00);
  \coordinate[label=above:$z$] (v5) at (-3.46, 2.00);
  \coordinate[label=above:$1$] (v6) at (3.46, 2.00);
  \draw[draw=black, fill=black, thin, solid] (v1) circle (1.42pt);
  \draw[draw=black, fill=black, thin, solid] (v2) circle (1.42pt);
  \draw[draw=black, fill=black, thin, solid] (v3) circle (1.42pt);
  \draw[draw=black, fill=black, thin, solid] (v4) circle (1.42pt);
  \draw[draw=black, fill=black, thin, solid] (v5) circle (1.42pt);
  \draw[draw=black, fill=black, thin, solid] (v6) circle (1.42pt);
  \draw[draw=black, thin, solid] (v1) -- (v2);
  \draw[draw=black, thin, solid] (v2) -- (v3);
  \draw[draw=black, thin, solid] (v3) -- (v1);
  \draw[draw=black, thin, solid] (v3) -- node[inner sep=0pt,left]{\tiny $1\!+\!\frac{1}{\lambda}$} (v5);
  \draw[draw=black, thin, solid] (v1) -- (v6);
  \draw[draw=black, thin, solid] (v2) -- (v6);
  \draw[draw=black, thin, solid] (v5) -- node[inner sep=0pt,above] {\tiny $1\!-\!\frac{1}{\lambda}$ }(v1);
  \draw[draw=black, thin, solid] (v3) -- node[inner sep=0pt,left] {\tiny $1\!-\!\frac{1}{\lambda}$}(v4);
\end{tikzpicture}
\end{minipage}
}

\newcommand{\EDExampleFive}{
\begin{minipage}{2.7cm}
\begin{tikzpicture}[scale=0.7, x=3ex, y = 3ex, baseline={([yshift=-4ex]current bounding box.center)}]
  \coordinate (v1) at (0.00, 2.00);
  \coordinate (v2) at (1.73, -1.00);
  \coordinate (v3) at (-1.73, -1.00);
  \coordinate[label=below:$0$] (v4) at (0.00, -4.00);
  \coordinate[label=above:$z$] (v5) at (-3.46, 2.00);
  \coordinate[label=above:$1$] (v6) at (3.46, 2.00);
  \draw[draw=black, fill=black, thin, solid] (v1) circle (1.42pt);
  \draw[draw=black, fill=black, thin, solid] (v2) circle (1.42pt);
  \draw[draw=black, fill=black, thin, solid] (v3) circle (1.42pt);
  \draw[draw=black, fill=black, thin, solid] (v4) circle (1.42pt);
  \draw[draw=black, fill=black, thin, solid] (v5) circle (1.42pt);
  \draw[draw=black, fill=black, thin, solid] (v6) circle (1.42pt);
  \draw[draw=black, thin, solid] (v1) -- (v2);
  \draw[draw=black, thin, solid] (v2) -- (v3);
  \draw[draw=black, thin, solid] (v3) -- (v1);
  \draw[draw=black, thin, solid] (v1) -- node[inner sep=0pt,above] {\tiny $1\!-\!\frac{1}{\lambda}$ } (v5);
  \draw[draw=black, thin, solid] (v3) -- node[inner sep=0pt,left] {\tiny $1\!+\!\frac{1}{\lambda}$ } (v5);
  \draw[draw=black, thin, solid] (v1) -- (v6);
  \draw[draw=black, thin, solid] (v2) -- (v6);
  \draw[draw=black, thin, solid] (v4) -- (v3);
\end{tikzpicture}
\end{minipage}
}

\newcommand{\lambdaLine}{
\begin{minipage}{1.2cm}
\begin{tikzpicture}[baseline={([yshift=-3ex]current bounding box.center)}]
  \coordinate (v1) at (0, 0);
  \coordinate (v2) at (1, 0);
  \draw[draw=black, fill=black, thin, solid] (v1) circle (1pt);
  \draw[draw=black, fill=black, thin, solid] (v2) circle (1pt);
  \draw[draw=black, thin, solid] (v1) -- node[inner sep=0pt,above] {\tiny $1\!-\!\frac{1}{\lambda}$ } (v2);
\end{tikzpicture}
\end{minipage}
}

\newcommand{\doubleLine}{
\begin{minipage}{2.3cm}
\begin{tikzpicture}
  \coordinate (v1) at (0, 0);
  \coordinate (v2) at (1, 0);
  \coordinate (v3) at (2, 0);
  \draw[draw=black, fill=black, thin, solid] (v1) circle (1pt);
  \draw[draw=black, fill=black, thin, solid] (v2) circle (1pt);
  \draw[draw=black, fill=black, thin, solid] (v3) circle (1pt);
  \draw[draw=black, thin, solid] (v1) -- (v2);
  \draw[draw=black, thin, solid] (v2) -- (v3);
\end{tikzpicture}
\end{minipage}
}

\newcommand{\nuLine}{
\begin{minipage}{1.2cm}
\begin{tikzpicture}[baseline={([yshift=-1ex]current bounding box.center)}]
  \coordinate (v1) at (0, 0);
  \coordinate (v2) at (1, 0);
  \draw[draw=black, fill=black, thin, solid] (v1) circle (1pt);
  \draw[draw=black, fill=black, thin, solid] (v2) circle (1pt);
  \draw[draw=black, thin, solid] (v1) -- node[inner sep=3pt,above] {\tiny $\nu$ } (v2);
\end{tikzpicture}
\end{minipage}
}

\newcommand{\nuLineLambda}{
\begin{minipage}{1.2cm}
\begin{tikzpicture}[baseline={([yshift=-3ex]current bounding box.center)}]
  \coordinate (v1) at (0, 0);
  \coordinate (v2) at (1, 0);
  \draw[draw=black, fill=black, thin, solid] (v1) circle (1pt);
  \draw[draw=black, fill=black, thin, solid] (v2) circle (1pt);
  \draw[draw=black, thin, solid] (v1) -- node[inner sep=3pt,above] {\tiny $\nu + \frac{1}{\lambda} $} (v2);
\end{tikzpicture}
\end{minipage}
}

\newcommand{\EDExampleTwoR}{
\begin{minipage}{2.5cm}
\begin{tikzpicture}[scale=0.7, x=3ex, y = 3ex, baseline={([yshift=-5ex]current bounding box.center)}]
  \coordinate (v1) at (0.00, 2.00);
  \coordinate (v2) at (1.73, -1.00);
  \coordinate (v3) at (-1.73, -1.00);
  \coordinate[label=below:$0$] (v4) at (0.00, -4.00);
  \coordinate[label=above:$z$] (v5) at (-3.46, 2.00);
  \coordinate[label=above:$1$] (v6) at (3.46, 2.00);
  \draw[draw=black, fill=black, thin, solid] (v1) circle (1.42pt);
  \draw[draw=black, fill=black, thin, solid] (v2) circle (1.42pt);
  \draw[draw=black, fill=black, thin, solid] (v3) circle (1.42pt);
  \draw[draw=black, fill=black, thin, solid] (v4) circle (1.42pt);
  \draw[draw=black, fill=black, thin, solid] (v5) circle (1.42pt);
  \draw[draw=black, fill=black, thin, solid] (v6) circle (1.42pt);
  \draw[draw=black, thin, solid] (v1) -- (v2);
  \draw[draw=black, thin, solid] (v2) -- (v3);
  \draw[draw=black, thin, solid] (v3) -- (v1);
  \draw[draw=black, thin, solid] (v1) -- (v5);
  \draw[draw=black, thin, solid] (v3) -- (v5);
  \draw[draw=black, thin, solid] (v1) -- (v6);
  \draw[draw=black, thin, solid] (v2) -- (v6);
  \draw[draw=black, thin, solid] (v3) -- (v4);
\end{tikzpicture}
\end{minipage}
}

\newcommand{\EDExampleThreeR}{
\begin{minipage}{2.5cm}
\begin{tikzpicture}[scale=0.7, x=3ex, y = 3ex, baseline={([yshift=-5ex]current bounding box.center)}]
  \coordinate (v1) at (0.00, 2.00);
  \coordinate (v2) at (1.73, -1.00);
  \coordinate (v3) at (-1.73, -1.00);
  \coordinate[label=below:$0$] (v4) at (0.00, -4.00);
  \coordinate[label=above:$z$] (v5) at (-3.46, 2.00);
  \coordinate[label=above:$1$] (v6) at (3.46, 2.00);
  \path (v5) -- (v1) coordinate[midway] (v7);

  \draw[draw=black, fill=black, thin, solid] (v1) circle (1.42pt);
  \draw[draw=black, fill=black, thin, solid] (v2) circle (1.42pt);
  \draw[draw=black, fill=black, thin, solid] (v3) circle (1.42pt);
  \draw[draw=black, fill=black, thin, solid] (v4) circle (1.42pt);
  \draw[draw=black, fill=black, thin, solid] (v5) circle (1.42pt);
  \draw[draw=black, fill=black, thin, solid] (v6) circle (1.42pt);
  \draw[draw=black, fill=black, thin, solid] (v7) circle (1.42pt);
  \draw[draw=black, thin, solid] (v1) -- (v2);
  \draw[draw=black, thin, solid] (v2) -- (v3);
  \draw[draw=black, thin, solid] (v3) -- (v1);
  \draw[draw=black, thin, solid] (v3) -- (v5);
  \draw[draw=black, thin, solid] (v1) -- (v6);
  \draw[draw=black, thin, solid] (v2) -- (v6);
  \draw[draw=black, thin, solid] (v5) -- (v1);
  \draw[draw=black, thin, solid] (v3) -- (v4);
\end{tikzpicture}
\end{minipage}
}

\newcommand{\EDExampleFourR}{
\begin{minipage}{2.5cm}
\begin{tikzpicture}[scale=0.7, x=3ex, y = 3ex, baseline={([yshift=-5ex]current bounding box.center)}]
  \coordinate (v1) at (0.00, 2.00);
  \coordinate (v2) at (1.73, -1.00);
  \coordinate (v3) at (-1.73, -1.00);
  \coordinate[label=below:$0$] (v4) at (0.00, -4.00);
  \coordinate[label=above:$z$] (v5) at (-3.46, 2.00);
  \coordinate[label=above:$1$] (v6) at (3.46, 2.00);
  \path (v5) -- (v1) coordinate[midway] (v7);
  \path (v4) -- (v3) coordinate[midway] (v8);
  \draw[draw=black, fill=black, thin, solid] (v1) circle (1.42pt);
  \draw[draw=black, fill=black, thin, solid] (v2) circle (1.42pt);
  \draw[draw=black, fill=black, thin, solid] (v3) circle (1.42pt);
  \draw[draw=black, fill=black, thin, solid] (v4) circle (1.42pt);
  \draw[draw=black, fill=black, thin, solid] (v5) circle (1.42pt);
  \draw[draw=black, fill=black, thin, solid] (v6) circle (1.42pt);
  \draw[draw=black, fill=black, thin, solid] (v7) circle (1.42pt);
  \draw[draw=black, fill=black, thin, solid] (v8) circle (1.42pt);
  \draw[draw=black, thin, solid] (v1) -- (v2);
  \draw[draw=black, thin, solid] (v2) -- (v3);
  \draw[draw=black, thin, solid] (v3) -- (v1);
  \draw[draw=black, thin, solid] (v3) -- node[inner sep=0pt,left]{\tiny $1\!+\!\frac{1}{\lambda}$} (v5);
  \draw[draw=black, thin, solid] (v1) -- (v6);
  \draw[draw=black, thin, solid] (v2) -- (v6);
  \draw[draw=black, thin, solid] (v5) -- (v1);
  \draw[draw=black, thin, solid] (v3) -- (v4);
\end{tikzpicture}
\end{minipage}
}

\newcommand{\EDExampleFiveR}{
\begin{minipage}{2.5cm}
\begin{tikzpicture}[scale=0.7, x=3ex, y = 3ex, baseline={([yshift=-5ex]current bounding box.center)}]
  \coordinate (v1) at (0.00, 2.00);
  \coordinate (v2) at (1.73, -1.00);
  \coordinate (v3) at (-1.73, -1.00);
  \coordinate[label=below:$0$] (v4) at (0.00, -4.00);
  \coordinate[label=above:$z$] (v5) at (-3.46, 2.00);
  \coordinate[label=above:$1$] (v6) at (3.46, 2.00);
  \path (v5) -- (v1) coordinate[midway] (v7);
  \draw[draw=black, fill=black, thin, solid] (v1) circle (1.42pt);
  \draw[draw=black, fill=black, thin, solid] (v2) circle (1.42pt);
  \draw[draw=black, fill=black, thin, solid] (v3) circle (1.42pt);
  \draw[draw=black, fill=black, thin, solid] (v4) circle (1.42pt);
  \draw[draw=black, fill=black, thin, solid] (v5) circle (1.42pt);
  \draw[draw=black, fill=black, thin, solid] (v6) circle (1.42pt);
  \draw[draw=black, fill=black, thin, solid] (v7) circle (1.42pt);
  \draw[draw=black, thin, solid] (v1) -- (v2);
  \draw[draw=black, thin, solid] (v2) -- (v3);
  \draw[draw=black, thin, solid] (v3) -- (v1);
  \draw[draw=black, thin, solid] (v3) -- node[inner sep=0pt,left]{\tiny $1\!+\!\frac{1}{\lambda}$} (v5);
  \draw[draw=black, thin, solid] (v1) -- (v6);
  \draw[draw=black, thin, solid] (v2) -- (v6);
  \draw[draw=black, thin, solid] (v5) -- (v1);
  \draw[draw=black, thin, solid] (v3) -- (v4);
\end{tikzpicture}
\end{minipage}
}

\newcommand{\approxtrigX}{
\begin{minipage}{2.5cm}
\begin{tikzpicture}
    \coordinate (v);
    \def \rad {1}
    \coordinate[label=above:$1$] (v1) at ([shift=(60:\rad)]v);
    \coordinate[label=left:$0$] (v2) at ([shift=(180:\rad)]v);
    \coordinate[label=below:$z$] (v3) at ([shift=(300:\rad)]v);

    \draw[draw=black, fill=black, thin, solid] (v1) circle (1pt);
    \draw[draw=black, fill=black, thin, solid] (v2) circle (1pt);
    \draw[draw=black, fill=black, thin, solid] (v3) circle (1pt);
    \draw[draw=black, thin, solid] (v1) -- node[midway, above] {\footnotesize $\tfrac{x}{\lambda}$} (v2);
    \draw[draw=black, thin, solid] (v1) --node[midway, right] {\footnotesize $\tfrac{\epsilon}{\lambda}$} (v3);
    \draw[draw=black, thin, solid] (v3) --node[midway, below] {\footnotesize $\tfrac{1 - 2\epsilon}{\lambda}$} (v2);

\end{tikzpicture}
\end{minipage}
}

\newcommand{\approxtrig}{
\begin{minipage}{2.5cm}
\begin{tikzpicture}
    \coordinate (v);
    \def \rad {1}
    \coordinate[label=above:$1$] (v1) at ([shift=(60:\rad)]v);
    \coordinate[label=left:$0$] (v2) at ([shift=(180:\rad)]v);
    \coordinate[label=below:$z$] (v3) at ([shift=(300:\rad)]v);

    \draw[draw=black, fill=black, thin, solid] (v1) circle (1pt);
    \draw[draw=black, fill=black, thin, solid] (v2) circle (1pt);
    \draw[draw=black, fill=black, thin, solid] (v3) circle (1pt);
    \draw[draw=black, thin, solid] (v1) -- node[midway, above] {\footnotesize $\tfrac{1 - 2\epsilon}{\lambda}$} (v2);
    \draw[draw=black, thin, solid] (v1) --node[midway, right] {\footnotesize $\tfrac{\epsilon}{\lambda}$} (v3);
    \draw[draw=black, thin, solid] (v3) --node[midway, below] {\footnotesize $\tfrac{1 - 2\epsilon}{\lambda}$} (v2);

\end{tikzpicture}
\end{minipage}
}

\newcommand{\reroutBubble}{
\begin{minipage}{2.5cm}
\begin{tikzpicture}
    \coordinate (v);
    \def \rad {1}
    \coordinate[label=above:$1$] (v1) at ([shift=(60:\rad)]v);
    \coordinate[label=left:$0$] (v2) at ([shift=(180:\rad)]v);
    \coordinate[label=below:$z$] (v3) at ([shift=(300:\rad)]v);

    \draw[draw=black, fill=black, thin, solid] (v) circle (1pt);
    \draw[draw=black, fill=black, thin, solid] (v1) circle (1pt);
    \draw[draw=black, fill=black, thin, solid] (v2) circle (1pt);
    \draw[draw=black, fill=black, thin, solid] (v3) circle (1pt);
    \draw[draw=black, thin, solid, bend left=30] (v) to (v2);
    \draw[draw=black, thin, solid, bend right=30] (v) to (v2);
    \draw[draw=black, thin, solid] (v) -- (v1);
    \draw[draw=black, thin, solid] (v2) -- (v3);

\end{tikzpicture}
\end{minipage}
}

\newcommand{\chainGen}{
\begin{minipage}{2.5cm}
\begin{tikzpicture}[baseline={([yshift=-4ex]current bounding box.center)}]
    \coordinate (v);
    \def \rad {1}
    \coordinate (v1) at ([shift=(0:\rad)]v);
    \coordinate (v2) at ([shift=(180:\rad)]v);

    \draw[draw=black, fill=black, thin, solid] (v) circle (1pt);
    \draw[draw=black, fill=black, thin, solid] (v1) circle (1pt);
    \draw[draw=black, fill=black, thin, solid] (v2) circle (1pt);
    \draw[draw=black, thin, solid] (v) --node[midway, above] {\footnotesize $1$} (v1);
    \draw[draw=black, thin, solid] (v) --node[midway, above] {\footnotesize $v$} (v2);

\end{tikzpicture}
\end{minipage}
}

\newcommand{\reroutLine}{
\begin{minipage}{2.5cm}
\begin{tikzpicture}[baseline={([yshift=-4ex]current bounding box.center)}]
    \coordinate (v);
    \def \rad {1}
    \coordinate (v1) at ([shift=(0:\rad)]v);
    \coordinate (v2) at ([shift=(180:\rad)]v);

    \draw[draw=black, fill=black, thin, solid] (v1) circle (1pt);
    \draw[draw=black, fill=black, thin, solid] (v2) circle (1pt);
    \draw[draw=black, thin, solid] (v1) --node[midway, above] {\footnotesize $v- \tfrac{1}{\lambda}$} (v2);

\end{tikzpicture}
\end{minipage}
}

\newcommand{\chainTwo}{
\begin{minipage}{2.5cm}
\begin{tikzpicture}[baseline={([yshift=-4ex]current bounding box.center)}]
    \coordinate (v);
    \def \rad {1}
    \coordinate (v1) at ([shift=(0:\rad)]v);
    \coordinate (v2) at ([shift=(180:\rad)]v);

    \draw[draw=black, fill=black, thin, solid] (v) circle (1pt);
    \draw[draw=black, fill=black, thin, solid] (v1) circle (1pt);
    \draw[draw=black, fill=black, thin, solid] (v2) circle (1pt);
    \draw[draw=black, thin, solid] (v) --node[midway, above] {\footnotesize $1$} (v1);
    \draw[draw=black, thin, solid] (v) --node[midway, above] {\footnotesize $2$} (v2);

\end{tikzpicture}
\end{minipage}
}

\newcommand{\reroutLineTwo}{
\begin{minipage}{2.5cm}
\begin{tikzpicture}[baseline={([yshift=-4ex]current bounding box.center)}]
    \coordinate (v);
    \def \rad {1}
    \coordinate (v1) at ([shift=(0:\rad)]v);
    \coordinate (v2) at ([shift=(180:\rad)]v);

    \draw[draw=black, fill=black, thin, solid] (v1) circle (1pt);
    \draw[draw=black, fill=black, thin, solid] (v2) circle (1pt);
    \draw[draw=black, thin, solid] (v1) --node[midway, above] {\footnotesize $2- \tfrac{1}{\lambda}$} (v2);

\end{tikzpicture}
\end{minipage}
}

\newcommand{\reroutLineTwoeps}{
\begin{minipage}{2.5cm}
\begin{tikzpicture}[baseline={([yshift=-4ex]current bounding box.center)}]
    \coordinate (v);
    \def \rad {1}
    \coordinate (v1) at ([shift=(0:\rad)]v);
    \coordinate (v2) at ([shift=(180:\rad)]v);

    \draw[draw=black, fill=black, thin, solid] (v1) circle (1pt);
    \draw[draw=black, fill=black, thin, solid] (v2) circle (1pt);
    \draw[draw=black, thin, solid] (v1) --node[midway, above] {\footnotesize $\tfrac{1 - 2\epsilon}{\lambda}$} (v2);

\end{tikzpicture}
\end{minipage}
}

\newcommand{\coproductI}{
\begin{minipage}{2.5cm}
\begin{tikzpicture}
    \coordinate[label=above:$x$] (v);
    \def \rad {1}
    \coordinate[label=above:$1$] (v1) at ([shift=(60:\rad)]v);
    \coordinate[label=left:$0$] (v2) at ([shift=(180:\rad)]v);
    \coordinate[label=below:$z$] (v3) at ([shift=(300:\rad)]v);

    \draw[draw=black, fill=black, thin, solid] (v) circle (1pt);
    \draw[draw=black, fill=black, thin, solid] (v1) circle (1pt);
    \draw[draw=black, fill=black, thin, solid] (v2) circle (1pt);
    \draw[draw=black, fill=black, thin, solid] (v3) circle (1pt);
    \draw[draw=black, thin, solid, bend left=30] (v) to (v2);
    \draw[draw=black, thin, solid, bend right=30] (v) to (v2);
    \draw[draw=black, thin, solid] (v) -- (v1);
    \draw[draw=black, thin, solid] (v) -- (v3);

\end{tikzpicture}
\end{minipage}
}

\newcommand{\coproductII}{
\begin{minipage}{2.5cm}
\begin{tikzpicture}
    \coordinate[label=above:$x$] (v);
    \def \rad {1}
    \coordinate[label=above:$1$] (v1) at ([shift=(60:\rad)]v);
    \coordinate[label=left:$0$] (v2) at ([shift=(180:\rad)]v);
    \coordinate[label=below:$z$] (v3) at ([shift=(300:\rad)]v);

    \draw[draw=black, fill=black, thin, solid] (v) circle (1pt);
    \draw[draw=black, fill=black, thin, solid] (v1) circle (1pt);
    \draw[draw=black, fill=black, thin, solid] (v2) circle (1pt);
    \draw[draw=black, fill=black, thin, solid] (v3) circle (1pt);
    \draw[draw=black, thin, solid, bend left=30] (v) to (v2);
    \draw[draw=black, thin, solid, bend right=30] (v) to (v2);
    \draw[draw=black, thin, solid] (v) -- (v1);
    \draw[draw=black, thin, solid] (v2) -- (v3);

\end{tikzpicture}
\end{minipage}
}

\newcommand{\coproductBubble}{
\begin{minipage}{2cm}
\begin{tikzpicture}
    \coordinate[label=right:$x$] (v);
    \def \rad {1}
    \coordinate[label=left:$0$] (v2) at ([shift=(180:\rad)]v);

    \draw[draw=black, fill=black, thin, solid] (v) circle (1pt);
    \draw[draw=black, fill=black, thin, solid] (v2) circle (1pt);
    \draw[draw=black, thin, solid, bend left=30] (v) to (v2);
    \draw[draw=black, thin, solid, bend right=30] (v) to (v2);

\end{tikzpicture}
\end{minipage}
}

\newcommand{\coproductTrig}{
\begin{minipage}{2.5cm}
\begin{tikzpicture}
    \coordinate[label=above left:{$x=0$}] (v);
    \def \rad {1}
    \coordinate[label=above:$1$] (v1) at ([shift=(60:\rad)]v);
    \coordinate[label=below:$z$] (v3) at ([shift=(300:\rad)]v);

    \draw[draw=black, fill=black, thin, solid] (v) circle (1pt);
    \draw[draw=black, fill=black, thin, solid] (v1) circle (1pt);
    \draw[draw=black, fill=black, thin, solid] (v3) circle (1pt);
    \draw[draw=black, thin, solid] (v) -- (v1);
    \draw[draw=black, thin, solid] (v) -- (v3);

\end{tikzpicture}
\end{minipage}
}

\newcommand{\coproductIII}{
\begin{minipage}{2.5cm}
\begin{tikzpicture}
     \def \eps {0.5}

    \coordinate (v);
    \def \rad {1}
    \coordinate[label=above:$1$] (v1) at ([shift=(60:\rad)]v);
    \coordinate[label=left:$0$] (v2) at ([shift=(180:\rad)]v);
    \coordinate[label=below:$z$] (v3) at ([shift=(300:\rad)]v);

     \coordinate[label=above:$x_1$] (x1) at ($(v2)!\eps!(v1)$);
    \coordinate[label=below:$x_2$] (x2) at ($(v2)!\eps!(v3)$);

    \draw[draw=black, fill=black, thin, solid] (v1) circle (1pt);
    \draw[draw=black, fill=black, thin, solid] (v2) circle (1pt);
    \draw[draw=black, fill=black, thin, solid] (v3) circle (1pt);
    \draw[draw=black, fill=black, thin, solid] (x1) circle (1pt);
    \draw[draw=black, fill=black, thin, solid] (x2) circle (1pt);

    \draw[draw=black, thin, solid, bend left=30] (x1) to (x2);
    \draw[draw=black, thin, solid, bend right=30] (x1) to (x2);
    
    \draw[draw=black, thin, solid] (v2) -- (x1) -- (v1);
    \draw[draw=black, thin, solid] (v2) -- (x2) -- (v3);

\end{tikzpicture}
\end{minipage}
}

\newcommand{\coproductIV}{
\begin{minipage}{2.5cm}
\begin{tikzpicture}
     \def \eps {0.5}

    \coordinate (v);
    \def \rad {1}
    \coordinate[label=above:$1$] (v1) at ([shift=(60:\rad)]v);
    \coordinate[label=left:$0$] (v2) at ([shift=(180:\rad)]v);
    \coordinate[label=below:$z$] (v3) at ([shift=(300:\rad)]v);

     \coordinate[label=above:$x_1$] (x1) at ($(v2)!\eps!(v1)$);
    \coordinate[label=below:$x_2$] (x2) at ($(v2)!\eps!(v3)$);

    \draw[draw=black, fill=black, thin, solid] (v1) circle (1pt);
    \draw[draw=black, fill=black, thin, solid] (v2) circle (1pt);
    \draw[draw=black, fill=black, thin, solid] (v3) circle (1pt);
    \draw[draw=black, fill=black, thin, solid] (x1) circle (1pt);
    \draw[draw=black, fill=black, thin, solid] (x2) circle (1pt);

    \draw[draw=black, thin, solid, bend left=30] (x1) to (x2);
    \draw[draw=black, thin, solid, bend right=30] (x1) to (x2);
    
    \draw[draw=black, thin, solid] (v2) -- (x1) -- (v1);
    \draw[draw=black, thin, solid] (v2) -- (x2);
    \draw[draw=black, thin, solid] (x1) -- (v3);

\end{tikzpicture}
\end{minipage}
}

\newcommand{\coproductV}{
\begin{minipage}{2.5cm}
\begin{tikzpicture}
     \def \eps {0.5}

    \coordinate (v);
    \def \rad {1}
    \coordinate[label=above:$1$] (v1) at ([shift=(60:\rad)]v);
    \coordinate[label=left:$0$] (v2) at ([shift=(180:\rad)]v);
    \coordinate[label=below:$z$] (v3) at ([shift=(300:\rad)]v);

     \coordinate[label=above:$x_1$] (x1) at ($(v2)!\eps!(v1)$);
    \coordinate[label=below:$x_2$] (x2) at ($(v2)!\eps!(v3)$);

    \draw[draw=black, fill=black, thin, solid] (v1) circle (1pt);
    \draw[draw=black, fill=black, thin, solid] (v2) circle (1pt);
    \draw[draw=black, fill=black, thin, solid] (v3) circle (1pt);
    \draw[draw=black, fill=black, thin, solid] (x1) circle (1pt);
    \draw[draw=black, fill=black, thin, solid] (x2) circle (1pt);

    \draw[draw=black, thin, solid, bend left=30] (x1) to (x2);
    \draw[draw=black, thin, solid, bend right=30] (x1) to (x2);
    
    \draw[draw=black, thin, solid] (v2) -- (x1) -- (v1);
    \draw[draw=black, thin, solid] (v2) -- (x2);
    \draw[draw=black, thin, solid, bend right=45] (v2) to (v3);

\end{tikzpicture}
\end{minipage}
}

\newcommand{\coproductIIInox}{
\begin{minipage}{2.5cm}
\begin{tikzpicture}
     \def \eps {0.5}

    \coordinate (v);
    \def \rad {1}
    \coordinate[label=above:$1$] (v1) at ([shift=(60:\rad)]v);
    \coordinate[label=left:$0$] (v2) at ([shift=(180:\rad)]v);
    \coordinate[label=below:$z$] (v3) at ([shift=(300:\rad)]v);

     \coordinate (x1) at ($(v2)!\eps!(v1)$);
    \coordinate (x2) at ($(v2)!\eps!(v3)$);

    \draw[draw=black, fill=black, thin, solid] (v1) circle (1pt);
    \draw[draw=black, fill=black, thin, solid] (v2) circle (1pt);
    \draw[draw=black, fill=black, thin, solid] (v3) circle (1pt);
    \draw[draw=black, fill=black, thin, solid] (x1) circle (1pt);
    \draw[draw=black, fill=black, thin, solid] (x2) circle (1pt);

    \draw[draw=black, thin, solid, bend left=30] (x1) to (x2);
    \draw[draw=black, thin, solid, bend right=30] (x1) to (x2);
    
    \draw[draw=black, thin, solid] (v2) -- (x1) -- (v1);
    \draw[draw=black, thin, solid] (v2) -- (x2) -- (v3);

\end{tikzpicture}
\end{minipage}
}

\newcommand{\coproductIVnox}{
\begin{minipage}{2.5cm}
\begin{tikzpicture}
     \def \eps {0.5}

    \coordinate (v);
    \def \rad {1}
    \coordinate[label=above:$1$] (v1) at ([shift=(60:\rad)]v);
    \coordinate[label=left:$0$] (v2) at ([shift=(180:\rad)]v);
    \coordinate[label=below:$z$] (v3) at ([shift=(300:\rad)]v);

     \coordinate (x1) at ($(v2)!\eps!(v1)$);
    \coordinate (x2) at ($(v2)!\eps!(v3)$);

    \draw[draw=black, fill=black, thin, solid] (v1) circle (1pt);
    \draw[draw=black, fill=black, thin, solid] (v2) circle (1pt);
    \draw[draw=black, fill=black, thin, solid] (v3) circle (1pt);
    \draw[draw=black, fill=black, thin, solid] (x1) circle (1pt);
    \draw[draw=black, fill=black, thin, solid] (x2) circle (1pt);

    \draw[draw=black, thin, solid, bend left=30] (x1) to (x2);
    \draw[draw=black, thin, solid, bend right=30] (x1) to (x2);
    
    \draw[draw=black, thin, solid] (v2) -- (x1) -- (v1);
    \draw[draw=black, thin, solid] (v2) -- (x2);
    \draw[draw=black, thin, solid] (x1) -- (v3);

\end{tikzpicture}
\end{minipage}
}

\newcommand{\coproductVnox}{
\begin{minipage}{2.5cm}
\begin{tikzpicture}
     \def \eps {0.5}

    \coordinate (v);
    \def \rad {1}
    \coordinate[label=above:$1$] (v1) at ([shift=(60:\rad)]v);
    \coordinate[label=left:$0$] (v2) at ([shift=(180:\rad)]v);
    \coordinate[label=below:$z$] (v3) at ([shift=(300:\rad)]v);

     \coordinate (x1) at ($(v2)!\eps!(v1)$);
    \coordinate (x2) at ($(v2)!\eps!(v3)$);

    \draw[draw=black, fill=black, thin, solid] (v1) circle (1pt);
    \draw[draw=black, fill=black, thin, solid] (v2) circle (1pt);
    \draw[draw=black, fill=black, thin, solid] (v3) circle (1pt);
    \draw[draw=black, fill=black, thin, solid] (x1) circle (1pt);
    \draw[draw=black, fill=black, thin, solid] (x2) circle (1pt);

    \draw[draw=black, thin, solid, bend left=30] (x1) to (x2);
    \draw[draw=black, thin, solid, bend right=30] (x1) to (x2);
    
    \draw[draw=black, thin, solid] (v2) -- (x1) -- (v1);
    \draw[draw=black, thin, solid] (v2) -- (x2);
    \draw[draw=black, thin, solid, bend right=45] (v2) to (v3);

\end{tikzpicture}
\end{minipage}
}

\newcommand{\coproductCap}{
\begin{minipage}{1.8cm}
\begin{tikzpicture}
     \def \eps {0.5}

    \coordinate (v);
    \def \rad {1}
    \coordinate[label=left:$0$] (v2) at ([shift=(180:\rad)]v);

     \coordinate[label=above:$x_1$] (x1) at ($(v2)!\eps!(v1)$);
    \coordinate[label=below:$x_2$] (x2) at ($(v2)!\eps!(v3)$);

    \draw[draw=black, fill=black, thin, solid] (v2) circle (1pt);
    \draw[draw=black, fill=black, thin, solid] (x1) circle (1pt);
    \draw[draw=black, fill=black, thin, solid] (x2) circle (1pt);

    \draw[draw=black, thin, solid, bend left=30] (x1) to (x2);
    \draw[draw=black, thin, solid, bend right=30] (x1) to (x2);
    
    \draw[draw=black, thin, solid] (v2) -- (x1);
    \draw[draw=black, thin, solid] (v2) -- (x2);
    
\end{tikzpicture}
\end{minipage}
}

\newcommand{\coproductTrigII}{
\begin{minipage}{2.5cm}
\begin{tikzpicture}
    \coordinate[label=above left:{\footnotesize $x_1=x_2=0$}] (v);
    \def \rad {1}
    \coordinate[label=above:$1$] (v1) at ([shift=(60:\rad)]v);
    \coordinate[label=below:$z$] (v3) at ([shift=(300:\rad)]v);

    \draw[draw=black, fill=black, thin, solid] (v) circle (1pt);
    \draw[draw=black, fill=black, thin, solid] (v1) circle (1pt);
    \draw[draw=black, fill=black, thin, solid] (v3) circle (1pt);
    \draw[draw=black, thin, solid] (v) -- (v1);
    \draw[draw=black, thin, solid] (v) -- (v3);

\end{tikzpicture}
\end{minipage}
}

\newcommand{\coproductBubbleTrigII}{
\begin{minipage}{2.55cm}
\begin{tikzpicture}
    \coordinate[label=right:{\footnotesize$x_1=x_2$}] (v);
    \def \rad {1}
    \coordinate[label=above:$1$] (v1) at ([shift=(60:\rad)]v);
    \coordinate[label=left:$0$] (v2) at ([shift=(180:\rad)]v);
    \coordinate[label=below:$z$] (v3) at ([shift=(300:\rad)]v);

    \draw[draw=black, fill=black, thin, solid] (v) circle (1pt);
    \draw[draw=black, fill=black, thin, solid] (v1) circle (1pt);
    \draw[draw=black, fill=black, thin, solid] (v2) circle (1pt);
    \draw[draw=black, fill=black, thin, solid] (v3) circle (1pt);
    \draw[draw=black, thin, solid, bend left=30] (v) to (v2);
    \draw[draw=black, thin, solid, bend right=30] (v) to (v2);
    \draw[draw=black, thin, solid] (v) -- (v1);
    \draw[draw=black, thin, solid] (v) -- (v3);

\end{tikzpicture}
\end{minipage}
}

\newcommand{\coproductBubbleTrigRedII}{
\begin{minipage}{2.5cm}
\begin{tikzpicture}
    \coordinate[label=right:{\footnotesize$x_1=x_2$}] (v);
    \def \rad {1}
    \coordinate[label=above:$1$] (v1) at ([shift=(60:\rad)]v);
    \coordinate[label=left:$0$] (v2) at ([shift=(180:\rad)]v);
    \coordinate[label=below:$z$] (v3) at ([shift=(300:\rad)]v);

    \draw[draw=black, fill=black, thin, solid] (v) circle (1pt);
    \draw[draw=black, fill=black, thin, solid] (v1) circle (1pt);
    \draw[draw=black, fill=black, thin, solid] (v2) circle (1pt);
    \draw[draw=black, fill=black, thin, solid] (v3) circle (1pt);
    \draw[draw=black, thin, solid, bend left=30] (v) to (v2);
    \draw[draw=black, thin, solid, bend right=30] (v) to (v2);
    \draw[draw=black, thin, solid] (v) -- (v1);
    \draw[draw=black, thin, solid] (v2) -- (v3);

\end{tikzpicture}
\end{minipage}
}

\newcommand{\completionKfivetwoDimreg}{
\begin{minipage}{5cm}
\begin{tikzpicture}[scale=1.5, x=3ex, y = 3ex, baseline={([yshift=-8.0ex]current bounding box.center)}]
  \coordinate[label=above:$0$] (v1) at (-1.00, 1.00);
  \coordinate[label=above right:$1$] (v2) at (1.00, 1.00);
  \coordinate[label=above:$z$] (v3) at (3.00, 1.00);
  \coordinate[label=above:$\infty$] (v4) at (1.00, 3.00);
  \coordinate (v5) at (0.00, -1.00);
  \coordinate (v6) at (2.00, -1.00);
  \draw[draw=black, fill=black, thin, solid] (v1) circle (0.67pt);
  \draw[draw=black, fill=black, thin, solid] (v2) circle (0.67pt);
  \draw[draw=black, fill=black, thin, solid] (v3) circle (0.67pt);
  \draw[draw=black, fill=black, thin, solid] (v4) circle (0.67pt);
  \draw[draw=black, fill=black, thin, solid] (v5) circle (0.67pt);
  \draw[draw=black, fill=black, thin, solid] (v6) circle (0.67pt);
  \draw[draw=black, thin, solid] (v1) -- (v5);
  \draw[draw=black, thin, solid] (v1) -- (v6);
  \draw[draw=black, thin, solid] (v2) -- (v5);
  \draw[draw=black, thin, solid] (v2) -- (v6);
  \draw[draw=black, thin, solid] (v3) -- (v5);
  \draw[draw=black, thin, solid] (v3) -- (v6);
  \draw[draw=black, thin, solid] (v5) -- (v6);
  \draw[draw=black, thin, solid] (v1) -- node[inner sep=0pt,above] {\tiny{$\frac{-3\!+\!5\ep}{\lambda}$}} (v2);
  \draw[draw=black, thin, solid] (v1) -- node[inner sep=0pt,above left] {\tiny{$\frac{1\!-\!3\ep}{\lambda}$}} (v4);
  \draw[draw=black, thin, solid] (v2) -- node[inner sep=0pt,right] {\tiny{$\frac{1\!-\!3\ep}{\lambda}$}} (v4);
  \draw[draw=black, thin, solid] (v3) -- node[inner sep=0pt,above right] {\tiny{$-2$}} (v4);
  \draw[draw=black, thin, solid, bend left=90, looseness=1.8] (v5) to node[inner sep=0pt,above left] {\tiny{$\frac{2\ep}{\lambda}$}} (v4);
  \draw[draw=black, thin, solid, bend right=90, looseness=1.8] (v6) to node[inner sep=0pt,above right] {\tiny{$\frac{2\ep}{\lambda}$}} (v4);

\end{tikzpicture}
\end{minipage}
}

\newcommand{\completionKfivetwoDimregExchanged}{
\begin{minipage}{5cm}
\begin{tikzpicture}[scale=1.5, x=3ex, y = 3ex, baseline={([yshift=-8.0ex]current bounding box.center)}]
  \coordinate[label=above:$1$] (v1) at (-1.00, 1.00);
  \coordinate[label=above right:$0$] (v2) at (1.00, 1.00);
  \coordinate[label=above:$\infty$] (v3) at (3.00, 1.00);
  \coordinate[label=above:$z$] (v4) at (1.00, 3.00);
  \coordinate (v5) at (0.00, -1.00);
  \coordinate (v6) at (2.00, -1.00);
  \draw[draw=black, fill=black, thin, solid] (v1) circle (0.67pt);
  \draw[draw=black, fill=black, thin, solid] (v2) circle (0.67pt);
  \draw[draw=black, fill=black, thin, solid] (v3) circle (0.67pt);
  \draw[draw=black, fill=black, thin, solid] (v4) circle (0.67pt);
  \draw[draw=black, fill=black, thin, solid] (v5) circle (0.67pt);
  \draw[draw=black, fill=black, thin, solid] (v6) circle (0.67pt);
  \draw[draw=black, thin, solid] (v1) -- (v5);
  \draw[draw=black, thin, solid] (v1) -- (v6);
  \draw[draw=black, thin, solid] (v2) -- (v5);
  \draw[draw=black, thin, solid] (v2) -- (v6);
  \draw[draw=black, thin, solid] (v3) -- (v5);
  \draw[draw=black, thin, solid] (v3) -- (v6);
  \draw[draw=black, thin, solid] (v5) -- (v6);
  \draw[draw=black, thin, solid] (v1) -- node[inner sep=0pt,above] {\tiny{$\frac{-3\!+\!5\ep}{\lambda}$}} (v2);
  \draw[draw=black, thin, solid] (v1) -- node[inner sep=0pt,above left] {\tiny{$\frac{1\!-\!3\ep}{\lambda}$}} (v4);
  \draw[draw=black, thin, solid] (v2) -- node[inner sep=0pt,right] {\tiny{$\frac{1\!-\!3\ep}{\lambda}$}} (v4);
  \draw[draw=black, thin, solid] (v3) -- node[inner sep=0pt,above right] {\tiny{$-2$}} (v4);
  \draw[draw=black, thin, solid, bend left=90, looseness=1.8] (v5) to node[inner sep=0pt,above left] {\tiny{$\frac{2\ep}{\lambda}$}} (v4);
  \draw[draw=black, thin, solid, bend right=90, looseness=1.8] (v6) to node[inner sep=0pt,above right] {\tiny{$\frac{2\ep}{\lambda}$}} (v4);

\end{tikzpicture}
\end{minipage}
}

\newcommand{\KfivetwoDimregDecompleted}{
\begin{minipage}{2cm}
\begin{tikzpicture}[scale=1.5, x=3ex, y = 3ex, baseline={([yshift=-8.0ex]current bounding box.center)}]
  \coordinate[label=above:$1$] (v1) at (0.00, 1.00);
  \coordinate[label=above right:$0$] (v2) at (2.00, 1.00);
  \coordinate[label=below:$z$] (v4) at (1.00, -3.00);
  \coordinate (v5) at (0.00, -1.00);
  \coordinate (v6) at (2.00, -1.00);
  \draw[draw=black, fill=black, thin, solid] (v1) circle (0.67pt);
  \draw[draw=black, fill=black, thin, solid] (v2) circle (0.67pt);
  \draw[draw=black, fill=black, thin, solid] (v4) circle (0.67pt);
  \draw[draw=black, fill=black, thin, solid] (v5) circle (0.67pt);
  \draw[draw=black, fill=black, thin, solid] (v6) circle (0.67pt);
  \draw[draw=black, thin, solid] (v1) -- (v5);
  \draw[draw=black, thin, solid] (v1) -- (v6);
  \draw[draw=black, thin, solid] (v2) -- (v5);
  \draw[draw=black, thin, solid] (v2) -- (v6);
  \draw[draw=black, thin, solid] (v5) -- (v6);
  \draw[draw=black, thin, solid] (v5) to node[inner sep=0pt,below left] {\tiny{$\frac{2\ep}{\lambda}$}} (v4);
  \draw[draw=black, thin, solid] (v6) to node[inner sep=0pt,below right] {\tiny{$\frac{2\ep}{\lambda}$}} (v4);

\end{tikzpicture}
\end{minipage}
}

\newcommand{\KfivetwoDimregDecompletedNo}{
\begin{minipage}{2cm}
\begin{tikzpicture}[scale=1.5, x=3ex, y = 3ex, baseline={([yshift=-8.0ex]current bounding box.center)}]
  \coordinate[label=above:$1$] (v1) at (0.00, 1.00);
  \coordinate[label=above right:$0$] (v2) at (2.00, 1.00);
  \coordinate[label=below:$z$] (v4) at (1.00, -3.00);
  \coordinate (v5) at (0.00, -1.00);
  \coordinate (v6) at (2.00, -1.00);
  \draw[draw=black, fill=black, thin, solid] (v1) circle (0.67pt);
  \draw[draw=black, fill=black, thin, solid] (v2) circle (0.67pt);
  \draw[draw=black, fill=black, thin, solid] (v4) circle (0.67pt);
  \draw[draw=black, fill=black, thin, solid] (v5) circle (0.67pt);
  \draw[draw=black, fill=black, thin, solid] (v6) circle (0.67pt);
  \draw[draw=black, thin, solid] (v1) -- (v5);
  \draw[draw=black, thin, solid] (v1) -- (v6);
  \draw[draw=black, thin, solid] (v2) -- (v5);
  \draw[draw=black, thin, solid] (v2) -- (v6);
  \draw[draw=black, thin, solid] (v5) -- (v6);

\end{tikzpicture}
\end{minipage}
}

\newcommand{\KfivetwoDimregDecompletedLeft}{
\begin{minipage}{2cm}
\begin{tikzpicture}[scale=1.5, x=3ex, y = 3ex, baseline={([yshift=-8.0ex]current bounding box.center)}]
  \coordinate[label=above:$1$] (v1) at (0.00, 1.00);
  \coordinate[label=above right:$0$] (v2) at (2.00, 1.00);
  \coordinate[label=below:$z$] (v4) at (1.00, -3.00);
  \coordinate (v5) at (0.00, -1.00);
  \coordinate (v6) at (2.00, -1.00);
  \draw[draw=black, fill=black, thin, solid] (v1) circle (0.67pt);
  \draw[draw=black, fill=black, thin, solid] (v2) circle (0.67pt);
  \draw[draw=black, fill=black, thin, solid] (v4) circle (0.67pt);
  \draw[draw=black, fill=black, thin, solid] (v5) circle (0.67pt);
  \draw[draw=black, fill=black, thin, solid] (v6) circle (0.67pt);
  \draw[draw=black, thin, solid] (v1) -- (v5);
  \draw[draw=black, thin, solid] (v1) -- (v6);
  \draw[draw=black, thin, solid] (v2) -- (v5);
  \draw[draw=black, thin, solid] (v2) -- (v6);
  \draw[draw=black, thin, solid] (v5) -- (v6);
  \draw[draw=black, thin, solid] (v5) to node[inner sep=0pt,below left] {\tiny{$-\frac{\ep}{\lambda}$}} (v4);

\end{tikzpicture}
\end{minipage}
}

\newcommand{\KfivetwoDimregDecompletedRight}{
\begin{minipage}{2cm}
\begin{tikzpicture}[scale=1.5, x=3ex, y = 3ex, baseline={([yshift=-8.0ex]current bounding box.center)}]
  \coordinate[label=above:$1$] (v1) at (0.00, 1.00);
  \coordinate[label=above right:$0$] (v2) at (2.00, 1.00);
  \coordinate[label=below:$z$] (v4) at (1.00, -3.00);
  \coordinate (v5) at (0.00, -1.00);
  \coordinate (v6) at (2.00, -1.00);
  \draw[draw=black, fill=black, thin, solid] (v1) circle (0.05);
  \draw[draw=black, fill=black, thin, solid] (v2) circle (0.05);
  \draw[draw=black, fill=black, thin, solid] (v4) circle (0.05);
  \draw[draw=black, fill=black, thin, solid] (v5) circle (0.05);
  \draw[draw=black, fill=black, thin, solid] (v6) circle (0.05);
  \draw[draw=black, thin, solid] (v1) -- (v5);
  \draw[draw=black, thin, solid] (v1) -- (v6);
  \draw[draw=black, thin, solid] (v2) -- (v5);
  \draw[draw=black, thin, solid] (v2) -- (v6);
  \draw[draw=black, thin, solid] (v5) -- (v6);
  \draw[draw=black, thin, solid] (v6) to node[inner sep=0pt,below right] {\tiny{$-\frac{\ep}{\lambda}$}} (v4);

\end{tikzpicture}
\end{minipage}
}

\newcommand{\KfivetwoDimregDecompletedLeftDecomp}{
\begin{minipage}{2cm}
\begin{tikzpicture}[scale=1.5, x=3ex, y = 3ex, baseline={([yshift=-8.0ex]current bounding box.center)}]
  \coordinate[label=above:$1$] (v1) at (0.00, 1.00);
  \coordinate[label=above right:$0$] (v2) at (2.00, 1.00);
  \coordinate[label=below:$z$] (v4) at (1.00, -3.00);
  \coordinate (v5) at (0.00, -1.00);
  \coordinate (v6) at (2.00, -1.00);
  \path (v5) -- (v4) coordinate[midway] (v7);
  \draw[draw=black, fill=black, thin, solid] (v1) circle (0.67pt);
  \draw[draw=black, fill=black, thin, solid] (v2) circle (0.67pt);
  \draw[draw=black, fill=black, thin, solid] (v4) circle (0.67pt);
  \draw[draw=black, fill=black, thin, solid] (v5) circle (0.67pt);
  \draw[draw=black, fill=black, thin, solid] (v6) circle (0.67pt);
  \draw[draw=black, fill=black, thin, solid] (v7) circle (0.67pt);
  \draw[draw=black, thin, solid] (v1) -- (v5);
  \draw[draw=black, thin, solid] (v1) -- (v6);
  \draw[draw=black, thin, solid] (v2) -- (v5);
  \draw[draw=black, thin, solid] (v2) -- (v6);
  \draw[draw=black, thin, solid] (v5) -- (v6);
  \draw[draw=black, thin, solid] (v5) -- (v4);

\end{tikzpicture}
\end{minipage}
}

\newcommand{\KfivetwoDimregDecompletedRightDecomp}{
\begin{minipage}{2cm}
\begin{tikzpicture}[scale=1.5, x=3ex, y = 3ex, baseline={([yshift=-8.0ex]current bounding box.center)}]
  \coordinate[label=above:$1$] (v1) at (0.00, 1.00);
  \coordinate[label=above right:$0$] (v2) at (2.00, 1.00);
  \coordinate[label=below:$z$] (v4) at (1.00, -3.00);
  \coordinate (v5) at (0.00, -1.00);
  \coordinate (v6) at (2.00, -1.00);
  \path (v6) -- (v4) coordinate[midway] (v7);
  \draw[draw=black, fill=black, thin, solid] (v1) circle (0.67pt);
  \draw[draw=black, fill=black, thin, solid] (v2) circle (0.67pt);
  \draw[draw=black, fill=black, thin, solid] (v4) circle (0.67pt);
  \draw[draw=black, fill=black, thin, solid] (v5) circle (0.67pt);
  \draw[draw=black, fill=black, thin, solid] (v6) circle (0.67pt);
  \draw[draw=black, fill=black, thin, solid] (v7) circle (0.67pt);
  \draw[draw=black, thin, solid] (v1) -- (v5);
  \draw[draw=black, thin, solid] (v1) -- (v6);
  \draw[draw=black, thin, solid] (v2) -- (v5);
  \draw[draw=black, thin, solid] (v2) -- (v6);
  \draw[draw=black, thin, solid] (v5) -- (v6);
  \draw[draw=black, thin, solid] (v6) -- (v4);

\end{tikzpicture}
\end{minipage}
}

\newcommand{\productAPeriod}{
\begin{minipage}{1.8cm}
\begin{tikzpicture}[x=2.5ex, y=2.5ex, baseline={([yshift=-4.0ex]current bounding box.center)}]
  \coordinate (v0);
  \coordinate[below=2 of v0] (v1);
  \coordinate[below=2 of v1] (v2);
  \filldraw (v0) circle (1pt);
  \filldraw (v1) circle (1pt);
  \filldraw (v2) circle (1pt);
  \filldraw[fill=black!20] (v0) to [bend left=90] (v1) to [bend left=90] (v2) to [bend right=90, looseness=1.5] (v0);
  \filldraw[fill=black!20] (v0) to [bend right=90] (v1) to [bend right=90] (v2) to [bend left=90, looseness=1.5] (v0);
\end{tikzpicture}
\end{minipage}
}

\newcommand{\productBPeriod}{
\begin{minipage}{1.3cm}
\begin{tikzpicture}[x=2.5ex, y=2.5ex, baseline={([yshift=-4.0ex]current bounding box.center)}]
  \coordinate (v0);
  \coordinate[below=2 of v0] (v1);
  \coordinate[below=2 of v1] (v2);

  \filldraw (v0) circle (1pt);
  \filldraw (v1) circle (1pt);
  \filldraw (v2) circle (1pt);
  \filldraw[fill=black!20] (v0) to [bend right=90] (v1) to [bend right=90] (v2) to [bend left=90, looseness=1.5] (v0);
  \draw[thin] (v0)--(v2);
  \draw[thin, bend left=40] (v0) to (v2);

\end{tikzpicture}
\end{minipage}
}

\newcommand{\productCPeriod}{
\begin{minipage}{1.3cm}
\begin{tikzpicture}[x=2.5ex, y=2.5ex, baseline={([yshift=-4.0ex]current bounding box.center)}]
  \coordinate (v0);
  \coordinate[below=2 of v0] (v1);
  \coordinate[below=2 of v1] (v2);

  \filldraw (v0) circle (1pt);
  \filldraw (v1) circle (1pt);
  \filldraw (v2) circle (1pt);
  \filldraw[fill=black!20] (v0) to [bend left=90] (v1) to [bend left=90] (v2) to [bend right=90, looseness=1.5] (v0);
  \draw[thin] (v0)--(v2);
  \draw[thin, bend right=40] (v0) to (v2);
\end{tikzpicture}

\end{minipage}
}

\newcommand{\productBUnfinished}{
\begin{minipage}{1.3cm}
\begin{tikzpicture}[x=2.5ex, y=2.5ex, baseline={([yshift=-4.0ex]current bounding box.center)}]
  \coordinate (v0);
  \coordinate[below=2 of v0] (v1);
  \coordinate[below=2 of v1] (v2);

  \filldraw (v0) circle (1pt);
  \filldraw (v1) circle (1pt);
  \filldraw (v2) circle (1pt);
  \filldraw[fill=black!20] (v0) to [bend right=90] (v1) to [bend right=90] (v2) to [bend left=90, looseness=1.5] (v0);

\end{tikzpicture}

\end{minipage}
}

\newcommand{\productCUnfinished}{
\begin{minipage}{1.3cm}
\begin{tikzpicture}[x=2.5ex, y=2.5ex, baseline={([yshift=-4.0ex]current bounding box.center)}]
  \coordinate (v0);
  \coordinate[below=2 of v0] (v1);
  \coordinate[below=2 of v1] (v2);

  \filldraw (v0) circle (1pt);
  \filldraw (v1) circle (1pt);
  \filldraw (v2) circle (1pt);
  \filldraw[fill=black!20] (v0) to [bend left=90] (v1) to [bend left=90] (v2) to [bend right=90, looseness=1.5] (v0);

\end{tikzpicture}

\end{minipage}
}

\newcommand{\boxMultiple}{
\begin{minipage}{1.5cm}
\begin{tikzpicture}
    \coordinate[label=left:$z$] (v);
    \def \rad {1}
    \coordinate[label=right:$x_1$] (v1) at ([shift=(45:\rad)]v);
    \coordinate[label=right:$x_n$] (v2) at ([shift=(315:\rad)]v);

    \draw[draw=black, fill=black, thin, solid] (v) circle (1pt);
    \draw[draw=black, fill=black, thin, solid] (v1) circle (1pt);
    \draw[draw=black, fill=black, thin, solid] (v2) circle (1pt);
    \draw[draw=black, thin, solid] (v) -- (v1);
    \draw[draw=black, thin, solid] (v) -- (v2);

    \path (v) -- (v1) coordinate[pos=0.8] (c1);
    \path (v) -- (v2) coordinate[pos=0.8] (c2);
    \draw[draw=black, thin, dotted, bend left=30] (c1) to (c2);

\end{tikzpicture}
\end{minipage}
}

\newcommand{\boxMultipleOne}{
\begin{minipage}{2cm}
\begin{tikzpicture}
    \coordinate[label=left:$z$] (v);
    \def \rad {1}
    \coordinate[label=right:$x_1$] (v1) at ([shift=(45:\rad)]v);
    \coordinate[label=right:$x_n$] (v2) at ([shift=(315:\rad)]v);
    \coordinate[label=right:$x_i$] (v3) at ([shift=(0:\rad)]v);

    \draw[draw=black, fill=black, thin, solid] (v) circle (1pt);
    \draw[draw=black, fill=black, thin, solid] (v1) circle (1pt);
    \draw[draw=black, fill=black, thin, solid] (v2) circle (1pt);
    \draw[draw=black, fill=black, thin, solid] (v3) circle (1pt);
    \draw[draw=black, thin, solid] (v) -- (v1);
    \draw[draw=black, thin, solid] (v) -- (v2);
    \draw[draw=black, thin, solid] (v) -- node[inner sep=0pt,above,xshift=2pt] {\tiny{$\nu_i + \frac{1}{\lambda}$}}(v3);

    \path (v) -- (v1) coordinate[pos=0.8] (c1);
    \path (v) -- (v2) coordinate[pos=0.8] (c2);
    \draw[draw=black, thin, dotted, bend left=30] (c1) to (c2);

\end{tikzpicture}
\end{minipage}
}

\newcommand{\boxMultipleTwo}{
\begin{minipage}{2.2cm}
\begin{tikzpicture}
    \coordinate[label=left:$z$] (v);
    \def \rad {1.4}
    \coordinate[label=right:$x_1$] (v1) at ([shift=(45:\rad)]v);
    \coordinate[label=right:$x_n$] (v2) at ([shift=(315:\rad)]v);
    \coordinate[label=right:$x_i$] (v3) at ([shift=(20:\rad)]v);
    \coordinate[label=right:$x_j$] (v4) at ([shift=(340:\rad)]v);

    \draw[draw=black, fill=black, thin, solid] (v) circle (1pt);
    \draw[draw=black, fill=black, thin, solid] (v1) circle (1pt);
    \draw[draw=black, fill=black, thin, solid] (v2) circle (1pt);
    \draw[draw=black, fill=black, thin, solid] (v3) circle (1pt);
    \draw[draw=black, fill=black, thin, solid] (v4) circle (1pt);
    \draw[draw=black, thin, solid] (v) -- (v1);
    \draw[draw=black, thin, solid] (v) -- (v2);

    \path (v) -- (v1) coordinate[pos=0.8] (c1);
    \path (v) -- (v2) coordinate[pos=0.8] (c2);
    \draw[draw=black, thin, dotted, bend left=30] (c1) to (c2);
    \draw[draw=black, thin, solid] (v) -- node[inner sep=0pt,above,xshift=2pt,sloped] {\tiny{$\nu_i + \frac{1}{\lambda}$}}(v3);
    \draw[draw=black, thin, solid] (v) -- node[inner sep=0pt,above,xshift=2pt,sloped] {\tiny{$\nu_j + \frac{1}{\lambda}$}}(v4);
    \draw[draw=black, thin, solid] (v3) -- node[inner sep=0pt,right] {\tiny{$-\frac{1}{\lambda}$}}(v4);

\end{tikzpicture}
\end{minipage}
}

\newcommand{\boxMultipleOneAttached}{
\begin{minipage}{2cm}
\begin{tikzpicture}
    \coordinate (v);
    \def \rad {1}
    \coordinate[label=right:$x_1$] (v1) at ([shift=(45:\rad)]v);
    \coordinate[label=right:$x_n$] (v2) at ([shift=(315:\rad)]v);
    \coordinate[label=right:$x_i$] (v3) at ([shift=(0:\rad)]v);
    \coordinate[label=left:$z$] (v4) at ([shift=(180:\rad)]v);

    \draw[draw=black, fill=black, thin, solid] (v) circle (1pt);
    \draw[draw=black, fill=black, thin, solid] (v1) circle (1pt);
    \draw[draw=black, fill=black, thin, solid] (v2) circle (1pt);
    \draw[draw=black, fill=black, thin, solid] (v3) circle (1pt);
    \draw[draw=black, fill=black, thin, solid] (v4) circle (1pt);
    \draw[draw=black, thin, solid] (v) -- (v1);
    \draw[draw=black, thin, solid] (v) -- (v2);
    \draw[draw=black, thin, solid] (v4) -- node[inner sep=1pt,above] {\tiny{$1$}}(v);
    \draw[draw=black, thin, solid] (v) -- node[inner sep=0pt,above,xshift=2pt] {\tiny{$\nu_i + \frac{1}{\lambda}$}}(v3);

    \path (v) -- (v1) coordinate[pos=0.8] (c1);
    \path (v) -- (v2) coordinate[pos=0.8] (c2);
    \draw[draw=black, thin, dotted, bend left=30] (c1) to (c2);

\end{tikzpicture}
\end{minipage}
}

\newcommand{\boxMultipleTwoAttached}{
\begin{minipage}{3.7cm}
\begin{tikzpicture}
    \coordinate (v);
    \def \rad {1.4}
    \coordinate[label=right:$x_1$] (v1) at ([shift=(45:\rad)]v);
    \coordinate[label=right:$x_n$] (v2) at ([shift=(315:\rad)]v);
    \coordinate[label=right:$x_i$] (v3) at ([shift=(20:\rad)]v);
    \coordinate[label=right:$x_j$] (v4) at ([shift=(340:\rad)]v);
    \coordinate[label=left:$z$] (v5) at ([shift=(180:\rad)]v);

    \draw[draw=black, fill=black, thin, solid] (v) circle (1pt);
    \draw[draw=black, fill=black, thin, solid] (v1) circle (1pt);
    \draw[draw=black, fill=black, thin, solid] (v2) circle (1pt);
    \draw[draw=black, fill=black, thin, solid] (v3) circle (1pt);
    \draw[draw=black, fill=black, thin, solid] (v4) circle (1pt);
    \draw[draw=black, fill=black, thin, solid] (v5) circle (1pt);
    \draw[draw=black, thin, solid] (v) -- (v1);
    \draw[draw=black, thin, solid] (v) -- (v2);

    \path (v) -- (v1) coordinate[pos=0.8] (c1);
    \path (v) -- (v2) coordinate[pos=0.8] (c2);
    \draw[draw=black, thin, dotted, bend left=30] (c1) to (c2);
    \draw[draw=black, thin, solid] (v) -- node[inner sep=0pt,above,xshift=2pt,sloped] {\tiny{$\nu_i + \frac{1}{\lambda}$}}(v3);
    \draw[draw=black, thin, solid] (v) -- node[inner sep=0pt,above,xshift=2pt,sloped] {\tiny{$\nu_j + \frac{1}{\lambda}$}}(v4);
    \draw[draw=black, thin, solid] (v3) -- node[inner sep=0pt,right] {\tiny{$-\frac{1}{\lambda}$}}(v4);
    \draw[draw=black, thin, solid] (v5) -- node[inner sep=1pt,above] {\tiny{$1$}}(v);

\end{tikzpicture}
\end{minipage}
}

\newcommand{\KfivetwoDimregDecompletedExtOne}{
\begin{minipage}{2cm}
\begin{tikzpicture}[scale=1.5, x=3ex, y = 3ex, baseline={([yshift=-8.0ex]current bounding box.center)}]
  \coordinate[label=above:$1$] (v1) at (0.00, 1.00);
  \coordinate[label=above right:$0$] (v2) at (2.00, 1.00);
  \coordinate (v4) at (1.00, -2.00);
  \coordinate (v5) at (0.00, -1.00);
  \coordinate (v6) at (2.00, -1.00);
  \coordinate[label=below:$z$] (v7) at (1.00, -3.00);
  \draw[draw=black, fill=black, thin, solid] (v1) circle (0.67pt);
  \draw[draw=black, fill=black, thin, solid] (v2) circle (0.67pt);
  \draw[draw=black, fill=black, thin, solid] (v4) circle (0.67pt);
  \draw[draw=black, fill=black, thin, solid] (v5) circle (0.67pt);
  \draw[draw=black, fill=black, thin, solid] (v6) circle (0.67pt);
  \draw[draw=black, fill=black, thin, solid] (v7) circle (0.67pt);
  \draw[draw=black, thin, solid] (v1) -- (v5);
  \draw[draw=black, thin, solid] (v1) -- (v6);
  \draw[draw=black, thin, solid] (v2) -- (v5);
  \draw[draw=black, thin, solid] (v2) -- (v6);
  \draw[draw=black, thin, solid] (v5) -- (v6);
  \draw[draw=black, thin, solid] (v7) -- (v4);
  \draw[draw=black, thin, solid] (v5) to node[inner sep=0pt,below left] {\tiny{$\frac{1 + 2\ep}{\lambda}$}} (v4);
  \draw[draw=black, thin, solid] (v6) to node[inner sep=0pt,below right] {\tiny{$\frac{2\ep}{\lambda}$}} (v4);

\end{tikzpicture}
\end{minipage}
}

\newcommand{\KfivetwoDimregDecompletedExtTwo}{
\begin{minipage}{2cm}
\begin{tikzpicture}[scale=1.5, x=3ex, y = 3ex, baseline={([yshift=-8.0ex]current bounding box.center)}]
  \coordinate[label=above:$1$] (v1) at (0.00, 1.00);
  \coordinate[label=above right:$0$] (v2) at (2.00, 1.00);
  \coordinate (v4) at (1.00, -2.00);
  \coordinate (v5) at (0.00, -1.00);
  \coordinate (v6) at (2.00, -1.00);
  \coordinate[label=below:$z$] (v7) at (1.00, -3.00);
  \draw[draw=black, fill=black, thin, solid] (v1) circle (0.67pt);
  \draw[draw=black, fill=black, thin, solid] (v2) circle (0.67pt);
  \draw[draw=black, fill=black, thin, solid] (v4) circle (0.67pt);
  \draw[draw=black, fill=black, thin, solid] (v5) circle (0.67pt);
  \draw[draw=black, fill=black, thin, solid] (v6) circle (0.67pt);
  \draw[draw=black, fill=black, thin, solid] (v7) circle (0.67pt);
  \draw[draw=black, thin, solid] (v1) -- (v5);
  \draw[draw=black, thin, solid] (v1) -- (v6);
  \draw[draw=black, thin, solid] (v2) -- (v5);
  \draw[draw=black, thin, solid] (v2) -- (v6);
  \draw[draw=black, thin, solid] (v7) -- (v4);
  \draw[draw=black, thin, solid] (v5) -- node[inner sep=0pt,above] {\tiny{$-\!\frac{\ep}{\lambda}$}} (v6);
  \draw[draw=black, thin, solid] (v5) to node[inner sep=0pt,below left] {\tiny{$\frac{1 + 2\ep}{\lambda}$}} (v4);
  \draw[draw=black, thin, solid] (v6) to node[inner sep=0pt,below right] {\tiny{$\frac{1 + 2\ep}{\lambda}$}} (v4);

\end{tikzpicture}
\end{minipage}
}

\newcommand{\KfivetwoDimregDecompletedExtThree}{
\begin{minipage}{2cm}
\begin{tikzpicture}[scale=1.5, x=3ex, y = 3ex, baseline={([yshift=-8.0ex]current bounding box.center)}]
  \coordinate[label=above:$1$] (v1) at (0.00, 1.00);
  \coordinate[label=above right:$0$] (v2) at (2.00, 1.00);
  \coordinate (v5) at (0.00, -1.00);
  \coordinate (v6) at (2.00, -1.00);
  \coordinate[label=below:$z$] (v4) at (1.00, -2.00);
  \draw[draw=black, fill=black, thin, solid] (v1) circle (0.67pt);
  \draw[draw=black, fill=black, thin, solid] (v2) circle (0.67pt);
  \draw[draw=black, fill=black, thin, solid] (v4) circle (0.67pt);
  \draw[draw=black, fill=black, thin, solid] (v5) circle (0.67pt);
  \draw[draw=black, fill=black, thin, solid] (v6) circle (0.67pt);
  \draw[draw=black, thin, solid] (v1) -- (v5);
  \draw[draw=black, thin, solid] (v1) -- (v6);
  \draw[draw=black, thin, solid] (v2) -- (v5);
  \draw[draw=black, thin, solid] (v2) -- (v6);
  \draw[draw=black, thin, solid] (v5) -- (v6);
  \draw[draw=black, thin, solid] (v5) to node[inner sep=0pt,below left] {\tiny{$\frac{1 + 2\ep}{\lambda}$}} (v4);
  \draw[draw=black, thin, solid] (v6) to node[inner sep=0pt,below right] {\tiny{$\frac{2\ep}{\lambda}$}} (v4);

\end{tikzpicture}
\end{minipage}
}

\newcommand{\threestarMunchhausen}{
\begin{minipage}{2.2cm}
\begin{tikzpicture}
    \coordinate (v);
    \def \rad {1}
    \coordinate[label=above:$1$] (v1) at ([shift=(60:\rad)]v);
    \coordinate[label=left:$0$] (v2) at ([shift=(180:\rad)]v);
    \coordinate[label=below:$z$] (v3) at ([shift=(300:\rad)]v);

    \draw[draw=black, fill=black, thin, solid] (v) circle (1pt);
    \draw[draw=black, fill=black, thin, solid] (v1) circle (1pt);
    \draw[draw=black, fill=black, thin, solid] (v2) circle (1pt);
    \draw[draw=black, fill=black, thin, solid] (v3) circle (1pt);
    \draw[draw=black, thin, solid] (v) -- node[inner sep=0pt,above] {\tiny $\frac{2 + a_0\ep}{\lambda}$} (v2);
    \draw[draw=black, thin, solid] (v) -- node[inner sep=0pt,below right] {\tiny $\frac{1+a_1\ep}{\lambda}$}(v1);
    \draw[draw=black, thin, solid] (v) -- node[inner sep=0pt,right] {\tiny $\frac{a_2\ep}{\lambda}$}(v3);

\end{tikzpicture}
\end{minipage}
}

\newcommand{\threestarMunchhausenExt}{
\begin{minipage}{2.2cm}
\begin{tikzpicture}
    \coordinate (v);
    \def \rad {1}
    \coordinate[label=above:$1$] (v1) at ([shift=(60:\rad)]v);
    \coordinate[label=left:$0$] (v2) at ([shift=(180:\rad)]v);
    \coordinate[label=below:$z$] (v3) at ([shift=(300:\rad)]v);

    \draw[draw=black, fill=black, thin, solid] (v) circle (1pt);
    \draw[draw=black, fill=black, thin, solid] (v1) circle (1pt);
    \draw[draw=black, fill=black, thin, solid] (v2) circle (1pt);
    \draw[draw=black, fill=black, thin, solid] (v3) circle (1pt);
    \draw[draw=black, thin, solid] (v) -- node[inner sep=0pt,above] {\tiny $\frac{2 + a_0\ep}{\lambda}$} (v2);
    \draw[draw=black, thin, solid] (v) -- node[inner sep=0pt,below right] {\tiny $\frac{1+a_1\ep}{\lambda}$}(v1);
    \draw[draw=black, thin, solid] (v) -- node[inner sep=0pt,right] {\tiny $\frac{1+a_2\ep}{\lambda}$}(v3);

\end{tikzpicture}
\end{minipage}
}

\newcommand{\threestarMunchhausenCompleted}{
\begin{minipage}{3.5cm}
\begin{tikzpicture}
    \coordinate (v) ;
    \def \rad {1}
    \coordinate[label=above:$1$] (v1) at ([shift=(90:\rad)]v);
    \coordinate[label=left:$0$] (v2) at ([shift=(180:\rad)]v);
    \coordinate[label=below:$\infty$] (v3) at ([shift=(270:\rad)]v);
    \coordinate[label=right:$z$] (v4) at ([shift=(0:\rad)]v);

    \draw[thin] (v) -- (v4);
    \draw[draw=black, thin, solid] (v) -- node[inner sep=0pt,above, xshift=4pt] {\tiny $\frac{2 + a_0\ep}{\lambda}$} (v2);
    \draw[draw=black, thin, solid] (v) -- node[inner sep=0pt,right] {\tiny $\frac{1+a_1\ep}{\lambda}$}(v1);
    \draw[draw=black, thin, solid] (v) -- node[inner sep=0pt,above] {\tiny $\frac{1+a_2\ep}{\lambda}$}(v4);
    \draw[draw=black, thin, solid] (v) -- node[inner sep=0pt,right] {\tiny $\nu_{x\infty}$}(v3);

    \draw[thin] (v1) -- node[inner sep=0pt,above left] {\tiny{$\nu_{01}$}} (v2);
    \draw[thin] (v3) -- node[inner sep=0pt,below right] {\tiny{$-\!\frac{1+a_2\ep}{\lambda}$}} (v4);
    \draw[thin] (v2) -- node[inner sep=0pt,below left] {\tiny{$\nu_{0\infty}$}} (v3);
    \draw[draw=black, thin, solid, bend right=90, looseness=2.5] (v1) to node[inner sep=0pt, pos=0.8, below left] {\tiny{$\nu_{1\infty}$}} (v3);

    \filldraw (v) circle (1pt);
    \filldraw (v1) circle (1pt);
    \filldraw (v2) circle (1pt);
    \filldraw (v3) circle (1pt);
    \filldraw (v4) circle (1pt);

\end{tikzpicture}
\end{minipage}
}

\newcommand{\threestarMunchhausenCompletedSwitched}{
\begin{minipage}{2.5cm}
\begin{tikzpicture}
    \coordinate (v) ;
    \def \rad {1}
    \coordinate[label=above:$0$] (v1) at ([shift=(90:\rad)]v);
    \coordinate[label=left:$1$] (v2) at ([shift=(180:\rad)]v);
    \coordinate[label=below:$z$] (v3) at ([shift=(270:\rad)]v);
    \coordinate[label=right:$\infty$] (v4) at ([shift=(0:\rad)]v);

    \draw[thin] (v) -- (v4);
    \draw[draw=black, thin, solid] (v) -- node[inner sep=0pt,above, xshift=4pt] {\tiny $\frac{2 + a_0\ep}{\lambda}$} (v2);
    \draw[draw=black, thin, solid] (v) -- node[inner sep=0pt,right] {\tiny $\frac{1+a_1\ep}{\lambda}$}(v1);
    \draw[draw=black, thin, solid] (v) -- node[inner sep=0pt,above] {\tiny $\frac{1+a_2\ep}{\lambda}$}(v4);
    \draw[draw=black, thin, solid] (v) -- node[inner sep=0pt,right] {\tiny $\nu_{x\infty}$}(v3);

    \draw[thin] (v1) -- node[inner sep=0pt,above left] {\tiny{$\nu_{01}$}} (v2);
    \draw[thin] (v3) -- node[inner sep=0pt,below right] {\tiny{$-\!\frac{1+a_2\ep}{\lambda}$}} (v4);
    \draw[thin] (v2) -- node[inner sep=0pt,below left] {\tiny{$\nu_{0\infty}$}} (v3);
    \draw[draw=black, thin, solid, bend right=90, looseness=2.5] (v1) to node[inner sep=0pt, pos=0.8, below left] {\tiny{$\nu_{1\infty}$}} (v3);

    \filldraw (v) circle (1pt);
    \filldraw (v1) circle (1pt);
    \filldraw (v2) circle (1pt);
    \filldraw (v3) circle (1pt);
    \filldraw (v4) circle (1pt);

\end{tikzpicture}
\end{minipage}
}

\newcommand{\threestarMunchhausenDecompleted}{
\begin{minipage}{3cm}
\begin{tikzpicture}
    \coordinate (v) ;
    \def \rad {1}
    \coordinate[label=above:$0$] (v1) at ([shift=(90:\rad)]v);
    \coordinate[label=left:$1$] (v2) at ([shift=(180:\rad)]v);
    \coordinate[label=below:$z$] (v3) at ([shift=(270:\rad)]v);

    \draw[draw=black, thin, solid] (v) -- node[inner sep=0pt,above] {\tiny $\frac{2 + a_0\ep}{\lambda}$} (v2);
    \draw[draw=black, thin, solid] (v) -- node[inner sep=0pt,right] {\tiny $\frac{1+a_1\ep}{\lambda}$}(v1);
    \draw[draw=black, thin, solid] (v) -- node[inner sep=0pt,right] {\tiny $\nu_{x\infty}$}(v3);

    \draw[thin] (v2) -- node[inner sep=0pt,below left] {\tiny{$\nu_{0\infty}$}} (v3);
    \draw[draw=black, thin, solid, bend left=90, looseness=2.5] (v1) to node[inner sep=0pt, pos=0.8, below right] {\tiny{$\nu_{1\infty}$}} (v3);

    \filldraw (v) circle (1pt);
    \filldraw (v1) circle (1pt);
    \filldraw (v2) circle (1pt);
    \filldraw (v3) circle (1pt);

\end{tikzpicture}
\end{minipage}
}

\newcommand{\threestarMunchhausenDecompletedTwo}{
\begin{minipage}{3cm}
\begin{tikzpicture}
    \coordinate (v) ;
    \def \rad {1}
    \coordinate[label=above:$0$] (v1) at ([shift=(90:\rad)]v);
    \coordinate[label=left:$1$] (v2) at ([shift=(180:\rad)]v);
    \coordinate[label=below:$z$] (v3) at ([shift=(270:\rad)]v);

    \draw[draw=black, thin, solid] (v) -- node[inner sep=0pt,above] {\tiny $\frac{2 + a_0\ep}{\lambda}$} (v2);
    \draw[draw=black, thin, solid] (v) -- node[inner sep=0pt,right] {\tiny $\frac{1+a_1\ep}{\lambda}$}(v1);
    \draw[draw=black, thin, solid] (v) -- node[inner sep=0pt,right] {\tiny $-\frac{(a_0+a_1+a_2+2)\ep}{\lambda}$}(v3);
    \filldraw (v) circle (1pt);
    \filldraw (v1) circle (1pt);
    \filldraw (v2) circle (1pt);
    \filldraw (v3) circle (1pt);

\end{tikzpicture}
\end{minipage}
}

\newcommand{\threestarMunchhausenDecompletedThree}{
\begin{minipage}{3cm}
\begin{tikzpicture}
    \coordinate (v) ;
    \def \rad {1}
    \coordinate[label=above:$1$] (v1) at ([shift=(90:\rad)]v);
    \coordinate[label=left:$0$] (v2) at ([shift=(180:\rad)]v);
    \coordinate[label=below:$1-z$] (v3) at ([shift=(270:\rad)]v);

    \draw[draw=black, thin, solid] (v) -- node[inner sep=0pt,above] {\tiny $\frac{2 + a_0\ep}{\lambda}$} (v2);
    \draw[draw=black, thin, solid] (v) -- node[inner sep=0pt,right] {\tiny $\frac{1+a_1\ep}{\lambda}$}(v1);
    \draw[draw=black, thin, solid] (v) -- node[inner sep=0pt,right] {\tiny $-\frac{(a_0+a_1+a_2+2)\ep}{\lambda}$}(v3);
    \filldraw (v) circle (1pt);
    \filldraw (v1) circle (1pt);
    \filldraw (v2) circle (1pt);
    \filldraw (v3) circle (1pt);

\end{tikzpicture}
\end{minipage}
}

\newcommand{\threestarMunchhausenDecompletedAttached}{
\begin{minipage}{3cm}
\begin{tikzpicture}
    \coordinate (v) ;
    \def \rad {1}
    \coordinate[label=above:$0$] (v1) at ([shift=(90:\rad)]v);
    \coordinate[label=left:$1$] (v2) at ([shift=(180:\rad)]v);
    \coordinate (v3) at ([shift=(270:\rad)]v);
    \coordinate[label=below:$z$] (v4) at ($(v3)+(-1, 0)$);

    \draw[draw=black, thin, solid] (v) -- node[inner sep=0pt,above] {\tiny $\frac{2 + a_0\ep}{\lambda}$} (v2);
    \draw[draw=black, thin, solid] (v) -- node[inner sep=0pt,right] {\tiny $\frac{1+a_1\ep}{\lambda}$}(v1);
    \draw[draw=black, thin, solid] (v) -- node[inner sep=0pt,right] {\tiny $\nu_{x\infty}$}(v3);
    \draw[draw=black, thin, solid] (v4) -- (v3);

    \draw[thin] (v2) -- node[inner sep=0pt,below left] {\tiny{$\nu_{0\infty}$}} (v3);
    \draw[draw=black, thin, solid, bend left=90, looseness=2.5] (v1) to node[inner sep=0pt, pos=0.8, below right] {\tiny{$\nu_{1\infty}$}} (v3);

    \filldraw (v) circle (1pt);
    \filldraw (v1) circle (1pt);
    \filldraw (v2) circle (1pt);
    \filldraw (v3) circle (1pt);
    \filldraw (v4) circle (1pt);

\end{tikzpicture}
\end{minipage}
}

\newcommand{\approxInvalidExample}{
\begin{minipage}{3cm}
\begin{tikzpicture}
    \coordinate (v1) at (0,0);
    \coordinate (v2) at (0,-1);
    \coordinate (v3) at (1, -2);
    \coordinate (v4) at (-1, -2);
    \coordinate (v5) at (2, -2);
    \coordinate (v6) at (-2, -2);

    \draw (0.4, -1) node{$v_3$};
    \draw (1, -2.3) node{$v_2$};
    \draw (-1, -2.3) node{$v_1$};
    
    \draw[draw=black, thin, solid, bend left=30] (v1) to (v2);
    \draw[draw=black, thin, solid, bend right=30] (v1) to (v2);
    \draw[draw=black, thin, solid, bend left=30] (v3) to (v4);
    \draw[draw=black, thin, solid, bend right=30] (v3) to (v4);
    \draw[draw=black, thin, solid] (v2) to (v3);
    \draw[draw=black, thin, solid] (v2) to (v4);
    \draw[draw=black, thin, solid] (v6) to (v4);
    \draw[draw=black, thin, solid] (v5) to (v3);

    \filldraw (v1) circle (1pt);
    \filldraw (v2) circle (1pt);
    \filldraw (v3) circle (1pt);
    \filldraw (v4) circle (1pt);
    \filldraw (v5) circle (1pt);
    \filldraw (v5) circle (1pt);
    \filldraw (v6) circle (1pt);

\end{tikzpicture}
\end{minipage}
}

\newcommand{\bubblethreestarone}{
\begin{minipage}{2.5cm}
\begin{tikzpicture}
    \coordinate (v);
    \def \rad {1}
    \coordinate[label=above:$1$] (v1) at ([shift=(60:\rad)]v);
    \coordinate[label=left:$0$] (v2) at ([shift=(180:\rad)]v);
    \coordinate[label=below:$z$] (v3) at ([shift=(300:\rad)]v);

    \draw[draw=black, fill=black, thin, solid] (v) circle (1pt);
    \draw[draw=black, fill=black, thin, solid] (v1) circle (1pt);
    \draw[draw=black, fill=black, thin, solid] (v2) circle (1pt);
    \draw[draw=black, fill=black, thin, solid] (v3) circle (1pt);
    \draw[draw=black, thin, solid] (v) to (v2);
    \draw[draw=black, thin, solid] (v) -- (v1);
    \draw[draw=black, thin, solid] (v) -- (v3);
    \draw[draw=black, thin, solid] (v2) -- (v3);

\end{tikzpicture}
\end{minipage}
}

\newcommand{\bubblethreestartwo}{
\begin{minipage}{2.5cm}
\begin{tikzpicture}
    \coordinate (v);
    \def \rad {1}
    \coordinate[label=above:$1$] (v1) at ([shift=(60:\rad)]v);
    \coordinate[label=left:$0$] (v2) at ([shift=(180:\rad)]v);
    \coordinate[label=below:$z$] (v3) at ([shift=(300:\rad)]v);

    \draw[draw=black, fill=black, thin, solid] (v) circle (1pt);
    \draw[draw=black, fill=black, thin, solid] (v1) circle (1pt);
    \draw[draw=black, fill=black, thin, solid] (v2) circle (1pt);
    \draw[draw=black, fill=black, thin, solid] (v3) circle (1pt);
    \draw[draw=black, thin, solid] (v) to (v2);
    \draw[draw=black, thin, solid] (v) -- (v1);
    \draw[draw=black, thin, solid] (v) -- (v3);
    \draw[draw=black, thin, solid] (v1) -- (v2);

\end{tikzpicture}
\end{minipage}
}
\newcommand{\bubblethreestarthree}{
\begin{minipage}{2.5cm}
\begin{tikzpicture}
    \coordinate (v);
    \def \rad {1}
    \coordinate[label=above:$1$] (v1) at ([shift=(60:\rad)]v);
    \coordinate[label=left:$0$] (v2) at ([shift=(180:\rad)]v);
    \coordinate[label=below:$z$] (v3) at ([shift=(300:\rad)]v);

    \draw[draw=black, fill=black, thin, solid] (v) circle (1pt);
    \draw[draw=black, fill=black, thin, solid] (v1) circle (1pt);
    \draw[draw=black, fill=black, thin, solid] (v2) circle (1pt);
    \draw[draw=black, fill=black, thin, solid] (v3) circle (1pt);
    \draw[draw=black, thin, solid] (v) to (v2);
    \draw[draw=black, thin, solid] (v) -- (v1);
    \draw[draw=black, thin, solid] (v) -- (v3);
    \draw[draw=black, thin, solid] (v1) -- (v2);
    \draw[draw=black, thin, solid] (v3) -- (v2);
    \draw[draw=black, thin, solid] (v1) -- node[inner sep=0pt, right] {\tiny $-1$} (v3);

\end{tikzpicture}
\end{minipage}
}

\newcommand{\bubblethreestarfour}{
\begin{minipage}{2.5cm}
\begin{tikzpicture}
    \coordinate (v);
    \def \rad {1}
    \coordinate[label=above:$1$] (v1) at ([shift=(60:\rad)]v);
    \coordinate[label=left:$0$] (v2) at ([shift=(180:\rad)]v);
    \coordinate[label=below:$z$] (v3) at ([shift=(300:\rad)]v);

    \draw[draw=black, fill=black, thin, solid] (v1) circle (1pt);
    \draw[draw=black, fill=black, thin, solid] (v2) circle (1pt);
    \draw[draw=black, fill=black, thin, solid] (v3) circle (1pt);
    \draw[draw=black, thin, solid] (v2) -- node[inner sep=0pt, below left] {\tiny ${1-\frac{\ep}{\lambda}}$} (v3);
    \draw[draw=black, thin, solid] (v2) -- (v1);

\end{tikzpicture}
\end{minipage}
}

\newcommand{\bubblethreestarfive}{
\begin{minipage}{2.5cm}
\begin{tikzpicture}
    \coordinate (v);
    \def \rad {1}
    \coordinate[label=above:$1$] (v1) at ([shift=(60:\rad)]v);
    \coordinate[label=left:$0$] (v2) at ([shift=(180:\rad)]v);
    \coordinate[label=below:$z$] (v3) at ([shift=(300:\rad)]v);

    \draw[draw=black, fill=black, thin, solid] (v1) circle (1pt);
    \draw[draw=black, fill=black, thin, solid] (v2) circle (1pt);
    \draw[draw=black, fill=black, thin, solid] (v3) circle (1pt);
    \draw[draw=black, thin, solid] (v2) -- node[inner sep=0pt, above left] {\tiny ${1-\frac{\ep}{\lambda}}$} (v1);
    \draw[draw=black, thin, solid] (v2) -- (v3);

\end{tikzpicture}
\end{minipage}
}

\newcommand{\bubblethreestarsix}{
\begin{minipage}{2.5cm}
\begin{tikzpicture}
    \coordinate (v);
    \def \rad {1}
    \coordinate[label=above:$1$] (v1) at ([shift=(60:\rad)]v);
    \coordinate[label=left:$0$] (v2) at ([shift=(180:\rad)]v);
    \coordinate[label=below:$z$] (v3) at ([shift=(300:\rad)]v);

    \draw[draw=black, fill=black, thin, solid] (v1) circle (1pt);
    \draw[draw=black, fill=black, thin, solid] (v2) circle (1pt);
    \draw[draw=black, fill=black, thin, solid] (v3) circle (1pt);
    \draw[draw=black, thin, solid] (v1) to (v2);
    \draw[draw=black, thin, solid] (v2) to (v3);
    \draw[draw=black, thin, solid] (v3) to (v1);
    \draw[draw=black, thin, solid] (v1) -- node[inner sep=0pt, right] {\tiny $-\frac{\ep}{\lambda}$} (v3);

\end{tikzpicture}
\end{minipage}
}

\newcommand{\approxInternalExample}{
\begin{minipage}{3cm}
\begin{tikzpicture}
    \coordinate[label=above:$0$] (v1) at (0,0);
    \coordinate (v2) at (-0.5,-1);
    \coordinate (v3) at (0.5,-1);
    \coordinate[label=below:$1$] (v4) at (-1, -2);
    \coordinate[label=below:$z$] (v5) at (1, -2);

    \draw[draw=black, thin, solid] (v1) to (v2);
    \draw[draw=black, thin, solid] (v1) to (v3);
    \draw[draw=black, thin, solid, bend left=30] (v3) to (v2);
    \draw[draw=black, thin, solid, bend right=30] (v3) to (v2);
    \draw[draw=black, thin, solid] (v2) to (v4);
    \draw[draw=black, thin, solid] (v1) to (v5);

    \filldraw (v1) circle (1pt);
    \filldraw (v2) circle (1pt);
    \filldraw (v3) circle (1pt);
    \filldraw (v4) circle (1pt);
    \filldraw (v5) circle (1pt);

\end{tikzpicture}
\end{minipage}
}

\newcommand{\approxInternalExampleReplaced}{
\begin{minipage}{3cm}
\begin{tikzpicture}
    \coordinate[label=above:$0$] (v1) at (0,0);
    \coordinate (v2) at (-0.5,-1);
    \coordinate (v3) at (0.5,-1);
    \coordinate[label=below:$1$] (v4) at (-1, -2);
    \coordinate[label=below:$z$] (v5) at (1, -2);

    \draw[draw=black, thin, solid] (v1) to node[inner sep=0pt, right] {\tiny {$\frac{2 - 3\ep}{\lambda}$}} (v3);
    \draw[draw=black, thin, solid] (v3) to (v5);
    \draw[draw=black, thin, solid] (v3) to node[inner sep=0pt, below right] {\tiny {$\frac{1 - 2\ep}{\lambda}$}} (v4);
    \draw[draw=black, thin, solid] (v1) to node[inner sep=0pt, above left] {\tiny {$\frac{\ep}{\lambda}$}} (v4);

    \filldraw (v1) circle (1pt);
    \filldraw (v3) circle (1pt);
    \filldraw (v4) circle (1pt);
    \filldraw (v5) circle (1pt);

\end{tikzpicture}
\end{minipage}
}

\newcommand{\completionKfivetwoP}{
\begin{minipage}{2.5cm}
\begin{tikzpicture}[
    x=3ex,
    y=3ex,
    yscale=-1,
    baseline={([yshift=-8.0ex]current bounding box.center)}
]
  \coordinate (v1) at (-1.00, 1.00);
  \coordinate (v2) at (1.00, 1.00);
  \coordinate (v3) at (3.00, 1.00);
  \coordinate (v4) at (1.00, 3.00);
  \coordinate (v5) at (0.00, -1.00);
  \coordinate (v6) at (2.00, -1.00);
  \coordinate (v7) at (-1.00, -1.00);
  \coordinate (v8) at (3.00, -1.00);

  \draw[draw=black, fill=black, thin, solid] (v1) circle (1pt);
  \draw[draw=black, fill=black, thin, solid] (v2) circle (1pt);
  \draw[draw=black, fill=black, thin, solid] (v3) circle (1pt);
  \draw[draw=black, fill=black, thin, solid] (v4) circle (1pt);
  \draw[draw=black, fill=black, thin, solid] (v5) circle (1pt);
  \draw[draw=black, fill=black, thin, solid] (v6) circle (1pt);

  \draw[draw=black, thin, solid] (v1) -- (v5);
  \draw[draw=black, thin, solid] (v1) -- (v6);
  \draw[draw=black, thin, solid] (v2) -- (v5);
  \draw[draw=black, thin, solid] (v2) -- (v6);
  \draw[draw=black, thin, solid] (v3) -- (v5);
  \draw[draw=black, thin, solid] (v3) -- (v6);
  \draw[draw=black, thin, solid] (v1) -- (v4);
  \draw[draw=black, thin, solid] (v2) -- (v4);
  \draw[draw=black, thin, solid] (v3) -- (v4);
  \draw[draw=black, thin, solid] (v5) -- (v7);
  \draw[draw=black, thin, solid] (v6) -- (v8);

\end{tikzpicture}
\end{minipage}
}

\newcommand{\Ksevenfactorized}{
\begin{minipage}{2.2cm}
\begin{tikzpicture}[x=3ex,y=3ex,
  baseline={([yshift=-7.5ex]current bounding box.center)}]

  \coordinate (C) at (-2.0, 0.0);
  \coordinate (D) at ( 2.0, 0.0);
  \coordinate (G) at ( 0.0, 0.0);
  \coordinate (C2) at (-2.0, -1.0);
  \coordinate (D2) at ( 2.0, -1.0);
  \coordinate (G2) at ( 0.0, -1.0);

  \coordinate (A) at (-1.0, 2.0);
  \coordinate (B) at ( 1.0, 2.0);

  \coordinate (E) at (-0.9,-3.0);
  \coordinate (F) at ( 0.9,-3.0);

  \draw[thin] (A)--(B);
  \draw[thin] (B)--(D);
  \draw[thin] (D2)--(F);
  \draw[thin] (C2)--(F);
  \draw[thin] (F)--(E);
  \draw[thin] (E)--(C2);
  \draw[thin] (E)--(D2);
  \draw[thin] (C)--(A);

  \draw[thin] (A)--(D);
  \draw[thin] (B)--(C);

  \draw[thin] (A)--(G);
  \draw[thin] (B)--(G);
  \draw[thin] (G2)--(E);
  \draw[thin] (G2)--(F);

  \draw[thin] (C)--(D)--(G);
  \draw[thin] (C2)--(D2)--(G2);
  \draw[thin, bend right=20] (C) to (D);
  \draw[thin, bend left=20] (C2) to (D2);
  \foreach \v in {A,B,C,D,E,F,G,C2,D2,G2}{
   \draw[draw=black, fill=black, thin, solid] (\v) circle (1pt);
  }
\end{tikzpicture}
\end{minipage}
}

\newcommand{\Psevenfive}{
\begin{minipage}{3cm}
\begin{tikzpicture}[scale=0.75]
    \coordinate (c) at (0,0);
    \coordinate (i1) at (0,1);
    \coordinate (i2) at (0,-1);
    \coordinate (i3) at (1,0);
    \coordinate (i4) at (-1,0);
    \coordinate (o1) at (2,2);
    \coordinate (o2) at (2,-2);
    \coordinate (o3) at (-2,2);
    \coordinate (o4) at (-2,-2);

    \filldraw (c) circle (1.33pt);
    \filldraw (i1) circle (1.33pt);
    \filldraw (i2) circle (1.33pt);
    \filldraw (i3) circle (1.33pt);
    \filldraw (i4) circle (1.33pt);
    \filldraw (o1) circle (1.33pt);
    \filldraw (o2) circle (1.33pt);
    \filldraw (o3) circle (1.33pt);
    \filldraw (o4) circle (1.33pt);

    \draw[thin] (c) -- (i1);
    \draw[thin] (c) -- (i2);
    \draw[thin] (c) -- (o1);
    \draw[thin] (c) -- (o4);
    \draw[thin] (i1) -- (i3) -- (i2) -- (i4) -- (i1);
    \draw[thin] (o1) -- (o2) -- (o4) -- (o3) -- (o1);
    \draw[thin] (o4) -- (i4);
    \draw[thin] (o1) -- (i3);
    \draw[thin] (o3) -- (i1);
    \draw[thin] (o3) -- (i4);
    \draw[thin] (o2) -- (i2);
    \draw[thin] (o2) -- (i3);

\end{tikzpicture}
\end{minipage}
}

\newcommand{\PsevenfiveDecompleted}{
\begin{minipage}{3cm}
\begin{tikzpicture}[scale=0.75]
    \coordinate (i1) at (0,1);
    \coordinate (i2) at (0,-1);
    \coordinate (i3) at (1,0);
    \coordinate (i4) at (-1,0);
    \coordinate (o1) at (2,2);
    \coordinate (o2) at (2,-2);
    \coordinate (o3) at (-2,2);
    \coordinate (o4) at (-2,-2);

    \filldraw (i1) circle (1.33pt);
    \filldraw (i2) circle (1.33pt);
    \filldraw (i3) circle (1.33pt);
    \filldraw (i4) circle (1.33pt);
    \filldraw (o1) circle (1.33pt);
    \filldraw (o2) circle (1.33pt);
    \filldraw (o3) circle (1.33pt);
    \filldraw (o4) circle (1.33pt);

    \draw[thin] (i1) -- (i3) -- (i2) -- (i4) -- (i1);
    \draw[thin] (o1) -- (o2) -- (o4) -- (o3) -- (o1);
    \draw[thin] (o4) -- (i4);
    \draw[thin] (o1) -- (i3);
    \draw[thin] (o3) -- (i1);
    \draw[thin] (o3) -- (i4);
    \draw[thin] (o2) -- (i2);
    \draw[thin] (o2) -- (i3);

\end{tikzpicture}
\end{minipage}
}

\newcommand{\PseventenDecompleted}{
\begin{minipage}{3.8cm}
\begin{tikzpicture}[scale=0.75]
    \coordinate (c) at (0, 0);
    \coordinate (i1) at (2,0);
    \coordinate (i2) at (-2,0);
    \coordinate (i3) at (1,1.5);
    \coordinate (i4) at (-1,1.5);
    \coordinate (i5) at (-1,-1.5);
    \coordinate (i6) at (1,-1.5);
    \coordinate (o) at (0,-3);

    \filldraw (c) circle (1.33pt);
    \filldraw (i1) circle (1.33pt);
    \filldraw (i2) circle (1.33pt);
    \filldraw (i3) circle (1.33pt);
    \filldraw (i4) circle (1.33pt);
    \filldraw (i5) circle (1.33pt);
    \filldraw (i6) circle (1.33pt);
    \filldraw (o) circle (1.33pt);

    \draw[thin] (i1) -- (i6) -- (i5) -- (i2) -- (i4) -- (i3) -- (i1);
    \draw[thin] (i6) -- (c) -- (i4); 
    \draw[thin] (i5) -- (c) -- (i3); 
    \draw[thin, bend right=10] (i5) to (o);
    \draw[thin, bend left=20] (i1) to (o);
    \draw[thin, bend right=20] (i2) to (o);
    \draw[thin, bend left=90, looseness=1.5] (i3) to (o);

\end{tikzpicture}
\end{minipage}
}

\newcommand{\Pseventen}{
\begin{minipage}{4cm}
\begin{tikzpicture}[scale=0.75]
    \coordinate (c) at (0, 0);
    \coordinate (i1) at (2,0);
    \coordinate (i2) at (-2,0);
    \coordinate (i3) at (1,1.5);
    \coordinate (i4) at (-1,1.5);
    \coordinate (i5) at (-1,-1.5);
    \coordinate (i6) at (1,-1.5);
    \coordinate (o) at (0,-3);
    \coordinate (o2) at (0,3);

    \filldraw (c) circle (1.33pt);
    \filldraw (i1) circle (1.33pt);
    \filldraw (i2) circle (1.33pt);
    \filldraw (i3) circle (1.33pt);
    \filldraw (i4) circle (1.33pt);
    \filldraw (i5) circle (1.33pt);
    \filldraw (i6) circle (1.33pt);
    \filldraw (o) circle (1.33pt);
    \filldraw (o2) circle (1.33pt);

    \draw[thin] (i1) -- (i6) -- (i5) -- (i2) -- (i4) -- (i3) -- (i1);
    \draw[thin] (i6) -- (c) -- (i4); 
    \draw[thin] (i5) -- (c) -- (i3); 
    \draw[thin, bend right=10] (i5) to (o);
    \draw[thin, bend left=20] (i1) to (o);
    \draw[thin, bend right=20] (i2) to (o);
    \draw[thin, bend left=90, looseness=1.5] (i3) to (o);
    \draw[thin, bend left=10] (i4) to (o2);
    \draw[thin, bend right=20] (i1) to (o2);
    \draw[thin, bend left=20] (i2) to (o2);
    \draw[thin, bend right=90, looseness=1.5] (i6) to (o2);

\end{tikzpicture}
\end{minipage}
}

\newcommand{\PTwistLeft}{
\begin{minipage}{4.5cm}
\begin{tikzpicture}[scale = 0.7, baseline={([yshift=-1.3ex]current bounding box.center)}]
    \coordinate[label=above:$a$] (va) at (0,3);
    \coordinate[label=above:$b$] (vb) at (0,2);
    \coordinate (vm) at (0, 2.15);
    \coordinate[label=above:$c$] (vc) at (0,1);
    \coordinate[label=above:$d$] (vd) at (0,0);
    \filldraw (va) circle (1.42pt);
    \filldraw (vb) circle (1.42pt);
    \filldraw (vc) circle (1.42pt);
    \filldraw (vd) circle (1.42pt);

    \draw[fill=black!20] (va) arc (90:270:3 and 1.5) arc (270:90:.5) arc (270:90:.5) arc (270:90:.5);
    \draw[fill=black!20] (va) arc (90:-90:3 and 1.5) arc (-90:90:.5) arc (-90:90:.5) arc (-90:90:.5);
    \node (G1) at (-1.5,1.5) {$\overline{G}_1$};
    \node (G2) at (+1.5,1.5) {$\overline{G}_2$};
    \node (G) at (-2,0) {$\overline{G}$};
\end{tikzpicture}
\end{minipage}
}

\newcommand{\PTwistRight}{
\begin{minipage}{4.5cm}
\begin{tikzpicture}[scale = 0.7, baseline={([yshift=-1.3ex]current bounding box.center)}]
    \coordinate[label=above:$a$] (va) at (0,3);
    \coordinate[label=above:$b$] (vb) at (0,2);
    \coordinate (vm) at (0, 2.15);
    \coordinate[label=below:$c$] (vc) at (0,1);
    \coordinate[label=above:$d$] (vd) at (0,0);
    \draw[fill=black!20] (vb) .. controls (-1,3) and (-3,3) .. (-3,1.5) .. controls (-3,0) and (-1,0) .. (vc) .. controls (-1,.5) .. (-1,.75) .. controls (-1,1) .. (vd) .. controls (-1,1.5) .. (va) .. controls (-1,2) .. (-1,2.25) .. controls (-1,2.5) .. (vb);
    \draw[fill=black!20] (-1,.75) .. controls (-1,1) .. (vd) .. controls (-1,1.5) .. (va) .. controls (-1,2) .. (-1,2.25);
    \draw[fill=black!20] (va) arc (90:-90:3 and 1.5) arc (-90:90:.5) arc (-90:90:.5) arc (-90:90:.5);
    \node (G1) at (-1.75,1.5) {$\overline{G}_1$};
    \node (G2) at (+2,1.5) {$\overline{G}_2$};
    \node (G) at (-2,0) {$\overline{G}'$};
    \filldraw (va) circle (1.42pt);
    \filldraw (vb) circle (1.42pt);
    \filldraw (vc) circle (1.42pt);
    \filldraw (vd) circle (1.42pt);
    \draw[bend left=90, looseness=1.5, dashed] (va) to (vd);
    \draw[bend left=90, looseness=1.5, dashed] (vb) to (vd);
    \draw[bend left=90, looseness=1.5, dashed] (va) to (vc);
\end{tikzpicture}
\end{minipage}
}

\newcommand{\Psevenfour}{
\begin{minipage}{3cm}
\begin{tikzpicture}[scale=0.75]
    \coordinate (c) at (0,0);
    \coordinate (i1) at (0,1);
    \coordinate (i2) at (0,-1);
    \coordinate (i3) at (1,0);
    \coordinate (i4) at (-1,0);
    \coordinate (o1) at (2,2);
    \coordinate (o2) at (2,-2);
    \coordinate (o3) at (-2,2);
    \coordinate (o4) at (-2,-2);

    \filldraw (c) circle (1.33pt);
    \filldraw (i1) circle (1.33pt);
    \filldraw (i2) circle (1.33pt);
    \filldraw (i3) circle (1.33pt);
    \filldraw (i4) circle (1.33pt);
    \filldraw (o1) circle (1.33pt);
    \filldraw (o2) circle (1.33pt);
    \filldraw (o3) circle (1.33pt);
    \filldraw (o4) circle (1.33pt);

    \draw[thin] (c) -- (i1);
    \draw[thin] (c) -- (i4);
    \draw[thin] (c) -- (o1);
    \draw[thin] (c) -- (o4);
    \draw[thin] (i1) -- (i3) -- (i2) -- (i4) -- (i1);
    \draw[thin] (o1) -- (o2) -- (o4) -- (o3) -- (o1);
    \draw[thin] (o4) -- (i2);
    \draw[thin] (o1) -- (i3);
    \draw[thin] (o3) -- (i1);
    \draw[thin] (o3) -- (i4);
    \draw[thin] (o2) -- (i2);
    \draw[thin] (o2) -- (i3);

\end{tikzpicture}
\end{minipage}
}

\newcommand{\PsevenfourSubgraph}{
\begin{minipage}{3cm}
\begin{tikzpicture}[scale=0.75]
    \coordinate[label=below:$2$] (c) at (0,0);
    \coordinate (i1) at (0,1);
    \coordinate[label=below:$3$] (i2) at (0,-1);
    \coordinate[label=right:$4$] (i3) at (1,0);
    \coordinate (i4) at (-1,0);
    \coordinate[label=above:$1$] (o3) at (-2,2);

    \filldraw (c) circle (1.33pt);
    \filldraw (i1) circle (1.33pt);
    \filldraw (i2) circle (1.33pt);
    \filldraw (i3) circle (1.33pt);
    \filldraw (i4) circle (1.33pt);
    \filldraw (o3) circle (1.33pt);

    \draw[thin] (c) -- (i1);
    \draw[thin] (c) -- (i4);
    \draw[thin] (i1) -- (i3) -- (i2) -- (i4) -- (i1);
    \draw[thin] (o3) -- (i1);
    \draw[thin] (o3) -- (i4);

\end{tikzpicture}
\end{minipage}
}

\newcommand{\PsevenfourSubgraphTwist}{
\begin{minipage}{3cm}
\begin{tikzpicture}[scale=0.75]
    \coordinate[label=below:$4$] (c) at (0,0);
    \coordinate (i1) at (0,1);
    \coordinate[label=below:$1$] (i2) at (0,-1);
    \coordinate[label=right:$2$] (i3) at (1,0);
    \coordinate (i4) at (-1,0);
    \coordinate[label=above:$3$] (o3) at (-2,2);

    \filldraw (c) circle (1.33pt);
    \filldraw (i1) circle (1.33pt);
    \filldraw (i2) circle (1.33pt);
    \filldraw (i3) circle (1.33pt);
    \filldraw (i4) circle (1.33pt);
    \filldraw (o3) circle (1.33pt);

    \draw[thin] (c) -- (i1);
    \draw[thin] (c) -- (i4);
    \draw[thin] (i1) -- (i3) -- (i2) -- (i4) -- (i1);
    \draw[thin] (o3) -- (i1);
    \draw[thin] (o3) -- (i4);

\end{tikzpicture}
\end{minipage}
}

\newcommand{\Psevenseven}{
\begin{minipage}{3cm}
\begin{tikzpicture}[scale=0.75]
    \coordinate (c) at (0,0);
    \coordinate (i2) at (0,-1);
    \coordinate (i3) at (1,0);
    \coordinate (o1) at (2,2);
    \coordinate (o2) at (2,-2);
    \coordinate (o3) at (-2,2);
    \coordinate (o4) at (-2,-2);

    \path (i2) -- (i3) coordinate[midway] (i1);
    \coordinate (i4) at (-0.5, 0.5);

    \filldraw (c) circle (1.33pt);
    \filldraw (i1) circle (1.33pt);
    \filldraw (i2) circle (1.33pt);
    \filldraw (i3) circle (1.33pt);
    \filldraw (i4) circle (1.33pt);
    \filldraw (o1) circle (1.33pt);
    \filldraw (o2) circle (1.33pt);
    \filldraw (o3) circle (1.33pt);
    \filldraw (o4) circle (1.33pt);

    \draw[thin] (c) -- (i1);
    \draw[thin] (c) -- (i2);
    \draw[thin] (c) -- (i3);
    \draw[thin] (c) -- (i4);
    \draw[thin] (o1) -- (o2) -- (o4) -- (o3) -- (o1);
    \draw[thin] (i2) -- (i1) -- (i3) -- (i2);
    \draw[thin] (o4) -- (i4);
    \draw[thin] (o4) -- (i2);
    \draw[thin] (o1) -- (i3);
    \draw[thin] (o1) -- (i4);
    \draw[thin] (o3) -- (i1);
    \draw[thin] (o3) -- (i4);
    \draw[thin] (o2) -- (i2);
    \draw[thin] (o2) -- (i3);
    \draw[bend right=30] (i1) to (o3);

\end{tikzpicture}
\end{minipage}
}

\newcommand{\FSplitLeft}{
\begin{minipage}{4.5cm}
\begin{tikzpicture}[scale = 0.7, baseline={([yshift=-1.3ex]current bounding box.center)}]
    \coordinate (va) at (0,3);
    \coordinate (vb) at (0,2);
    \coordinate (vm) at (0, 2.15);
    \coordinate (vc) at (0,1);
    \coordinate (vd) at (0,0);
    \filldraw (va) circle (1.42pt);
    \filldraw (vb) circle (1.42pt);
    \filldraw (vc) circle (1.42pt);
    \filldraw (vd) circle (1.42pt);

    \draw[fill=black!20] (va) arc (90:270:3 and 1.5) arc (270:90:.5) arc (270:90:.5) arc (270:90:.5);
    \draw[fill=black!20] (va) arc (90:-90:3 and 1.5) arc (-90:90:.5) arc (-90:90:.5) arc (-90:90:.5);
    \node (G1) at (-1.5,1.5) {$\overline{G}_1$};
    \node (G2) at (+1.5,1.5) {$\overline{G}_2$};
    \node (G) at (-2,-0.2) {$\overline{G}$};
\end{tikzpicture}
\end{minipage}
}

\newcommand{\FSplitRight}{
\begin{minipage}{4.5cm}
\begin{tikzpicture}[scale = 0.7, baseline={([yshift=-1.3ex]current bounding box.center)}]
    \coordinate (va) at (0,3);
    \coordinate (vb) at (0,2);
    \coordinate (vm) at (0, 2.15);
    \coordinate (vc) at (0,1);
    \coordinate (vd) at (0,0);
    \filldraw (va) circle (1.42pt);
    \filldraw (vb) circle (1.42pt);
    \filldraw (vc) circle (1.42pt);
    \filldraw (vd) circle (1.42pt);

    \draw[fill=black!20] (va) arc (90:270:3 and 1.5) arc (270:90:.5) arc (270:90:.5) arc (270:90:.5);
    \draw[fill=black!20] (va) arc (90:-90:3 and 1.5) arc (-90:90:.5) arc (-90:90:.5) arc (-90:90:.5);
    \node (G1) at (-1.5,1.5) {$\overline{G}_1$};
    \node (G2) at (+1.5,1.5) {$\overline{G}_2^*$};
    \node (G) at (-2,-0.2) {$\overline{G}'$};
\end{tikzpicture}
\end{minipage}
}

\newcommand{\FiveTwistComplete}{
\begin{minipage}{4.5cm}
\begin{tikzpicture}[scale = 0.7, baseline={([yshift=-1.3ex]current bounding box.center)}]
    \filldraw[thick,fill=black!10] (0,0) ellipse (3 and 2);
    \filldraw[thick,fill=white] (0,0) ellipse (2 and 1.16);
    \filldraw[thick,fill=black!10] (2,0) -- (1,1) -- (-1,1) -- (-1,-1) -- (1,-1) -- (2,0);
    \fill (2,0) circle (3pt) node[anchor=east] {$\infty\,$};
    \fill (1,1) circle (3pt);
    \fill (-1,1) circle (3pt);
    \fill (1,-1) circle (3pt);
    \fill (-1,-1) circle (3pt);
    \node[below=0.2 of current bounding box] {$\overline{G}$};
\end{tikzpicture}
\end{minipage}
}

\newcommand{\FiveTwistDecompleted}{
\begin{minipage}{4.5cm}
\begin{tikzpicture}[scale = 0.7, baseline={([yshift=-1.3ex]current bounding box.center)}]
    \filldraw[thick,fill=black!10] (0,0) ellipse (3 and 2);
    \filldraw[thick,fill=white] (0,0) ellipse (2 and 1.16);
    \filldraw[thick,fill=black!10] (-1,-1) rectangle (1,1);
    \draw[thick,dashed] (-1,-1) -- (1,1);
    \draw[thick,dashed] (-1,1) -- (1,-1);
    \fill (1,1) circle (3pt);
    \fill (-1,1) circle (3pt);
    \fill (1,-1) circle (3pt);
    \fill (-1,-1) circle (3pt);
    \node at (0.6,0) {$G_1$};
    \node at (-2.42,0) {$G_2$};
    \node[below=0.2 of current bounding box] {$G=\overline{G}\setminus\infty$};
\end{tikzpicture}
\end{minipage}
}

\newcommand{\coproductBubbleZeroX}{
\begin{minipage}{2.9cm}
\begin{tikzpicture}
    \coordinate[label=right:{\footnotesize $x_1 = x_2$}] (v);
    \def \rad {1}
    \coordinate[label=left:$0$] (v2) at ([shift=(180:\rad)]v);

    \draw[draw=black, fill=black, thin, solid] (v) circle (1pt);
    \draw[draw=black, fill=black, thin, solid] (v2) circle (1pt);
    \draw[draw=black, thin, solid, bend left=30] (v) to (v2);
    \draw[draw=black, thin, solid, bend right=30] (v) to (v2);

\end{tikzpicture}
\end{minipage}
}

\newcommand{\coproductXOneXTwo}{
\begin{minipage}{0.6cm}
\begin{tikzpicture}
     \def \eps {0.5}

    \coordinate (v);
    \def \rad {1}
    \coordinate (v2) at ([shift=(180:\rad)]v);

    \coordinate[label=above:$x_1$] (x1) at ($(v2)!\eps!(v1)$);
    \coordinate[label=below:$x_2$] (x2) at ($(v2)!\eps!(v3)$);

    \draw[draw=black, fill=black, thin, solid] (x1) circle (1pt);
    \draw[draw=black, fill=black, thin, solid] (x2) circle (1pt);

    \draw[draw=black, thin, solid, bend left=30] (x1) to (x2);
    \draw[draw=black, thin, solid, bend right=30] (x1) to (x2);
    
\end{tikzpicture}
\end{minipage}
}

\newcommand{\CKExampleII}{
\begin{minipage}{3cm}
\begin{tikzpicture}
     \def \eps {0.5}

    \coordinate (v);
    \def \rad {1}
    \coordinate[label=above:$1$] (v1) at ([shift=(60:\rad)]v);
    \coordinate[label=left:$0$] (v2) at ([shift=(180:\rad)]v);
    \coordinate[label=below:$x_3$] (v3) at ([shift=(300:\rad)]v);

     \coordinate[label=above:$x_1$] (x1) at ($(v2)!\eps!(v1)$);
    \coordinate[label=below:$x_2$] (x2) at ($(v2)!\eps!(v3)$);
    \coordinate[label=below:$z$] (z) at ($(v3)+(1,0)$);

    \draw[draw=black, fill=black, thin, solid] (v1) circle (1pt);
    \draw[draw=black, fill=black, thin, solid] (v2) circle (1pt);
    \draw[draw=black, fill=black, thin, solid] (v3) circle (1pt);
    \draw[draw=black, fill=black, thin, solid] (x1) circle (1pt);
    \draw[draw=black, fill=black, thin, solid] (x2) circle (1pt);
    \draw[draw=black, fill=black, thin, solid] (z) circle (1pt);

    \draw[draw=black, thin, solid, bend left=30] (x1) to (x2);
    \draw[draw=black, thin, solid, bend right=30] (x1) to (x2);
    \draw[draw=black, thin, solid, bend left=30] (v3) to (z);
    \draw[draw=black, thin, solid, bend right=30] (v3) to (z);
    
    \draw[draw=black, thin, solid] (v2) -- (x1) -- (v1);
    \draw[draw=black, thin, solid] (v2) -- (x2) -- (v3);
    \draw[draw=black, thin, solid] (v1) -- (v3);

\end{tikzpicture}
\end{minipage}
}

\newcommand{\coproductBubbleXZ}{
\begin{minipage}{2cm}
\begin{tikzpicture}
    \coordinate[label=right:{$z$}] (v);
    \def \rad {1}
    \coordinate[label=left:$x_3$] (v2) at ([shift=(180:\rad)]v);

    \draw[draw=black, fill=black, thin, solid] (v) circle (1pt);
    \draw[draw=black, fill=black, thin, solid] (v2) circle (1pt);
    \draw[draw=black, thin, solid, bend left=30] (v) to (v2);
    \draw[draw=black, thin, solid, bend right=30] (v) to (v2);

\end{tikzpicture}
\end{minipage}
}

\newcommand{\coproductTrigOneZeroZ}{
\begin{minipage}{2.5cm}
\begin{tikzpicture}
    \coordinate[label=above left:{\footnotesize $x_1=x_2=0$}] (v);
    \def \rad {1}
    \coordinate[label=above:$1$] (v1) at ([shift=(60:\rad)]v);
    \coordinate[label=below:{\footnotesize $x_3 = z$}] (v3) at ([shift=(300:\rad)]v);

    \draw[draw=black, fill=black, thin, solid] (v) circle (1pt);
    \draw[draw=black, fill=black, thin, solid] (v1) circle (1pt);
    \draw[draw=black, fill=black, thin, solid] (v3) circle (1pt);
    \draw[draw=black, thin, solid] (v) -- (v1);
    \draw[draw=black, thin, solid] (v) -- (v3);
    \draw[draw=black, thin, solid] (v1) -- (v3);

\end{tikzpicture}
\end{minipage}
}

\newcommand{\reroutthreestar}{
\begin{minipage}{2.2cm}
\begin{tikzpicture}
    \coordinate (v);
    \def \rad {1}
    \coordinate[label=above:$1$] (v1) at ([shift=(60:\rad)]v);
    \coordinate[label=left:$0$] (v2) at ([shift=(180:\rad)]v);
    \coordinate[label=below:$z$] (v3) at ([shift=(300:\rad)]v);

    \draw[draw=black, fill=black, thin, solid] (v) circle (1pt);
    \draw[draw=black, fill=black, thin, solid] (v1) circle (1pt);
    \draw[draw=black, fill=black, thin, solid] (v2) circle (1pt);
    \draw[draw=black, fill=black, thin, solid] (v3) circle (1pt);
    \draw[draw=black, thin, solid] (v) -- node[inner sep=0pt,above] {\tiny $\frac{2-3\ep}{\lambda}$} (v2);
    \draw[draw=black, thin, solid] (v) -- (v1);
    \draw[draw=black, thin, solid] (v) -- (v3);

\end{tikzpicture}
\end{minipage}
}

\newcommand{\coprodDecompletedEvaled}{
\begin{minipage}{1.8cm}
\begin{tikzpicture}
    \coordinate (v) ;
    \def \rad {1}
    \coordinate[label=above:$z$] (v1) at ([shift=(90:\rad)]v);
    \coordinate[label=below:$0$] (v3) at ([shift=(270:\rad)]v);
    \coordinate[label=right:$1$] (v4) at ([shift=(0:\rad)]v);

    \draw[thin] (v) -- (v4);
    \draw[draw=black, thin, solid] (v) -- (v1);
    \draw[draw=black, thin, solid] (v) -- (v4);
    \draw[draw=black, thin, solid] (v) -- node[inner sep=0pt,right] {\tiny $\frac{3\ep}{\lambda}$}(v3);

    \filldraw (v) circle (1pt);
    \filldraw (v1) circle (1pt);
    \filldraw (v3) circle (1pt);
    \filldraw (v4) circle (1pt);

\end{tikzpicture}
\end{minipage}
}

\newcommand{\CKExampleIIxxcontracted}{
\begin{minipage}{3cm}
\begin{tikzpicture}
     \def \eps {0.5}

    \coordinate (v);
    \def \rad {1}
    \coordinate[label=above:$1$] (v1) at ([shift=(60:\rad)]v);
    \coordinate[label=left:$0$] (v2) at ([shift=(180:\rad)]v);
    \coordinate[label=below:$x_3$] (v3) at ([shift=(300:\rad)]v);

    \coordinate (x1) at ($(v2)!\eps!(v1)$);
    \coordinate (x2) at ($(v2)!\eps!(v3)$);
    \coordinate[label=above:{\footnotesize $x_1 = x_2$}] (x3) at ($(x1)!\eps!(x2)$);
    \coordinate[label=below:$z$] (z) at ($(v3)+(1,0)$);

    \draw[draw=black, fill=black, thin, solid] (v1) circle (1pt);
    \draw[draw=black, fill=black, thin, solid] (v2) circle (1pt);
    \draw[draw=black, fill=black, thin, solid] (v3) circle (1pt);
    \draw[draw=black, fill=black, thin, solid] (x3) circle (1pt);
    \draw[draw=black, fill=black, thin, solid] (z) circle (1pt);

    \draw[draw=black, thin, solid, bend left=30] (v2) to (x3);
    \draw[draw=black, thin, solid, bend right=30] (v2) to (x3);
    \draw[draw=black, thin, solid, bend left=30] (v3) to (z);
    \draw[draw=black, thin, solid, bend right=30] (v3) to (z);
    
    \draw[draw=black, thin, solid] (x3) -- (v1);
    \draw[draw=black, thin, solid] (x3) -- (v3);
    \draw[draw=black, thin, solid] (v1) -- (v3);

\end{tikzpicture}
\end{minipage}
}

\newcommand{\CKExampleIIxzcontracted}{
\begin{minipage}{2.5cm}
\begin{tikzpicture}
     \def \eps {0.5}

    \coordinate (v);
    \def \rad {1}
    \coordinate[label=above:$1$] (v1) at ([shift=(60:\rad)]v);
    \coordinate[label=left:$0$] (v2) at ([shift=(180:\rad)]v);
    \coordinate[label=below:{$x_3=z$}] (v3) at ([shift=(300:\rad)]v);

     \coordinate[label=above:$x_1$] (x1) at ($(v2)!\eps!(v1)$);
    \coordinate[label=below:$x_2$] (x2) at ($(v2)!\eps!(v3)$);

    \draw[draw=black, fill=black, thin, solid] (v1) circle (1pt);
    \draw[draw=black, fill=black, thin, solid] (v2) circle (1pt);
    \draw[draw=black, fill=black, thin, solid] (v3) circle (1pt);
    \draw[draw=black, fill=black, thin, solid] (x1) circle (1pt);
    \draw[draw=black, fill=black, thin, solid] (x2) circle (1pt);

    \draw[draw=black, thin, solid, bend left=30] (x1) to (x2);
    \draw[draw=black, thin, solid, bend right=30] (x1) to (x2);
    
    \draw[draw=black, thin, solid] (v2) -- (x1) -- (v1);
    \draw[draw=black, thin, solid] (v2) -- (x2) -- (v3);
    \draw[draw=black, thin, solid] (v1) -- (v3);

\end{tikzpicture}
\end{minipage}
}

\newcommand{\CKExampleIIxxzerocontracted}{
\begin{minipage}{4cm}
\begin{tikzpicture}
     \def \eps {0.5}

    \coordinate (v);
    \def \rad {1}
    \coordinate[label=above:$1$] (v1) at ([shift=(60:\rad)]v);
    \coordinate(v2) at ([shift=(180:\rad)]v);
    \coordinate[label=below:$x_3$] (v3) at ([shift=(300:\rad)]v);

     \coordinate (x1) at ($(v2)!\eps!(v1)$);
    \coordinate (x2) at ($(v2)!\eps!(v3)$);
    \coordinate[label=below:$z$] (z) at ($(v3)+(1,0)$);
    \coordinate[label=left:{\footnotesize $x_1 = x_2 = 0$}] (x3) at ($(x1)!\eps!(x2)$);

    \draw[draw=black, fill=black, thin, solid] (v1) circle (1pt);
    \draw[draw=black, fill=black, thin, solid] (v3) circle (1pt);
    \draw[draw=black, fill=black, thin, solid] (x3) circle (1pt);
    \draw[draw=black, fill=black, thin, solid] (z) circle (1pt);

    \draw[draw=black, thin, solid, bend left=30] (v3) to (z);
    \draw[draw=black, thin, solid, bend right=30] (v3) to (z);
    
    \draw[draw=black, thin, solid] (x3) -- (v1);
    \draw[draw=black, thin, solid] (x3) -- (v3);
    \draw[draw=black, thin, solid] (v1) -- (v3);

\end{tikzpicture}
\end{minipage}
}

\newcommand{\coproductDoubleBubble}{
\begin{minipage}{2.2cm}
\begin{tikzpicture}
     \def \eps {0.5}

    \coordinate (v);
    \def \rad {1}
    \coordinate (v1) at ([shift=(60:\rad)]v);
    \coordinate (v2) at ([shift=(180:\rad)]v);
    \coordinate[label=below:$x_3$] (v3) at ([shift=(300:\rad)]v);

     \coordinate[label=above:$x_1$] (x1) at ($(v2)!\eps!(v1)$);
    \coordinate[label=below:$x_2$] (x2) at ($(v2)!\eps!(v3)$);
    \coordinate[label=below:$z$] (z) at ($(v3)+(1,0)$);

    \draw[draw=black, fill=black, thin, solid] (v3) circle (1pt);
    \draw[draw=black, fill=black, thin, solid] (x1) circle (1pt);
    \draw[draw=black, fill=black, thin, solid] (x2) circle (1pt);
    \draw[draw=black, fill=black, thin, solid] (z) circle (1pt);

    \draw[draw=black, thin, solid, bend left=30] (x1) to (x2);
    \draw[draw=black, thin, solid, bend right=30] (x1) to (x2);
    \draw[draw=black, thin, solid, bend left=30] (v3) to (z);
    \draw[draw=black, thin, solid, bend right=30] (v3) to (z);

\end{tikzpicture}
\end{minipage}
}

\newcommand{\CKExampleIIxxxzcontracted}{
\begin{minipage}{3cm}
\begin{tikzpicture}
     \def \eps {0.5}

    \coordinate (v);
    \def \rad {1}
    \coordinate[label=above:$1$] (v1) at ([shift=(60:\rad)]v);
    \coordinate[label=left:$0$] (v2) at ([shift=(180:\rad)]v);
    \coordinate[label=below:{$x_3=z$}] (v3) at ([shift=(300:\rad)]v);

    \coordinate (x1) at ($(v2)!\eps!(v1)$);
    \coordinate (x2) at ($(v2)!\eps!(v3)$);
    \coordinate[label=above:{\footnotesize $x_1 = x_2$}] (x3) at ($(x1)!\eps!(x2)$);

    \draw[draw=black, fill=black, thin, solid] (v1) circle (1pt);
    \draw[draw=black, fill=black, thin, solid] (v2) circle (1pt);
    \draw[draw=black, fill=black, thin, solid] (v3) circle (1pt);
    \draw[draw=black, fill=black, thin, solid] (x3) circle (1pt);

    \draw[draw=black, thin, solid, bend left=30] (v2) to (x3);
    \draw[draw=black, thin, solid, bend right=30] (v2) to (x3);
    
    \draw[draw=black, thin, solid] (x3) -- (v1);
    \draw[draw=black, thin, solid] (x3) -- (v3);
    \draw[draw=black, thin, solid] (v1) -- (v3);

\end{tikzpicture}
\end{minipage}
}

\newcommand{\coproductFull}{
\begin{minipage}{3cm}
\begin{tikzpicture}
     \def \eps {0.5}

    \coordinate (v);
    \def \rad {1}
    \coordinate (v1) at ([shift=(60:\rad)]v);
    \coordinate[label=left:$0$] (v2) at ([shift=(180:\rad)]v);
    \coordinate[label=below:$x_3$] (v3) at ([shift=(300:\rad)]v);

     \coordinate[label=above:$x_1$] (x1) at ($(v2)!\eps!(v1)$);
    \coordinate[label=below:$x_2$] (x2) at ($(v2)!\eps!(v3)$);
    \coordinate[label=below:$z$] (z) at ($(v3)+(1,0)$);

    \draw[draw=black, fill=black, thin, solid] (v2) circle (1pt);
    \draw[draw=black, fill=black, thin, solid] (v3) circle (1pt);
    \draw[draw=black, fill=black, thin, solid] (x1) circle (1pt);
    \draw[draw=black, fill=black, thin, solid] (x2) circle (1pt);
    \draw[draw=black, fill=black, thin, solid] (z) circle (1pt);

    \draw[draw=black, thin, solid, bend left=30] (x1) to (x2);
    \draw[draw=black, thin, solid, bend right=30] (x1) to (x2);
    \draw[draw=black, thin, solid, bend left=30] (v3) to (z);
    \draw[draw=black, thin, solid, bend right=30] (v3) to (z);
    
    \draw[draw=black, thin, solid] (v2) -- (x1);
    \draw[draw=black, thin, solid] (v2) -- (x2);

\end{tikzpicture}
\end{minipage}
}

\newcommand{\CKExampleIIxxzeroxzcontracted}{
\begin{minipage}{3cm}
\begin{tikzpicture}
     \def \eps {0.5}

    \coordinate (v);
    \def \rad {1}
    \coordinate[label=above:$1$] (v1) at ([shift=(60:\rad)]v);
    \coordinate(v2) at ([shift=(180:\rad)]v);
    \coordinate[label=below:{$x_3=z$}] (v3) at ([shift=(300:\rad)]v);

     \coordinate (x1) at ($(v2)!\eps!(v1)$);
    \coordinate (x2) at ($(v2)!\eps!(v3)$);
    \coordinate[label=left:{\footnotesize $x_1 = x_2 = 0$}] (x3) at ($(x1)!\eps!(x2)$);

    \draw[draw=black, fill=black, thin, solid] (v1) circle (1pt);
    \draw[draw=black, fill=black, thin, solid] (v3) circle (1pt);
    \draw[draw=black, fill=black, thin, solid] (x3) circle (1pt);
    
    \draw[draw=black, thin, solid] (x3) -- (v1);
    \draw[draw=black, thin, solid] (x3) -- (v3);
    \draw[draw=black, thin, solid] (v1) -- (v3);

\end{tikzpicture}
\end{minipage}
}

\newcommand{\CKExampleIIRR}{
\begin{minipage}{3cm}
\begin{tikzpicture}
     \def \eps {0.5}

    \coordinate (v);
    \def \rad {1}
    \coordinate[label=above:$1$] (v1) at ([shift=(60:\rad)]v);
    \coordinate[label=left:$0$] (v2) at ([shift=(180:\rad)]v);
    \coordinate[label=below:$x_3$] (v3) at ([shift=(300:\rad)]v);

     \coordinate[label=above:$x_1$] (x1) at ($(v2)!\eps!(v1)$);
    \coordinate[label=below:$x_2$] (x2) at ($(v2)!\eps!(v3)$);
    \coordinate[label=below:$z$] (z) at ($(v3)+(1,0)$);

    \draw[draw=black, fill=black, thin, solid] (v1) circle (1pt);
    \draw[draw=black, fill=black, thin, solid] (v2) circle (1pt);
    \draw[draw=black, fill=black, thin, solid] (v3) circle (1pt);
    \draw[draw=black, fill=black, thin, solid] (x1) circle (1pt);
    \draw[draw=black, fill=black, thin, solid] (x2) circle (1pt);
    \draw[draw=black, fill=black, thin, solid] (z) circle (1pt);

    \draw[draw=black, thin, solid, bend left=30] (x1) to (x2);
    \draw[draw=black, thin, solid, bend right=30] (x1) to (x2);
    \draw[draw=black, thin, solid, bend left=30] (v3) to (z);
    \draw[draw=black, thin, solid, bend right=30] (v3) to (z);
    
    \draw[draw=black, thin, solid] (v2) -- (x1);
    \draw[draw=black, thin, solid] (v2) -- (x2) -- (v3);
    \draw[draw=black, thin, solid] (v1) -- (v3);
    \draw[draw=black, thin, solid] (v1) -- (x2);

\end{tikzpicture}
\end{minipage}
}

\newcommand{\CKExampleIIRRxzcontracted}{
\begin{minipage}{3cm}
\begin{tikzpicture}
     \def \eps {0.5}

    \coordinate (v);
    \def \rad {1}
    \coordinate[label=above:$1$] (v1) at ([shift=(60:\rad)]v);
    \coordinate[label=left:$0$] (v2) at ([shift=(180:\rad)]v);
    \coordinate[label=below:{$x_3=z$}] (v3) at ([shift=(300:\rad)]v);

     \coordinate[label=above:$x_1$] (x1) at ($(v2)!\eps!(v1)$);
    \coordinate[label=below:$x_2$] (x2) at ($(v2)!\eps!(v3)$);

    \draw[draw=black, fill=black, thin, solid] (v1) circle (1pt);
    \draw[draw=black, fill=black, thin, solid] (v2) circle (1pt);
    \draw[draw=black, fill=black, thin, solid] (v3) circle (1pt);
    \draw[draw=black, fill=black, thin, solid] (x1) circle (1pt);
    \draw[draw=black, fill=black, thin, solid] (x2) circle (1pt);

    \draw[draw=black, thin, solid, bend left=30] (x1) to (x2);
    \draw[draw=black, thin, solid, bend right=30] (x1) to (x2);
    
    \draw[draw=black, thin, solid] (v2) -- (x1);
    \draw[draw=black, thin, solid] (v2) -- (x2) -- (v3);
    \draw[draw=black, thin, solid] (v1) -- (v3);
    \draw[draw=black, thin, solid] (v1) -- (x2);

\end{tikzpicture}
\end{minipage}
}

\newcommand{\CKExampleIIapprox}{
\begin{minipage}{2.3cm}
\begin{tikzpicture}
     \def \eps {0.5}

    \coordinate (v);
    \def \rad {1}
    \coordinate[label=above:$1$] (v1) at ([shift=(60:\rad)]v);
    \coordinate[label=left:$0$] (v2) at ([shift=(180:\rad)]v);
    \coordinate[label=below:$z$] (v3) at ([shift=(300:\rad)]v);

     \coordinate[label=above:$x_1$] (x1) at ($(v2)!\eps!(v1)$);
    \coordinate[label=below:$x_2$] (x2) at ($(v2)!\eps!(v3)$);

    \draw[draw=black, fill=black, thin, solid] (v1) circle (1pt);
    \draw[draw=black, fill=black, thin, solid] (v2) circle (1pt);
    \draw[draw=black, fill=black, thin, solid] (v3) circle (1pt);
    \draw[draw=black, fill=black, thin, solid] (x1) circle (1pt);
    \draw[draw=black, fill=black, thin, solid] (x2) circle (1pt);

    \draw[draw=black, thin, solid, bend left=30] (x1) to (x2);
    \draw[draw=black, thin, solid, bend right=30] (x1) to (x2);
    
    \draw[draw=black, thin, solid] (v2) -- (x1) -- (v1);
    \draw[draw=black, thin, solid] (v2) -- (x2) -- (v3);
    \draw[draw=black, thin, solid] (v1) -- (v3);

\end{tikzpicture}
\end{minipage}
}

\newcommand{\CKExampleIIRRapprox}{
\begin{minipage}{2.3cm}
\begin{tikzpicture}
     \def \eps {0.5}

    \coordinate (v);
    \def \rad {1}
    \coordinate[label=above:$1$] (v1) at ([shift=(60:\rad)]v);
    \coordinate[label=left:$0$] (v2) at ([shift=(180:\rad)]v);
    \coordinate[label=below:$z$] (v3) at ([shift=(300:\rad)]v);

     \coordinate[label=above:$x_1$] (x1) at ($(v2)!\eps!(v1)$);
    \coordinate[label=below:$x_2$] (x2) at ($(v2)!\eps!(v3)$);

    \draw[draw=black, fill=black, thin, solid] (v1) circle (1pt);
    \draw[draw=black, fill=black, thin, solid] (v2) circle (1pt);
    \draw[draw=black, fill=black, thin, solid] (v3) circle (1pt);
    \draw[draw=black, fill=black, thin, solid] (x1) circle (1pt);
    \draw[draw=black, fill=black, thin, solid] (x2) circle (1pt);

    \draw[draw=black, thin, solid, bend left=30] (x1) to (x2);
    \draw[draw=black, thin, solid, bend right=30] (x1) to (x2);
    
    \draw[draw=black, thin, solid] (v2) -- (x1);
    \draw[draw=black, thin, solid] (v2) -- (x2) -- (v3);
    \draw[draw=black, thin, solid] (v1) -- (v3);
    \draw[draw=black, thin, solid] (v1) -- (x2);

\end{tikzpicture}
\end{minipage}
}

\newcommand{\CKExampleIIRRapproxTwo}{
\begin{minipage}{3cm}
\begin{tikzpicture}
     \def \eps {0.5}

    \coordinate (v);
    \def \rad {1}
    \coordinate[label=above:$1$] (v1) at ([shift=(60:\rad)]v);
    \coordinate[label=left:$0$] (v2) at ([shift=(180:\rad)]v);
    \coordinate[label=below:$x_3$] (v3) at ([shift=(300:\rad)]v);

     \coordinate (x1) at ($(v2)!\eps!(v1)$);
    \coordinate (x2) at ($(v2)!\eps!(v3)$);
    \coordinate[label=below:$z$] (z) at ($(v3)+(1,0)$);
    \coordinate[label=above:{$x_2$}] (x3) at ($(x1)!\eps!(x2)$);

    \draw[draw=black, fill=black, thin, solid] (v1) circle (1pt);
    \draw[draw=black, fill=black, thin, solid] (v2) circle (1pt);
    \draw[draw=black, fill=black, thin, solid] (v3) circle (1pt);
    \draw[draw=black, fill=black, thin, solid] (x3) circle (1pt);
    \draw[draw=black, fill=black, thin, solid] (z) circle (1pt);

    \draw[draw=black, thin, solid, bend left=30] (v3) to (z);
    \draw[draw=black, thin, solid, bend right=30] (v3) to (z);
    
    \draw[draw=black, thin, solid] (v2) --  node[inner sep=2pt,below] {\tiny $\frac{2-3\ep}{\lambda}$} (x3);
    \draw[draw=black, thin, solid] (x3) -- (v3);
    \draw[draw=black, thin, solid] (v1) -- (v3);
    \draw[draw=black, thin, solid] (v1) -- (x3);

\end{tikzpicture}
\end{minipage}
}

\newcommand{\CKExampleIIRRapproxTwoRR}{
\begin{minipage}{3cm}
\begin{tikzpicture}
     \def \eps {0.5}

    \coordinate (v);
    \def \rad {1}
    \coordinate[label=above:$1$] (v1) at ([shift=(60:\rad)]v);
    \coordinate[label=left:$0$] (v2) at ([shift=(180:\rad)]v);
    \coordinate[label=below:$x_3$] (v3) at ([shift=(300:\rad)]v);

     \coordinate (x1) at ($(v2)!\eps!(v1)$);
    \coordinate (x2) at ($(v2)!\eps!(v3)$);
    \coordinate[label=below:$z$] (z) at ($(v3)+(1,0)$);
    \coordinate[label=above:{$x_2$}] (x3) at ($(x1)!\eps!(x2)$);

    \draw[draw=black, fill=black, thin, solid] (v1) circle (1pt);
    \draw[draw=black, fill=black, thin, solid] (v2) circle (1pt);
    \draw[draw=black, fill=black, thin, solid] (v3) circle (1pt);
    \draw[draw=black, fill=black, thin, solid] (x3) circle (1pt);
    \draw[draw=black, fill=black, thin, solid] (z) circle (1pt);

    \draw[draw=black, thin, solid, bend left=30] (v3) to (z);
    \draw[draw=black, thin, solid, bend right=30] (v3) to (z);
    
    \draw[draw=black, thin, solid] (v2) --  node[inner sep=2pt,below] {\tiny $\frac{2-3\ep}{\lambda}$} (x3);
    \draw[draw=black, thin, solid] (x3) -- (v3);
    \draw[draw=black, thin, solid] (v1) -- (v3);
    \draw[draw=black, thin, solid] (v1) -- (v2);

\end{tikzpicture}
\end{minipage}
}

\newcommand{\CKExampleIIRRapproxTwoapprox}{
\begin{minipage}{2.3cm}
\begin{tikzpicture}
     \def \eps {0.5}

    \coordinate (v);
    \def \rad {1}
    \coordinate[label=above:$1$] (v1) at ([shift=(60:\rad)]v);
    \coordinate[label=left:$0$] (v2) at ([shift=(180:\rad)]v);
    \coordinate[label=below:$z$] (v3) at ([shift=(300:\rad)]v);

     \coordinate (x1) at ($(v2)!\eps!(v1)$);
    \coordinate (x2) at ($(v2)!\eps!(v3)$);
    \coordinate[label=above:{$x_2$}] (x3) at ($(x1)!\eps!(x2)$);

    \draw[draw=black, fill=black, thin, solid] (v1) circle (1pt);
    \draw[draw=black, fill=black, thin, solid] (v2) circle (1pt);
    \draw[draw=black, fill=black, thin, solid] (v3) circle (1pt);
    \draw[draw=black, fill=black, thin, solid] (x3) circle (1pt);

    \draw[draw=black, thin, solid] (v2) --  node[inner sep=0pt,below] {\tiny $\frac{2-3\ep}{\lambda}$} (x3);
    \draw[draw=black, thin, solid] (x3) -- node[inner sep=0pt,above right] {\tiny $\frac{1 - 2\ep}{\lambda}$} (v3);
    \draw[draw=black, thin, solid] (v1) -- node[inner sep=0pt,below right] {\tiny $\frac{1}{\lambda}$} (x3);

\end{tikzpicture}
\end{minipage}
}

\newcommand{\CKExampleIIRRapproxTwoRRapprox}{
\begin{minipage}{2.3cm}
\begin{tikzpicture}
     \def \eps {0.5}

    \coordinate (v);
    \def \rad {1}
    \coordinate[label=above:$1$] (v1) at ([shift=(60:\rad)]v);
    \coordinate[label=left:$0$] (v2) at ([shift=(180:\rad)]v);
    \coordinate[label=below:$z$] (v3) at ([shift=(300:\rad)]v);

     \coordinate (x1) at ($(v2)!\eps!(v1)$);
    \coordinate (x2) at ($(v2)!\eps!(v3)$);
    \coordinate[label=above:{$x_2$}] (x3) at ($(x1)!\eps!(x2)$);

    \draw[draw=black, fill=black, thin, solid] (v1) circle (1pt);
    \draw[draw=black, fill=black, thin, solid] (v2) circle (1pt);
    \draw[draw=black, fill=black, thin, solid] (v3) circle (1pt);
    \draw[draw=black, fill=black, thin, solid] (x3) circle (1pt);

    \draw[draw=black, thin, solid] (v2) --  node[inner sep=0pt,below] {\tiny $\frac{2-3\ep}{\lambda}$} (x3);
    \draw[draw=black, thin, solid] (x3) -- node[inner sep=0pt,above right] {\tiny $\frac{1 - 2\ep}{\lambda}$} (v3);
    \draw[draw=black, thin, solid] (v1) -- node[inner sep=0pt,below right] {\tiny $\frac{\ep}{\lambda}$} (x3);

\end{tikzpicture}
\end{minipage}
}

\newcommand{\CKExampleIIRRapproxTwoRRconvolution}{
\begin{minipage}{3cm}
\begin{tikzpicture}
     \def \eps {0.5}

    \coordinate (v);
    \def \rad {1}
    \coordinate[label=above:$1$] (v1) at ([shift=(60:\rad)]v);
    \coordinate[label=left:$0$] (v2) at ([shift=(180:\rad)]v);
    \coordinate[label=below:$x_3$] (v3) at ([shift=(300:\rad)]v);

     \coordinate (x1) at ($(v2)!\eps!(v1)$);
    \coordinate (x2) at ($(v2)!\eps!(v3)$);
    \coordinate[label=below:$z$] (z) at ($(v3)+(1,0)$);

    \draw[draw=black, fill=black, thin, solid] (v1) circle (1pt);
    \draw[draw=black, fill=black, thin, solid] (v2) circle (1pt);
    \draw[draw=black, fill=black, thin, solid] (v3) circle (1pt);
    \draw[draw=black, fill=black, thin, solid] (z) circle (1pt);

    \draw[draw=black, thin, solid, bend left=30] (v3) to (z);
    \draw[draw=black, thin, solid, bend right=30] (v3) to (z);
    
    \draw[draw=black, thin, solid] (v2) -- node[inner sep=0pt, below left] {\tiny $\frac{1 - 3\ep}{\lambda}$} (v3);
    \draw[draw=black, thin, solid] (v1) -- (v3);

\end{tikzpicture}
\end{minipage}
}

\newcommand{\momentumthreepointp}{
\begin{minipage}{2cm}
\begin{tikzpicture}[rotate=-90]
    \def\circleradius{0.5cm}
    \def\linelength{1cm}
    \def\arcradius{0.35cm}
    \def\shrink{0.4cm} 
    \def\offset{-0.35cm} 

    \coordinate (Afull) at (120:{\circleradius+\linelength});
    \coordinate (Bfull) at (240:{\circleradius+\linelength});
    \coordinate (Cfull) at (0:{\circleradius+\linelength});
    
    \coordinate (A) at ($ (120:{\circleradius+\linelength-\shrink}) + (30:\offset) $);
    \coordinate (B) at ($ (240:{\circleradius+\linelength-\shrink}) + (330:\offset)$);
    \coordinate (C) at ($ (0:{\circleradius+\linelength-\shrink}) + (270:\offset) $);
    \coordinate (AS) at ($ (120:{\circleradius+\linelength-\shrink}) - (30:\offset) $);

    \filldraw[black!20] (0,0) circle (\circleradius);
    \draw[thick] (0,0) circle (\circleradius);
    
    \foreach \angle in {0,120,240} {
        \filldraw
            (\angle:\circleradius) circle (0.05);
    }

    \draw[thick, ->, bend right=30] (B) to (A);

    \draw[thick, ->, bend right=30] (AS) to (C);

    \node at (0,0) {$G^p$};
    \node at (C) [above right =0.5 and 0.2] {$\scriptstyle p_1$};
    \node at (A) [left = 0.4] {$\scriptstyle p_2$};
\end{tikzpicture}
\end{minipage}}

\newcommand{\momentumthreepointza}{
\begin{minipage}{2cm}
\begin{tikzpicture}[rotate=-90]
    \def\circleradius{0.5cm}
    \def\linelength{1cm}
    \def\arcradius{0.35cm}
    \def\shrink{0.4cm} 
    \def\offset{-0.35cm} 

    \coordinate (Afull) at (120:{\circleradius+\linelength});
    \coordinate (Bfull) at (240:{\circleradius+\linelength});
    \coordinate (Cfull) at (0:{\circleradius+\linelength});
    
    \coordinate (A) at ($ (120:{\circleradius+\linelength-\shrink}) + (30:\offset) $);
    \coordinate (B) at ($ (240:{\circleradius+\linelength-\shrink}) + (330:\offset)$);
    \coordinate (C) at ($ (0:{\circleradius+\linelength-\shrink}) + (270:\offset) $);
    \coordinate (AS) at ($ (120:{\circleradius+\linelength-\shrink}) - (30:\offset) $);

    \filldraw[black!20] (0,0) circle (\circleradius);
    \draw[thick] (0,0) circle (\circleradius);
    
    \foreach \angle in {0,120,240} {
        \filldraw
            (\angle:\circleradius) circle (0.05);
    }

    \draw[thick, ->, bend right=30] (B) to (A);

    \draw[thick, ->, bend right=30] (AS) to (C);

    \node at (0,0) {$G^p$};
    \node at (C) [above right =0.5 and 0.2] {$\scriptstyle z_1$};
    \node at (A) [left = 0.4] {$\scriptstyle z_2$};
\end{tikzpicture}
\end{minipage}}

\newcommand{\momentumthreepointzb}{
\begin{minipage}{2.2cm}
\begin{tikzpicture}[rotate=-90]
    \def\circleradius{0.5cm}
    \def\linelength{1cm}
    \def\arcradius{0.35cm}
    \def\shrink{0.4cm} 
    \def\offset{-0.35cm} 

    \coordinate (Afull) at (120:{\circleradius+\linelength});
    \coordinate (Bfull) at (240:{\circleradius+\linelength});
    \coordinate (Cfull) at (0:{\circleradius+\linelength});
    
    \coordinate (A) at ($ (120:{\circleradius+\linelength-\shrink}) + (30:\offset) $);
    \coordinate (B) at ($ (240:{\circleradius+\linelength-\shrink}) + (330:\offset)$);
    \coordinate (C) at ($ (0:{\circleradius+\linelength-\shrink}) + (270:\offset) $);
    \coordinate (AS) at ($ (120:{\circleradius+\linelength-\shrink}) - (30:\offset) $);

    \filldraw[black!20] (0,0) circle (\circleradius);
    \draw[thick] (0,0) circle (\circleradius);
    
    \draw[thick] (120:\circleradius) -- (120:{\circleradius + \linelength});
    
    \node at (110:{\circleradius + \linelength/2}) {$\scriptstyle \nu^p$};

    \foreach \angle in {0,240} {
        \filldraw
            (\angle:\circleradius) circle (0.05);
    }

    \draw[thick, ->, bend right=30] (B) to (A);

    \draw[thick, ->, bend right=30] (AS) to (C);

    \node at (0,0) {$G^p$};
    \node at (C) [above right =0.5 and 0.2] {$\scriptstyle z_1$};
    \node at (A) [left = 0.4] {$\scriptstyle z_2$};
\end{tikzpicture}
\end{minipage}}

\newcommand{\momentumthreepointzc}{
\begin{minipage}{2.7cm}
\begin{tikzpicture}[rotate=-90]
    \def\circleradius{0.5cm}
    \def\linelength{1cm}
    \def\arcradius{0.35cm}
    \def\shrink{0.4cm} 
    \def\offset{-0.35cm} 

    \coordinate (Afull) at (120:{\circleradius+\linelength});
    \coordinate (Bfull) at (240:{\circleradius+\linelength});
    \coordinate (Cfull) at (0:{\circleradius+\linelength});
    
    \coordinate (A) at ($ (120:{\circleradius+\linelength-\shrink}) + (30:\offset) $);
    \coordinate (B) at ($ (240:{\circleradius+\linelength-\shrink}) + (330:\offset)$);
    \coordinate (C) at ($ (0:{\circleradius+\linelength-\shrink}) + (270:\offset) $);
    \coordinate (AS) at ($ (120:{\circleradius+\linelength-\shrink}) - (30:\offset) $);

    \filldraw[black!20] (0,0) circle (\circleradius);
    \draw[thick] (0,0) circle (\circleradius);
    \draw[thick] (240:\circleradius) -- (240:{\circleradius + \linelength});
    \node at (252:{\circleradius + \linelength/2}) {$\scriptstyle \nu^p$};

    \foreach \angle in {0,120} {
        \filldraw
            (\angle:\circleradius) circle (0.05);
    }

    \draw[thick, ->, bend right=30] (B) to (A);

    \draw[thick, ->, bend right=30] (AS) to (C);

    \node at (0,0) {$G^p$};
    \node at (C) [above right =0.5 and 0.2] {$\scriptstyle z_1$};
    \node at (A) [left = 0.4] {$\scriptstyle z_2$};
\end{tikzpicture}
\end{minipage}}

\newcommand{\momentumthreepointzd}{
\begin{minipage}{2cm}
\begin{tikzpicture}[rotate=-90]
    \def\circleradius{0.5cm}
    \def\linelength{1cm}
    \def\arcradius{0.35cm}
    \def\shrink{0.4cm} 
    \def\offset{-0.35cm} 

    \coordinate (Afull) at (120:{\circleradius+\linelength});
    \coordinate (Bfull) at (240:{\circleradius+\linelength});
    \coordinate (Cfull) at (0:{\circleradius+\linelength});
    
    \coordinate (A) at ($ (120:{\circleradius+\linelength-\shrink}) + (30:\offset) $);
    \coordinate (B) at ($ (240:{\circleradius+\linelength-\shrink}) + (330:\offset)$);
    \coordinate (C) at ($ (0:{\circleradius+\linelength-\shrink}) + (270:\offset) $);
    \coordinate (AS) at ($ (120:{\circleradius+\linelength-\shrink}) - (30:\offset) $);

    \filldraw[black!20] (0,0) circle (\circleradius);
    \draw[thick] (0,0) circle (\circleradius);
    \draw[thick] (0:\circleradius) -- (0:{\circleradius + \linelength});
    \node at (-12:{\circleradius + \linelength/2}) {$\scriptstyle \nu^p$};

    \foreach \angle in {120,240} {
        \filldraw
            (\angle:\circleradius) circle (0.05);
    }

    \draw[thick, ->, bend right=30] (B) to (A);

    \draw[thick, ->, bend right=30] (AS) to (C);

    \node at (0,0) {$G^p$};
    \node at (C) [above right =0.5 and 0.2] {$\scriptstyle z_1$};
    \node at (A) [left = 0.4] {$\scriptstyle z_2$};
\end{tikzpicture}
\end{minipage}}

\newcommand{\positionthreepointzb}{
\begin{minipage}{3.8cm}
\begin{tikzpicture}[rotate=-90]
    \def\circleradius{0.5cm}
    \def\linelength{1cm}
    \def\labels{0.8cm}

    \coordinate (Afull) at (120:{\circleradius+\linelength});
    \coordinate (Bfull) at (240:{\circleradius+\linelength});
    \coordinate (Cfull) at (0:{\circleradius+\linelength});
    
    \filldraw[black!20] (0,0) circle (\circleradius);
    \draw[thick] (0,0) circle (\circleradius);
    \draw[thick] (120:\circleradius) -- (120:{\circleradius + \linelength});
    \draw[thick] (240:\circleradius) .. controls (280:\linelength) and (320:\linelength) .. (0:\circleradius);
    \node at (110:\circleradius + \linelength/2) {$\scriptstyle \nu$};
    \node at (300:1.23cm) {$\scriptstyle \nu-\frac{\lambda+1}\lambda$};

    \foreach \angle in {0,240} {
        \filldraw
            (\angle:\circleradius) circle (0.05);
    }
    \filldraw(120:\circleradius+\linelength) circle (0.05);

    \node at (0,0) {$G$};
    \node at (0:\labels) {$\scriptstyle z_2$};
    \node at (120:\labels + \linelength) {$\scriptstyle 0$};
    \node at (240:\labels) {$\scriptstyle z_1$};

\end{tikzpicture}
\end{minipage}}

\newcommand{\positionthreepointzc}{
\begin{minipage}{3.8cm}
\begin{tikzpicture}[rotate=-90]
    \def\circleradius{0.5cm}
    \def\linelength{1cm}
    \def\labels{0.8cm}

    \coordinate (Afull) at (120:{\circleradius+\linelength});
    \coordinate (Bfull) at (240:{\circleradius+\linelength});
    \coordinate (Cfull) at (0:{\circleradius+\linelength});
    
    \filldraw[black!20] (0,0) circle (\circleradius);
    \draw[thick] (0,0) circle (\circleradius);
    \draw[thick] (240:\circleradius) -- (240:{\circleradius + \linelength});
    \draw[thick] (0:\circleradius) .. controls (40:\linelength) and (80:\linelength) .. (120:\circleradius);
    \node at (250:\circleradius + \linelength/2) {$\scriptstyle \nu$};
    \node at (60:1.23cm) {$\scriptstyle \nu-\frac{\lambda+1}\lambda$};

    \foreach \angle in {0,120} {
        \filldraw
            (\angle:\circleradius) circle (0.05);
    }
    \filldraw(240:\circleradius+\linelength) circle (0.05);

    \node at (0,0) {$G$};
    \node at (0:\labels) {$\scriptstyle z_2$};
    \node at (120:\labels) {$\scriptstyle 0$};
    \node at (240:\labels + \linelength) {$\scriptstyle z_1$};

\end{tikzpicture}
\end{minipage}}

\newcommand{\positionthreepointzd}{
\begin{minipage}{2cm}
\begin{tikzpicture}[rotate=-90]
    \def\circleradius{0.5cm}
    \def\linelength{1cm}
    \def\labels{0.8cm}

    \coordinate (Afull) at (120:{\circleradius+\linelength});
    \coordinate (Bfull) at (240:{\circleradius+\linelength});
    \coordinate (Cfull) at (0:{\circleradius+\linelength});
    
    \filldraw[black!20] (0,0) circle (\circleradius);
    \draw[thick] (0,0) circle (\circleradius);
    \draw[thick] (0:\circleradius) -- (0:{\circleradius + \linelength});
    \draw[thick] (120:\circleradius) .. controls (160:\linelength) and (200:\linelength) .. (240:\circleradius);
    \node at (-10:\circleradius + \linelength/2) {$\scriptstyle \nu$};
    \node at (180:1cm) {$\scriptstyle \nu-\frac{\lambda+1}\lambda$};

    \foreach \angle in {120,240} {
        \filldraw
            (\angle:\circleradius) circle (0.05);
    }
    \filldraw(0:\circleradius+\linelength) circle (0.05);

    \node at (0,0) {$G$};
    \node at (0:\labels + \linelength) {$\scriptstyle z_2$};
    \node at (120:\labels) {$\scriptstyle 0$};
    \node at (240:\labels) {$\scriptstyle z_1$};

\end{tikzpicture}
\end{minipage}}

\newcommand{\rotationExTwo}{
\begin{minipage}{2.3cm}
\begin{tikzpicture}[x=3ex, y = 3ex, baseline={([yshift=-8.0ex]current bounding box.center)}]
  \coordinate[label=above:$0$] (v1) at (-1.00, 1.00);
  \coordinate[label=above:$1$] (v2) at (3.00, 1.00);
  \coordinate[label=below:$z$] (v4) at (2.00, -3.00);
  \coordinate (v5) at (0.00, -1.00);
  \coordinate (v6) at (2.00, -1.00);
  \draw[draw=black, fill=black, thin, solid] (v1) circle (1pt);
  \draw[draw=black, fill=black, thin, solid] (v2) circle (1pt);
  \draw[draw=black, fill=black, thin, solid] (v4) circle (1pt);
  \draw[draw=black, fill=black, thin, solid] (v5) circle (1pt);
  \draw[draw=black, fill=black, thin, solid] (v6) circle (1pt);
  \draw[draw=black, thin, solid] (v1) -- (v5);
  \draw[draw=black, thin, solid] (v1) -- (v6);
  \draw[draw=black, thin, solid] (v2) -- (v5);
  \draw[draw=black, thin, solid] (v2) -- (v6);
  \draw[draw=black, thin, solid] (v5) -- (v6);
  \draw[draw=black, thin, solid] (v4) -- node[inner sep=0pt, right] {\tiny $\frac{1 + 2\ep}{\lambda}$} (v6);

\end{tikzpicture}
\end{minipage}
}

\newcommand{\rotationExA}{
\begin{minipage}{2.3cm}
\begin{tikzpicture}[x=3ex, y = 3ex, baseline={([yshift=-8.0ex]current bounding box.center)}]
  \coordinate[label=above:$0$] (v1) at (-1.00, 1.00);
  \coordinate[label=above:$1$] (v2) at (3.00, 1.00);
  \coordinate[label=below:$z$] (v4) at (2.00, -3.00);
  \coordinate (v5) at (0.00, -1.00);
  \coordinate (v6) at (2.00, -1.00);
  \draw[draw=black, fill=black, thin, solid] (v1) circle (1pt);
  \draw[draw=black, fill=black, thin, solid] (v2) circle (1pt);
  \draw[draw=black, fill=black, thin, solid] (v4) circle (1pt);
  \draw[draw=black, fill=black, thin, solid] (v5) circle (1pt);
  \draw[draw=black, fill=black, thin, solid] (v6) circle (1pt);
  \draw[draw=black, thin, solid] (v1) -- (v5);
  \draw[draw=black, thin, solid] (v1) -- (v6);
  \draw[draw=black, thin, solid] (v2) -- (v5);
  \draw[draw=black, thin, solid] (v2) -- (v6);
  \draw[draw=black, thin, solid] (v5) -- (v6);
  \draw[draw=black, thin, solid] (v4) -- node[inner sep=0pt, right] {\tiny $\frac{1 + a\ep}{\lambda}$} (v6);

\end{tikzpicture}
\end{minipage}
}

\newcommand{\rotationExTwoAttached}{
\begin{minipage}{2.3cm}
\begin{tikzpicture}[x=3ex, y = 3ex, baseline={([yshift=-8.0ex]current bounding box.center)}]
  \coordinate[label=above:$0$] (v1) at (-1.00, 1.00);
  \coordinate[label=above:$1$] (v2) at (3.00, 1.00);
  \coordinate[label=below:$z$] (v4) at (2.00, -3.00);
  \coordinate (v5) at (0.00, -1.00);
  \coordinate (v6) at (2.00, -1.00);
  \draw[draw=black, fill=black, thin, solid] (v1) circle (1pt);
  \draw[draw=black, fill=black, thin, solid] (v2) circle (1pt);
  \draw[draw=black, fill=black, thin, solid] (v4) circle (1pt);
  \draw[draw=black, fill=black, thin, solid] (v5) circle (1pt);
  \draw[draw=black, fill=black, thin, solid] (v6) circle (1pt);
  \draw[draw=black, thin, solid] (v1) -- (v5);
  \draw[draw=black, thin, solid] (v1) -- (v6);
  \draw[draw=black, thin, solid] (v2) -- (v5);
  \draw[draw=black, thin, solid] (v2) -- (v6);
  \draw[draw=black, thin, solid] (v5) -- (v6);
  \draw[draw=black, thin, solid] (v1) -- node[inner sep=0pt, above] {\tiny $\frac{-1 + 3\ep}{\lambda}$} (v2);
  \draw[draw=black, thin, solid] (v4) -- node[inner sep=0pt, right] {\tiny $\frac{1 + 2\ep}{\lambda}$} (v6);

\end{tikzpicture}
\end{minipage}
}

\newcommand{\rotationExTwoRotated}{
\begin{minipage}{3cm}
\begin{tikzpicture}[x=3ex, y = 3ex, baseline={([yshift=-8.0ex]current bounding box.center)}]
  \coordinate[label=above:$0$] (v1) at (-1.00, 1.00);
  \coordinate (v2) at (3.00, 1.00);
  \coordinate[label=above:$1$] (v4) at (3.00, 3.00);
  \coordinate (v5) at (0.00, -1.00);
  \coordinate[label=below:$z$] (v6) at (2.00, -1.00);
  \draw[draw=black, fill=black, thin, solid] (v1) circle (1pt);
  \draw[draw=black, fill=black, thin, solid] (v2) circle (1pt);
  \draw[draw=black, fill=black, thin, solid] (v5) circle (1pt);
  \draw[draw=black, fill=black, thin, solid] (v6) circle (1pt);
  \draw[draw=black, fill=black, thin, solid] (v4) circle (1pt);
  \draw[draw=black, thin, solid] (v1) -- (v5);
  \draw[draw=black, thin, solid] (v1) -- (v6);
  \draw[draw=black, thin, solid] (v2) -- (v5);
  \draw[draw=black, thin, solid] (v2) -- (v6);
  \draw[draw=black, thin, solid] (v5) -- (v6);
  \draw[draw=black, thin, solid] (v4) -- node[inner sep=0pt,right] {\tiny $\frac{1 + 2\ep}{\lambda}$} (v2);
  \draw[draw=black, thin, solid,bend right = 90, looseness=1.7] (v1) to node[inner sep=0pt,below left] {\tiny $\frac{-1 + 3\ep}{\lambda}$} (v6);

\end{tikzpicture}
\end{minipage}
}

\newcommand{\rotationExTwoCombined}{
\begin{minipage}{3cm}
\begin{tikzpicture}[x=3ex, y = 3ex, baseline={([yshift=-8.0ex]current bounding box.center)}]
  \coordinate[label=above:$0$] (v1) at (-1.00, 1.00);
  \coordinate (v2) at (3.00, 1.00);
  \coordinate[label=above:$1$] (v4) at (3.00, 3.00);
  \coordinate (v5) at (0.00, -1.00);
  \coordinate[label=below:$z$] (v6) at (2.00, -1.00);
  \draw[draw=black, fill=black, thin, solid] (v1) circle (1pt);
  \draw[draw=black, fill=black, thin, solid] (v2) circle (1pt);
  \draw[draw=black, fill=black, thin, solid] (v4) circle (1pt);
  \draw[draw=black, fill=black, thin, solid] (v5) circle (1pt);
  \draw[draw=black, fill=black, thin, solid] (v6) circle (1pt);
  \draw[draw=black, thin, solid] (v1) -- (v5);
  \draw[draw=black, thin, solid] (v2) -- (v5);
  \draw[draw=black, thin, solid] (v2) -- (v6);
  \draw[draw=black, thin, solid] (v5) -- (v6);
  \draw[draw=black, thin, solid] (v4) -- node[inner sep=0pt, right] {\tiny $\frac{1 + 2\ep}{\lambda}$} (v2);
  \draw[draw=black, thin, solid,bend right = 90, looseness=1.7] (v1) to node[inner sep=0pt,below left] {\tiny $\frac{2\ep}{\lambda}$} (v6);

\end{tikzpicture}
\end{minipage}
}

\newcommand{\rotationExTwoDetached}{
\begin{minipage}{3cm}
\begin{tikzpicture}[x=3ex, y = 3ex, baseline={([yshift=-8.0ex]current bounding box.center)}]
  \coordinate[label=above:$0$] (v1) at (-1.00, 1.00);
  \coordinate (v2) at (3.00, 1.00);
  \coordinate[label=above:$1$] (v4) at (3.00, 3.00);
  \coordinate (v5) at (0.00, -1.00);
  \coordinate[label=below:$z$] (v6) at (2.00, -1.00);
  \draw[draw=black, fill=black, thin, solid] (v1) circle (1pt);
  \draw[draw=black, fill=black, thin, solid] (v2) circle (1pt);
  \draw[draw=black, fill=black, thin, solid] (v4) circle (1pt);
  \draw[draw=black, fill=black, thin, solid] (v5) circle (1pt);
  \draw[draw=black, fill=black, thin, solid] (v6) circle (1pt);
  \draw[draw=black, thin, solid] (v1) -- (v5);
  \draw[draw=black, thin, solid] (v2) -- (v5);
  \draw[draw=black, thin, solid] (v2) -- (v6);
  \draw[draw=black, thin, solid] (v5) -- (v6);
  \draw[draw=black, thin, solid] (v4) -- node[inner sep=0pt, right] {\tiny $\frac{1 + 2\ep}{\lambda}$} (v2);

\end{tikzpicture}
\end{minipage}
}

\newcommand{\rotationExNone}{
\begin{minipage}{2.3cm}
\begin{tikzpicture}[x=3ex, y = 3ex, baseline={([yshift=-8.0ex]current bounding box.center)}]
  \coordinate[label=above:$0$] (v1) at (-1.00, 1.00);
  \coordinate[label=above:$1$] (v2) at (3.00, 1.00);
  \coordinate[label=below:$z$] (v4) at (2.00, -3.00);
  \coordinate (v5) at (0.00, -1.00);
  \coordinate (v6) at (2.00, -1.00);
  \draw[draw=black, fill=black, thin, solid] (v1) circle (1pt);
  \draw[draw=black, fill=black, thin, solid] (v2) circle (1pt);
  \draw[draw=black, fill=black, thin, solid] (v4) circle (1pt);
  \draw[draw=black, fill=black, thin, solid] (v5) circle (1pt);
  \draw[draw=black, fill=black, thin, solid] (v6) circle (1pt);
  \draw[draw=black, thin, solid] (v1) -- (v5);
  \draw[draw=black, thin, solid] (v1) -- (v6);
  \draw[draw=black, thin, solid] (v2) -- (v5);
  \draw[draw=black, thin, solid] (v2) -- (v6);
  \draw[draw=black, thin, solid] (v5) -- (v6);
  \draw[draw=black, thin, solid] (v4) -- (v6);

\end{tikzpicture}
\end{minipage}
}

\newcommand{\graphicalfunctiona}{
\begin{minipage}{2.5cm}
\begin{tikzpicture}[rotate=-90]
    \def\circleradius{0.5cm}
    \def\linelength{1cm}
    \def\labels{0.8cm}

    \coordinate (Afull) at (120:{\circleradius+\linelength});
    \coordinate (Bfull) at (240:{\circleradius+\linelength});
    \coordinate (Cfull) at (0:{\circleradius+\linelength});
    
    \filldraw[black!20] (0,0) circle (\circleradius);
    \draw[thick] (0,0) circle (\circleradius);
    \draw[thick] (0:\circleradius) -- (0:{\circleradius + \linelength});
    \node at (-10:\circleradius + \linelength/2) {$\scriptstyle \nu$};

    \foreach \angle in {120,240} {
        \filldraw
            (\angle:\circleradius) circle (0.05);
    }
    \filldraw(0:\circleradius+\linelength) circle (0.05);

    \node at (180:\linelength) {$G$};
    \node at (16:1.65cm) {$\scriptstyle z_2=z$};
    \node at (120:\labels) {$\scriptstyle 0$};
    \node at (240:\labels+0.1cm) {$\scriptstyle z_1=1$};

\end{tikzpicture}
\end{minipage}}

\newcommand{\graphicalfunctionb}{
\begin{minipage}{4cm}
\begin{tikzpicture}[rotate=-90]
    \def\circleradius{0.5cm}
    \def\linelength{1cm}
    \def\labels{0.8cm}

    \coordinate (Afull) at (120:{\circleradius+\linelength});
    \coordinate (Bfull) at (240:{\circleradius+\linelength});
    \coordinate (Cfull) at (0:{\circleradius+\linelength});
    
    \filldraw[black!20] (0,0) circle (\circleradius);
    \draw[thick] (0,0) circle (\circleradius);
    \draw[thick] (120:\circleradius) -- (120:{\circleradius + \linelength});
    \draw[thick] (240:\circleradius) .. controls (280:\linelength) and (320:\linelength) .. (0:\circleradius);
    \draw[thick] (120:\circleradius) .. controls (160:\linelength) and (200:\linelength) .. (240:\circleradius);
    \draw[thick] (120:\circleradius+\linelength) .. controls (160:2cm) and (220:2cm) .. (240:\circleradius);
    \node at (110:\circleradius + \linelength/2) {$\scriptstyle \nu$};
    \node at (300:1.23cm) {$\scriptstyle \nu-\frac{\lambda+1}\lambda$};
    \node at (170:1cm) {$\scriptstyle \frac{\lambda+1}\lambda-N_G$};
    \node at (220:1.6cm) {$\scriptstyle N_G-\nu$};
    
    \foreach \angle in {0,240} {
        \filldraw
            (\angle:\circleradius) circle (0.05);
    }
    \filldraw(120:\circleradius+\linelength) circle (0.05);
   \filldraw(0:\circleradius+2*\linelength) circle (0.0);

    \node at (0:\labels) {$\scriptstyle z_2=z$};
    \node at (120:\labels + \linelength) {$\scriptstyle 0$};
    \node at (250:\labels+0.15cm) {$\scriptstyle z_1=1$};

\end{tikzpicture}
\end{minipage}}

\newcommand{\graphicalfunctionc}{
\begin{minipage}{4cm}
\begin{tikzpicture}[rotate=-90]
    \def\circleradius{0.5cm}
    \def\linelength{1cm}
    \def\labels{0.8cm}

    \coordinate (Afull) at (120:{\circleradius+\linelength});
    \coordinate (Bfull) at (240:{\circleradius+\linelength});
    \coordinate (Cfull) at (0:{\circleradius+\linelength});
    
    \filldraw[black!20] (0,0) circle (\circleradius);
    \draw[thick] (0,0) circle (\circleradius);
    \draw[thick] (240:\circleradius) -- (240:{\circleradius + \linelength});
    \draw[thick] (0:\circleradius) .. controls (40:\linelength) and (80:\linelength) .. (120:\circleradius);
    \draw[thick] (120:\circleradius) .. controls (160:\linelength) and (200:\linelength) .. (240:\circleradius);
    \draw[thick] (120:\circleradius) .. controls (140:2cm) and (200:2cm) .. (240:\circleradius+\linelength);
    \node at (250:\circleradius + \linelength/2) {$\scriptstyle \nu$};
    \node at (60:1.23cm) {$\scriptstyle \nu-\frac{\lambda+1}\lambda$};
    \node at (190:1cm) {$\scriptstyle \frac{\lambda+1}\lambda-N_G$};
    \node at (140:1.6cm) {$\scriptstyle N_G-\nu$};

    \foreach \angle in {0,120} {
        \filldraw
            (\angle:\circleradius) circle (0.05);
    }
    \filldraw(240:\circleradius+\linelength) circle (0.05);
   \filldraw(0:\circleradius+2*\linelength) circle (0.0);

    \node at (0:\labels) {$\scriptstyle z_2=z$};
    \node at (117:\labels) {$\scriptstyle 0$};
    \node at (240:\labels + \linelength+0.1cm) {$\scriptstyle z_1=1$};

\end{tikzpicture}
\end{minipage}}

\newcommand{\NewTwist}{
\begin{minipage}{4.5cm}
\begin{tikzpicture}[scale = 0.7, baseline={([yshift=-1.3ex]current bounding box.center)}]
    \coordinate[label=above:$0$] (va) at (0,3);
    \coordinate[label=above:$1$] (vb) at (0,2);
    \coordinate (vm) at (0, 2.15);
    \coordinate[label=above:$z$] (vc) at (0,1);
    \coordinate[label=above:$\infty$] (vd) at (0,0);
    \filldraw (va) circle (1.42pt);
    \filldraw (vb) circle (1.42pt);
    \filldraw (vc) circle (1.42pt);
    \filldraw (vd) circle (1.42pt);

    \draw[fill=black!20] (va) arc (90:270:3 and 1.5) arc (270:90:.5) arc (270:90:.5) arc (270:90:.5);
    \draw[fill=black!20] (va) arc (90:-90:3 and 1.5) arc (-90:90:.5) arc (-90:90:.5) arc (-90:90:.5);
    \node (G1) at (-1.5,1.5) {$\overline{G}_1$};
    \node (G2) at (+1.5,1.5) {$\overline{G}_2$};
    \node (G) at (-2,0) {$\overline{G}$};
\end{tikzpicture}
\end{minipage}
}

\newcommand{\momentumthreepointpPositionized}{
\begin{minipage}{2cm}
\begin{tikzpicture}[rotate=-90,baseline={([yshift=6.5ex]current bounding box.center)}]
    \def\circleradius{0.5cm}
    \def\linelength{1cm}
    \def\arcradius{0.35cm}
    \def\shrink{0.4cm} 
    \def\offset{-0.35cm} 

    \coordinate (Afull) at (120:{\circleradius+\linelength});
    \coordinate (Bfull) at (240:{\circleradius+\linelength});
    \coordinate (Cfull) at (0:{\circleradius+\linelength});
    
    \coordinate (A) at ($ (120:{\circleradius+\linelength-\shrink}) + (30:\offset) $);
    \coordinate (B) at ($ (240:{\circleradius+\linelength-\shrink}) + (330:\offset)$);
    \coordinate (C) at ($ (0:{\circleradius+\linelength-\shrink}) + (270:\offset) $);
    \coordinate (AS) at ($ (120:{\circleradius+\linelength-\shrink}) - (30:\offset) $);

    \filldraw[black!20] (0,0) circle (\circleradius);
    \draw[thick] (0,0) circle (\circleradius);
    
    \foreach \angle in {0,120,240} {
        \filldraw
            (\angle:\circleradius) circle (0.05);
    }
    \coordinate[label=below:$p_2$] (l1) at (0:\circleradius);
    \coordinate[label=above right:$0$] (l2) at (120:\circleradius);
    \coordinate[label=above left:$p_1$] (l3) at (240:\circleradius);

    \node at (0,0) {$G$};
\end{tikzpicture}
\end{minipage}}

\newcommand{\KfivetwoDimregDecompletedExtTwoConst}{
\begin{minipage}{2cm}
\begin{tikzpicture}[scale=1.5, x=3ex, y = 3ex, baseline={([yshift=-8.0ex]current bounding box.center)}]
  \coordinate[label=above:$1$] (v1) at (0.00, 1.00);
  \coordinate[label=above right:$0$] (v2) at (2.00, 1.00);
  \coordinate (v4) at (1.00, -2.00);
  \coordinate (v5) at (0.00, -1.00);
  \coordinate (v6) at (2.00, -1.00);
  \coordinate[label=below:$z$] (v7) at (1.00, -3.00);
  \draw[draw=black, fill=black, thin, solid] (v1) circle (0.67pt);
  \draw[draw=black, fill=black, thin, solid] (v2) circle (0.67pt);
  \draw[draw=black, fill=black, thin, solid] (v4) circle (0.67pt);
  \draw[draw=black, fill=black, thin, solid] (v5) circle (0.67pt);
  \draw[draw=black, fill=black, thin, solid] (v6) circle (0.67pt);
  \draw[draw=black, fill=black, thin, solid] (v7) circle (0.67pt);
  \draw[draw=black, thin, solid] (v1) -- (v5);
  \draw[draw=black, thin, solid] (v1) -- (v6);
  \draw[draw=black, thin, solid] (v2) -- (v5);
  \draw[draw=black, thin, solid] (v2) -- (v6);
  \draw[draw=black, thin, solid] (v7) -- (v4);
  \draw[draw=black, thin, solid] (v5) to (v4);
  \draw[draw=black, thin, solid] (v6) to (v4);

\end{tikzpicture}
\end{minipage}
}

\newcommand{\KfivetwoDimregDecompletedExtOneConstruct}{
\begin{minipage}{2cm}
\begin{tikzpicture}[scale=1.5, x=3ex, y = 3ex, baseline={([yshift=-8.0ex]current bounding box.center)}]
  \coordinate[label=above:$1$] (v1) at (0.00, 1.00);
  \coordinate[label=above right:$0$] (v2) at (2.00, 1.00);
  \coordinate (v5) at (0.00, -1.00);
  \coordinate (v6) at (2.00, -1.00);
  \coordinate[label=below:$z$] (v4) at (1.00, -2.00);
  \draw[draw=black, fill=black, thin, solid] (v1) circle (0.67pt);
  \draw[draw=black, fill=black, thin, solid] (v2) circle (0.67pt);
  \draw[draw=black, fill=black, thin, solid] (v4) circle (0.67pt);
  \draw[draw=black, fill=black, thin, solid] (v5) circle (0.67pt);
  \draw[draw=black, fill=black, thin, solid] (v6) circle (0.67pt);
  \draw[draw=black, thin, solid] (v1) -- (v5);
  \draw[draw=black, thin, solid] (v1) -- (v6);
  \draw[draw=black, thin, solid] (v2) -- (v5);
  \draw[draw=black, thin, solid] (v2) -- (v6);
  \draw[draw=black, thin, solid] (v5) -- (v6);
  \draw[draw=black, thin, solid] (v5) to node[inner sep=0pt,below left] {\tiny{$\frac{1 + 2\ep}{\lambda}$}} (v4);
  \draw[draw=black, thin, solid] (v6) to node[inner sep=0pt,below right] {\tiny{$\frac{2\ep}{\lambda}$}} (v4);

\end{tikzpicture}
\end{minipage}
}

\newcommand{\KfivetwoDimregDecompletedExtOneConComplete}{
\begin{minipage}{3.4cm}
\begin{tikzpicture}[scale=1.5, x=3ex, y = 3ex, baseline={([yshift=-8.0ex]current bounding box.center)}]
  \coordinate[label=above:$1$] (v1) at (0.00, 1.00);
  \coordinate[label=above right:$0$] (v2) at (2.00, 1.00);
  \coordinate (v5) at (0.00, -1.00);
  \coordinate (v6) at (2.00, -1.00);
  \coordinate[label=below:$z$] (v4) at (1.00, -2.00);
  \coordinate[label=right:$\infty$] (v7) at (4.00, 0.00);
  \draw[draw=black, fill=black, thin, solid] (v1) circle (0.67pt);
  \draw[draw=black, fill=black, thin, solid] (v2) circle (0.67pt);
  \draw[draw=black, fill=black, thin, solid] (v4) circle (0.67pt);
  \draw[draw=black, fill=black, thin, solid] (v5) circle (0.67pt);
  \draw[draw=black, fill=black, thin, solid] (v6) circle (0.67pt);
  \draw[draw=black, fill=black, thin, solid] (v7) circle (0.67pt);
  \draw[draw=black, thin, solid] (v1) -- (v5);
  \draw[draw=black, thin, solid] (v1) -- (v6);
  \draw[draw=black, thin, solid] (v2) -- (v5);
  \draw[draw=black, thin, solid] (v2) -- (v6);
  \draw[draw=black, thin, solid] (v5) -- (v6);
  \draw[draw=black, thin, solid] (v5) to node[inner sep=0pt,right] {\tiny{$\frac{2\ep}{\lambda}$}} (v4);
  \draw[draw=black, thin, solid] (v6) to node[inner sep=-1pt,below right] {\tiny{$\frac{1 + 2\ep}{\lambda}$}} (v4);
  \draw[draw=black, thin, solid] (v1) to node[inner sep=0pt,above ] {\tiny{$\frac{-2 - \ep}{\lambda}$}} (v2);
  \draw[draw=black, thin, solid] (v2) to node[inner sep=0pt,above] {\tiny{$\frac{3\ep}{\lambda}$}} (v7);
  \draw[draw=black, thin, solid] (v6) to node[inner sep=0pt,above left] {\tiny{$-\frac{\ep}{\lambda}$}} (v7);
  \draw[draw=black, thin, solid] (v1) to [bend left=90, looseness=1.5] node[inner sep=0pt, below left ] {\tiny{$\frac{3\ep}{\lambda}$}} (v7);
  \draw[draw=black, thin, solid] (v4) to [bend right=90, looseness=0.8] node[inner sep=6pt, above] {\tiny{$\frac{-1-4\ep}{\lambda}$}} (v7);
  \draw[draw=black, thin, solid] (v5) to [bend right=90, looseness=1.9] node[inner sep=2pt, above]
  {\tiny{$1$}} (v7);
\end{tikzpicture}
\end{minipage}
}

\newcommand{\KfivetwoDimregDecompletedExtOneConCDC}{
\begin{minipage}{3.4cm}
\begin{tikzpicture}[scale=1.5, x=3ex, y = 3ex, baseline={([yshift=-8.0ex]current bounding box.center)}]
  \coordinate[label=above:$z$] (v1) at (0.00, 1.00);
  \coordinate (v5) at (0.00, -1.00);
  \coordinate (v6) at (2.00, -1.00);
  \coordinate[label=below:$1$] (v4) at (1.00, -2.00);
  \coordinate[label=right:$0$] (v7) at (4.00, 0.00);
  \draw[draw=black, fill=black, thin, solid] (v1) circle (0.67pt);
  \draw[draw=black, fill=black, thin, solid] (v4) circle (0.67pt);
  \draw[draw=black, fill=black, thin, solid] (v5) circle (0.67pt);
  \draw[draw=black, fill=black, thin, solid] (v6) circle (0.67pt);
  \draw[draw=black, fill=black, thin, solid] (v7) circle (0.67pt);
  \draw[draw=black, thin, solid] (v1) -- (v5);
  \draw[draw=black, thin, solid] (v1) -- (v6);
  \draw[draw=black, thin, solid] (v5) -- (v6);
  \draw[draw=black, thin, solid] (v5) to node[inner sep=0pt,right] {\tiny{$\frac{2\ep}{\lambda}$}} (v4);
  \draw[draw=black, thin, solid] (v6) to node[inner sep=0pt,below right] {\tiny{$\frac{1 + 2\ep}{\lambda}$}} (v4);
  \draw[draw=black, thin, solid] (v6) to node[inner sep=0pt,above left] {\tiny{$-\frac{\ep}{\lambda}$}} (v7);
  \draw[draw=black, thin, solid] (v1) to node[inner sep=0pt, above right] {\tiny{$\frac{3\ep}{\lambda}$}} (v7);
  \draw[draw=black, thin, solid] (v5) to [bend right=90, looseness=1.9] node[inner sep=2pt, above]
  {\tiny{$1$}} (v7);
\end{tikzpicture}
\end{minipage}
}
\newcommand{\KfivetwoDimregDecompletedExtOneConCDCR}{
\begin{minipage}{2.2cm}
\begin{tikzpicture}[scale=1.5, x=3ex, y = 3ex, baseline={([yshift=-8.0ex]current bounding box.center)}]
  \coordinate[label=above:$z$] (v1) at (0.00, 1.00);
  \coordinate (v5) at (0.00, -1.00);
  \coordinate (v6) at (2.00, -1.00);
  \coordinate[label=below:$1$] (v4) at (1.00, -2.00);
  \coordinate[label=right:$0$] (v7) at (2.00, 1.00);
  \draw[draw=black, fill=black, thin, solid] (v1) circle (0.67pt);
  \draw[draw=black, fill=black, thin, solid] (v4) circle (0.67pt);
  \draw[draw=black, fill=black, thin, solid] (v5) circle (0.67pt);
  \draw[draw=black, fill=black, thin, solid] (v6) circle (0.67pt);
  \draw[draw=black, fill=black, thin, solid] (v7) circle (0.67pt);
  \draw[draw=black, thin, solid] (v1) -- (v5);
  \draw[draw=black, thin, solid] (v1) -- (v6);
  \draw[draw=black, thin, solid] (v5) -- (v6);
  \draw[draw=black, thin, solid] (v5) to node[inner sep=0pt,below left] {\tiny{$\frac{2\ep}{\lambda}$}} (v4);
  \draw[draw=black, thin, solid] (v6) to node[inner sep=0pt,below right] {\tiny{$\frac{1 + 2\ep}{\lambda}$}} (v4);
  \draw[draw=black, thin, solid] (v6) to node[inner sep=0pt,right] {\tiny{$-\frac{\ep}{\lambda}$}} (v7);
  \draw[draw=black, thin, solid] (v5) to (v7);
\end{tikzpicture}
\end{minipage}
}

\newcommand{\kfivetwoexpandedone}{
\begin{minipage}{2.2cm}
\begin{tikzpicture}[scale=1.5, x=3ex, y = 3ex, baseline={([yshift=-8.0ex]current bounding box.center)}]
  \coordinate[label=above:$z$] (v1) at (0.00, 1.00);
  \coordinate (v5) at (0.00, -1.00);
  \coordinate (v6) at (2.00, -1.00);
  \coordinate[label=below:$1$] (v4) at (1.00, -2.00);
  \coordinate[label=right:$0$] (v7) at (2.00, 1.00);
  \draw[draw=black, fill=black, thin, solid] (v1) circle (0.67pt);
  \draw[draw=black, fill=black, thin, solid] (v4) circle (0.67pt);
  \draw[draw=black, fill=black, thin, solid] (v5) circle (0.67pt);
  \draw[draw=black, fill=black, thin, solid] (v6) circle (0.67pt);
  \draw[draw=black, fill=black, thin, solid] (v7) circle (0.67pt);
  \draw[draw=black, thin, solid] (v1) -- (v5);
  \draw[draw=black, thin, solid] (v1) -- (v6);
  \draw[draw=black, thin, solid] (v5) -- (v6);
  \draw[draw=black, thin, solid] (v6) to (v4);
  \draw[draw=black, thin, solid] (v6) to node[inner sep=0pt,right] {\tiny{$-\frac{\ep}{\lambda}$}} (v7);
  \draw[draw=black, thin, solid] (v5) to (v7);
\end{tikzpicture}
\end{minipage}
}

\newcommand{\kfivetwoexpandedtwo}{
\begin{minipage}{2.2cm}
\begin{tikzpicture}[scale=1.5, x=3ex, y = 3ex, baseline={([yshift=-8.0ex]current bounding box.center)}]
  \coordinate[label=above:$z$] (v1) at (0.00, 1.00);
  \coordinate (v5) at (0.00, -1.00);
  \coordinate (v6) at (2.00, -1.00);
  \coordinate[label=below:$1$] (v4) at (1.00, -2.00);
  \coordinate[label=right:$0$] (v7) at (2.00, 1.00);
  \draw[draw=black, fill=black, thin, solid] (v1) circle (0.67pt);
  \draw[draw=black, fill=black, thin, solid] (v4) circle (0.67pt);
  \draw[draw=black, fill=black, thin, solid] (v5) circle (0.67pt);
  \draw[draw=black, fill=black, thin, solid] (v6) circle (0.67pt);
  \draw[draw=black, fill=black, thin, solid] (v7) circle (0.67pt);
  \draw[draw=black, thin, solid] (v1) -- (v5);
  \draw[draw=black, thin, solid] (v1) -- (v6);
  \draw[draw=black, thin, solid] (v5) -- (v6);
  \draw[draw=black, thin, solid] (v5) to node[inner sep=0pt,below left] {\tiny{$\frac{2\ep}{\lambda}$}} (v4);
  \draw[draw=black, thin, solid] (v6) to (v4);
  \draw[draw=black, thin, solid] (v5) to (v7);
\end{tikzpicture}
\end{minipage}
}

\newcommand{\kfivetwoexpandedthree}{
\begin{minipage}{2.2cm}
\begin{tikzpicture}[scale=1.5, x=3ex, y = 3ex, baseline={([yshift=-8.0ex]current bounding box.center)}]
  \coordinate[label=above:$z$] (v1) at (0.00, 1.00);
  \coordinate (v5) at (0.00, -1.00);
  \coordinate (v6) at (2.00, -1.00);
  \coordinate[label=below:$1$] (v4) at (1.00, -2.00);
  \coordinate[label=right:$0$] (v7) at (2.00, 1.00);
  \draw[draw=black, fill=black, thin, solid] (v1) circle (0.67pt);
  \draw[draw=black, fill=black, thin, solid] (v4) circle (0.67pt);
  \draw[draw=black, fill=black, thin, solid] (v5) circle (0.67pt);
  \draw[draw=black, fill=black, thin, solid] (v6) circle (0.67pt);
  \draw[draw=black, fill=black, thin, solid] (v7) circle (0.67pt);
  \draw[draw=black, thin, solid] (v1) -- (v5);
  \draw[draw=black, thin, solid] (v1) -- (v6);
  \draw[draw=black, thin, solid] (v5) -- (v6);
  \draw[draw=black, thin, solid] (v6) to node[inner sep=0pt,below right] {\tiny{$\frac{1 + 2\ep}{\lambda}$}} (v4);
  \draw[draw=black, thin, solid] (v5) to (v7);
\end{tikzpicture}
\end{minipage}
}

\newcommand{\kfivetwoexpandedzero}{
\begin{minipage}{2.2cm}
\begin{tikzpicture}[scale=1.5, x=3ex, y = 3ex, baseline={([yshift=-8.0ex]current bounding box.center)}]
  \coordinate[label=above:$z$] (v1) at (0.00, 1.00);
  \coordinate (v5) at (0.00, -1.00);
  \coordinate (v6) at (2.00, -1.00);
  \coordinate[label=below:$1$] (v4) at (1.00, -2.00);
  \coordinate[label=right:$0$] (v7) at (2.00, 1.00);
  \draw[draw=black, fill=black, thin, solid] (v1) circle (0.05);
  \draw[draw=black, fill=black, thin, solid] (v4) circle (0.05);
  \draw[draw=black, fill=black, thin, solid] (v5) circle (0.05);
  \draw[draw=black, fill=black, thin, solid] (v6) circle (0.05);
  \draw[draw=black, fill=black, thin, solid] (v7) circle (0.05);
  \draw[draw=black, thin, solid] (v1) -- (v5);
  \draw[draw=black, thin, solid] (v1) -- (v6);
  \draw[draw=black, thin, solid] (v5) -- (v6);
  \draw[draw=black, thin, solid] (v6) to (v4);
  \draw[draw=black, thin, solid] (v5) to (v7);
\end{tikzpicture}
\end{minipage}
}

\newcommand{\threestarMunchhausenCompletedVs}{
\begin{minipage}{3.5cm}
\begin{tikzpicture}
    \coordinate (v) ;
    \def \rad {1}
    \coordinate[label=above:$1$] (v1) at ([shift=(90:\rad)]v);
    \coordinate[label=left:$0$] (v2) at ([shift=(180:\rad)]v);
    \coordinate[label=below:$\infty$] (v3) at ([shift=(270:\rad)]v);
    \coordinate[label=right:$z$] (v4) at ([shift=(0:\rad)]v);

    \draw[thin] (v) -- (v4);
    \draw[draw=black, thin, solid] (v) -- node[inner sep=0pt,above, xshift=4pt] {\tiny $\frac{2 - 3\ep}{\lambda}$} (v2);
    \draw[draw=black, thin, solid] (v) -- node[inner sep=0pt,right] {\tiny $\frac{1 -\ep}{\lambda}$}(v1);
    \draw[draw=black, thin, solid] (v) -- node[inner sep=0pt,above] {\tiny $\frac{1 - \ep}{\lambda}$}(v4);
    \draw[draw=black, thin, solid] (v) -- node[inner sep=0pt,right] {\tiny $\nu_{x\infty}$}(v3);

    \draw[thin] (v1) -- node[inner sep=0pt,above left] {\tiny{$\nu_{01}$}} (v2);
    \draw[thin] (v3) -- node[inner sep=0pt,below right] {\tiny{$-\!\frac{1 -\ep}{\lambda}$}} (v4);
    \draw[thin] (v2) -- node[inner sep=0pt,below left] {\tiny{$\nu_{0\infty}$}} (v3);
    \draw[draw=black, thin, solid, bend right=90, looseness=2.5] (v1) to node[inner sep=0pt, pos=0.8, below left] {\tiny{$\nu_{1\infty}$}} (v3);

    \filldraw (v) circle (1pt);
    \filldraw (v1) circle (1pt);
    \filldraw (v2) circle (1pt);
    \filldraw (v3) circle (1pt);
    \filldraw (v4) circle (1pt);

\end{tikzpicture}
\end{minipage}
}

\newcommand{\coprodDecompletedVs}{
\begin{minipage}{3cm}
\begin{tikzpicture}
    \coordinate (v) ;
    \def \rad {1}
    \coordinate[label=above:$z$] (v1) at ([shift=(90:\rad)]v);
    \coordinate[label=below:$0$] (v3) at ([shift=(270:\rad)]v);
    \coordinate[label=right:$1$] (v4) at ([shift=(0:\rad)]v);

    \draw[thin] (v) -- (v4);
    \draw[draw=black, thin, solid] (v) -- node[inner sep=0pt,right] {\tiny $\frac{1-\ep}{\lambda}$}(v1);
    \draw[draw=black, thin, solid] (v) -- node[inner sep=0pt,above] {\tiny $\frac{1-\ep}{\lambda}$}(v4);
    \draw[draw=black, thin, solid] (v) -- node[inner sep=0pt,right] {\tiny $\nu_{x\infty}$}(v3);

    \draw[thin] (v3) -- node[inner sep=0pt,below right] {\tiny{$-\!\frac{1-\ep}{\lambda}$}} (v4);
    \draw[draw=black, thin, solid, bend right=90, looseness=1.5] (v1) to node[inner sep=0pt, left] {\tiny{$\nu_{1\infty}$}} (v3);

    \filldraw (v) circle (1pt);
    \filldraw (v1) circle (1pt);
    \filldraw (v3) circle (1pt);
    \filldraw (v4) circle (1pt);

\end{tikzpicture}
\end{minipage}
}

\newcommand{\productWithPlanar}{
\begin{minipage}{3.5cm}
\begin{tikzpicture}[x=3.75ex, y=3.75ex, baseline={([yshift=-4.0ex]current bounding box.center)}]
  \coordinate[label={above:$u$}] (v0);
  \coordinate[below=2 of v0, label={above:$v$}] (v1);
  \coordinate[below=2 of v1, label={below:$w$}] (v2);
  \coordinate[below=1 of v0, label={[red]above:$w$}] (u1);
  \coordinate[below=1 of v1, label={[red]below:$u$}] (u2);
  \coordinate[right=2.5 of v1, label={[red]right:$v$}] (u3);
  \filldraw (v0) circle (1pt);
  \filldraw (v1) circle (1pt);
  \filldraw (v2) circle (1pt);
  \filldraw[draw=red, fill=red] (u1) circle (1pt);
  \filldraw[draw=red, fill=red] (u2) circle (1pt);
  \filldraw[draw=red, fill=red] (u3) circle (1pt);
  \filldraw[fill=black!20] (v0) to [bend left=90] (v1) to [bend left=90] (v2) to [bend right=90, looseness=1.5] (v0);
  \filldraw[fill=black!20] (v0) to [bend right=90] (v1) to [bend right=90] (v2) to [bend left=90, looseness=1.5] (v0);
  \filldraw[fill=red!20, fill opacity=0.5,draw=red] (u3) to [bend right=90] (u1) to [bend left=90] (u2) to [bend right=90] (u3);

  \node[below=0.75 of v2] (G) {$G$};
  \node[right=0.8 of v2] (p) {planar};
  \node[right=0.8 of v0] (p) {\color{red} planar dual};
    
\end{tikzpicture}

\end{minipage}
}

\newcommand{\productFourierTwist}{
\begin{minipage}{2.5cm}
\begin{tikzpicture}[x=3.75ex, y=3.75ex, baseline={([yshift=-4.0ex]current bounding box.center)}]
  \coordinate[label={above:$u$}] (v0);
  \coordinate[below=2 of v0, label={below:$v$}] (v1);
  \coordinate[below=2 of v1, label={below:$w$}] (v2);
  \filldraw (v0) circle (1pt);
  \filldraw (v1) circle (1pt);
  \filldraw (v2) circle (1pt);
  \filldraw[fill=red!20,draw=red] (v0) to [bend left=90] (v1) to [bend left=90] (v2) to [bend right=90, looseness=1.5] (v0);
  \filldraw[fill=black!20] (v0) to [bend right=90] (v1) to [bend right=90] (v2) to [bend left=90, looseness=1.5] (v0);

  \node[below=0.75 of v2] (G) {$G'$};

\end{tikzpicture}

\end{minipage}
}

\newcommand{\asymptoticExpansion}{
\begin{tikzpicture}[
  every node/.style={font=\small},
  every path/.style={very thick},
  blob/.style={circle, fill=black!20, draw=black, minimum size=0.85cm, thin}
]

\coordinate[label=left:$z$] (z) at (-0.3,0) {};
\coordinate[label=below:$0$] (zero) at (1.5,-1.0) {};
\coordinate[label=right:$1$] (one) at (3.3,0) {};

\coordinate (n1) at (0.8,0) {};
\coordinate (n2) at (2.2,0) {};

\draw (z) to[bend left=65] (n2);
\draw (n1) to[bend left=65] (one);

\draw (z) -- (n1);
\draw (n1) -- (n2);
\draw (n2) -- (one);
\draw (n1) -- (zero);
\draw (n2) -- (zero);

\filldraw (z) circle (1.3pt);
\filldraw (zero) circle (1.3pt);
\filldraw (one) circle (1.3pt);

\filldraw[fill=black!20, draw=black, thin] (n1) circle (0.45cm);
\filldraw[fill=black!20, draw=black, thin] (n2) circle (0.45cm);

\node at (4.2,0) {$\sim\,\displaystyle\sum$};

\coordinate[label=left:$z$] (z2) at (5.1,0) {};
\coordinate[label=below:$0$] (zeroA) at (5.9,-1) {};
\coordinate (na) at (5.9,0) {};

\draw (z2) -- (na);
\draw (na) -- (zeroA);

\filldraw (z2) circle (1.3pt);
\filldraw (zeroA) circle (1.3pt);

\filldraw[fill=black!20, draw=black, thin] (na) circle (0.45cm);

\coordinate[label=below:$0$] (zeroB) at (7.7,-0.9) {};
\coordinate[label=right:$1$] (oneB) at (8.8,0) {};
\coordinate (nb) at (7.7,0) {};

\draw (zeroB) to[bend left=110, looseness=2.75] ($(nb) + (0, 0.45cm)$);
\draw (zeroB) to[bend left=60] ($(nb) + (-0.45cm, 0)$);

\draw (zeroB) -- (nb);
\draw (oneB) -- (zeroB);
\draw (nb) -- (oneB);

\filldraw (zeroB) circle (1.3pt);
\filldraw (oneB) circle (1.3pt);

\filldraw[fill=black!20, draw=black, thin] (nb) circle (0.45cm);

\end{tikzpicture}
}

\definecolor{darkgreen}{rgb}{0.0, 0.5, 0.0}

\def\bnon{\begin{equation*}}
\def\enon{\end{equation*}}
\def\bnu{\begin{equation}}
\def\enu{\end{equation}}
\def\bqa{\begin{eqnarray*}}
\def\eqa{\end{eqnarray*}}

\def\li{\mathcal{L}_{1}(z)}
\def\lo{\mathcal{L}_{0}(z)}
\def\loi{\mathcal{L}_{01}(z)}
\def\lio{\mathcal{L}_{10}(z)}
\def\lii{\mathcal{L}_{11}(z)}
\def\loo{\mathcal{L}_{00}(z)}
\def\lioi{\mathcal{L}_{101}(z)}
\def\liio{\mathcal{L}_{110}(z)}
\def\loii{\mathcal{L}_{011}(z)}
\def\looi{\mathcal{L}_{001}(z)}
\def\loio{\mathcal{L}_{010}(z)}
\def\lioo{\mathcal{L}_{100}(z)}

\def\loiz{\mathcal{L}_{01\bar{z}}(z)}
\def\lioz{\mathcal{L}_{10\bar{z}}(z)}
\def\trilog{\loi-\lio}
\def\trilogbk{\left(\loi-\lio\right)}
\def\imz{z-\bar{z}}

\def\diffz{\partial_z-\partial_{\bar{z}}}
\def\fgi{f_{G_1}}
\def\fg{f_{G}}
\def\inv{^{-1}}

\def\invz{\frac{1}{z\bar{z}}}
\def\invzi{\frac{1}{(z-1)(\bar{z}-1)}}
\def\invmz{\frac{1}{\imz}}
\def\modz{z\bar{z}}
\def\modzi{(z-1)(\bar{z}-1)}
\def\den{\modz\modzi}
\def\invall{\frac{1}{\modz\modzi}}
\def\ep{\varepsilon}
\def\Li{\operatorname{Li}}
\newcommand{\IntLoop}[2]{\int \frac{\dd[#2]{#1}}{\pi^{2}}}
\def\sv{\int_{\text{sv}}\dd z}
\def\svb{\int_{\text{sv}}\dd\bar{z}}
\newtheorem{theo}{Theorem}
\newtheorem{lem}[theo]{Lemma}
\newtheorem{defin}[theo]{Definition}

\tikzfeynmanset{warn luatex=false}

\begin{document}
\begin{titlepage}
\thispagestyle{empty}
\noindent
PUBDB-2026-01399
\hfill
April 2026 \\
\vspace{1.0cm}

\begin{center}
  {\bf \Large
  Graphical Functions by Examples
  }
  \vspace{1.5cm}

 {\large
   Mrigankamauli Chakraborty,
   Marco Klann,
   Sven-Olaf Moch,
   Pooja Mukherjee, \\[0.5ex] 
   Tobias Porsche,
   Oliver Schnetz
   and
   Leonid A. Shumilov
   \\
 }
 \vspace{0.5cm}
 {\footnotesize{
     \begin{flushleft}       
 {\it
   II. Institut f\"ur Theoretische Physik, Universit\"at Hamburg,
   Luruper Chaussee 149, D-22761 Hamburg, Germany \\
 }
     \end{flushleft}
}}
  \vspace{2.0cm}
  \large {\bf Abstract}
\end{center}
  
\noindent 
Graphical functions have emerged as a powerful framework for evaluating multi‑loop Feynman integrals in perturbative quantum field theory.
Defined as massless three‑point position‑space integrals, they reveal rich analytic structures and have enabled major advances, including the highest‑loop results currently known in several quantum field theories.
Their role extends to conformal field theory, and recent algorithmic developments now allow many graphical functions to be computed automatically.
This review, based on graduate-level lectures held by O.S. in 2025/26 at the University of Hamburg, introduces the central ideas behind graphical functions, covering periods, Feynman residues, and the treatment of regular and singular cases in both integer and non‑integer dimensions.
It also discusses connections to momentum space and self‑duality, and provides guidance for further study, offering a coherent entry point into a topic not addressed in standard textbooks.

\end{titlepage}
\newpage

\setcounter{footnote}{0}
\setcounter{page}{1}

\tableofcontents
\bigskip
\section{Introduction}
\label{sec:realintro}

The evaluation of Feynman integrals at high loop orders is a central task in perturbative quantum field theory (QFT).
Analytic methods are especially valuable, as they clarify the mathematical structures underlying perturbation theory.
In particle physics, high‑order calculations are essential for precision predictions of observables at collider experiments, where multi‑loop corrections often determine the theoretical uncertainty and are crucial for testing the Standard Model or uncovering hints of new physics.
In condensed‑matter physics, similarly, accurate determinations of critical exponents within the renormalization‑group framework rely on perturbative expansions computed to very high loop order. 

Progress in mathematics has significantly expanded our ability to perform such calculations and has led to the development of graphical functions~\cite{Schnetz:2013hqa} (see also~\cite{Borinsky:2021gkd}).
These are massless position‑space Feynman integrals in QFT that depend on three vectors $z_0$, $z_1$, and $z_2$ in $D$‑dimensional Euclidean space.
They form a relatively new framework for evaluating Feynman integrals, making use of the special structure of three‑point functions in position space, and they provide a powerful tool for calculations in perturbative QFT.
Their application has led to major advances in the renormalization of QFTs at very high loop orders, including landmark results for six-loop calculations in $\phi^3$ theory~\cite{Schnetz:2025wtu} and seven-loop calculations in $\phi^4$ theory~\cite{Schnetz:2022nsc}. 
Graphical functions also play an important role in four‑dimensional conformal QFTs, where they were developed independently, see, for example, \cite{Drummond:2013nda}.
In recent years, significant progress has been achieved, enabling many graphical functions to be computed automatically with the computer algebra system \texttt{MAPLE}, using tools developed by one of the authors~\cite{hyperlogprocedures}.

This review grew out of lecture notes from a series of graduate-level courses taught by Oliver Schnetz at the University of Hamburg. Its purpose is to introduce the core ideas behind this method and to show how to apply the techniques needed for automated computations.
The material presented here is not covered in standard textbooks on Feynman integrals (such as~\cite{Weinzierl:2022eaz}), and this review is intended to help close this gap in the literature.
Proofs are mostly omitted; the aim is to present a pedagogical introduction based on fully worked out examples.
Readers interested in the detailed mathematics should consult the original sources. 

In this article, we use the convention that the dimension is
\begin{align}
    D=2\lambda+2 =  d-2\ep \quad \text{with} \quad d \in \{ 4, 6, 8, ... \}
\end{align}
in dimensional regularisation.
We use this convention because it is commonly used in
physics. Note that in previous articles on graphical functions and also in \cite{hyperlogprocedures} the convention is $D=d-\epsilon$. For comparison of the results, it is necessary to replace $\ep$ by $\epsilon/2$ in this article. Likewise, we construct hyperlogarithms by adding letters to the right (as in normal writing) which is mathematically the most consistent notation.
So,
\begin{equation}\label{Lidef}
\Li_{a_1,\ldots a_n}(z)=\int_{0<t_1<\ldots<t_n<z} 
\frac{\dd t_1}{t_1-a_1}\wedge\ldots\wedge \frac{\dd t_n}{t_n-a_n}.
\end{equation}
Likewise, multiple zeta values (MZVs) have increasing indices,
\begin{equation}\label{MZVdef}
\zeta(n_1,\ldots n_r)=\sum_{1\leq k_1<\ldots<k_r}\frac{1}{k_1^{n_1}\cdots k_r^{n_r}}.
\end{equation}
This notation is opposite to the default notation in \cite{hyperlogprocedures}. However, the software has an option to switch to the left-to-right convention, as can be found in the documentation. Otherwise, one has to reverse all words in hyperlogarithms and in MZVs before comparing results.

The review begins with an introduction and the necessary theoretical background on complex functions
(functions on $\mathbb{C}$) in Section \ref{sec:intro}, then develops the concepts of (scalar) $p$‑integrals, Feynman residues, and Feynman periods in Section \ref{sec:periods}. 
The article presents the calculation of the $K_{3,4}$‑period as an illustrative example (Section \ref{sec:k34}) and explains how to attach a leg in dimensional regularisation, distinguishing between regular and singular cases 
in Sections \ref{sec:leg_dimreg_reg} and \ref{sec:leg_dimreg_sing}.
An overview of key identities follows in Sections \ref{sec:identities} - \ref{sec:gf_dimreg_sing}, along with a discussion of periods and graphical functions in both even integer and general integer dimensions.
The review then examines the calculation of three-point functions in momentum space via self-duality in Section \ref{sec:self_duality}.
It concludes with a perspective on future developments in the field (Section \ref{sec:future}) and recommendations for further reading in Section~\ref{sec:further}.

\section{Setting the stage}
\label{sec:intro}
To illustrate the main ideas of graphical functions we will first work through a simple convergent integral in $D = 4$ dimensions.
Consider the well-known three-point one-loop integral
\begin{align}
\label{three-point-mom}
    \momentumthreepoint\qquad = \IntLoop{q}{4}\dfrac{1}{q^2(p_1 - q)^2(p_2 - q)^2}.
\end{align}
The method of graphical functions is designed for Feynman graphs whose propagators are massless. Therefore, it applies only to graphs consisting entirely of massless edges.
Another feature of the graphical function method is that we consider integrals in position space.
By renaming the momentum variables, we see that the integral is equivalent to a graph with one internal vertex in position space:
\begin{align}
\label{three-point-pos}
    \threepointdual \quad\qquad \Rightarrow \qquad \thetreeResc \quad = \quad\IntLoop{x}{4}\dfrac{1}{x^2(z_1 - x)^2(z_2 - x)^2}.  
\end{align}
As we can see, this corresponds to the dual graph.

The integral in Equation \eqref{three-point-pos} depends on three external vectors $0, z_1, z_2 \in \mathbb{R}^4$.
Note that we will be working in Euclidean space in these notes.

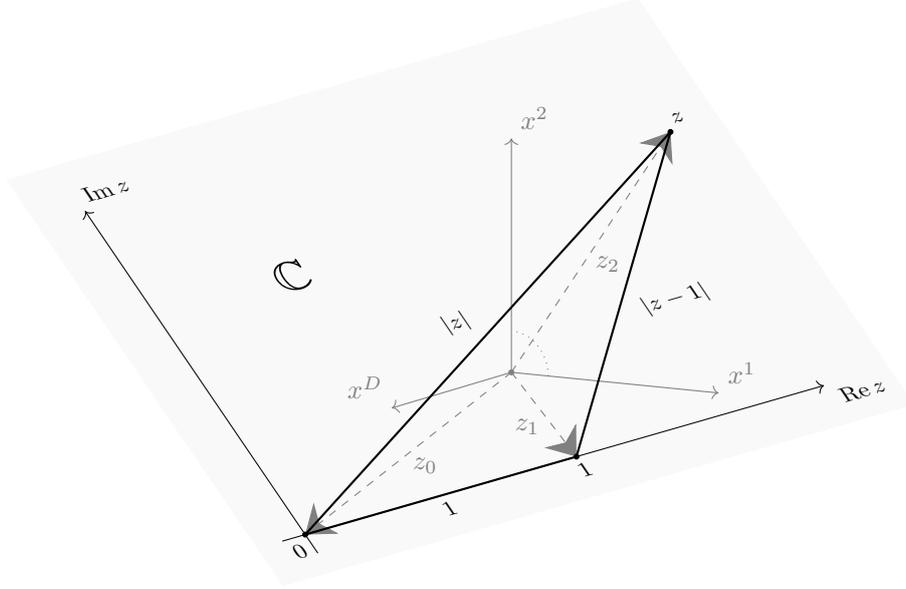
\begin{figure}
\tdplotsetmaincoords{80}{120}
\centering
\begin{tikzpicture}[tdplot_main_coords,scale=.7]
\draw[thin, ->,black!50] (0,0,0) -- (4.5,0,0) node[anchor=south east,opacity=1]{$x^D$};

\draw[thin, ->,black!50] (0,0,0) -- (0,4.5,0)  node[anchor=south west,opacity=1]{$x^1$};

\draw[thin, ->,black!50] (0,0,0) -- (0,0,4.5)  node[anchor=south west,opacity=1]{$x^2$};

\tdplotsetrotatedcoords{0}{90}{0};
\draw[dotted,black!50,tdplot_rotated_coords] (0,.8,0) arc (90:180:.8);

\pgfmathsetmacro{\Ox}{32}
\pgfmathsetmacro{\Oy}{14}
\pgfmathsetmacro{\Oz}{3}
\pgfmathsetmacro{\Onex}{-5}
\pgfmathsetmacro{\Oney}{3}
\pgfmathsetmacro{\Onez}{1}
\pgfmathsetmacro{\Zx}{-12}
\pgfmathsetmacro{\Zy}{1}
\pgfmathsetmacro{\Zz}{6}

\tdplotcrossprod(\Onex,\Oney,\Onez)(\Zx,\Zy,\Zz)
\pgfmathsetmacro{\rx}{\tdplotresx}
\pgfmathsetmacro{\ry}{\tdplotresy}
\pgfmathsetmacro{\rz}{\tdplotresz}
\tdplotcrossprod(\rx,\ry,\rz)(\Onex,\Oney,\Onez)
\pgfmathsetmacro{\tx}{\tdplotresx/100}
\pgfmathsetmacro{\ty}{\tdplotresy/100}
\pgfmathsetmacro{\tz}{\tdplotresz/100}

\pgfmathsetmacro{\dplane}{\rx*\Ox + \ry*\Oy + \rz*\Oz}

\pgfmathsetmacro{\OneLen}{sqrt(\Onex*\Onex+\Oney*\Oney+\Onez*\Onez)};
\pgfmathsetmacro{\iLen}{sqrt(\tx*\tx+\ty*\ty+\tz*\tz)};

\tikzset{perspective/.style= {canvas is plane={O(0,0,0)x(\Onex/\OneLen,\Oney/\OneLen,\Onez/\OneLen)y(\tx/\iLen,\ty/\iLen,\tz/\iLen)}} }
\pgfmathsetmacro{\axisscale}{2};
\tikzset{perspective2/.style= {canvas is plane={O(0,0,0)x(\axisscale*\Onex/\OneLen,\axisscale*\Oney/\OneLen,\axisscale*\Onez/\OneLen)y(\axisscale*\tx/\iLen,\axisscale*\ty/\iLen,\axisscale*\tz/\iLen)}} }

\pgfmathsetmacro{\axisovershoot}{2};
\pgfmathsetmacro{\axisundershoot}{.5};

\coordinate (v0) at (\Ox,\Oy,\Oz);
\coordinate (v1) at ([shift={(\Onex,\Oney,\Onez)}]v0);
\coordinate (vz) at ([shift={(\Zx,\Zy,\Zz)}]v0);
\coordinate (vi) at ([shift={(\tx,\ty,\tz)}]v0);

\filldraw[black!50] (0,0,0) circle (1.3pt);

\draw[dashed,-{Stealth[length=10pt, width=15pt]},black!50] (0,0,0) -- node[inner sep=1.5pt,below right] {$z_0$} (v0);
\draw[dashed,-{Stealth[length=10pt, width=15pt]},black!50] (0,0,0) -- node[inner sep=1.5pt,below left] {$z_1$} (v1);
\draw[dashed,-{Stealth[length=10pt, width=15pt]},black!50] (0,0,0) -- node[inner sep=1.5pt,below right] {$z_2$} (vz);
\draw[thin,->] ($(v0)!-\axisundershoot/\OneLen!(v1)$) -- ($(v0)!{1+2.7*\axisovershoot/\OneLen}!(v1)$) node[perspective,anchor=north west,opacity=1]{$\Re z$};
\draw[thin,->] ($(v0)!-\axisundershoot/\iLen!(vi)$) -- ($(v0)!{\OneLen/\iLen+1.4*\axisovershoot/\iLen}!(vi)$) node[perspective,anchor=south west,opacity=1]{$\Im z$};

\coordinate (C1) at ($(v0)!{2*\OneLen/\iLen}!(vi)$);
\node[perspective2] (C) at ($(C1)!{.5}!(v1)$) {$\mathbb{C}$};

\pgfmathsetmacro{\xscale}{1.7};
\coordinate (R0) at (${1 + 2.1*\axisundershoot/\OneLen + 2*\axisundershoot/\iLen}*(v0) - 2.1*\axisundershoot/\OneLen*(v1)- 2*\axisundershoot/\iLen*(vi)$);
\coordinate (R1) at (${ - \xscale*2*\axisovershoot/\OneLen + 2*\axisundershoot/\iLen}*(v0) + {1 + 2*\xscale*\axisovershoot/\OneLen}*(v1)- 2*\axisundershoot/\iLen*(vi)$);
\coordinate (R2) at (${ - \OneLen/\iLen - 2*\xscale*\axisovershoot/\OneLen - 2*\axisovershoot/\iLen}*(v0) + {1 + 2*\xscale*\axisovershoot/\OneLen}*(v1)+ {\OneLen/\iLen + 2*\axisovershoot/\iLen}*(vi)$);
\coordinate (R3) at (${1 - \OneLen/\iLen + 2.1*\axisundershoot/\OneLen - 2*\axisovershoot/\iLen}*(v0) - {2.1*\axisundershoot/\OneLen}*(v1)+ {\OneLen/\iLen + 2*\axisovershoot/\iLen}*(vi)$);

\draw[opacity = .05,fill,black!50]  (R0) -- (R1) -- (R2) -- (R3);

\filldraw (v0) circle(1.3pt) node[below left,perspective] {$0$};
\filldraw (v1) circle(1.3pt) node[below,perspective] {$1$};
\filldraw (vz) circle(1.3pt) node[above right,perspective] {$z$};

\draw[thick] (v0) -- node[perspective,below]{$1$} (v1);
\draw[thick] (v0) -- node[perspective,above left]{$|z|$} (vz);
\draw[thick] (vz) -- node[perspective,below right]{$|z-1|$} (v1);

\end{tikzpicture}
\caption{The three external vertices $z_0,z_1,z_2$ span the complex plane (picture by M. Borinsky).}
\label{fig:gfC}
\end{figure}

Three distinct points in $\mathbb{R}^4$ span a plane $\mathbb{R}^2$, which can be identified with the complex plane $\mathbb{C}$, see Figure \ref{fig:gfC}.
We can always rescale our coordinate system such that one variable (say $z_1$) acquires unit norm.
This allows us to identify these points with points on the complex plane
\begin{align}
\label{complex-param}
    z_1 \to 1 \in \mathbb{C} && z_2 \to z \in \mathbb{C},
\end{align}
such that the integral becomes a function of one complex variable $z$.
The reduction to $\mathbb{C}$ is only possible for the external variables $z_1$ and $z_2$; 
the integration variable remains in $\mathbb{R}^4$. There is no direct way to project the integration
variable down to $\mathbb{C}$ -- we will have to solve a differential equation to effectively do this.

In Euclidean space, graphical functions are real-analytic functions depending on $z$ and its complex conjugate
$\bar{z}$. Any result, however, can be interpreted in Minkowski space if one considers $z$ and $\bar{z}$ as two independent real variables.

Note that during the rescaling step we collect a constant factor $\norm{z_1}^{-(D-2)N_G}$. For a graph $G$ with $E_G$ edges and $V^{\rm int}_G$
internal vertices in four dimensions, we have
\begin{align}
 N_G= E_G - 2V^{\rm int}_G,
\end{align}
where the first term is due to the propagators and the second term comes from the integral measure.

A more general definition of the scaling weight $N_G$ (the superficial degree of divergence)
is given in definition \ref{NGdef}.
In our example, we find $N_G=1$.
Therefore, the original graph is related to the graphical function via
\begin{align}
  \thetreeResc = \frac{1}{\norm{z_1}^{2}}\thetree.
\end{align}
One of the key ideas in working with graphical functions is that the propagator in $D$ dimensions is the solution of the Klein-Gordon equation
\begin{align}
\label{box-app-1}
    \Box_D\dfrac{1}{x^{D-2}} = - \dfrac{4}{\Gamma\left(\frac{D - 2}{2}\right)}\delta^{(D)}(x) \overset{D = 4}{=} -4\delta^{(4)}(x).
\end{align}
Applying the operator $\Box_4$ to the external vertex identified with $z$ in the graph~\eqref{three-point-pos} and using complex variables~\eqref{complex-param} we
get the effective Laplace equation (proposition 3.22 in \cite{Schnetz:2013hqa})
\begin{align}
\label{box-app-2}
    \dfrac{1}{z - \bar{z}}\partial_z\partial_{\bar{z}}(z - \bar{z})\quad \thetreeX \quad = -\quad\thetwo. 
\end{align}

The graph obtained on the right hand side of~\eqref{box-app-2} is a trivial graphical function
with no integration
\begin{align}
    \thetwo \quad = \quad \dfrac{1}{|z|^2|z - 1|^2}  \overset{z \in \mathbb{C}}{=} \dfrac{1}{z\bar{z}(z - 1)(\bar{z} - 1)}.
\end{align}
Note that in the above equation, we used the fact that the square of the norm factorizes in $\mathbb{C}$.

This leads to the \emph{key idea}: instead of integration in the variable $x$, we can use~\eqref{box-app-2} as a differential equation and solve it to obtain the initial graph~\eqref{three-point-pos}.
Note that this gives us only the rescaled graph and we need to take care of a scaling factor as well, which we will discuss later.

Let us try to find such a solution. In order to simplify the differential equation, we perform the partial fraction decomposition
\begin{align}
\label{equ:pfrac}
    \dfrac{z-\bar{z}}{z\bar{z}(z- 1)(\bar{z} - 1)} = \dfrac{1}{z(\bar{z} - 1)} - \dfrac{1}{\bar{z}(z - 1)}
\end{align}
to write the differential equation~\eqref{box-app-2} in the form
\begin{align}
    \partial_{\bar{z}}\partial_{z}(z - \bar{z})\quad\thetree = -\dfrac{1}{z(\bar{z}-1)} + \dfrac{1}{\bar{z}(z - 1)}. 
\end{align}
The advantage of using the complex parametrisation~\eqref{complex-param} is that the operator $\Box_4$ factorizes into $\partial_{z}\partial_{\bar{z}}$.
This makes it possible to solve the partial differential equation by taking primitives
\begin{align}
    \partial_z(z- \bar{z})\thetree\quad = -\dfrac{1}{z}\log(\bar{z}-1) + \dfrac{1}{z- 1}\log\bar{z} + C(z).
\end{align}
There is an ambiguity that leads to an integration constant $C(z)$.
The initial integral~\eqref{three-point-pos} is not a generic function but is associated with a Feynman graph.
This constrains its shape.
One restriction is that~\eqref{three-point-pos} should be a single-valued function.
Using the ambiguity in choosing $C(z)$, we can rewrite the solution as
\begin{align}
\label{diff-red}
    \partial_z(z - \bar{z})\quad\thetree = -\dfrac{1}{z}\underbrace{\log(\bar{z} - 1)(z - 1)}_{\li} + \dfrac{1}{z- 1}\underbrace{\log\modz}_{\lo} + \tilde{C}(z),
\end{align}
with
\begin{align}
    \tilde{C}(z) = C(z) + \frac{1}{z}\log(z - 1) - \frac{1}{z-1}\log(z).
\end{align}
The functions on the right hand side are single-valued logarithms.
\begin{align}
  \mathcal{L}_0(z) &= \log z\bar{z},\notag\\
  \mathcal{L}_{a}(z) &= \log\left(1 - \frac{z}{a}\right)\left(1 - \frac{\bar{z}}{a}\right),\quad\text{if }\mathbb{C}\ni a\neq0, \notag\\
    \partial_z\mathcal{L}_{a}(z) &= \dfrac{1}{z - a}, \notag \\
    \mathcal{L}_a(0) &= 0.
\end{align}
Note that in the case of $a = 0$ we used the regularisation prescription
\begin{align}
    \mathcal{L}_0(0) = \log 0 = 0.
\end{align}
This choice of regularisation is arbitrary;
we could have chosen any constant as the value of $\mathcal{L}(0)$.
The reason is that all such contributions cancel out in the final answer.
This regularisation prescription (using the notion of a tangential base point)
 was introduced and analyzed by Deligne \cite{Deligne1989}.

Single-valued logarithms generalize to {\em single-valued multiple polylogarithms} that are single-valued analogs to Goncharov polylogarithms,
\begin{align}
\label{polylog-def}
     \sv\dfrac{\mathcal{L}_w(z)}{z - a} :=\int\limits_{0, \text{sv}}^{z}\dd t~\dfrac{\mathcal{L}_w(t)}{t - a} = \mathcal{L}_{wa}(z),
\end{align}
where $w$ is a word consisting of letters $a_1, a_2,...\in\{0,1\}$ and $wa$ is the word $w$ with a concatenated letter $a\in\{0,1\}$. \
The subscript $\text{sv}$ has been used to indicate that we perform the integration in a way that produces a single-valued function as the result.
Subsequent examples may help to understand how this is achieved. Further details on the construction of single-valued graphical functions can be found in (e.g.) \cite{Schnetz:2013hqa}.

The obvious generalization to letters $a_1, a_2,...\in\mathbb{C}$ leads to
{\em single-valued hyperlogarithms}. The theory of single-valued hyperlogarithms
was established by Brown~\cite{CRMATH_2004__338_7_527_0, BrSVMPII}.

Single-valued hyperlogarithms are further generalized in \cite{Schnetz:2021ebf}.
This generalization will be the content of lecture notes to appear in 2027. Here, we
take the results from this theory as a black box that is implemented in the Maple package
{\tt HyperlogProcedures} \cite{hyperlogprocedures}.

The recursive definition of single-valued multiple polylogarithms allows us to perform the second integration in~\eqref{diff-red}
\begin{align}
\label{three-point-res}
    \thetree \quad &= \dfrac{1}{z- \bar{z}}\Big(- \sv\dfrac{\mathcal{L}_{1}(z)}{z} + \sv\dfrac{\mathcal{L}_{0}(z)}{z - 1}\Big) \notag \\
    &=\dfrac{\loi - \lio}{z - \bar{z}} = \dfrac{4\mathrm{i}D(z)}{z - \bar{z}} = \dfrac{2D(z)}{\operatorname{Im}(z)},
\end{align}
where $D(z) = \operatorname{Im}(\Li_{2}(z) + \log(1 - z)\log |z|)$ is the Bloch-Wigner-dilogarithm (see e.g.\ \cite{Zagier2007}) and $\Li_k(z) = \sum_{n = 1}^{\infty}\frac{z^n}{n^k}$, see \eqref{Lidef}.

Note that the constant $\tilde{C}(z)$ in Equation~\eqref{three-point-res} has been omitted, as it is constrained by the single-valuedness, analytic structure, and symmetry of the initial integral~\eqref{three-point-pos}. 
Indeed, the kernel of the integration takes the form
\begin{align}\label{kernel}
    \dfrac{f(z) - g(\bar{z})}{z - \bar{z}} .
\end{align}
By symmetry $z\mapsto\bar{z}$ and by single-valuedness, $f$ and $g$ have to be rational
functions. From analysing the pole structure at $z=\bar{z}$ in general and
at $z=0,1,\infty$ in particular, we find $f=g\in\mathbb{R}$, so that $f(z) - g(\bar{z})=0$.
We conclude that \eqref{three-point-res} is the unique solution of the effective
Laplace equation in the space of graphical functions.
This uniqueness of the solution is a universal property in the theory of graphical functions; see Theorem 36 in \cite{Borinsky:2021gkd}.

\vspace{5pt}
Before we continue with more calculations, let us state some general points.
Note that the above definition of graphical functions is valid for all massless graphs with three external points in position space.
We do not need to start with the dual of a momentum space integral (see Section \ref{sec:self_duality} for momentum space integrals).
In any dimension $D$, the external vertices $z_0, z_1, z_2\in\mathbb{R}^D$ span a two-dimensional plane in $\mathbb{R}^D$.
Graphical functions, as we treat them here, do not work in two dimensions, since \eqref{box-app-1} no longer holds.
Instead, the Green's function is, up to constants, $\log{x^2}$.
For this reason, and because there is yet no theory of graphical functions in odd dimensions, we need to work in $D \geq 4$ even dimensional spacetime.
In these notes, we will work in $D = 4$ or $D = 4 - 2\ep$ dimensions, where we are interested in the limit as $\ep \rightarrow 0$. We will not consider the generalisation to higher dimensions, but refer the reader to \cite{Borinsky:2021gkd}. In general, at higher
dimensions, the theory of graphical functions is more complicated than in four dimensions, although the general structure is quite analogous.

We can always shift the coordinate system to obtain $z_0=0$.
Then, we are in the same position as in our previous example and can scale to $\norm{z_1}\to 1$.
Thereafter, we identify $z_2$ with the complex coordinate $z$.

Lastly, let us mention that the identification of the plane with $\mathbb{C}$ is not unique.
Instead of identifying $z_2$ with $z$, we could also identify $z_2$ with $\bar{z}$.
This ambiguity gives rise to the symmetry $z\mapsto\bar{z}$ in graphical functions. Note that the symmetry $z\mapsto\bar{z}$ does
not follow from the fact that graphical functions are real by definition
(for example, $(z-\bar{z})/2\mathrm{i} =\mathrm{Im} z\in\mathbb{R}$ and anti-symmetric).

\section{\texorpdfstring{$p$}{p}-Integrals, Feynman residues, Feynman periods}
\label{sec:periods}
A problem that can be solved efficiently with graphical functions is the calculation of numbers associated with the evaluation of Feynman graphs.
For example, in $D=4$ dimensions we consider
\begin{align}
   \periodA = \quad\dfrac{1}{p^2}P_3, && \periodB = \quad\dfrac{1}{\ep}P_3 + O(1), && \quad \periodC = \quad P_3,
\end{align}
where $P_3$ is the same constant in all three cases.
Note that the second graph is a tetrahedron with external lines attached.
In the third graph, we assume that one vertex is $0\in\mathbb{R}^4$ and another
vertex is $z_1\in\mathbb{R}^4$ with $\norm{z_1}=1$.

Let us show how to calculate $P_3$ given by the following Feynman period
\begin{align}
    P_3 = \quad\periodX = \IntLoop{x_1}{4}\IntLoop{x_2}{4}\dfrac{1}{x_1^2(x_1 - x_2)^2x_2^2(x_1 - z_1)^2(x_2 - z_1)^2}.
\end{align}
The name `period' comes from number theory, where it defines a class of numbers that are algebraic integrals \cite{Kontsevich2001}.

By renaming integration variables, we can see that $P_3$ can be expressed as an integral over the graphical function \eqref{three-point-res} in the previous section
\begin{align}
\label{period-int-1}
    P_3 = \quad\periodZ = \IntLoop{z_2}{4}\underbrace{\dfrac{1}{z_2^2(z_2 - z_1)^2}}_{\to\dfrac{1}{z\bar{z}(z-1)(\bar{z}-1)}}\underbrace{\IntLoop{x}{4}\dfrac{1}{x^2(z_1- x)^2(z_2 - x)^2}}_{\to \dfrac{\loi - \lio}{z - \bar{z}}}.
\end{align}
To simplify the integral, observe that an $\mathrm{SO}(3)$ subgroup of
$\mathrm{SO}(4)$ leaves $z_1$ invariant.
Each $z \in \mathbb{R}^4$ can be mapped to two different points in the complex plane via an element of this subgroup.
Therefore, the integral factors into half of the volume of $\mathrm{SO}(3)$, i.e. $2\pi$, and an integral over the complex plane (see Lemma 3.34 in \cite{Schnetz:2013hqa}).
\begin{align}
    \dd^4z_2 \rightarrow 2 \pi \sin^2\theta r^3\dd r\dd\theta = \frac{\pi}{4i}(z - \bar{z})^2 \dd z \wedge \dd\bar{z},
\end{align}
where we used $r\sin\theta = \frac{1}{2i}(z - \bar{z})$ and $r \dd r \dd\theta = \frac{i}{2} \dd z\wedge \dd\bar{z}$.
Substituting this into~\eqref{period-int-1} yields
\begin{align}
\label{period-int-2}
    P_3 &= \dfrac{1}{4\pi i}\int_{\mathbb{C}}\dfrac{z - \bar{z}}{z\bar{z}(z - 1)(\bar{z} - 1)}\Big(\loi - \lio\Big)\dd z \wedge \dd\bar{z} \notag \\
    & = \dfrac{1}{4\pi i}\int_{\mathbb{C}}\Biggl(\dfrac{1}{z(\bar{z} - 1)} - \dfrac{1}{\bar{z}(z - 1)}\Biggr)\Big(\loi -\lio\Big)\dd z \wedge \dd\bar{z},
\end{align}
where we made use of the decomposition \eqref{equ:pfrac}.

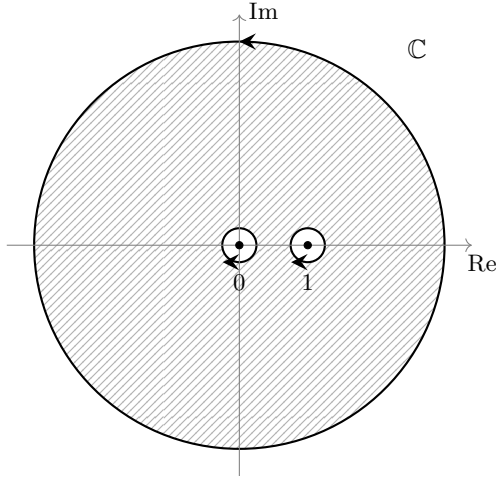
\begin{figure}[ht]
\centering

\begin{tikzpicture}[scale=0.9]
 
\begin{scope}
  \clip (0,0) circle (3cm);
  \begin{scope}[even odd rule]
    \clip (0,0) circle (3cm)
          (0,0) circle (0.25cm)
          (1,0) circle (0.25cm);
    \fill[pattern=north east lines, pattern color=gray!60] (0,0) circle (3cm);
  \end{scope}
\end{scope}
 
\draw[thick,
  decoration={markings, mark=at position 0.25 with {\arrow[scale=1.5]{stealth}}},
  postaction={decorate}
] (0,0) circle (3cm);
 
\draw[thick,
  decoration={markings, mark=at position 0.75 with {\arrow[scale=1.5]{stealth reversed}}},
  postaction={decorate}
] (0,0) circle (0.25cm);
 
\draw[thick,
  decoration={markings, mark=at position 0.75 with {\arrow[scale=1.5]{stealth reversed}}},
  postaction={decorate}
] (1,0) circle (0.25cm);
 
\draw[->, thin, gray] (-3.4,0) -- (3.4,0);
\draw[->, thin, gray] (0,-3.4) -- (0,3.4);
 
\node[below, font=\small] at (0,-0.3)  {$0$};
\node[below, font=\small] at (1,-0.3)  {$1$};
 
\node[below right, font=\small] at (3.2,0) {$\text{Re}$};
\node[above right, font=\small] at (0,3.2) {$\text{Im}$};
 
\filldraw[black] (0,0) circle (1.5pt);
\filldraw[black] (1,0) circle (1.5pt);
 
\node[font=\normalsize] at (2.6, 2.9) {$\mathbb{C}$};
 
\end{tikzpicture}
\caption{Integration Contours}
\label{fig:contour}
\end{figure}

The definition of single-valued multiple polylogarithms allows us to take a single-valued primitive of the integrand, yielding
\begin{align}
    F(z)=\sv\Biggl(\dfrac{1}{z(\bar{z} - 1)} - \dfrac{1}{\bar{z}(z - 1)}\Biggr)\Big(\loi -\lio\Big) = \dfrac{\loio -\lioo}{\bar{z} - 1} - \dfrac{\loii - \lioi}{\bar{z}}.
\end{align}
This does not immediately solve the initial problem of integration over the complex plane, but it allows us to present the period in the following form (note that $\dd\bar{z}\wedge\dd\bar{z}=0$);
\begin{align}
\label{period-int-3}
    P_3 = \dfrac{1}{4\pi i}\int_{\mathbb{C}} \dd F(z)\dd\bar{z}.
\end{align}
This integral can be computed using Stokes' theorem. For any regular differential form $\omega$ in some region $X$, we have
\begin{align}
    \int_{X}\dd\omega = \int_{\partial X}\omega.
\end{align}
At first glance, this seems to vanish, since $\partial \mathbb{C} = 0$.
This is in conflict with integrating a positive function.
In fact, in our example, the 1-form $\omega$ is not regular on $\mathbb{C}$ but has poles at $\{0, 1, \infty\}$ when considering our integrand as a function on the Riemann sphere $\mathbb{C}\cup\{\infty\}$.
We must remove infinitesimal neighbourhoods around each of these points from the domain of integration
$X$. Thus
\begin{align}
    \partial\Big(\mathbb{C}\char`\\ \{0, 1, \infty\}\Big) = S_\infty - S_0 - S_1, 
\end{align}
where $S_a$ is a circle around $a$ with radius $\ep$ (with radius $1/\ep$ in the case $a = \infty$), where the limit $\ep \to 0$ is implied; see Figure~\ref{fig:contour}. 

Using the parametrisation $z = a + \ep e^{i\varphi}$, the integral over $S_a$ corresponds to the anti-residue at $a$
\begin{align}
    \dfrac{1}{2\pi i}\int_{S_a}\dfrac{\dd\bar{z}}{\bar{z} - \bar{a}} = \dfrac{1}{2\pi i}\int_0^{2\pi}\dd\varphi~\dfrac{-i\ep e^{-i\varphi}}{\ep e^{-i\varphi}} = -1,
\end{align}
while all other terms vanish.
Likewise, with $z = \frac1\ep e^{i\varphi}$ for $S_\infty$, we obtain
\begin{align}
    \dfrac{1}{2\pi i}\int_{S_\infty}\dfrac{\dd\bar{z}}{\bar{z}} = \dfrac{1}{2\pi i}\int_0^{2\pi}\dd\varphi~\dfrac{-i\ep^{-1} e^{-i\varphi}}{\ep^{-1} e^{-i\varphi}} = -1.
\end{align}
Applying this to the function $F(z)$ we get (see Theorem 2.29 in \cite{Schnetz:2013hqa})
\begin{align}
    \dfrac{1}{2\pi i}\int_{\mathbb{C}\char`\\ \{0, 1, \infty\}}\dd F(z)\dd\bar{z} = \underbrace{\overline{\operatorname{res}}_{0}F + \overline{\operatorname{res}}_{1}F-\overline{\operatorname{res}}_{\infty}F}_{\text{anti-residues}}.
\end{align}
The standard definition of residues requires that the function is analytic away from isolated singularities.
Since single-valued multiple polylogarithms are not analytic (they depend on both $z$ and $\bar{z}$), we must be careful in principle, but it turns out that at the singular points $a=0,1$ every graphical function admits an expansion
that is a polynomial in the single-valued logarithm at $a$ with meromorphic coefficients in $z$ and $\bar{z}$;
see Theorem 5 in \cite{Borinsky:2021gkd}. In these single-valued log-Laurent expansions, the anti-residue
is the coefficient of $(\mathcal{L}_a(z))^0(z-a)^0(\bar{z}-a)^{-1}$.

At the singular point $\infty$ the situation is analogous.
The anti-residue $\overline{\operatorname{res}}_{\infty}F$ can be extracted from the expansion of $F$ in
$1/z$ as the coefficient of $(\mathcal{L}_0(z))^0z^0\bar{z}^{-1}$.

For the integral~\eqref{period-int-3}, we get
\begin{align}
\label{period-polylogs}
    P_3 = \dfrac{1}{2}\Biggl(\underbrace{-\mathcal{L}_{011}(0) + \mathcal{L}_{101}(0)}_{= 0\text{ by definition}} + \mathcal{L}_{010}(1) - \mathcal{L}_{100}(1) - \mathcal{L}_{010}(\infty) + \mathcal{L}_{100}(\infty) + \mathcal{L}_{011}(\infty) - \mathcal{L}_{101}(\infty)\Biggr).
\end{align}
The final step is to understand how to evaluate multiple polylogarithms at the point $\infty$ by translating them to the multiple polylogarithms at 1.
We use the variable transformation
\begin{align}
    z \mapsto \dfrac{z}{z - 1}
\end{align}
that maps $1 \mapsto \infty$, $\infty \mapsto 1$ and $0 \mapsto 0$.
The 1-forms $\dd z/z$ and $\dd z/(z-1)$, corresponding to the letters 0 and 1, respectively, transform as
\begin{subequations}
\begin{align}
    \dfrac{\dd z}{z} &\mapsto \dfrac{\dd\frac{z}{z- 1}}{\frac{z}{z - 1}} = \Biggl(\dfrac{1}{z} - \dfrac{1}{z - 1}\Biggr)\dd z, \\
    \dfrac{\dd z}{z - 1} &\mapsto \dfrac{\dd\frac{z}{z - 1}}{\frac{z}{z - 1} - 1} = \Biggl(-\dfrac{1}{z - 1}\Biggr)\dd z.
\end{align}
\end{subequations}
Substituting these transformations into the definition of single-valued multiple polylogarithms implies the map of the letters $0 \mapsto 0 - 1$ and $1 \mapsto -1$ . Here, we assume that words are linear with respect the substitution of letters
\begin{align}
u(\lambda a+\mu b)v=\lambda\, uav+\mu\, ubv\quad\text{for any words $u,v$, letter $a,b$ and constants $\lambda,\mu\in\mathbb{C}$.}
\end{align}
In the above transformations, we consider 0 and 1 as letters and the sign $-$ as
constant $-1\in\mathbb{C}$. Likewise, single-valued hyperlogarithms are defined to be
linear in words,
\begin{align}\label{Lsha}
\mathcal{L}_{\lambda u+\mu v}(z)=\lambda\mathcal{L}_u(z)+\mu\mathcal{L}_v(z)\quad
\text{for any words $u,v$ and constants $\lambda,\mu\in\mathbb{C}$.}
\end{align}

The substitutions act on all letters in the words of each multiple polylogarithm.

Concretely, in~\eqref{period-polylogs} we get
\begin{subequations}\label{eq:limit}
\begin{align}
    \mathcal{L}_{010}(\infty) &= -\mathcal{L}_{010}(1) + \mathcal{L}_{110}(1) + \mathcal{L}_{011}(1) - \mathcal{L}_{111}(1),  \label{eq:limitA} \\
    \mathcal{L}_{100}(\infty) &= -\mathcal{L}_{100}(1) + \mathcal{L}_{110}(1) + \mathcal{L}_{101}(1) - \mathcal{L}_{111}(1), \label{eq:limitB} \\ 
    \mathcal{L}_{011}(\infty) &= \mathcal{L}_{011}(1) - \mathcal{L}_{111}(1), \label{eq:limitC} \\
    \quad \mathcal{L}_{101}(\infty) &= \mathcal{L}_{101}(1) - \mathcal{L}_{111}(1). \label{eq:limitD} 
\end{align}
\end{subequations}
Because the multiple polylogarithms are single-valued, we do not have to investigate
in which (tangential) direction the singularities are approached.

Gathering all contributions, we obtain
\begin{align}
   -\mathcal{L}_{010}(\infty)+\mathcal{L}_{100}(\infty)+\mathcal{L}_{011}(\infty)-\mathcal{L}_{101}(\infty) = \mathcal{L}_{010}(1) - \mathcal{L}_{100}(1).
\end{align}
This particular example has the symmetry that all residues sum up to zero.
In more complicated examples, this is not the case.

We obtain
\begin{align}
    P_3 = \mathcal{L}_{010}(1) - \mathcal{L}_{100}(1).
\end{align}

The evaluations of the multiple polylogarithms can be calculated \cite{Schnetz:2013hqa,hyperlogprocedures,CRMATH_2004__338_7_527_0,Schnetz:2021ebf},
\begin{align}
    \mathcal{L}_{010}(1) = 4\zeta(3),\quad\mathcal{L}_{100}(1) = -2\zeta(3),
\end{align}
where $\zeta(n)$ is the Riemann zeta function. Eventually we obtain the
classical result \cite{Chetyrkin:1980pr}
\begin{align}\label{zeta3}
    P_3 = 6\zeta(3).
\end{align}
\section{The \texorpdfstring{$K_{3,4}$}{K34}-period}
\label{sec:k34}
The calculation of $P_3$ in the previous section is quite simple; it can also be done using other methods and techniques. Let us now present an example in which the use of graphical functions provides a significant advantage. We consider the following 4-point graph in $D=4$, referring to the $K_{3,4}$-period, which was first calculated analytically in 1999~\cite{Schnetz:1999if}
\begin{equation}
\kperiod
\end{equation}
We may place $0, 1, z$ on any of the vertices (see definition and Theorem 2.7 in \cite{Schnetz:2008mp}); all choices yield the same result. We choose
\begin{equation}
\kperiodlem \quad = \quad \left(\thetree\right)^4 = \quad\left(\frac{\loi-\lio}{\imz}\right)^4.
\label{eq:productPeriod}
\end{equation}
This follows because we do not integrate over external vertices.
We can calculate the period by integrating over the entire complex plane
\begin{equation}
\mathcal{P}_{K_{3,4}} = \frac{1}{4\pi i}\int_{\mathbb{C}}\left(\frac{\loi-\lio}{\imz}\right)^4\;\left(\imz\right)^2\,\dd z\wedge \dd\bar{z}.
\end{equation}
As in the previous section, we want to apply Stokes' theorem.
To this end, we first compute the single-valued integral with respect to $z$,
\begin{align}
    \sv&\;\frac{\trilogbk^4}{(\imz)^2}\notag \\[6pt]
    &\overset{\text{\small{IBP}}}{=} \;\underbrace{-\,\frac{\trilogbk^4}{\imz}}_{(\text{I})}\; + \underbrace{\,4\sv\; \frac{\trilogbk^3}{z-\bar{z}}\left(\frac{\lo}{z-1}-\frac{\li}{z}\right)}_{(\text{II})}.
\end{align}
Term (I) is finite at any point, as $\trilog\sim\imz$ close to the real axis, where the poles would be. Thus, it has no anti-residues rendering the integral zero.
For (II), we obtain
\begin{align}
\text{(II)} &= \; \frac{4}{\bar{z}-1} \sv\;\left(\frac{1}{\imz}-\frac{1}{z-1}\right)\underbrace{\lo\trilogbk^3}_{\mathcal{L}_{\text{IIa}}(z)}  \notag \\[5pt]
&  -\frac{4}{\bar{z}} \sv\;\left(\frac{1}{\imz}-\frac{1}{z}\right)\underbrace{\li\trilogbk^3 }_{\mathcal{L}_{\text{IIb}}(z)}
\end{align}
after a partial fraction decomposition.
In general, (single-valued) hyperlogarithms shuffle ($\shuffle$) their words upon multiplication
\begin{equation}
    \mathcal{L}_u(z)\mathcal{L}_v(z)=\mathcal{L}_{u\, \shuffle\, v}(z).
\end{equation}
The shuffle product of two words is an operation that produces the sum of all possible interleavings of their letters, without changing the relative order with respect to each word. More formally, the shuffle product is best defined recursively.
For any letters $a$, $b$ and words $w$, $v$ we have
\begin{equation}
    aw \shuffle bv = a(w \shuffle bv) + b(aw \shuffle v).
\end{equation}
The empty word shuffles trivially. For letters $a,b,c,d$, we e.g.\ get
\begin{equation}
    ab \shuffle cd = abcd+acbd+acdb+cabd+cadb+cdab.
\end{equation}
Conveniently, one can think of two decks of cards. One deck is inserted into the other, so that the internal order of each deck is preserved. The shuffle product represents the sum
over all possible outcomes of such a shuffle.

In hyperlogarithms, the shuffled expression is linearly expanded (see \eqref{Lsha}).
This leads to expressions for $\mathcal{L}_{\text{IIa}}$ and $\mathcal{L}_{\text{IIb}}$ in terms of 26 single-valued multiple polylogarithms with words of length 8. Concretely, the words IIa and IIb are given by
\begin{align}\label{eqK34}
\text{IIa} &=
144\,{\scriptstyle 0000111} + 72\,{\scriptstyle 0001011} - 72\,{\scriptstyle 0001110} - 24\,{\scriptstyle 0010101} - 48\,{\scriptstyle 0010110} - 48\,{\scriptstyle 0011001} - 24\,{\scriptstyle 0011010}\notag\\
&\quad-72\,{\scriptstyle 0100011} - 48\,{\scriptstyle 0100101} - 24\,{\scriptstyle 0100110} - 24\,{\scriptstyle 0101001} + 24\,{\scriptstyle 0101100} + 24\,{\scriptstyle 0110010} + 48\,{\scriptstyle 0110100}\notag\\
&\quad+ 72\,{\scriptstyle 0111000} - 144\,{\scriptstyle 1000011} - 72\,{\scriptstyle 1000101} + 24\,{\scriptstyle 1001010} + 48\,{\scriptstyle 1001100} + 72\,{\scriptstyle 1010001} + 48\,{\scriptstyle 1010010}\notag\\
&\quad+ 24\,{\scriptstyle 1010100} + 144\,{\scriptstyle 1100001} + 72\,{\scriptstyle 1100010} - 72\,{\scriptstyle 1101000} - 144\,{\scriptstyle 1110000},\notag\\
\text{IIb} &=
144\,{\scriptstyle 0001111} + 72\,{\scriptstyle 0010111} - 72\,{\scriptstyle 0011101} - 144\,{\scriptstyle 0011110} - 24\,{\scriptstyle 0101011} - 48\,{\scriptstyle 0101101} - 72\,{\scriptstyle 0101110}\notag\\
&\quad- 48\,{\scriptstyle 0110011} - 24\,{\scriptstyle 0110101} + 72\,{\scriptstyle 0111010} + 144\,{\scriptstyle 0111100} - 72\,{\scriptstyle 1000111} - 48\,{\scriptstyle 1001011} - 24\,{\scriptstyle 1001101}\notag\\
&\quad- 24\,{\scriptstyle 1010011} + 24\,{\scriptstyle 1010110} + 24\,{\scriptstyle 1011001} + 48\,{\scriptstyle 1011010} + 72\,{\scriptstyle 1011100} + 24\,{\scriptstyle 1100101} + 48\,{\scriptstyle 1100110}\notag\\
&\quad+ 48\,{\scriptstyle 1101001} + 24\,{\scriptstyle 1101010} + 72\,{\scriptstyle 1110001} - 72\,{\scriptstyle 1110100} - 144\,{\scriptstyle 1111000},
\end{align}
where we write words in small font and the numerical prefactors denote their coefficients. As in the previous section, we get
\begin{flalign}\label{K34formula}
   \mathcal{P}_{K_{3,4}} & = \frac{1}{2} (\overline{\text{res}}_0+\overline{\text{res}}_1 - \overline{\text{res}}_\infty)\left(\frac{4}{\bar{z}-1}\mathcal{L}_{\text{IIa}\bar{z}-\text{IIa}1}(z)-\frac{4}{\bar{z}}\mathcal{L}_{\text{IIb}\bar{z}-\text{IIb}0}(z)\right)\notag \\[5pt]
   & = 2\left(\mathcal{L}_{\text{IIa}\bar{z}-\text{IIa}1}(1)-\mathcal{L}_{\text{IIa}\bar{z}-\text{IIa}1}(\infty)+\mathcal{L}_{\text{IIb}\bar{z}-\text{IIb}0}(\infty)\right)
\end{flalign}
upon applying Stokes' theorem.
Note that neither $\mathcal{L}_{\text{IIa}\bar{z}}$ nor $\mathcal{L}_{\text{IIb}\bar{z}}$ are single-valued multiple polylogarithms because their words have a non-constant letter $\bar{z}$.
They are members of a wider class of functions called generalized single-valued hyperlogarithms (GSVHs) (formally, we have only added the letter
$\bar{z}$).

There exists a fully developed theory of GSVHs which is quite general and complex \cite{Schnetz:2021ebf}. Explaining GSVHs is the topic of an ongoing series
of lectures also held by OS. at the University of Hamburg. The lectures will be made publicly available in 2027 by lecture notes in the style of these notes on graphical functions.

The core theorem in the theory of GSVHs is the existence of a commutative hexagon from which efficient
algorithms for all operations (taking (anti-)primitives, (anti-)derivatives, evaluations, complex conjugation)
can be extracted. Even for classical single-valued hyperlogarithms, algorithms most efficiently
use the commutative hexagon (treating them as special cases of GSVHs).

With this theory of GSVHs, the evaluations in \eqref{K34formula} at 1 and at $\infty$ can easily be calculated \cite{hyperlogprocedures},
\begin{subequations}
\begin{align}
\mathcal{L}_{\text{IIa}\bar{z}-\text{IIa}1}(1) &= -\frac{1728}{5}\zeta(5,3) + \frac{232}{2625}\pi^8 - 648\zeta(3)\zeta(5)\;, \\[2pt]
\mathcal{L}_{\text{IIa}\bar{z}-\text{IIa}1}(\infty) &= 0, \quad
\mathcal{L}_{\text{IIb}\bar{z}-\text{IIb}0}(\infty) = \mathcal{L}_{\text{IIa}\bar{z}-\text{IIa}1}(1).
\end{align}
\end{subequations}
Combining these results, we obtain the period
\begin{equation}
\mathcal{P}_{K_{3,4}} = 4\mathcal{L}_{\text{IIa}\bar{z}-\text{IIa}1}(1) = -\frac{6912}{5}\zeta(5,3) + \frac{928}{2625}\pi^8 - 2592\zeta(3)\zeta(5) , 
\end{equation}
where $\zeta(5,3) = \sum_{1\leq k_1< k_2}\frac{1}{k_1^5k_2^3}$ is an MZV; see \eqref{MZVdef}.

\section{Attaching a leg in dimensional regularisation, regular case}
\label{sec:leg_dimreg_reg}
We have now seen how graphical functions can be used to calculate 
(finite) Feynman integrals and periods in four dimensions.
To perform more general computations in QFT, we will need a method to calculate
regularized Feynman integrals.
Often, the preferred method of regularisation is dimensional regularisation~\cite{tHooft:1972tcz,Bollini:1972ui} working in $D = 4 - 2\ep$ dimensions. In general, graphical functions generalize to dimensionally regularized integrals
if one uses a specific internal subtraction scheme. The use of this subtraction scheme
can be fully automatized, so that dimensionally regularized integrals pose no obstruction to
the theory of graphical functions.

To explore the situation, let us start with non-divergent graphs in $D$ dimensions. \\
Contrary to momentum space, the explicit form of the Laplacian changes with dimension. 
After switching to the complex parametrization, we now get the following differential equation for removing an edge:
\begin{equation}
    \left[\frac{1}{\imz}\partial_z\partial_{\bar{z}}(\imz)+\frac{\ep}{\imz}(\diffz)\right]\blobline = \; -\frac{1}{\Gamma(1-\ep)}\;\blob.
\end{equation}
Note that in the limit $\ep \rightarrow 0$ this reduces to the four-dimensional case. \\
We want to solve for $f_{G_1}(z) = \blobline$ as an expansion in $\ep$
\begin{equation}
    f_{G_1}(z) = \underbrace{\ldots}_{\text{poles in $\ep$}} + \;\fgi^0(z)\; +\;\ep\fgi^1(z)\; + \;\mathcal{O}(\ep^2).
\end{equation} 

To invert the Laplacian, we use the following identity for linear operators $A$ and $B$:
\begin{flalign}
    A+\ep B &= A(1+\ep A^{-1}B)\notag \\
    \implies(A+\ep B)\inv &= (1+\ep A^{-1}B)\inv A\inv=\sum_{k = 0}^\infty (-\ep A\inv B)^kA\inv.
\end{flalign}
In our case, $A= \frac{1}{z- \bar{z}}\partial_z\partial_{\bar{z}}(z- \bar{z})=:\Delta_0$ is the four-dimensional effective Laplacian, and $B=\frac{1}{\imz}(\diffz)=:\delta_0$.
Thus we obtain the relation
\begin{equation}
\fgi = -\frac{1}{\Gamma(1-\ep)}\sum_{k = 0}^\infty (-\ep \Delta_0\inv \delta_0)^k\Delta_0\inv\fg \label{sum}.
\end{equation}
for appending an edge.

Let us show how this works on a specific example, namely the graphs
\begin{equation}
f_{G_1}(z) = \thetree, \qquad f_G(z) = \thetwo,
\label{eq:fsupers}
\end{equation}
which we already evaluated in four dimensions. Let us now calculate it up to the order $\mathcal{O}(\ep)$. 

Using superscripts for orders in $\ep$, we get from expanding~\eqref{sum}
\begin{subequations}
\begin{align}
\label{sum-orders-1}
    f^0_{G_1} &= -\Delta_0^{-1}f^0_{G}, \\
    f^1_{G_1} &= \gamma_{E}\Delta^{-1}_0f^0_G-\Delta_0^{-1}f^1_{G} + \Delta_0^{-1}\delta_0\Delta_0^{-1}f_G^0,
\label{sum-orders-2}
\end{align}
\end{subequations}
where $\gamma_{E}$ is the Euler–Mascheroni constant. 
Note that the definition of the operator $\Delta_0^{-1}$ is somewhat subtle, since the Laplace operator has the non-trivial kernel \eqref{kernel}. In this section, we consider the action of $\Delta_0^{-1}$ on regular (for $\ep\to0$) graphical functions. In this case, the result can be constructed
from any inverse of $\Delta_0$ by adding the element in the kernel so that the sum has the general properties of a graphical function.

It is convenient to define
\begin{equation}
\label{Laplace-inv}
    \Delta_0\inv = \invmz\sv\svb\;(\imz),
\end{equation}
where both sides are understood as operators acting on single-valued functions.

Equation~\eqref{sum-orders-1} then reproduces the four-dimensional calculation
\begin{equation}\label{fG1}
    f^0_{G_1}(z) = \dfrac{\loi - \lio}{z - \bar{z}}.
\end{equation}
Let us work out the three terms in~\eqref{sum-orders-2}. The first term is $\gamma_Ef^0_{G_1}(z)$.

The propagator in $D$ dimensions is $\norm{x}^{-2\lambda}$ (where $\lambda:=1-\ep$).
Therefore, the graphical function for $G$ is
\begin{flalign}
    \fg(z) &= \frac{1}{(\modz)^\lambda(\modzi)^\lambda}&&\nonumber \\
    &= \invz\invzi(\modz\modzi)^\ep \;=\; \frac{1+\ep\log(\den)}{\den} + \mathcal{O}(\ep^2)&&\nonumber \\[4pt]
    &= \frac{1+\ep(\lo +\li)}{\den}+ \mathcal{O}(\ep^2),
\end{flalign}
which gives
\begin{equation}
    f_G^{1}(z) = \dfrac{\lo + \li}{z\bar{z}(z- 1)(\bar{z} - 1)}.
\end{equation}
Applying the inverted Laplacian~\eqref{Laplace-inv} we arrive at
\begin{equation}
\label{Laplace-inv-applied}
    \Delta_0\inv\fg^{1}(z) = \invmz\sv\svb\left(\frac{1}{z(\bar{z}-1)}-\frac{1}{\bar{z}(z-1)}\right)(\lo +\li).
\end{equation}
To integrate with respect to $\bar{z}$ we use the fact that for a single-valued multiple polylogarithm with word $w$
\begin{equation}\label{antiprim}
  \partial_{\bar{z}} \mathcal{L}_{aw}(z) = \frac{\mathcal{L}_w(z)}{\bar{z} - a}\quad\text{if $w$ has at most two letters.}
\end{equation}
Note that \eqref{antiprim} is false for words $w$ with length $\geq3$.
Inverting \eqref{antiprim}, we easily see
\begin{align}
    \svb\;\left(\frac{1}{z(\bar{z}-1)}-\frac{1}{\bar{z}(z-1)}\right)\big(\lo +\li\big) =\dfrac{\lio + \lii}{z} - \dfrac{\loo + \loi}{z - 1}.
\end{align}
An integration with respect to $z$ gives
\begin{equation}
\Delta_0\inv\fg^1(z)=\frac{\lioo+\liio-\looi-\loii}{\imz}.
\end{equation}

For the third term in \eqref{sum-orders-2}, the first application of the inverse Laplace operator was already performed in~\eqref{sum-orders-1} and \eqref{fG1}. Application of $\delta_0$ gives
\begin{align}
    \delta_0\Delta_0^{-1}f^0_G(z) &= \dfrac{1}{z - \bar{z}}(\partial_z - \partial_{\bar{z}})\dfrac{\lio - \loi}{z - \bar{z}} \nonumber \\
    &= \frac{1}{(\imz)^2}\left(\frac{\li}{z}-\frac{\lo}{z-1}-\frac{\lo}{\bar{z} - 1}+\frac{\li}{\bar{z}}\right)+2\,\frac{\loi-\lio}{(\imz)^3}.
\end{align}
As a final step, we apply the inverse Laplace operator once again using formula~\eqref{Laplace-inv}. Integrating with respect to $\bar{z}$ we obtain
\begin{flalign}
    \svb&\;\bigg[\underbrace{\frac{2}{(\imz)^2}(\loi-\lio)}_{\text{IBP}}+\underbrace{\frac{1}{\imz}\bigg[\left(\frac{1}{z}+\frac{1}{\bar{z}}\right)\li-\left(\frac{1}{z-1}+\frac{1}{\bar{z}-1}\right)\lo\bigg]}_{\text{(I)}}\bigg] && \notag \\[7pt]
    &=\;2\frac{\loi-\lio}{\imz}-\svb\;\frac{2}{\imz}\left(\frac{\li}{\bar{z}}-\frac{\lo}{\bar{z}-1}\right) + \text{(I)} && \notag \\[4pt]
     &=\;2\frac{\loi-\lio}{\imz}+\svb\frac{1}{\imz}\left[\left(\frac{1}{z}-\frac{1}{\bar{z}}\right)\li-\left(\frac{1}{z-1}-\frac{1}{\bar{z}-1}\right)\lo\right]&&\notag \\
    &=\;2\frac{\loi-\lio}{\imz}-\svb\left(\frac{\li}{z\bar{z}}-\frac{\lo}{(z-1)(\bar{z}-1)}\right)&&\notag \\
    &=\;2\frac{\loi-\lio}{\imz}-\frac{\loi}{z}+\frac{\lio}{z-1}.
\end{flalign}
The primitive (with respect to $z$) is not a single-valued multiple polylogarithm but a
GSVH (introduced in the previous section; see~\cite{Schnetz:2021ebf} for details).
Nonetheless, we can (formally) perform the final integration over $z$ and obtain
\begin{equation}
\Delta_0\inv\delta_0\Delta_0\inv \fg^0(z)=\frac{2(\loiz-\lioz)-\loio+\lioi}{\imz}.
\end{equation}

Combining all three terms, we finally arrive at the result
\begin{align}
    f^1_{G_1}(z) &= \dfrac{1}{z - \bar{z}}\Big(\gamma_{E}\big(\lio - \loi\big) + 2\big(\loiz - \lioz\big) \nonumber \\
    &\quad + \looi + \lioi - \lioo - \liio + \loii - \loio\Big).
\end{align}
Note that the result is not singular on the real axis; the only singularities are at $0, 1, \text{and}~\infty$. This is a general property of graphical functions \cite{Golz:2015rea}.
The result is explicitly single-valued, symmetric under $z\to\bar{z}$ (this can be checked
with \cite{hyperlogprocedures}) and has the right pole structure at $0$, $1$ and $\infty$.
This determines the result uniquely as graphical function and we do not need to add
a correction term from the kernel of $\Delta_0^{-1}$.

To conclude this section, let us mention that the use of formula~\eqref{sum} makes it possible to compute the expansion in $\ep$ in a very straightforward and algorithmic way. As a result, attaching a leg to the graphical functions is very efficient. For example, using \texttt{HyperlogProcedures}~\cite{hyperlogprocedures} within the computer algebra system \texttt{MAPLE}, the result for $f_{G_1}(z)$ can be obtained up to $\mathcal{O}(\ep^{10})$ on a single core.

\section{Attaching a leg in dimensional regularisation, singular case}
\label{sec:leg_dimreg_sing}
Now that we understand the main idea behind appending an edge in the regular case, let us look at divergent integrals.
Here we will find the problem that $\Delta_0$ has a kernel:
\begin{align}
    \Delta_0 \frac{1}{z\bar{z}} = \frac{1}{z-\bar{z}} \partial_z \partial_{\bar{z}} \frac{z-\bar{z}}{z \bar{z}} = 0.
\end{align}
A convergent graphical function $f_{G_1}(z)$ never has a term $(z\bar{z})^{-1}$.
It is of order $\abs{z}^{0}$ in the limit $z \to 0$, but it may have a term of the form $\log(z \bar{z})$. 
For example:
\begin{align}
    \thetree = \frac{\mathcal{L}_{01}(z)-\mathcal{L}_{10}(z)}{z-\bar{z}} = 2 - \log (z \bar{z}) + \mathcal{O}(\abs{z}).
\end{align}
On the contrary, a graphical function $f_{G_1}(z)$ that is singular for $\ep\to0$ may have terms proportional to $(z \bar{z})^{-1}$.
For example:
\begin{align}
    f_{G_1}(z) = \bubblethreestar = \frac{1}{\ep z \bar{z}}+ \mathcal{O}(|z|^0)+ \mathcal{O}(\ep^0).
\end{align}
There are two possible solutions to this problem.
We can either subtract the singularities before performing the integration or correct the result after integration.
Both solutions are very similar, but the first has the advantage of providing a valuable cross-check.

Let us explore this approach by calculating $g_{G_1}(z)$ in the above example to order $\ep^0$.
The graph $G_1$ has a sub-divergence due to its bubble subgraph.
We consider the effective Laplace equation with $\lambda = 1 - \ep$,
\begin{equation}
     \left[ \frac{1}{z-\bar{z}} \partial_z \partial_{\bar{z}} (z-\bar{z}) + \frac{\ep }{z-\bar{z}}( \partial_z - \partial_{\bar{z}}) \right] f_{G_1}(z) = - \frac{1}{\Gamma(\lambda)} f_G(z),
\end{equation}
where
\begin{align}
    f_G(z) = \bubblethreestarremoved = \frac{1}{(z \bar{z})^{2 \lambda} ((z-1)(\bar{z}-1))^\lambda} \overset{\abs{z}\rightarrow 0}{\longrightarrow}  \abs{z}^{-4}
\end{align}
\begin{figure}
\centering
\asymptoticExpansion
\caption{The asymptotic expansion of graphical functions at $z=0$. Bold lines stand for sets of edges.}
\label{fig:asympt}
\end{figure}

leads to a logarithmic divergence in $f_{G_1}(z)$ where $z$ is replaced by the internal
variable $x\in\mathbb{R}^D$.
We subtract this divergence using the expansion of $f_G(z)$ at $z=0$ depicted in Figure \ref{fig:asympt},
\begin{align}\label{asex}
    f_G (z) = \sum_{V \subset V^{\mathrm{int}}_G} f_{G[V \cup \{0,z\}]}(z) f_{G[V \cup \{0,z\} = 0]}(z) (1 + \mathcal{O}(\abs{z})).
\end{align}
We denote by $G[V \cup \{0,z\}]$ the induced graph consisting of all edges in $G$ with vertices of $V \cup \{0,z\}$, while $G[V \cup \{0,z\} = 0]$ represents the factor graph where all edges and vertices in $V \cup \{0,z\}$ are contracted to the vertex $0$. 
$V_G^{\text{int}}$ is the set of internal vertices of $G$.
The result is only valid if the right hand side exists (in exotic situations, it can happen that some expression on the right hand side does not exist in dimensional regularization). The expansion formula (\ref{asex}) is a position space version of the method of expansion by regions in momentum space (see e.g.\ \cite{Smirnov:2021dkb} and the references therein). It assumes that all contributions to the expansion can be obtained by simple scalings of the 
integration variables. Mathematically, it has the status of a conjecture that has been tested in many thousands of cases. Still, we know from tropical geometry that the fully generic situation can be much more complicated. It seems to be a property of QFTs that it is particularly well-behaved in such expansions.

In our example, $V_G^{\mathrm{int}}=\emptyset$ and the induced and factor graphs are
\begin{equation}
  \label{equ:inducedFactorGraphs}
    f_{G[\{0,z\}]} = \bubblethreestarinduced = \frac{1}{(z \bar{z})^{2\lambda}} \quad\text{and}\quad
    f_{G[\{0,z\}=0]} = \bubblethreestarfactor = 1,
\end{equation}
respectively.
After subtracting their product, we get the regulated expression
\begin{align}
    f^{\text{reg}}_G(z) = \frac{1-((z-1)(\bar{z}-1))^\lambda}{(z \bar{z})^{2 \lambda} ((z-1)(\bar{z}-1))^\lambda} \overset{\abs{z}\rightarrow 0}{\longrightarrow} \abs{z}^{-3}.
\end{align}
This is no longer divergent upon integration over $x\in\mathbb{R}^D$ in $f_{G_1}(z)$.
Expanding in $ \ep $ using $ \lambda = 1 - \ep $ gives to zeroth order
\begin{align}
    f^{\text{reg}}_G(z) = \frac{z + \bar{z} - z \bar{z}}{(z \bar{z})^2 (z-1) (\bar{z}-1)} + \mathcal{O} (\ep).
\end{align}
The Laplace operator to zeroth order in $\ep$ reads
\begin{align}
    \Delta = \Delta_0 + \mathcal{O}(\ep)= \frac{1}{z-\bar{z}} \partial_z \partial_{\bar{z}} (z-\bar{z})  + \mathcal{O}(\ep).
\end{align}
We obtain
\begin{align}
    \Delta f^{\text{reg}}_{G_1} (z) &= \Delta_0 f^{\text{reg}}_{G_1} (z) + \mathcal{O}(\ep)
    = - \frac{1}{\Gamma(1)} f^{\text{reg}}_{G} (z) + \mathcal{O}(\ep).
\end{align}
Since we only took the constant term in $\ep$, this is the same expression as in $D=4$ dimensions.
We again follow our integration procedure.
First, we multiply by $z-\bar{z}$ and do a partial fraction decomposition to get the differential equation
\begin{equation}
     \partial_{\bar{z}}  \partial_z (z-\bar{z}) f^{\text{reg}}_{G_1} (z)  = - \frac{1}{z^2(\bar{z}-1)} +  \frac{1}{\bar{z}^2(z-1)} + \mathcal{O}(\ep).
\end{equation}
Next, we integrate with respect to $\bar{z}$. Modulo higher terms in $\ep$ and ignoring integration constants, we obtain
\begin{align}
    \partial_z (z-\bar{z}) f^{\text{reg}}_{G_1} (z)  &= 
    \svb  \left( - \frac{1}{z^2(\bar{z}-1)} +  \frac{1}{\bar{z}^2(z-1)} \right)
    = -\frac{1}{z^2} \mathcal{L}_1(z) -  \frac{1}{\bar{z}(z-1)}
\end{align}
and with respect to $z$
\begin{align}
    ( z-\bar{z}) f^{\text{reg}}_{G_1} (z)  &= \sv  \left( -\frac{1}{z^2} \mathcal{L}_1(z) -  \frac{1}{\bar{z}(z-1)}  \right) \notag  \\
    &= \frac{1}{z} \mathcal{L}_1(z)  - \frac{1}{\bar{z}} \mathcal{L}_1(z)  -\sv  \left( \frac{1}{z-1} - \frac{1}{z} \right) \notag  \\
    &= -\frac{z-\bar{z}}{z \bar{z}} \mathcal{L}_1(z) - \mathcal{L}_1(z) + \mathcal{L}_0(z).
\end{align}
Dividing by $z-\bar{z}$ yields the expression
\begin{align}
     f^{\text{reg}}_{G_1} (z)  = -\frac{\mathcal{L}_1(z)}{z \bar{z}} + \frac{\mathcal{L}_0(z) - \mathcal{L}_1(z)}{z-\bar{z}}+C(z) + \mathcal{O}(\ep)
\end{align}
for some function $c(z)$ in the kernel of $\Delta_0$.
The first term is symmetric under the exchange $ z \leftrightarrow \bar{z} $ and exhibits singularities at $ z = 0 $, $ z = 1 $, and $ z = \infty $, whereas the second term is antisymmetric under this exchange and singular along the real axis.
Furthermore, the second term is in the kernel of $\Delta_0$
\begin{align*}
    \Delta_0  \frac{\mathcal{L}_0(z) - \mathcal{L}_1(z)}{z-\bar{z}} &= \frac{1}{z-\bar{z}} \partial_{\bar{z}} \partial_z (\mathcal{L}_0(z) - \mathcal{L}_1(z)) = \frac{1}{z-\bar{z}} \partial_{\bar{z}}  \left( \frac{1}{z} - \frac{1}{z-1} \right) = 0.
\end{align*}
The function $C(z)$ has to cancel this term because it cannot appear in a graphical function.
(There exists an algorithm to subtract these spurious terms in general.)
Finally, the regularised part is
\begin{align}
    f^{\text{reg}}_{G_1} (z) = -\frac{\mathcal{L}_1(z)}{z\bar{z}} + \mathcal{O}(\ep).
\end{align}
Looking at the graphs \eqref{equ:inducedFactorGraphs}, we see that we have subtracted the bubble from $G$. From the graphical function of $G_1$, we have subtracted a bubble with
a leg attached.
This subtraction is the two-point function
\begin{align}
    f_{G_1}^\text{sing}(z) = \bubblethreestarfactoredattached =\int \frac{\dd[D]{x} }{\pi^\frac{D}{2}} \frac{1}{\norm{x}^{4\lambda}} \frac{1}{\norm{x-z}^{2\lambda}}.
\end{align}
Solving for this, we can recover the singular part.
Using the formula
\begin{equation}
  \label{equ:dimRegConvolution}
    \int \frac{\dd[D]{x} }{\pi^\frac{D}{2}} \frac{1}{\norm{x}^{2\lambda \nu}} \frac{1}{\norm{x-z}^{2\lambda \mu}} = \frac{\Gamma((1-\nu)\lambda + 1)\Gamma((1-\mu)\lambda + 1) \Gamma((\nu + \mu -1)\lambda -1)}{\Gamma(\lambda \nu)\Gamma(\lambda \mu)\Gamma((2-\nu-\mu)\lambda+2)} \frac{1}{\norm{z}^{2\lambda(\mu+\nu)-D}}
\end{equation}
with $\mu=1$, $\nu = 2$ and $\lambda = 1-\ep$, we find
\begin{equation}
  \begin{aligned}
    f_{G_1}^\text{sing}(z) &= \int \frac{\dd[D]{x} }{\pi^\frac{D}{2}} \frac{1}{\norm{x}^{4\lambda}} \frac{1}{\norm{x-z}^{2\lambda}}
    =  \frac{\Gamma(\ep)\Gamma(1)\Gamma(1-2\ep)}{\Gamma(2-2\ep)\Gamma(1-\ep)\Gamma(1+\ep)} \frac{1}{\norm{z}^{2-4\ep}} \\ 
   &= \frac{1}{\ep (1-2\ep)\Gamma(1-\ep) (z \bar{z})^{1-2\ep}}
   = \frac{1}{\ep z \bar{z}} + \frac{2-\gamma_E + 2 \mathcal{L}_0(z)}{z \bar{z}} + \mathcal{O} (\ep).
  \end{aligned}
\end{equation}
Combining the regular and singular contributions yields
\begin{align}\label{threestarsing}
    f_{G_1}(z) = f_{G_1}^\text{reg}(z) + f_{G_1}^\text{sing}(z) = \frac{1}{\ep z \bar{z}} + \frac{2-\gamma_E + 2 \mathcal{L}_0(z) - \mathcal{L}_1(z)}{z \bar{z}} + \mathcal{O} (\ep).
\end{align}
Note that we did not need to analyse the singularity structure in advance.
We also did not have to start from order $\ep^1$ because we did not lose any order in $\ep$ during the calculation.

In general, the function $f_{G_1}^\text{reg}(z)$ contains most of the structure of $f_{G_1}(z)$ missing poles in $\ep$, while $f_{G_1}^\text{sing}(z)$ has the poles but is simpler because it only demands the calculation of two-point functions.
The scaling behaviour in the first term of (\ref{equ:inducedFactorGraphs}) follows
from power counting. We need the exact scaling weight (here $2\lambda$),
it must not be expanded in $\ep$.
The prefactor and the second term in (\ref{equ:inducedFactorGraphs}) have to
be calculated to higher order in $\ep$ because (\ref{equ:dimRegConvolution}) will generate a pole in $\ep$ (here we need the order $\ep^1$, but both terms are
exactly $1$).
This can be done using graphical functions, where the simplicity of two-point functions typically (over-)compensates the difficulty due to the extra order in $\ep$.

Singularities at $z=1$ can be treated analogously, while singularities at $z=\infty$ are infrared singularities and require a slightly different subtraction.
In this case, the correct expansion is
\begin{align}\label{eq:expinfty}
    f_G (z) = \sum_{V \subset V^{\mathrm{int}}_G} f_{G[V \cup \{0,1\}]}(z) f_{G[V \cup \{0,1\} = 0]}(z) (1 + \mathcal{O}(1/\abs{z})).
\end{align}
We exemplify this case with the infrared singular graphical function
\begin{equation}
\label{equ:IRthreestar}
  f_{G_1}(z) = \IRthreestar.
\end{equation}
Here, the graphical function $G$ without a leg that is attached to $z$ scales like
\begin{align}
    f_G(z) = \frac{1}{(z \bar{z})^{a \ep} ((z-1)(\bar{z}-1))^\lambda} \overset{\abs{z}\rightarrow \infty}{\longrightarrow}  \abs{z}^{-2}.
\end{align}
This leads to an infrared divergence in the graphical function $f_{G_1}$ in the
limit $\ep\to0$ because with the attached leg $1/\norm{z-x}^{2\lambda}$ the integrand scales as $\norm{x}^4$ as $x\rightarrow\infty$.
So, a $\abs{z}^{-2}$ behaviour at infinity of $f_G(z)$ always implies a logarithmic infrared divergence in $f_{G_1}(z)$ (in $D = 4$ dimensions).

In our case, the only contribution to (\ref{eq:expinfty}) is
\begin{equation}
    f_{G[\{0,1\}]} = 1 \quad\text{and}\quad
    f_{G[\{0,1\}=0]}(z) = \IRthreestarfactor = \frac{1}{(z \bar{z})^{a\ep+\lambda}}.
\end{equation}
After subtracting the singularity, we get the regularised expression
\begin{equation}
    f^\text{reg}_G(z) = \frac{1-((\frac{1}{z}-1)(\frac{1}{\bar{z}}-1)^\lambda}{(z \bar{z})^{a \ep} ((z-1)(\bar{z}-1)^\lambda} \overset{\abs{z}\rightarrow \infty}{\longrightarrow} \abs{z}^{-3}.
\end{equation}
The calculation for the regularised part is performed in direct analogy to the previous example.
We expand in $\ep$,
\begin{align}
    f^\text{reg}_G(z) = \frac{1}{(z-1)(\bar{z}-1)} - \frac{1}{z \bar{z}} + \mathcal{O(\ep)},
\end{align}
multiply by $-(z-\bar{z})$,
\begin{align}
   \partial_{\bar{z}} \partial_z (z- \bar{z}) f^\text{reg}_{G_1}(z)  = \frac{1}{z-1} - \frac{1}{\bar{z}-1}+ \frac{1}{\bar{z}} - \frac{1}{z}+\mathcal{O(\ep)},
\end{align}
integrate with respect to $\bar{z}$,
\begin{align}
   \partial_z (z- \bar{z}) f^\text{reg}_{G_1}(z)  =  \frac{\bar{z}}{z-1} -\mathcal{L}_1(z) + \mathcal{L}_0(z) - \frac{\bar{z}}{z}+C_1(z)+\mathcal{O(\ep)},
\end{align}
integrate with respect to $z$ using integration by parts,
\begin{align}
    (z- \bar{z}) f^\text{reg}_{G_1}(z) = (\mathcal{L}_0(z) -  \mathcal{L}_1(z)) (z- \bar{z}) + \mathcal{L}_1(z)+C_2(z)+\mathcal{O(\ep)},
\end{align}
and divide by $z-\bar{z}$,
\begin{align}
    f^\text{reg}_{G_1}(z) = \mathcal{L}_0(z) -  \mathcal{L}_1(z) + \frac{\mathcal{L}_1(z)}{z - \bar{z}}+C(z)+\mathcal{O(\ep)},
\end{align}
where $C_1,C_2,C$ are integration `constants'.
As before, $C(z)\in\text{ker}\,\Delta_0$ kills the term $\mathcal{L}_1(z)/(z - \bar{z})$.

To obtain the singular contribution, we consider the graph subtracted from $G$ and attach an edge,
\begin{equation}
    f_{G_1}^\text{sing}(z) = \IRthreestarfactoredattached = \int \frac{\dd[D]{x} }{\pi^\frac{D}{2}} \frac{1}{\norm{x}^{2\lambda + 2a\ep}} \frac{1}{\norm{x-z}^{2\lambda}}.
\end{equation}
Using \eqref{equ:dimRegConvolution} with $\mu = 1$ and $\nu = 1 + a\ep/\lambda$, we get
\begin{align}
    f^\text{sing}_{G_1}(z) &= \frac{1}{(-a\ep+1)(a\ep -\ep) \Gamma(1-\ep)} \frac{1}{\abs{z}^{2(a-1)\ep}} \notag \\
    &= \frac{1}{(a-1)\ep} + \frac{a-\gamma_E}{a-1} -\mathcal{L}_0(z) + \mathcal{O}(\ep).
\end{align}
Adding the regular and singular contributions yields
\begin{align}
\label{equ:IRthreestarSol}
    f_{G_1}(z)  = f^\text{reg}_{G_1}(z)  + f^\text{sing}_{G_1}(z) = \frac{1}{(a-1)\ep} + \frac{a-\gamma_E}{a-1} -\mathcal{L}_1(z) + \mathcal{O}(\ep).
\end{align}
\section{Identities for periods and graphical functions (overview)}
\label{sec:identities}

Graphical functions are powerful because they are specialised to three‑point functions in quantum field theory, which satisfy many non-trivial identities originating from symmetry, conformal properties, and the underlying graph structure. 
Three‑point functions are particularly valuable because they form the basic building blocks for more complicated amplitudes, and their constrained structure allows one to extract analytic information that is otherwise difficult to obtain.

In many applications, however, the ultimate goal is the computation of two‑point functions. These also obey important identities and are central objects in the renormalization of quantum field theory. 
In renormalization procedures, two‑, three‑, and/or four-point functions need to be renormalized to determine propagator and vertex counterterms. In practice, one reduces higher $n$-point
functions to two-point functions by taking limits. This reduction does not alter the renormalization of the $n$-point functions.

In Section \ref{sec:periods}, we saw how graphical functions can be used in integer dimensions to calculate two-point functions. An alternative to this integration over $z$ using the residue theorem is taking the limit $z\to0$ (or $1$, $\infty$) \cite{Brown:2015ztw}.
This method also works in non-integer dimensions where the residue method is tricky to use
because the integration measure acquires powers of $\log(z-\bar{z})$ that lead out
of our preferred function space of GSVHs \cite{Schnetz:2021ebf}.

With the possibility to integrate over $z$, three‑point graphical functions are an efficient tool for obtaining renormalization data across different types of Green functions.

\begin{table}[H]
    \centering
    \begin{tabular}{c|c|c|c}
         &even integer dimensions   & \makecell{$D = 4 - 2\ep$\\ (convergent for $\ep$ = 0)} & \makecell{$D = 4 - 2\ep$ \\ (div.\ for $\ep$ = 0)} \\
         \hline
        \makecell{numbers,\\periods,\\2-point functions } &  \makecell{classical Feynman periods;\\completion,\\factorisation, twist} &\makecell{taking the limit $z\to0$}& \makecell{$\Delta$-Y factors\\(uniqueness)}\\
        \hline
         \makecell{3-point functions}&\makecell{classical graphical functions;\\ adding edges, transformations\\IBP, external differentiation\\ \texttt{HyperInt}, \texttt{HyperForm}}  & \makecell{approximation} & \makecell{rerouting}
    \end{tabular}
    \caption{Some techniques applied in the computation of two- and three point functions.}
    \label{tab:GFoverview}
\end{table}
In Table \ref{tab:GFoverview} we list some techniques and structural relations that are introduced to handle two‑ and three‑point functions in three different settings: even integer dimensions, dimensional regularisation without divergences, and fully general dimensional regularisation. Each of these regimes presents its own challenges, and different identities or analytic tools become available depending on the underlying QFT.

\begin{figure}
\centering
\begin{tikzpicture}[x=2.3ex,y=2.3ex]

    \begin{scope}[local bounding box=fullcateye]
        \coordinate (vm);
        \coordinate [above=1 of vm] (vo);
        \coordinate [left =2 of vm] (vl);
        \coordinate [right=2 of vm] (vr);
        \coordinate [left =1 of vl] (vll);
        \coordinate [right=1 of vr] (vrr);

        \draw (vm) to [bend left =90] (vo);
        \draw (vm) to [bend right=90] (vo);
        \draw (vl) to [bend left =20] (vo);
        \draw (vr) to [bend right=20] (vo);
        \draw (vl) to [bend right=50] (vr);

        \draw (vl) -- (vm);
        \draw (vr) -- (vm);
        \draw (vll) -- (vl);
        \draw (vrr) -- (vr);

        \node [left =.2 of vll] {$0$};
        \node [right=.2 of vrr] {$1$};

        \filldraw (vm) circle (1pt);
        \filldraw (vo) circle (1pt);
        \filldraw (vl) circle (1pt);
        \filldraw (vr) circle (1pt);
        \filldraw (vll) circle (1pt);
        \filldraw (vrr) circle (1pt);
    \end{scope}

    \begin{scope}[xshift=130,local bounding box=fullcateyez]
        \coordinate (vm);
        \coordinate [above=1 of vm] (vo);
        \coordinate [left =2 of vm] (vl);
        \coordinate [right=2 of vm] (vr);
        \coordinate [left =1 of vl] (vll);
        \coordinate [right=1 of vr] (vrr);

        \draw (vm) to [bend left =90] (vo);
        \draw (vm) to [bend right=90] (vo);
        \draw (vl) to [bend left =20] (vo);
        \draw (vr) to [bend right=20] (vo);
        \draw (vl) to [bend right=50] (vr);

        \draw (vl) -- (vm);
        \draw (vr) -- (vm);
        \draw (vll) -- (vl);
        \draw (vrr) -- (vr);

        \node [left =.2 of vll] {$0$};
        \node [right=.2 of vrr] {$1$};
        \node [above=.2 of vo ] {$z$};

        \filldraw (vm) circle (1pt);
        \filldraw (vo) circle (1pt);
        \filldraw (vl) circle (1pt);
        \filldraw (vr) circle (1pt);
        \filldraw (vll) circle (1pt);
        \filldraw (vrr) circle (1pt);
    \end{scope}

    \draw[-latex,thick] (fullcateye-|fullcateyez.west) -- (fullcateye) node[midway,above]{$\int \dd^2 z$};
    \draw[-latex,thick,dashed]  ([xshift=30]fullcateye-|fullcateyez.east) -- (fullcateye-|fullcateyez.east)  node[midway,above]{};

    \begin{scope}[yshift=-40,local bounding box=fullcateyez2]
        \coordinate (vm);
        \coordinate [above=1 of vm] (vo);
        \coordinate [below=1 of vm] (vu);
        \coordinate [left =2 of vm] (vl);
        \coordinate [right=2 of vm] (vr);
        \coordinate [left =1 of vl] (vll);
        \coordinate [right=1 of vr] (vrr);

        \draw (vm) to [bend left =90] (vo);
        \draw (vm) to [bend right=90] (vo);
        \draw (vl) to [bend left =20] (vo);
        \draw (vr) to [bend right=20] (vo);
        \draw (vl) to [bend right=50] (vr);

        \draw (vl) -- (vm);
        \draw (vr) -- (vm);
        \draw (vll) -- (vl);
        \draw (vrr) -- (vr);

        \node [left =.2 of vll] {$z$};
        \node [right=.2 of vrr] {$1$};
        \node [above=.2 of vo ] {$0$};

        \node [below=.2 of vu ] {\phantom{$0$}};

        \filldraw (vm) circle (1pt);
        \filldraw (vo) circle (1pt);
        \filldraw (vl) circle (1pt);
        \filldraw (vr) circle (1pt);
        \filldraw (vll) circle (1pt);
        \filldraw (vrr) circle (1pt);
    \end{scope}

    \draw[-latex,thick]  (fullcateyez2) -- ([xshift=-30]fullcateyez2-|fullcateyez2.west)  node[midway,above]{$0 \leftrightarrow z$};

    \begin{scope}[yshift=-40,xshift=130,local bounding box=cateyeredl]
        \coordinate (vm);
        \coordinate [above=1 of vm] (vo);
        \coordinate [below=1 of vm] (vu);
        \coordinate [left =2 of vm] (vl);
        \coordinate [right=2 of vm] (vr);
        \coordinate [left =1 of vl] (vll);
        \coordinate [right=1 of vr] (vrr);

        \draw (vm) to [bend left =90] (vo);
        \draw (vm) to [bend right=90] (vo);
        \draw (vl) to [bend left =20] (vo);
        \draw (vr) to [bend right=20] (vo);
        \draw (vl) to [bend right=50] (vr);

        \draw (vl) -- (vm);
        \draw (vr) -- (vm);
        \draw (vrr) -- (vr);

        \node [left =.2 of vl] {$z$};
        \node [right=.2 of vrr] {$1$};
        \node [above=.2 of vo ] {$0$};
        \node [below=.2 of vu ] {\phantom{$0$}};
        \node [right=.2 of vrr ] {\phantom{$0$}};
        \node [left=.2 of vll ] {\phantom{$0$}};

        \filldraw (vm) circle (1pt);
        \filldraw (vo) circle (1pt);
        \filldraw (vl) circle (1pt);
        \filldraw (vr) circle (1pt);
        \filldraw (vrr) circle (1pt);
    \end{scope}

    \draw[-latex,thick] (fullcateyez2-|cateyeredl.west) -- (fullcateyez2) node[midway,above]{$\Delta_0^{-1}$};
    \draw[-latex,thick,dashed]  ([xshift=30]fullcateyez2-|cateyeredl.east) -- (fullcateyez2-|cateyeredl.east)  node[midway,above]{};

    \begin{scope}[yshift=-80,xshift=0,local bounding box=cateyeredl2]
        \coordinate (vm);
        \coordinate [above=1 of vm] (vo);
        \coordinate [below=1 of vm] (vu);
        \coordinate [left =2 of vm] (vl);
        \coordinate [right=2 of vm] (vr);
        \coordinate [left =1 of vl] (vll);
        \coordinate [right=1 of vr] (vrr);

        \draw (vm) to [bend left =90] (vo);
        \draw (vm) to [bend right=90] (vo);
        \draw (vr) to [bend right=20] (vo);
        \draw (vl) to [bend right=50] (vr);

        \draw (vl) -- (vm);
        \draw (vr) -- (vm);
        \draw (vrr) -- (vr);

        \node [left =.2 of vl] {$z$};
        \node [right=.2 of vrr] {$1$};
        \node [above=.2 of vo ] {$0$};
        \node [below=.2 of vu ] {\phantom{$0$}};
        \node [right=.2 of vrr ] {\phantom{$0$}};
        \node [left=.2 of vll ] {\phantom{$0$}};

        \filldraw (vm) circle (1pt);
        \filldraw (vo) circle (1pt);
        \filldraw (vl) circle (1pt);
        \filldraw (vr) circle (1pt);
        \filldraw (vrr) circle (1pt);
    \end{scope}

    \draw[-latex,thick]  (cateyeredl2) -- ([xshift=-30]cateyeredl2-|cateyeredl2.west)  node[midway,above]{\text{ext.~edges}};

    \begin{scope}[yshift=-80,xshift=130,local bounding box=cateyeredl3]
        \coordinate (vm);
        \coordinate [above=1 of vm] (vo);
        \coordinate [below=1 of vm] (vu);
        \coordinate [left =2 of vm] (vl);
        \coordinate [right=2 of vm] (vr);
        \coordinate [left =1 of vl] (vll);
        \coordinate [right=1 of vr] (vrr);

        \draw (vm) to [bend left =90] (vo);
        \draw (vm) to [bend right=90] (vo);
        \draw (vr) to [bend right=20] (vo);
        \draw (vl) to [bend right=50] (vr);

        \draw (vl) -- (vm);
        \draw (vr) -- (vm);
        \draw (vrr) -- (vr);

        \node [left =.2 of vl] {$1$};
        \node [right=.2 of vrr] {$z$};
        \node [right=.2 of vrr ] {\phantom{$0$}};
        \node [left=.2 of vll ] {\phantom{$0$}};
        \node [above=.2 of vo ] {$0$};
        \node [below=.2 of vu ] {\phantom{$0$}};

        \filldraw (vm) circle (1pt);
        \filldraw (vo) circle (1pt);
        \filldraw (vl) circle (1pt);
        \filldraw (vr) circle (1pt);
        \filldraw (vrr) circle (1pt);
    \end{scope}

    \draw[-latex,thick]  (cateyeredl2-|cateyeredl3.west) -- (cateyeredl2)  node[midway,above]{$1 \leftrightarrow z$};

    \draw[-latex,thick,dashed]  ([xshift=30]cateyeredl3-|cateyeredl3.east) -- (cateyeredl3)  node[midway,above]{};

    \begin{scope}[yshift=-120,local bounding box=cateyeredlr]
        \coordinate (vm);
        \coordinate [above=1 of vm] (vo);
        \coordinate [below=1 of vm] (vu);
        \coordinate [left =2 of vm] (vl);
        \coordinate [right=2 of vm] (vr);
        \coordinate [left =1 of vl] (vll);
        \coordinate [right=1 of vr] (vrr);

        \draw (vm) to [bend left =90] (vo);
        \draw (vm) to [bend right=90] (vo);
        \draw (vr) to [bend right=20] (vo);
        \draw (vl) to [bend right=50] (vr);

        \draw (vl) -- (vm);
        \draw (vr) -- (vm);

        \node [left =.2 of vl] {$1$};
        \node [right=.2 of vr] {$z$};
        \node [above=.2 of vo ] {$0$};
        \node [below=.2 of vu ] {\phantom{$0$}};
        \node [right=.2 of vrr ] {\phantom{$0$}};
        \node [left=.2 of vll ] {\phantom{$0$}};

        \filldraw (vm) circle (1pt);
        \filldraw (vo) circle (1pt);
        \filldraw (vl) circle (1pt);
        \filldraw (vr) circle (1pt);
    \end{scope}

    \draw[-latex,thick]  (cateyeredlr) -- ([xshift=-30]cateyeredlr-|cateyeredlr.west)  node[midway,above]{$\Delta_0^{-1}$};

    \begin{scope}[yshift=-120,xshift=130,local bounding box=cateyeredlr2]
        \coordinate (vm);
        \coordinate [above=1 of vm] (vo);
        \coordinate [below=1 of vm] (vu);
        \coordinate [left =2 of vm] (vl);
        \coordinate [right=2 of vm] (vr);
        \coordinate [left =1 of vl] (vll);
        \coordinate [right=1 of vr] (vrr);

        \draw (vm) to [bend left =90] (vo);
        \draw (vm) to [bend right=90] (vo);

        \draw (vl) -- (vm);
        \draw (vr) -- (vm);

        \node [left =.2 of vl] {$1$};
        \node [right=.2 of vr] {$z$};
        \node [above=.2 of vo ] {$0$};
        \node [below=.2 of vu ] {\phantom{$0$}};
        \node [right=.2 of vrr ] {\phantom{$0$}};
        \node [left=.2 of vll ] {\phantom{$0$}};

        \filldraw (vm) circle (1pt);
        \filldraw (vo) circle (1pt);
        \filldraw (vl) circle (1pt);
        \filldraw (vr) circle (1pt);
    \end{scope}

    \draw[-latex,thick]  (cateyeredlr-|cateyeredlr2.west) -- (cateyeredlr)  node[midway,above]{\text{ext.~edges}};
    \draw[-latex,thick,dashed]  ([xshift=30]cateyeredlr2-|cateyeredlr2.east) -- (cateyeredlr2)  node[midway,above]{};

    \begin{scope}[yshift=-160,xshift=0,local bounding box=cateyeredlr3]
        \coordinate (vm);
        \coordinate [above=1 of vm] (vo);
        \coordinate [below=1 of vm] (vu);
        \coordinate [left =2 of vm] (vl);
        \coordinate [right=2 of vm] (vr);
        \coordinate [left =1 of vl] (vll);
        \coordinate [right=1 of vr] (vrr);

        \draw (vm) to [bend left =90] (vo);
        \draw (vm) to [bend right=90] (vo);

        \draw (vl) -- (vm);

        \node [left =.2 of vl] {$1$};
        \node [below right=.2 of vm] {$z$};
        \node [above=.2 of vo ] {$0$};
        \node [below=.2 of vu ] {\phantom{$0$}};
        \node [right=.2 of vrr ] {\phantom{$0$}};
        \node [left=.2 of vll ] {\phantom{$0$}};

        \filldraw (vm) circle (1pt);
        \filldraw (vo) circle (1pt);
        \filldraw (vl) circle (1pt);
    \end{scope}

    \draw[-latex,thick]  (cateyeredlr3) -- ([xshift=-30]cateyeredlr3-|cateyeredlr3.west)  node[midway,above]{$\Delta_0^{-1}$};


    \begin{scope}[yshift=-160,xshift=130,local bounding box=one]
        \coordinate (vm);
        \coordinate [above=1 of vm] (vo);
        \coordinate [below=1 of vm] (vu);
        \coordinate [left =2 of vm] (vl);
        \coordinate [right=2 of vm] (vr);
        \coordinate [left =1 of vl] (vll);
        \coordinate [right=1 of vr] (vrr);



        \node [left =.2 of vl] {$1$};
        \node [below right=.2 of vm] {$z$};
        \node [above=.2 of vo ] {$0$};
        \node [below=.2 of vu ] {\phantom{$0$}};
        \node [right=.2 of vrr ] {\phantom{$0$}};
        \node [left=.2 of vll ] {\phantom{$0$}};

        \filldraw (vm) circle (1pt);
        \filldraw (vo) circle (1pt);
        \filldraw (vl) circle (1pt);
    \end{scope}

    \draw[-latex,thick]  (one) -- ([xshift=-42]one-|one.west)  node[midway,above]{\text{ext.~edges}};
\end{tikzpicture}
\caption{Construction of the cat-eye graph in four dimensions (picture by M. Borinsky).}
\end{figure}

It is useful to distinguish two classes of tools. The first class comprises tools
that work to all orders in $\ep$, only limited by moderate constraints
from time and memory consumption. All calculations from the previous sections
fall into this class. We do not include tools in class one that work to all orders
in $\ep$ but are time-consuming to apply (twist, IBP).
The first class has adding edges, transformations, completion
(in integer dimensions), factorization, products, and taking the limit $z\to0$.
If a Feynman integral (a period, a two-point function or a graphical function)
can be calculated only using tools in class one, we call it `constructible'
(for a definition see Section 7 in \cite{Borinsky:2021gkd}).
It is a very easy and quick task to check if a Feynman graph has a constructible integral.
At low loop orders most (up to three loops typically all) Feynman integrals
are constructible and hence easy to calculate with the method of graphical functions.
At moderate loop orders, constructible graphs (integrals) are still
a sizeable fraction of all graphs.

The second class comprises all other tools. They are much less straightforward to
use and some (approximation, parametric integration) only work for low orders in $\ep$.
Some of these tools are standard techniques such as integration‑by‑parts (IBP) identities, introduced in \cite{Chetyrkin:1981qh}
and formalized in the Laporta approach~\cite{Laporta:2000dsw}. The parametric integration method was developed in \cite{Brown:2008um}. It exploits the linear reducibility of Feynman integrals in terms of hyperlogarithms, whenever such reducibility is present. It has been fully automated in the {\texttt{HyperInt}} package~\cite{Panzer:2014caa} for {\texttt{MAPLE}} and in {\texttt{HyperFORM}}~\cite{Kardos:2025klp} under \texttt{FORM}~\cite{Davies:2026cci}. 

The combination of all tools provide an efficient framework for performing parametric integrations and expressing the results in terms of iterated integrals, see also Section \ref{sec:further}.

\begin{figure}[ht!]
  \centering
\includegraphics{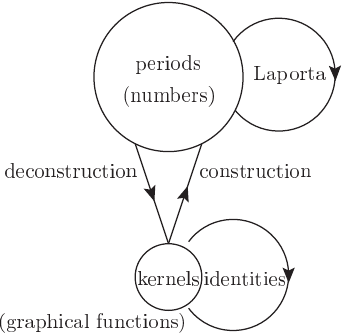}
\caption{(De-)construction relating  graphical functions and periods together with additional relations derived from IBP (Laporta algorithm) and identities among graphical functions.}
\label{fig:de-construction}
\end{figure}

The power of the theory of graphical functions lies in the ability to extend results for graphs
with a moderate number of edges to more complex graphs by further adding edges. 
Moreover, the interplay of graphical functions and periods via (de-)construction is illustrated in Figure~\ref{fig:de-construction}.

Table~\ref{tab:GFoverview} serves as a roadmap for the methods we will develop. 
In the remainder of this review, we will proceed through its entries systematically, moving from top to bottom and from left to right. The tools developed in previous steps still apply to
later steps; the entries in Table~\ref{tab:GFoverview} only show augmentations of the toolbox.  
We will build up the necessary toolkit step by step, starting from the most constrained situations and gradually extending to the general framework required for renormalization of QFT.

\section{Periods in even integer dimensions}
\label{sec:periodsEven}
Let us continue our discussion of periods from Sections \ref{sec:periods} and \ref{sec:k34}.
As a reminder, the numeric structure of the graphs
\begin{equation}\label{K4period}
  \WheelDiv \underline{\sim} ~\periodA \underline{\sim} ~\periodB \underline{\sim} ~6 \zeta_3
\end{equation}
includes the same number, the Feynman period $6\zeta(3)$ \cite{Kontsevich2001}.
Conjecturally, Feynman periods are a comodule under the motivic Galois coaction \cite{Panzer:2016snt,Brown:2015fyf,Brown:2015ylf}.

The motivic coaction is a mathematically interesting and restrictive structure,
but not yet very useful for the calculation of Feynman periods.
We will now go through some tools that we can use to facilitate an effective calculation.

A good review on Feynman periods for mathematically inclined readers is also \cite{Borinsky:2022lds}.

\subsection{Completion}\label{sectcomp}
Consider a primitive graph, i.e. one that is only logarithmically divergent and has no subdivergences.
An example is given by the tetrahedron in (\ref{K4period}).
Note that by scale invariance, a primitive graph can be opened at any edge to get a $p$-integral.
The corresponding period is the evaluation of the $p$-integral at $p = 1$.
All $p$-integrals obtained by opening any edges are identical.
The underlying mathematical structure is projective geometry, which allows one the transition
to different affine charts.

\begin{center}
$p$-integral, affine chart $\leftrightarrow$ primitive graph, projective space
\end{center}
It is also possible to calculate Feynman periods in position space where the evaluation
$p=1$ is replaced by choosing two vertices $z_0=0$ and $z_1$ with $\norm{z_1}=1$.
We use labels 0 and 1 for $z_0$ and $z_1$, respectively.

For completion, we use conformal symmetry that is generated by Euclidean symmetry plus
the inversion
\begin{equation}
    x \rightarrow \tilde{x} = \frac{x}{\norm{x}^2},
\end{equation}
which keeps the direction of $x$ but inverts the length of $x$.

The corresponding operation on the graph amounts to adding a vertex
with label $\infty$ and connecting all external (half-)edges to it.
The result is a vacuum graph, the completion $\overline{G}$ of $G$.
Some examples from various QFTs are:

\begin{alignat}{2}
\phi^4 :& \mathop{\periodBwithoutexternal}\limits_{K_4} &&\rightarrow \quad\periodBconnectedtoInfinity\quad=\quad\mathop{\kfive}\limits_{K_5}\quad=\;6\zeta_3,\notag\\
  &\thefourwexternal &&\rightarrow\quad  \TriangleWithOuterCurves \quad= \;1,\notag\\
  &\completionKfivetwoinverted ~ &&\rightarrow \quad\completionKseven=\;(6\zeta(3))^2,\notag\\
  &\mathop{\kperiodlem}\limits_{K_{3,4}} &&\rightarrow \quad \mathop{\kperiodfour}\limits_{K_{4,4}}\quad=\;P_{6,4}\;\text{ in \cite{Schnetz:2008mp}}\notag\\
\phi^3 :& \mathop{\TriangleWithExternals}\limits_{K_3} &&\rightarrow\quad\Trianglecompletion = \quad\mathop{\WheelDivrotated}\limits_{K_{4}}\hspace{-7mm}=\;1\notag\\
\text{Yukawa-}\phi^4:& \YukawaTriangle &&\rightarrow ~(-1)\;\BoxWithDiagonals =\; -1\notag\\
&\BoxWithExternals &&\rightarrow\quad\BoxWithDiagonalsVertex =\; 1\notag\\
&\Yukawavertex &&\rightarrow (-1)\;\BoxWithDiagonalsVertex =\;-1
\end{alignat}
In Yukawa-$\phi^4$ theory (the Gross-Neveu-Yukawa model~\cite{Gross:1974jv}), we obtain a factor of $-1$
by completing a Yukawa vertex \cite{Schnetz:2025opm}. 
In the last two examples, we saw that the same completed graph contributes to the Yukawa $\beta$-function and to the $\phi^4$ $\beta$-function in Yukawa-$\phi^4$ theory. Two more examples from the three-loop Yukawa $\beta$-function:
\begin{align}
    &\threeloopYukawa\quad \rightarrow \quad(-1)~\CompletionthreeLoopYukawa =\;-1,
    \qquad \CompletionthreeLoopYukawatwo = \;3\zeta_3 -1
\end{align}
It is proved in \cite{Schnetz:2008mp} that every decompletion of a completed graph
has the same Feynman period. This implies that the vertices $0,1,\infty$ can be chosen
freely in completed graphs, so that we consider completed graphs as unlabelled.

In particular, convergence of Feynman periods can be defined in terms of completed graphs.
\begin{defin}
A completed graph is primitive if and only if its internal weighted edge-connectivity is strictly greater than $\frac{D}{\lambda} = \frac{2 D}{D-2}$.
That means that in every cut of a primitive completed graph into two graphs with two or more vertices each, the weights of the cut edges add up to a total weight $>\frac{D}\lambda$.
\end{defin}

\begin{theo}
Every decompletion of a completed primitive graph has the same Feynman period (even
if decompletions are non-isomorphic graphs). If a completed graph is not primitive,
then all decompletions are divergent.
\end{theo}
Physically, a non-trivial edge-cut with total weight
$\leq\frac{D}{\lambda}$ corresponds to a subdivergence in the decompleted graph.

\begin{defin}
The Feynman period of a completed primitive $\bar{G}$ is defined as
\begin{equation}
     P_{\overline{G}} = P_{\overline{G}\backslash v}  \hspace{3mm} \text{for any vertex $v \in \overline{G}$}.
\end{equation}
\end{defin}
To calculate $P_{\overline{G}}$ we choose any three vertices in $\overline{G}$ as $\{0, 1, \infty\}$.
Then we can decomplete by removing $\infty$ and calculate the Feynman integral of $\overline{G}$ with $0=0\in\mathbb{R}^D$ and $1=z_1\in\mathbb{R}^D$ with $\norm{z_1}=1$.

A completed primitive graph can be considered as equivalence class of primitive graphs with the same Feynman period.
Since many non-isomorphic graphs can have the same completion, the number of completed
primitive graphs is significantly smaller than the number of primitive graphs.
See \cite{Schnetz:2008mp,Schnetz:2016fhy} for a list of completed primitive graphs in $\phi^4$ theory. More results are in
\cite{hyperlogprocedures}.

\subsection{Factorisation}\label{sec:fac}
If the completed primitive graph $\overline{G}$ has a three-vertex cut, we can choose these three vertices to be the external vertices $\{0,1,\infty\}$.
Then $P_{\overline{G}}$ factorizes into the product of the cut periods.
The cut graphs are not yet vacuum graphs although they have a vertex $\infty$.
We thus have to add edges of suitable weights between all cut vertices to lift the
cut graphs to completed primitive graphs.
Graphically, this is of the form
\begin{equation}
  P_{\bar{G}} = \underset{\bar{G}}{\productAPeriod} \rightarrow
  \productBUnfinished, \quad \productCUnfinished \rightarrow 
  \underset{\overline{G}_1}{\productBPeriod} \times
  \underset{\overline{G}_2}{\productCPeriod} = P_{\overline{G}_1} \cdot P_{\overline{G}_2}.
\end{equation}
In the first step, we split the graph at the three cut vertices.
In the second step, we attach three edges (a triangle) to the cut vertices
to restore the vertex degrees of the cut vertices.
Then, we multiply the Feynman periods of the two completed graphs.
In $\phi^4$ theory, the extra edges in the cut graphs have weight 1.

In Yukawa-$\phi^4$ theory factorisation is slightly different. We have 
\begin{equation}
  P_{\overline{G}} = \underset{\overline{G}}{\YukawaG} \rightarrow
  \underset{\overline{G}_1}{\YukawaGone} \times
  \underset{\overline{G}_2}{\YukawaGtwo} = P_{\overline{G}_1} \cdot P_{\overline{G}_2}.
\end{equation}
We can apply factorisation to determine the period of the associated $p$-integral
\begin{equation}
  \completionKfivetwoP \rightarrow\qquad \completionKfivetwoinverted.
\end{equation}
We first complete this graph and then factorize along the three vertices in the middle
\begin{equation}
  \completionKfivetwoinverted \rightarrow\qquad \completionKseven \rightarrow\qquad \Ksevenfactorized =\; (6\zeta(3))^2
\end{equation}
 Conversely, factorisation gives a multiplication of Feynman periods whose graphs have
 triangles.

 Factorisation generalizes the method of uniqueness \cite{KAZAKOV1983406}. It is particularly powerful in
 theories that have a three-valent vertex and conformal symmetry in convergent subgraphs, so that completion is possible. Examples are delta-wye identities in six-dimensional $\phi^3$ theory \cite{Schnetz:2025wtu} and in four-dimensional Yukawa-$\phi^4$ theory \cite{Schnetz:2025opm}.
 
\subsection{Planar duality}
Periods of planar dual graphs are related.
\begin{theo}\label{thmplanarP}
   Let $G^{*}$ be the planar dual of the (non-completed) scalar graph $G$ with edge-weights $\nu_e^* = \frac{\lambda+1}{\lambda}-\nu_e$, where we identify the labels in $G^*$ and $G$ if the edges cross. The Feynman periods $P_{G^*}$ and $P_G$ are related by
   \begin{equation}
     P_{G} = \left(\prod_e \frac{\Gamma(\lambda\nu_e^*)}{\Gamma(\lambda\nu_e)}\right)P_{G^*}.
   \end{equation}
\end{theo}
The proof uses Fourier transformations, similar to the duality we used in the introduction~\cite{Broadhurst:1995km}. 
A trivial example is the self-dual graph $K_4$ in four dimensions,
\begin{equation}
  \WheelDivrotatedwdots \rightarrow\quad  ~\WheelDivrotatedwselfdual =\quad ~\SelfDualKFour
\end{equation}
with $\nu_e = 1$ for all edges $e$. The graph $K_4$ is also self-dual in dimensions
$\neq4$, but in this case the weights transform non-trivially, so that the dual
weighted graph may not be isomorphic to the original weighted graph.

Note that for any graph with $\nu_e=1$ for all edges $e$
we get $P_G = P_{G^*}$ in four dimensions.
A non-trivial example comes from the graph $P_{7,5}$ in \cite{Schnetz:2008mp}
which decompletes as
\begin{equation}
  P_{7,5} =\quad \Psevenfive\quad = \quad\PsevenfiveDecompleted\quad.
\end{equation}
Taking the dual and completing again, we get the equality of periods
\begin{equation}
  P_{7,5} = \quad\PseventenDecompleted = \quad\Pseventen = P_{7,10}\,.
\end{equation}
Note that there exists no planar duality in Yukawa-$\phi^4$ theory \cite{Schnetz:2025opm}.

\subsection{The twist}\label{thetwist}
If a completed primitive graph has a three vertex cut, we already saw the product identity.
For a four vertex cut there exists the (Schnetz) twist identity.
The twist is a pairwise exchange at the four split vertices on one side of the split graph, followed by suitably moving edges to render the twisted graph completed primitive \cite{Schnetz:2008mp}.
Graphically
\begin{equation}
  \PTwistLeft = \quad\PTwistRight,
\end{equation}
where $a,b,c,d$ are the split vertices and the dashed lines represent (potentially) moved edges.

These twists are known as ``magic identities'' in Super Yang-Mills theory \cite{Drummond:2006rz}.
Let us for example consider the graph $P_{7,4}$ in $\phi^4$ theory
\begin{equation}
  P_{7,4} =\quad \Psevenfour\quad.
\end{equation}
We twist the subgraph depicted below at the vertices $1-3$ and $2-4$.
\begin{equation}
  \PsevenfourSubgraph \rightarrow \PsevenfourSubgraphTwist
\end{equation}
Because every vertex in the subgraph has valence two, no edges have to be moved to
obtain a completed graph. Inserting the twisted graph back at the cut vertices $1,2,3,4$, we get the identity
\begin{equation}
  P_{7,4} =\quad\Psevenseven\quad = P_{7,7}.
\end{equation}

\subsection{The Fourier split}
The Fourier split combines the twist with the Fourier identity between planar duals
\cite{Fouriersplit}.
Schematically,
\begin{equation}\label{Fsplit}
  \FSplitLeft =\quad \FSplitRight,
\end{equation}
where the subgraph $\overline{G}_2^*$ is obtained as follows.

We decomplete the left hand side of (\ref{Fsplit}) at one of the four split vertices.
Then, we assume that one side is externally planar (planar with all three remaining
cut vertices on the outer face). If the weight of the planar side (say $G_2$)
satisfies $N_{G_2}=\frac{\lambda+1}\lambda$ (see Theorem \ref{thmplanar}), then
we exchange $G_2$ for its planar dual and glue it back into the graph, yielding
(the drawing is adopted from the PhD thesis of J. Reynes~\cite{reynesExtensionsTutteInterlace})

\begin{equation}
    \raisebox{-1.4ex}{\productWithPlanar} = \raisebox{-1.4ex}{\productFourierTwist}\;.
\end{equation}
Afterwards, we complete the graph to obtain the right hand side of (\ref{Fsplit}).
Note that in dimensions $\geq6$, the edge weights in the dual graph change according
to Theorems \ref{thmplanarP} and \ref{thmplanar}.

The Fourier split is rather rare; it has only been studied in four dimensions.

\subsection{The Five-twist}
The five-twist is a more complicated combination of twisting and taking planar duals.
The general idea is to decomplete, four vertex cut, dualize, twist, dualize again, glue, and complete.

The five-twist is the first identity acting on five vertex splits in the completed
graph. It was only found quite recently \cite{Schnetz:2025mjw}.

\begin{figure}
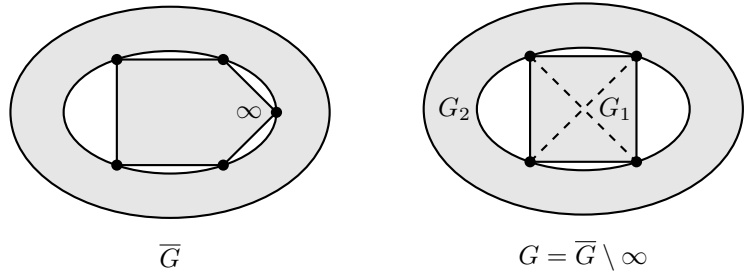

  \centering
  \FiveTwistComplete \qquad \FiveTwistDecompleted
  \caption{General form of the five-twist. The left graph is completed, the right graph represents a decompletion of the left graph where the five-twist acts
  as reflections of the inserted square along one or both of the dashed diagonals.}
  \label{fig:5twist}
\end{figure}
Schematically, the five-twist is depicted in Figure \ref{fig:5twist}.
Note that the five-twist requires a list of conditions on the insertions in Figure
\ref{fig:5twist}.

\subsection{Other identities}
Even more recently, a new identity was derived from the self-duality of massless
scalar three-point integrals \cite{Schnetz:2026mht}. 
This identity is explained in detail in Section \ref{sec:self_duality}.
It is quite powerful and leads at low loop orders to similar identities as the twist.

From combinatorial invariants, we expect that there exist more identities
for $\phi^4$ periods (see e.g.\ \cite{Panzer:2019yxl}). 
At eight loops, the first of these conjectured identities are
\begin{equation}\label{conjid}
P_{8,30}=P_{8,36}\qquad\text{and}\qquad P_{8,31}=P_{8,35}
\end{equation}
in the notation of \cite{Schnetz:2008mp}. There exists no chain of known identities
that establishes (\ref{conjid}). This lack of understanding of some identities shows
that the theory of periods and graphical functions is under ongoing development, with many important properties yet to be found.
\section{Graphical functions in integer dimensions}
\label{sec:gf_integer}
Now that we have seen multiple identities for periods, let us have a look at the more complicated case of graphical functions.

\subsection{External Edges}
We do not integrate over external vertices in graphical functions.
Therefore, all edges between them are trivial factors.
An edge of weight $\nu$ between the external vertices $0$ and $1$ contributes the factor $|1-0|^{-2\lambda\nu} = 1$ to the graphical function.
The edges $0$-$z$ and $1$-$z$ contribute $(z\bar{z})^{-\lambda\nu}$ and $((z-1)(\bar{z}-1))^{-\lambda\nu}$, respectively.
Graphically, this is
\begin{equation}
  \externalA = \externalB =(z\bar{z})^{\lambda\nu}\externalC =((z-1)(\bar{z}-1))^{\lambda\nu} \externalD.
\end{equation}

\subsection{Products}\label{sec:prod}
If the graph $G$ consists of two disjoint graphs that are connected only through the vertices $0$, $1$ and $z$, then the graphical function $f_G(z)$ factors into the product of the two subgraphs.
This also follows from the fact that we do not integrate over external vertices.
Graphically,
\begin{equation}\label{eqprod}
\productA = \productB \times \productC.
\end{equation}
We have used this product formula in \eqref{eq:productPeriod} to calculate the $K_{3,4}$-period.

The simplest non-trivial example is the graph with two non-adjacent internal vertices.
The graph splits into two identical graphs with one internal vertex whose graphical function was calculated in (\ref{three-point-res}),
\begin{equation}
  \productK = \left(\frac{\trilog}{\imz}\right)^2.
\end{equation}

\subsection{Completion}\label{sec:completion}
Completion (conformal symmetry) is an important tool also for the calculation of graphical functions, although
it is not quite as powerful as it is for Feynman periods.
The graph $G$ is completed by adding an external vertex with label $\infty$ that connects
to internal vertices with edges whose weights are chosen so that all internal vertices have weighted degree $\frac{D}{\lambda} = \frac{2D}{D - 2}$.
We also add edges $z$-$\infty$, $0$-$1$, $0$-$\infty$, $1$-$\infty$ so that all external vertices have a weighted degree of $0$.

\begin{lem}
Completion is always possible and unique. If the weights are (half-)integers in ($\phi^3$) $\phi^4$ theory, then the completed graph also has (half-)integer weights \cite{Schnetz:2008mp}.
\end{lem}

Decompletion is done by removing the vertex $\infty$ from the graph. Because the edge
$0$-$1$ that may have been added upon completion is trivial, completion does not
change the value of the graphical function.
Like in the case of Feynman periods, we denote the completion of the graph $G$ by $\overline{G}$.

\begin{defin}
The graphical function of a completed graph is the graphical function of its decompletion.
\begin{equation}
f_{\overline{G}}(z) \coloneqq f_{G}(z)
\end{equation}
\end{defin}

\begin{figure}
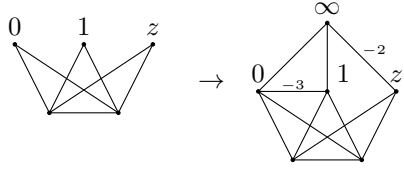

  \center
  \Kfivetwo
  $\rightarrow$
  \completionKfivetwo
  \caption{The completion of the complete graph with two internal vertices and unit weights.
  Weights $\neq1$ are denoted by small numbers in the completed graph.}
  \label{fig:k52completion}
\end{figure}

The simplest example is the three-star with unit weights. Its completion is
\begin{equation}
  f_{\bar G}(z) = \completionthreestar,
\end{equation}
where the edges $0$-$1$ and $z$-$\infty$ have weight $-1$.
A more complicated example is given in Figure \ref{fig:k52completion}.

The external vertices $0,1,z$ can be chosen freely in Feynman periods, allowing us to
drop labels in completed graphs for Feynman periods. This is not possible in the case
of graphical functions. The residual symmetry is to permute external labels. 
\begin{theo}
The graphical function $f_{\overline{G}}(z)$ is invariant under a simultaneous transformation of the four external vertices and the argument $z\rightarrow z'$ such that the cross-ratio of $(0, 1, z, \infty)$ is preserved.
By this condition, $z'$ is uniquely determined as a M\"obius transformation that permutes
the singularities $0,1,\infty$.
\end{theo}

We give some examples of transformations.
The permutation $\pi_{01} = (01)$ leads to the identity
\begin{equation}
\frac{(0-z)(1-\infty)}{(1-z)(0-\infty)} = \frac{(1-z')(0-\infty)}{(0-z')(1-\infty)}
\end{equation}
that implies $z' = 1 - z$. For $\pi_{0z} = (0z)$,
\begin{equation}
\frac{(0-z)(1-\infty)}{(1-z)(0-\infty)} = \frac{(z'-0)(1-\infty)}{(1-0)(z'-\infty)}
\end{equation}
yielding $z' = -\frac{z}{1-z}$.
The transposition $\pi_{1z} = (1z)$ leads to
\begin{equation}
\frac{(0-z)(1-\infty)}{(1-z)(0-\infty)} = \frac{(0-1)(z'-\infty)}{(z'-1)(0-\infty)}.
\end{equation}
This is solved by $z' = z^{-1}$.

(Two of) the above three cases generate the $S_3$ group of M\"obius transformations $\frac{az+b}{cz+d}$ that permute the singularities $0, 1, \infty$ on the Riemann sphere $\mathbb{C}\cup\{\infty\}$.

The transformation group of the external vertices is the permutation group $S_4$.
The stabilizer of the identity transformation in $z$ is the Kleinian group of double transpositions id, $(01)(z\infty)$, $(0z)(1\infty)$, and  $(0\infty)(1z)$.

As an example of how to use completion for the calculation of graphical functions, we consider the graph in Figure \ref{fig:k52completion}.
\begin{equation}
  f_{\overline{G}} = \completionKfivetwo = \completionKfivetwoPermuted = \KfivetwoPermutedDecompleted = \invall\times\KfivetwoPermutedDecompletedExternalRemoved.
\end{equation}
The first identity is the double transposition $\pi = (01)(z\infty)$.
In the second step, we decompleted the graph and removed the edge between $0$ and $1$ because it does not contribute to the value of the graphical function.
Finally, we made the contribution of the edges between $z$ and $0,$, $1,$ explicit.
Note that the resulting graph has an isolated vertex $z$.
Whenever this is the case, the graphical function does not depend on $z$; it becomes
a Feynman period that is much easier to calculate.

With a trivial edge $0$-$1$ of weight $1$ we recognize the $K_4$ graph whose period was
calculated in (\ref{zeta3}). We obtain
\begin{equation}
  f_{\overline{G}} = \frac{6\zeta(3)}{z\bar{z}(z-1)(\bar{z}-1)}.
\end{equation}

We summarize some remarks
\begin{enumerate}
\item Completion of graphical functions is weaker than completion of Feynman periods, because one can only permute external vertices in graphical functions while one can also swap internal and external vertices in periods; see Section \ref{sectcomp}.
\item Without completion, one can still permute the three external vertices $0,1,z$. This follows from the freedom to assign the three external vertices $z_0,z_1,z_2$ to $0,1,z$ in Section \ref{sec:realintro}; see (\ref{AGdef}).
A detailed discussion of this property is given in the less trivial case
of graphical functions with spin (numerator structure); see \cite{Schnetz:2025opm}.
\item With double transpositions, one can move the external vertex $z$ to any other external vertex in the completed graph without transformation of $z$.
\item Infrared convergence of an uncompleted graphical function is ultraviolet convergence at $\infty$ of the completion. So, convergence of graphical functions is best formulated for completed graphs; see Theorem \ref{thmconv}.
\item Completion exists for theories with spin 0 and spin 1/2, but is more complicated in this case. Completion does not exist if particles have spin $\geq1$; see \cite{Schnetz:2025opm}.
\item If one deletes external edges (edges between external vertices) from a completed graph, one gets ``internally completed'' graphs.
Internally completed graphs are bipartite graphs with external and internal vertices that classify graphical functions of equal complexity.
\item Double transpositions prove the twist identity for Feynman periods; see Section \ref{thetwist}.
\end{enumerate}

To elaborate on Remark 4, we define the weight of a (sub-)graph $g$ of $G$.
\begin{defin}\label{NGdef}
  For any subgraph $g$ of the graph $G$ with edges $\mathcal{E}_g$ of weights $\nu_e$ and internal vertices $\mathcal{V}^{\mathrm{int}}_g$, the weight $N_g$ is defined as
  \begin{equation}
    N_g = \sum_{e \in \mathcal{E}_g} \nu_e - \frac{\lambda + 1}{\lambda} \abs{\mathcal{V}^{\mathrm{int}}_g}.
  \end{equation}
\end{defin}
A necessary and sufficient condition for the convergence of the graphical function $f_G(z)$
can now be formulated.
\begin{theo}\label{thmconv}
  Let $\overline{G}$ be a completed graph with external vertices $0, 1, z, \infty$. The graphical function $f_{\overline{G}}(z)$ is convergent if and only if the weight $N_{\overline{G}(V)}$ for the induced subgraph $\overline{G}(V)$ fulfils
\begin{equation}\label{convgfineq}
  N_{\overline{G}\left[V\right]} < \left(\abs{V^{\mathrm{ext}}} - 1\right) \frac{\lambda + 1}{\lambda}
\end{equation}
for all $V \subset\mathcal{V}_G$ with $\abs{V} \geq 2$ vertices from which $\abs{V^{\mathrm{ext}}} \leq 1$ are external.
\end{theo}
The above theorem corrects the trivial case of one internal vertex in Proposition 11
of \cite{Borinsky:2021gkd}.

\begin{figure}
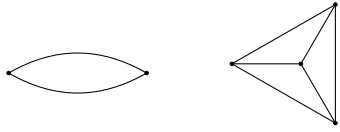

  \center
  \TwoPointDiv
  \qquad
  \WheelDiv
  \caption{Two examples of divergent subgraphs in $\phi^4$ theory. One vertex can be external,
  all other vertices are internal.}
  \label{fig:divergences}
\end{figure}

As an example, consider the graphs in Figure \ref{fig:divergences} in $D=4$ dimensions.
The first graph has the weight
\begin{equation}
  N = 2 - 2\abs{\mathcal{V}^{\text{int}}} = 2\left( \abs{\mathcal{V}^{\text{ext}}} - 1\right).
\end{equation}
Therefore, it is a divergent subgraph.
Similarly, the second subgraph has the weight
\begin{equation}
  N = 6 - 2\abs{\mathcal{V}^{\text{int}}} = 2\left( \abs{\mathcal{V}^{\text{ext}}} - 1\right).
\end{equation}
Hence, this graph is also a divergent subgraph.

\vspace{5pt}
We can classify graphs that are difficult to calculate in four dimensions.
First, the weights of the edges must be either 1 or negative.
Otherwise, the graph has the left graph in Figure \ref{fig:divergences} as subgraph.

An external vertex with a single edge can be reduced.
We may assume that the vertex has the label $z$.
If the weight of the single edge is 1,
we can solve the effective Laplace equation; see Section \ref{sec:intro}.
If the weight of the single edge is negative, a reduction to periods is possible;
see \cite{Borinsky:2021gkd}.

\begin{figure}
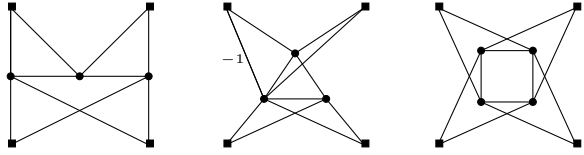

  \centering
  \HardExampleOne \qquad \HardExampleTwo \qquad \HardExampleThree
  \caption{\label{fig:hardExamples} Three classes of internally completed graphical functions that are difficult to calculate. The outside square vertices are external and carry the labels $0, 1, z, \infty$ in some order.}
\end{figure}

Thus, difficult graphical functions have at least two edges connected to each external vertex.
The smallest examples have three internal vertices; see the left two graphs in Figure \ref{fig:hardExamples}.
The left graph can be calculated using Gegenbauer methods \cite{Borinsky:2021gkd}, the second graph by external differentiation at the external vertex with a numerator; see Section \ref{extdiff}.
We find that all convergent graphical functions with $\leq3$ internal vertices can be calculated
in four dimensions (and also in higher even dimensions) \cite{hyperlogprocedures}.

Some difficult graphical functions can be calculated with \texttt{HyperInt} \cite{Panzer:2014caa,hyperint}
or \texttt{HyperFORM} \cite{kardos_2025_17706909} if they are linearly reducible.
Still, there are graphs with four and more internal vertices that are beyond reach
of today's methods. The third graph in Figure \ref{fig:hardExamples}, e.g., is conjectured to have a non-hyperlogarithmic graphical function.

\subsection{Planar Duality}
Planar duality is a consequence of self-duality of massless scalar three-point integrals;
see Section \ref{sec:self_duality}. It was proved in parametric space in \cite{Golz:2015rea}.
\begin{theo}\label{thmplanar}
  Let $G$ be an uncompleted externally planar graph, i.e.\ $G \cup\{0$-$1, 1$-$z, 0$-$z\}$ is planar. Let $G$ have the weight $N_G = \frac{\lambda + 1}{\lambda}$.
  Label the edges of the external dual $G^*$ of $G$ by corresponding edges in $G$.
  The labels of the external vertices in $G^*$ are opposite to the external labels in $G$;
  see (\ref{explandual}).

  Moreover, the weights $\nu_e^*$ of the edges in $G^*$ are related to the weights $\nu_e$ of
  the edges in $G$ according to $\nu_{e^*} = \frac{\lambda +1}{\lambda} - \nu_e$.
  Then, the graphical functions of $G^*$ and $G$ are related by
  \begin{equation}
    \label{equ:planarDuality}
    f_{G^*}(z) = \Big(\prod_{e \in E_G} \frac{\Gamma(\lambda\nu_e)}{\Gamma(\lambda\nu_e^*)}\Big) f_G(z).
  \end{equation}
\end{theo}

Note that, contrary to the example in the introduction, the theorem does not map from momentum space into position space. It is a position space relation that is specific to massless, scalar integrals with three external vertices.

As an example, consider the following calculation of a non-trivial graphical function in four dimensions.
\begin{equation}\label{explandual}
  \begin{split}
    \Triangle \qquad &=  \TriangleWithOneZero \qquad = \TriangleDual \qquad = -4\Delta_0^{-1}\TriangleDualWOEdge \\
                     &= -4\Delta_0^{-1} \TriangleDualCompleted \quad\qquad = -4\Delta_0^{-1} \TriangleDualCompletedPermuted \\
                     &= -4\Delta_0^{-1} \TriangleDualCompletedPermutedDecompleted \qquad =  \TriangleDualCompletedPermutedDecompletedAttached \\
                     &= \TreeSlim \times \quad \TriangleRightPart = 20\zeta(5)\frac{\trilog}{\imz}.
  \end{split}
\end{equation}

The original graph $G$ has weight $N_G = 1 \neq 2$, so that we cannot apply planar duality directly.
After adding an edge of weight 1 between $0$ and $1$,
the total weight increases to $2$ without changing the value of the graphical function.
Then, we apply Theorem \ref{thmplanar} where the dual weights remain at $\nu_e^* = 1$,
so that the gamma factors in (\ref{equ:planarDuality}) are trivial.

Now, we have a single edge of weight 1 attached to $z$, which we remove by inverting
the effective Laplace operator $\Delta_0$.
In the next three steps we complete the graph, apply the double transposition $(01)(z\infty)$, and decomplete again.
We graphically invert $\Delta_0$ by appending an edge of weight 1 at $z$.
The resulting graph is a product (see Section \ref{sec:prod}) that
factorizes into the tree graph calculated in the introduction (see (\ref{three-point-res}))
and a Feynman period that completes to the graph $P_4$ in \cite{Schnetz:2008mp}.

\subsection{The twist}
Like in the case of Feynman periods, we can twist (internally) completed graphical functions \cite{Borinsky:2021gkd}.
\begin{equation}
  \GFTwistLeft = \quad\GFTwistRight,
\end{equation}
where $0,1,z,\infty$ can be split vertices, but have to be on one side of the split
(here in $\overline{G}_2$). The right graph must be an (internally) completed graphical function.
This means that edges may have to be moved as in the case of the twist of the
Feynman period. Alternatively, one can generate pairs of edges of opposite weight
(with total weight zero) between the split vertices and associate one edge to $\overline{G}_1$
while the other edge goes to $\overline{G}_2$ such that the twist does not change the valence of
the split vertices.

\subsection{External differentiation}\label{extdiff}
It is possible to apply the differential operators $\Delta$ and $z_2^\mu \partial_{z_2^\mu}$
to a position space three-point integral.
Here, we only consider the second case.

If $z_2$ connects to $N$ internal vertices $x_i$ by edges of weights $\nu_i$, $i=1,\ldots,N$,
then the integrand of the graphical function has the factor
\begin{equation}
  \prod_{i = 1}^{N}\norm{z_2 - x_i}^{-2\lambda\nu_i}.
\end{equation}
Applying $z_2^\mu \partial_{z^\mu_2}$, we get
\begin{equation}\label{eqextdiff}
  \begin{aligned}
    z_2^\mu \frac{\partial}{\partial z_2^\mu} \prod_{i = 1}^{N}\norm{z_2 - x_i}^{-2\lambda\nu_i} &= z_2^\mu \sum_{j=1}^{N} (-2\lambda\nu_j)\frac{(z_2 - x_j)_\mu}{\norm{z_2 - x_j}^2}\prod_{i=1}^{N}\norm{z_2 - x_i}^{-2\lambda\nu_i} \\
    &= \sum_{j=1}^{N} \lambda\nu_j\left(\frac{x_j^2 - z_2^2}{\norm{z_2 - x_j}^2} - 1\right)\prod_{i=1}^{N}\norm{z_2 - x_i}^{-2\lambda\nu_i}.
  \end{aligned}
\end{equation}
All terms on the right hand side can be interpreted as factors in the integrand of a graphical function.
The factor $x_j^2$ decreases the weight of the edge connecting $0$ and $x_j$ by $1/\lambda$,
the factor $z_2^2$ equals $z\bar{z}$, while the denominator increases the weight
of the edge connecting $z_2$ and $x_j$ by $1/\lambda$.

Applying the differential operator $z_2^\mu \partial_{z^\mu_2}$ to the only invariants
$z_1^2\sim 1$ $z_2^2\sim z\bar{z}$ $(z_2-z_1)^2\sim (z-1)(\bar{z}-1)$, we find that
\begin{equation}
z_2^\mu \frac{\partial}{\partial z_2^\mu}\sim z\partial_z+\bar{z}\partial_{\bar{z}}
\end{equation}
is the effective action of the differential on a graphical function.

Of particular interest is the case where $z_2$ connects to $N=2$ internal vertices, where
we can use (\ref{eqextdiff}) to eliminate the numerator $x_1^2$. We have
\begin{equation}
  \begin{aligned}
    &\left(\lambda\nu_1 + \lambda\nu_2 + z\partial_z+\bar{z}\partial_{\bar{z}}\right) \frac{1}{\norm{z_2-x_1}^{2\lambda\nu_1}\norm{z_2-x_2}^{2\lambda\nu_2}}\\
    &\quad= \lambda\nu_1\frac{x_1^2 - z\bar{z}}{\norm{z_2-x_1}^{2\lambda\nu_1 + 2}\norm{z_2-x_2}^{2\lambda\nu_2}}+ \lambda\nu_2\frac{x_2^2 - z\bar{z}}{\norm{z_2-x_1}^{2\lambda\nu_1}\norm{z_2-x_2}^{2\lambda\nu_2+2}}.
  \end{aligned}
\end{equation}
An example with $\nu_1=1-1/\lambda$, $\nu_2=1$ in $D = 4-2\epsilon$ dimensions is
\begin{equation}
  \begin{aligned}
    \EDExample &= z\bar{z}\EDExampleTwo + \frac{2\lambda - 1 +z\partial_z+\bar{z}\partial_{\bar{z}}}{\lambda - 1}\EDExampleThree \\
    &- \frac{\lambda}{\lambda - 1} \left( \EDExampleFour - z\bar{z}\EDExampleFive \right).
  \end{aligned}
\end{equation}
Although we are ultimately interested in $D=4$ dimensions, we had to use dimensional regularisation
in order to get a meaningful expression (otherwise $\nu_1=0$ and the term with $x_1^2$ does not exist).
On the right hand side this is reflected by a $1-\lambda=\epsilon$ pole that stops us from immediately taking the limit $\lambda \rightarrow 1$.

To proceed we use $D$-dimensional convolution \eqref{equ:dimRegConvolution} backwards,
\begin{equation}
  \label{equ:decompLine}
  \lambdaLine = (\lambda - 1)\Gamma(\lambda)~\doubleLine.
\end{equation}
The $\epsilon^{-1}$ poles in the coefficients cancel, but we can still not take the limit $\lambda\to1$
because the individual graphical functions on the right hand side become divergent in $D=4$ dimensions.
We obtain
\begin{equation}
\begin{aligned}
  \EDExample &= z\bar{z}\left(\EDExampleTwoR + \lambda\Gamma(\lambda)\EDExampleFiveR\right)\\
  &+ (2\lambda - 1 + z\partial_z + \bar{z}\partial_{\bar{z}})\Gamma(\lambda)\EDExampleThreeR + \lambda(1 - \lambda)\Gamma(\lambda)^2\EDExampleFourR.
\end{aligned}
\end{equation}
The graphs on the right hand side are constructible; they can be solved using techniques described before.
With $\textsc{HyperlogProcedures}$ we obtain
\begin{equation}
    f_G(z)=\frac{\mathcal{L}_5(z)}{((z - 1)(\bar{z} - 1)(z - \bar{z}))},
\end{equation}
where $\mathcal{L}_5(z)$ is a single-valued multiple polylogarithm of weight five.
We obtain the explicit expression
\begin{align}
\mathcal{L}_5=&(z - 1)(\bar{z} - 1)\,(\mathcal{L}_{10011}-\mathcal{L}_{11001}+\mathcal{L}_{1101}-\mathcal{L}_{1011})
+z\bar{z}\,(\mathcal{L}_{00010}-\mathcal{L}_{01000}+\mathcal{L}_{0100}-\mathcal{L}_{0010})\notag\\
&-(z - 1)\bar{z}\,(\mathcal{L}_{00011}-\mathcal{L}_{01001}+\mathcal{L}_{0101}-\mathcal{L}_{0011})
-z(\bar{z} - 1)\,
(\mathcal{L}_{10010}-\mathcal{L}_{11000}+\mathcal{L}_{1100}-\mathcal{L}_{1010})\notag\\
&+(z - 1)\,(\mathcal{L}_{01011}-\mathcal{L}_{01101}+6\zeta(3)\,\mathcal{L}_{1})
+(\bar{z} - 1)\,(\mathcal{L}_{10110}-\mathcal{L}_{11010})+
z\,\mathcal{L}_{01100}-\bar{z}\,\mathcal{L}_{00110}\\
&-(z-\bar{z})\,(\mathcal{L}_{01010}+6\zeta(3)\,\mathcal{L}_{0}+6\zeta(3))
-2\zeta(3)(z\bar{z} - 3z - 3\bar{z} + 6)\,\mathcal{L}_{10}+2\zeta(3)(z - 1)(\bar{z}-3)\,\mathcal{L}_{11}.\notag
\end{align}

\subsection{Other identities and methods}
Factorization of Feynman periods, see Section \ref{sec:fac}, lifts to a factor identity for (internally) completed graphical functions that have a three-vertex cut.

There exist numerous other tools, most prominently integration-by-parts and the Gegenbauer identity \cite{Borinsky:2021gkd}.

As mentioned in Section \ref{sec:completion}, a tool outside the theory of graphical
functions that can be used to calculate some hard cases with not too many edges
is parametric integration \cite{Brown:2008um,Brown:2009ta} with implementations \texttt{HyperInt} \cite{Panzer:2014caa,hyperint} or \texttt{HyperFORM} \cite{kardos_2025_17706909}.

It is expected that more identities for graphical functions will be found in the future.
\section{Regular graphical functions in non-integer dimensions}
\label{sec:gf_dimreg_reg}
The methods presented so far generalize to non-integer dimensions. Most prominently, it was shown in Section \ref{sec:leg_dimreg_reg} how to append an edge of weight 1 in this setup.

The main addition to our toolbox is the method of approximation.
The idea is to replace a graphical function by a sum of graphical functions such that the expansions in $\ep$ coincide to some order in $\ep$. Typically, this is done by replacing
a suitable subgraph by a sum of simpler graphs that
approximate the subgraph insertion to some order in $\ep$.

As an example, let us look again at the graph in Figure \ref{fig:k52completion}, for which we found the result in $D = 4$ dimensions using completion (see Section \ref{sec:completion}).
Completion can also be used to calculate the $\mathcal{O}(\ep)$ correction in $D = 4 - \ep$ dimensions.

All internal vertices have total weight four which is not conformal in $D=4-2\ep$ dimensions where internal vertices need to have the weight $\frac{4 - 2\ep}{1 - \ep}$.
We have to connect the internal vertices to $\infty$ with edges of weight $\frac{2\ep}{1 - \ep}$ and obtain using the double transposition $(01)(z\infty)$,
\begin{equation}
  \label{equ:KfivetwoDimRegDecompleted}
  \begin{aligned}
  \Kfivetwo &= \completionKfivetwoDimreg = \completionKfivetwoDimregExchanged \\
    &= \frac{1}{(z\bar{z})^{1-3\ep}((z-1)(\bar{z}-1))^{1-3\ep}} \KfivetwoDimregDecompleted.
  \end{aligned}
\end{equation}
Note that the edges connected to $z$ vanish in the limit $\ep \rightarrow 0$.
If we expand their propagators to $\mathcal{O}(\ep)$, we get three contributions (we label the internal vertices by $x_1$ and $x_2$)
\begin{equation}
  \frac{1}{\norm{z_2-x_1}^{4\ep}\norm{z_2-x_2}^{4\ep}} = 1 - 4\ep\log(\norm{z_2-x_1}) - 4\ep\log(\norm{z_2-x_2}) + \mathcal{O}(\ep^2).
\end{equation}
We also have
\begin{equation}
  2\ep\log(\norm{z_2-x_1}) = 1-\frac{1}{\norm{z_2-x_1}^{2\ep}} + \mathcal{O}(\ep^2)
\end{equation}
with the same formula for $z_2-x_2$.
Because the graphical function is convergent for $\ep\to0$, we are free to expand under the integration sign;
an expansion of the integrand lifts to an expansion of the integral.
Graphically, we get the approximation
\begin{equation}
  \KfivetwoDimregDecompleted = 5\KfivetwoDimregDecompletedNo - 2\KfivetwoDimregDecompletedLeft -2\KfivetwoDimregDecompletedRight + \mathcal{O}(\ep^2).
\end{equation}
We have carefully chosen the weight $-\ep/\lambda = 1 - 1/\lambda$ in the approximation, so that we can apply the identity \eqref{equ:decompLine} to the two graphs on the right. We get
\begin{equation}
  \label{equ:KfivetwoDimRegDecompletedDecomposition}
  \KfivetwoDimregDecompleted = 5\KfivetwoDimregDecompletedNo + 2\ep\Gamma(\lambda)\KfivetwoDimregDecompletedLeftDecomp + 2\ep\Gamma(\lambda)\KfivetwoDimregDecompletedRightDecomp + \mathcal{O}(\ep^2).
\end{equation}
The first term on the right hand side is a $z$-independent period. It can be constructed by appending edges and integration over $z$ (or rather taking the limit $z\to0$ of a graphical function with and extra edges $x_1$-$0$ of weight $-1$ and $x_1$-$z$ of weight 1, see \cite{Schnetz:2022nsc}). We obtain
\begin{equation}
  \KfivetwoDimregDecompletedNo = 6\zeta(3) + \left(12\zeta(3)(1-\gamma_E) + \frac{\pi^4}{10}\right)\ep + \mathcal{O}(\ep^2).
\end{equation}
The latter two graphs are isomorphic and also constructible. We get
\begin{equation}
  \begin{aligned}
    \KfivetwoDimregDecompletedLeftDecomp &= -\frac{6\zeta(3)}{\ep} - 6(1-3\gamma_E)\zeta(3) - \frac{\pi^4}{10} - \frac{6(z-1)\zeta(3)\li}{\imz} \\[-40pt]
                                         &+\frac{z\bar{z}(\mathcal{L}_{0010}(z)-\mathcal{L}_{0100}(z))+(z-1)(\bar{z} - 1)(\mathcal{L}_{1011}(z)-\mathcal{L}_{1101}(z))}{\imz} \\
    &+ \frac{z(\bar{z}-1)(\mathcal{L}_{1100}(z) - \mathcal{L}_{1010}(z))-\bar{z}(z-1)(\mathcal{L}_{0011}(z)-\mathcal{L}_{0101}(z))}{\imz} + \mathcal{O}(\ep).
  \end{aligned}
\end{equation}
Note that the pole in $\ep$ is cancelled by the coefficient in Equation \eqref{equ:KfivetwoDimRegDecompletedDecomposition}, which is also the reason we only need this graph to $\mathcal{O}(1)$. \\
Combining these expressions with \eqref{equ:KfivetwoDimRegDecompleted} and \eqref{equ:KfivetwoDimRegDecompletedDecomposition} and expanding the $z$-dependence of the prefactor in $\ep$, we obtain
\begin{equation}\label{K52oeps}
  \begin{aligned}
    \Kfivetwo &=  \frac{6\zeta(3)}{z\bar{z}(z-1)(\bar{z} - 1)}  + \Biggl( \frac{\pi^4 + 120(3 - \gamma_E)\zeta(3)}{10 z \bar{z} (z - 1) (\bar{z} - 1)}- \frac{6\zeta(3)(z + 3 \bar{z} - 4)\mathcal{L}_{1}(z)}{z \bar{z} (z-1)(\bar{z} - 1)(\imz)}\\
    &+ \frac{18\zeta(3)\mathcal{L}_{0}(z)}{z \bar{z} (z-1)(\bar{z} - 1)} +4\frac{\mathcal{L}_{0010}(z) - \mathcal{L}_{0100}(z)}{(z - 1) (\bar{z} - 1) (\imz)}+4\frac{\mathcal{L}_{1011}(z) - \mathcal{L}_{1101}(z)}{z\bar{z}(\imz)}\\
    &+4\frac{\mathcal{L}_{1100}(z) - \mathcal{L}_{1010}(z)}{(z - 1)\bar{z}(\imz)}-4\frac{\mathcal{L}_{0011}(z) - \mathcal{L}_{0101}(z)}{z (\bar{z} - 1) (\imz)}   \Biggr)\ep  + \mathcal{O}(\ep^2).
  \end{aligned}
\end{equation}
\section{Singular graphical functions in non-integer dimensions}
\label{sec:gf_dimreg_sing}
In this section, we discuss the application of various methods for the calculation of dimensionally regularised graphical functions  
that are singular in the limit $\ep\to0$.

\subsection{Approximation}
The method of approximation was introduced in the previous section;
it is very powerful but needs to be refined in the context of singular graphical functions.

We start with a basic example and consider a graphical function that we already computed in Section \ref{sec:leg_dimreg_sing}
\begin{align}\label{approx1}
    \bubblethreestar = \dfrac{1}{\ep z\bar{z}} + \dfrac{2 - \gamma_E + 2\lo - \li}{z\bar{z}} + O(\ep).
\end{align}
We don't want to redo the calculation but rather use the result to derive an approximation rule.
The single-valued logarithms can be written as the first terms in an expansion of propagators, namely,
\begin{subequations}
\label{equ:edgeApprox}
\begin{align}
    \dfrac{1}{\left(z\bar{z}\right)^{-2\ep}} &= 1 +2\ep\,\underbrace{\log(z\bar{z})}_{\lo} + O(\ep^2), \\
    \dfrac{1}{\big((z - 1)(\bar{z} - 1)\big)^{\ep}} &= 1 - \ep\,\underbrace{\log\big((\bar{z} - 1)(z - 1)\big)}_{\li} + O(\ep^2).
\end{align}
\end{subequations}
Therefore, we have
\begin{align}
    \bubblethreestar = \left(\dfrac{1}{\ep} + 2 - \gamma_E  \right)\dfrac{1}{(z\bar{z})^{1-2\ep}\big((\bar{z} - 1)(z - 1)\big)^\ep} + O(\ep).
\end{align}
The benefit of the above version of (\ref{approx1})
is that the right hand side is a graphical function,
\begin{align}
  \label{equ:approxbubble}
    \bubblethreestar = \left(\dfrac{1}{\ep} + 2 - \gamma_E\right)\approxtrigX + O(\ep).
\end{align}
We introduced an edge between the vertices $0$ and $1$
of weight $x/\lambda$. This edge does not change the formula,
but it changes the total scaling weight of the graph on the right hand side.
We adjust $x$, such that both sides have equal scaling weight,
\begin{align}\label{scalingweights}
    \text{l.h.s.:}& ~4 - \dfrac{D}{2\lambda} = \dfrac{4 - 4\ep - 2 + \ep}{\lambda} = \dfrac{2 - 3\ep}{\lambda},\notag \\
    \text{r.h.s.:}& ~\dfrac{x + 1 - 2\ep + \ep}{\lambda} = \dfrac{x + 1 -\ep}{\lambda},
\end{align}
yielding $x = 1 - 2\ep$. The final version of the approximation is symmetric,
\begin{equation}
    \label{equ:threestarApprox}
    \bubblethreestar = \left(\dfrac{1}{\ep} + 2 - \gamma_E\right)\approxtrig + O(\ep).
\end{equation}
Both sides in (\ref{equ:threestarApprox}) having the same scaling weight allow us to substitute the subgraph on the left by the expression on the right if only two orders in $\ep$ are needed.
Such substitutions considerably simplify the calculation of hard graphical functions.

If the full graph is singular in the limit $\ep\to0$, there 
is an important restriction.
The approximation must not change the scaling behaviour of subdivergences (other than the bubble that is substituted).

If, for example, we try to approximate the top bubble in the graph
\begin{equation}
  \approxInvalidExample\qquad\qquad,
\end{equation}
we run into the problem that the scaling of the bottom bubble
(between the vertices $v_1$ and $v_2$) changes from $2$ to $2 + \ep/\lambda$. Such a substitution gives a false result even in the
first two orders in $\ep$.

If, on the other hand, we substitute the bottom bubble (either with
the left or right pair of attached edges), we find that
the scaling of the top bubble is unaffected. More critical
is the scaling of the singular triangle $v_1,v_2,v_3$ that
collapses to a double edge after substitution. However, the scaling weight
does not change. The triangle has the scaling weight $(2-3\epsilon)/\lambda$ (see the top line in (\ref{scalingweights})) which equals
the weight $1+(1-2\ep)/\lambda$ of the double edge after substitution.
In this case, the approximation is possible.

We can approximate the graph on the left hand side of \eqref{equ:threestarApprox} to one more order in $\ep$ if we allow more terms
on the right hand side,
\begin{equation}
  \begin{aligned}
    \bubblethreestar& = -\ep\left(\bubblethreestarone + \bubblethreestartwo - \bubblethreestarthree\right)\\
    &\mkern-72mu+ \frac{1}{\ep(1-2\ep)\Gamma(1-\ep)} \left(\bubblethreestarfour + \bubblethreestarfive - \bubblethreestarsix\right) + \mathcal{O}(\ep^2).
  \end{aligned}
\end{equation}
We conclude this section with the following example.
\begin{equation}
  \begin{aligned}
 \approxInternalExample =& \left(\frac{1}{\ep}+2-\gamma_E\right) \approxInternalExampleReplaced + \mathcal{O}(1) \\
    =& \frac{1}{z\bar{z}}\left(\frac{1}{\ep^2} + \frac{5 - 2\gamma_E + 3\lo - 2\li }{\ep}\right) + \mathcal{O}(1).
  \end{aligned}
\end{equation}
As explained above, the approximation \eqref{equ:approxbubble} is valid for the two
lowest orders in $\ep$ although the shape of the singular
triangle is affected (the scaling is preserved). 
Because the edge attached to $z$ has weight 1, the graph on
the right hand side can be constructed as described in Section \ref{sec:leg_dimreg_sing}.

\subsection{External differentiation}\label{extdiffII}
As explained in Section \ref{extdiff}, we get relations between graphical functions by differentiating with respect to the external variable $z$.
In this section, we consider the effective Laplace operator $\Delta$.
Acting on a single edge connected to $z$ gives
\begin{equation}\label{lapedge}
  \Delta\;\nuLine = \Delta_{z_2}\,\norm{z_2 - x_i}^{-2\lambda\nu} = 4\lambda^2\nu(\nu - 1)\;\nuLineLambda.
\end{equation}
If $z$ is connected to multiple edges, we also get mixed terms from the derivatives acting on different edges.
The general expression is
\begin{equation}
  \Delta\boxMultiple = 4\lambda^2\left[ \left(\sum_{j=1}^N \nu_j - 1 \right)\sum_{i=1}^N \nu_i\boxMultipleOne - \sum_{1\leq i < j \leq N}\nu_i\nu_j\boxMultipleTwo\right], 
\end{equation}
where $\nu_i$ is the weight of the edge connecting $z$ and $x_i$. \\
We append an edge of weight 1 to the external vertex $z$ (creating the internal vertex $x$).
The Laplace operator $\Delta$ acts on this
edge by generating the $\delta$ factor $-4\delta^{(D)}(z-x)/\Gamma(\lambda)$. We obtain
\begin{equation}\label{extdifappended}
  \begin{aligned}
    \boxMultiple = \lambda^2\Gamma(\lambda)\Biggl[& \left(1 - \sum_{j=1}^N \nu_j \right)\sum_{i=1}^N \nu_i\boxMultipleOneAttached\\
    &+ \sum_{1\leq i < j \leq N}\nu_i\nu_j\boxMultipleTwoAttached\Biggr].
  \end{aligned}
\end{equation}
The formula is useful in two setups.
If all $\nu_i$ vanish in the limit $\ep\to0$ or if
one $\nu_k$ is $1$ in the limit $\ep\to0$ while all other $\nu_i$ vanish in this limit, then
each coefficient on the right hand side has a low degree of order $\geq1$ in $\ep$. This implies that
we need to calculate the corresponding graphical
function to lower order in $\ep$.
The first setup has the additional advantage that
the second sum (which may be more complicated because the graphs may have an additional edge connecting $x_i$ and $x_j$) is suppressed by two orders in $\ep$.

We illustrate this setup by presenting an alternative
way to calculate the complete graphical function
with two internal vertices in four dimensions to order $\ep$ (see Section \ref{sec:gf_dimreg_reg}). From (\ref{equ:KfivetwoDimRegDecompleted}), we have
\begin{equation}
\Kfivetwo = \frac{1}{(z\bar{z})^{1-3\ep}((z-1)(\bar{z}-1))^{1-3\ep}} \KfivetwoDimregDecompleted.
\end{equation}
External differentiation (\ref{extdifappended}) gives
\begin{equation}
    \label{K52Exp}
    \KfivetwoDimregDecompleted = \lambda^2\Gamma(\lambda)\left[\left(1 - \frac{4\ep}{\lambda}\right)\frac{4\ep}{\lambda} \KfivetwoDimregDecompletedExtOne + \frac{4\ep^2}{\lambda^2} \KfivetwoDimregDecompletedExtTwo \right],
\end{equation}
where we collected the contributions of two isomorphic graphs.
The second graph on the right hand side is regular in $\ep$, so we can drop it in an expansion to order $\ep$,
\begin{equation}\label{K52eq}
  \Kfivetwo = \frac{(1 - 5\ep)4\ep\Gamma(1-\ep)}{\left(z\bar{z}(z-1)(\bar{z}-1)\right)^{1-3\ep}}\KfivetwoDimregDecompletedExtOne + \mathcal{O}(\ep^2).
\end{equation}
The graph on the right hand side is still not constructible,
but we can derive it from the graph
\begin{equation}\label{K52eqgraph}
\KfivetwoDimregDecompletedExtThree
\end{equation}
by appending an edge. The graphical function (\ref{K52eqgraph}) is regular in the limit $\ep\to0$.
The pole in $\ep$ in the graphical function on the right hand side
of (\ref{K52eq}) originates from the infrared
singularity (which is constructible) when appending the edge (see Section \ref{sec:leg_dimreg_sing}).
Because the coefficient on the right hand side of (\ref{K52eq}) has
a factor $\ep$, we need the graphical function (\ref{K52eqgraph})
only to order $\ep^0$. Hence, we can set $\ep=0$ in (\ref{K52eqgraph}) which renders the graph constructible. We recover the result (\ref{K52oeps}).\\[3pt]

With the same technique, we can calculate the graph to order $\ep^2$.
From \eqref{equ:KfivetwoDimRegDecompleted} and \eqref{K52Exp}, we get
\begin{equation}
\label{K52epssqu}
\Kfivetwo = \frac{4\ep\Gamma(\lambda)}{(z\bar{z}(z-1)(\bar{z}-1))^{1-3\ep}}\left[\left(1 - 5\ep\right) \KfivetwoDimregDecompletedExtOne + \ep \KfivetwoDimregDecompletedExtTwo \right]+\mathcal{O}(\ep^3).
\end{equation}
The second graph is regular in the limit $\ep\to0$. We only need the leading order, where it becomes
\begin{equation}
    \KfivetwoDimregDecompletedExtTwo = \KfivetwoDimregDecompletedExtTwoConst + \mathcal{O}(\ep).
\end{equation}
The graph on the right hand side is constructible.
For the first graph on the right hand side of \eqref{K52epssqu}, we need an expansion
to order $\ep^1$. We first observe that the graph is constructible from the graph
\begin{equation}\label{eps2ex}
\KfivetwoDimregDecompletedExtOneConstruct = \KfivetwoDimregDecompletedExtOneConComplete,
\end{equation}
by completion. We use the double transposition $(0\infty)(1z)$ and decomplete, yielding
\begin{equation}
\KfivetwoDimregDecompletedExtOneConstruct = \KfivetwoDimregDecompletedExtOneConCDC = \frac{1}{(z\bar{z})^{3\ep}} \KfivetwoDimregDecompletedExtOneConCDCR.
\end{equation}
With the double transposition, we managed to create a second edge that vanishes in the limit
$\ep\to0$. The lower right edge in the rightmost graph has weight $(1+2\ep)/\lambda=1+3\ep/\lambda$
that reduces to an edge of weight
$1$ in the limit $\ep\to0$. We need the first two orders in
$\ep$ (the left hand side is finite in the limit $\ep\to0$). The generic setup of the expansion is
\begin{equation}
f(\ep,\ep,\ep)=f(\ep,0,0)+f(0,\ep,0)+f(0,0,\ep)-2f(0,0,0)+\mathcal{O}(\ep^2)
\end{equation}
for any analytic function $f:\mathbb{R}^3\rightarrow\mathbb{R}$. Concretely, we obtain
\begin{equation}
\KfivetwoDimregDecompletedExtOneConCDCR = \kfivetwoexpandedone + \kfivetwoexpandedtwo + \kfivetwoexpandedthree - 2\kfivetwoexpandedzero + \mathcal{O}(\ep^2). 
\end{equation}
All of the graphs on the right hand side are constructible,
which allows us to calculate the leading three orders of \eqref{K52epssqu}.
Note that a direct expansion of the left hand side
of \eqref{eps2ex} does not suffice for a reduction to constructible graphs.
\\[3pt]

Our last example in this subsection is the solution of the generic
vertex with three edges using a ``M\"unchhausen'' (bootstrap) method.
In Sections \ref{sec:leg_dimreg_reg} and \ref{sec:leg_dimreg_sing},
we saw how to calculate the three-star if one of the edges has weight 1.
Here, we assume that the $\ep$ coefficient of all weights are
variables. We will see that we can still calculate the graph
to any order in $\ep$ (subject to memory and time constraints).

We start with
\begin{equation}
  \begin{aligned}
  \label{equ:munchhausenDiff}
  \Delta\threestarMunchhausen =&~4\lambda^2\frac{a_2\ep}{\lambda}\left(\frac{a_2\ep}{\lambda} -1\right)\threestarMunchhausenExt\\ =& -4a_2\ep\left(1 - (a_2 + 1)\ep \right ) \threestarMunchhausenCompleted,
  \end{aligned}
\end{equation}
where we have completed the graph in the second step.
The additional $\nu_{ij}$ have to be chosen according to the completion condition that the internal vertex has total weight $D/\lambda$ and
the external vertices have total weight and zero.
For the internal vertex $x$ we get
\begin{equation}
  \label{equ:nuxinf}
  \nu_{x\infty} = \frac{D}{\lambda} - \frac{4 + (a_0 + a_1 + a_2)}{\lambda} = -\frac{(a_0 + a_1 + a_2 + 2)\ep}{\lambda}.
\end{equation}
For the external edges we find the unique solution
\begin{equation}
  \label{equ:nurest}
    \nu_{0\infty} = \frac{(a_1 + a_2 + 1)\ep}{\lambda},\quad
    \nu_{01} = -\frac{2 + (a_0 + a_1 + a_2 + 1)\ep}{\lambda},\quad
    \nu_{1\infty} = \frac{1 + (a_0 + a_2 + 1)\ep}{\lambda}.
\end{equation}
We doubly transpose the external vertices with $(01)(z\infty)$ and decomplete, yielding
\begin{equation}
  \begin{aligned}
    \Delta\threestarMunchhausen =& -4a_2\ep\left(1 - (a_2 + 1)\ep \right ) \threestarMunchhausenCompletedSwitched\\
    =& -4a_2\ep\left(1 - (a_2 + 1)\ep \right ) \threestarMunchhausenDecompleted.
  \end{aligned}
\end{equation}
We invert the effective Laplacian by attaching an edge to the external vertex $z$,
\begin{equation}\label{Munch}
    \threestarMunchhausen = a_2\ep(1 - (a_2 + 1)\ep )\Gamma(1-\ep)\threestarMunchhausenDecompletedAttached.
\end{equation}
Altogether, we managed to construct the original graph from
the graph
\begin{equation}\label{Munch2}
\threestarMunchhausenDecompletedTwo\quad=\quad
\threestarMunchhausenDecompletedThree
\end{equation}
with a coefficient in (\ref{Munch}) that has a factor of $\ep$.
On the right hand side of (\ref{Munch2}) we used that the transposition (01) is equivalent to a substitution $z \rightarrow 1 - z$ (see Section \ref{sec:completion}).

Because appending an edge does not
lose an order in $\ep$, it suffices to know the graphical function
on the right hand side of (\ref{Munch2}) to one order less in $\ep$.
This graph, however, equals the original graph with $a_2$ replaced by $-(a_0+a_1+a_2+2)$.

We can iterate these steps to lower the order in $\ep$ until
the graphical function is zero (at order $\ep^{-2}$).
This constructs the generic three-vertex from its
(trivial) divergences by repeatedly inverting the effective Laplace
operator.

By taking derivatives (see (\ref{lapedge})) and using completion, we can calculate generic three-stars
with different weights in the limit $\ep\to0$.

In practice, one reaches
the order $\ep^6$ with this method. Higher orders (up to $\ep^9$)
can be obtained using IBP and contiguous relations.

\subsection{Rerouting}
The idea behind rerouting is the subtraction of simpler graphs that carry the same subdivergences as the original graph. This lowers the pole in $\ep$, so that less orders in $\ep$ are necessary to acquire the wanted order.
With less orders in $\ep$ to cover, approximation becomes more powerful.

The method of rerouting was devised in the context of renormalization by Brown and Kreimer \cite{Brown:2011pj}.

Consider a graph $X$ whose graphical function has a pole of order $N$ in $\ep$, $X \sim \ep^{-N}$. We subtract and add a sum of graphs $Y_i$ to get the relation 
\begin{equation}
    \label{equ:re-routing}
    X = \underbrace{\Big(X - \sum_i Y_i\Big)}_{\sim \ep^{-n}, ~~ n < N} + \underbrace{\sum_i Y_i}_{\sim \ep^{-N}}. 
\end{equation}
The subtraction is advantageous if all $Y_i$ are graphs that are easier to calculate than the graph $X$.

Let us illustrate the application of this method with a simple example.
\begin{equation}
    \bubblethreestar = \left(\bubblethreestar - \reroutBubble\right) + \reroutBubble.
\end{equation}
The difference in brackets is regular in $\ep$. If we are only interested in calculations only up to $\mathcal{O}(\ep^{-1})$ we can omit the term.
The remaining graph is simpler and can be calculated applying the convolution relation \eqref{equ:dimRegConvolution}. In general, we have
\begin{align}
    \label{equ:generalChain}
    \chainGen =& \quad \dfrac{\Gamma(1 + \lambda(1 - v))\Gamma(-1 + \lambda v)}{\Gamma(\lambda)\Gamma(\lambda v)\Gamma(2 +  \lambda(1- v))\Gamma(1)}\quad\reroutLine  \notag\\
            =& \quad \dfrac{1}{\big(1 + \lambda(1 - v)\big)(\lambda v - 1)\Gamma(\lambda)}\quad \reroutLine.
\end{align}
With $\nu=2$, we obtain
\begin{equation}
    \label{equ:twoChain}
    \chainTwo \!\!= \;\;\dfrac{1}{(1-\lambda)(2\lambda - 1)\Gamma(\lambda)}\quad\reroutLineTwo
    \!\!=\;\; \dfrac{1}{\ep(1 - 2\ep)\Gamma(1 - \ep)}\quad\reroutLineTwoeps
\end{equation}
for the chain connecting $0$ and $1$ in (\ref{equ:generalChain}).
We can read off the leading term in the Laurent expansion in $\ep$
(see (\ref{threestarsing})),
\begin{align}
    \bubblethreestar =&\quad \reroutBubble + O(1) \nonumber \\
    =&\quad \dfrac{1}{\ep(1 - 2\ep)\Gamma(1 - \ep)}\dfrac{1}{z\bar{z}} + O(1) =\quad \dfrac{1}{\ep z\bar{z}} + O(1).
\end{align}

To use rerouting for more complicated graphs, we need a tool
that calculates the singularity structure of $X-\sum_i Y_i$.
This is solved by a variant of the Connes-Kreimer coproduct of renormalization
\cite{Kreimer:1997dp, Connes:1999zw, Connes:1999yr}.
The Connes-Kreimer coproduct reveals the singularity structure of Feynman
integrals by pure combinatorics without performing integrations. For a detailed review of Hopf algebras in the context of renormalization, we recommend \cite{Panzer:2012gp}.\\[3pt]
In our context, we need a modification of the combinatorial structure.
Because we want to apply approximation to parts of the subtraction
$X-\sum_i Y_i$ that are not affected by modifications of subgraphs,
we need to lift all graphs to graphs with labelled vertices (internal and
external). In some sense, this is a simplification because graph isomorphy
becomes trivial in this context. Moreover, we want to keep the external
vertices of the graph which causes an asymmetry between subgraphs and
factor graphs.

We will also restrict ourselves to the main case of interest
in QFT: graphs with only logarithmic subdivergences and no two-point insertions.
Higher divergences come from two-point insertions and two-point insertions can be
integrated out. This means, they can be replaced by propagators with the (non-integer)
scaling weight of the insertion times a numerical factor that only depends on $\ep$ 
(their Feynman period).

We call a non-empty graph $g$ log-divergent if we have equality in (\ref{convgfineq}),
i.e.\ $|\mathcal{V}_g|\geq2$, $|\mathcal{V}_g^{\mathrm{ext}}|\leq1$ and
\begin{equation}\label{conv2}
N_{g} = \left(\abs{\mathcal{V}_g^{\mathrm{ext}}} - 1\right) \frac{\lambda + 1}{\lambda}.
\end{equation}

The $\mathbb{C}$-span of labelled log-divergent graphs together with the empty graph
$1$ (the unit) and disjoint union as multiplication is a Hopf-algebra under the coproduct defined in \eqref{equ:coaction}.

We consider graphs $G$ with only logarithmic subdivergences, which means that
no induced subgraph of $G$ is allowed to have a `$>$' sign in \eqref{conv2}.
The $\mathbb{C}$ span of such graphs forms an algebra with a coaction.
For a more precise discussion of the Connes-Kreimer coaction, see Section \ref{sectcoact}.

\begin{defin}\label{def:coaction}
    The Hopf algebra of labelled log-divergent graphs with the empty graph $1$
    coacts to the left on the $\mathbb{C}$-span of labelled graphs with only
    logarithmic subdivergences. The (reduced) Connes-Kreimer coaction ($\Delta'$) $\Delta$ is defined by
    \begin{equation}\label{equ:coaction}
        \Delta(G) = 1\otimes G+\Delta'(G),\quad
        \Delta'(G) =\sum_{\substack{g \subsetneq G \\ g~\text{log div.}}} g \otimes G / g,
    \end{equation}
    where $G / g$ denotes the factor graph of $G$ in which the subgraph $g$ is contracted. \end{defin}

We obtain the following theorem.
\begin{theo}\label{lem:regularity}
A linear combination of graphs is regular in the limit $\ep\to0$ if and only if its reduced Connes-Kreimer coaction is trivial. That is,
\begin{equation}
    X \text{ is regular} \quad \Longleftrightarrow \quad \Delta'(X) = 0.
\end{equation}
\end{theo}
The algorithm is as follows. We calculate the reduced coaction on $X$
and on a suitable set of rerouted graphs $Y_i$.
Searching for a linear combination that gives $\Delta'(X-\sum Y_i)=0$
is equivalent to solving a linear system.
Then, we know by Theorem \ref{lem:regularity} that $X-\sum Y_i$ is of the order $\ep^0$.

Later we will generalize the setup to subtractions that only partially regularise
the singularities in $X$ (in the sense of \eqref{equ:re-routing} with $n>0$).

Let us apply the reduced Connes-Kreimer coaction to the above example for rerouting.
The reduced coactions on the two graphs are
\begin{subequations}
\begin{equation}
    \Delta'\left(\coproductI\right) = \coproductBubble \otimes \coproductTrig
\end{equation}
\begin{equation}
    \Delta'\left(\coproductII\right) = \coproductBubble \otimes \coproductTrig.
\end{equation}
\end{subequations}
In the factor graph we label contracted vertices by equations. By linearity, we obtain
\begin{equation}
    \Delta'\left(\coproductI - \coproductII\right) = 0, 
\end{equation}
which proves our statement that this difference is regular. 

For graphs with subdivergences we have a wider choice of subtractions.
To illustrate this, we apply the reduced coaction to the following three graphs with
poles of order two in $\ep$.
\begin{subequations}
\begin{equation}
    \Delta'\left(\coproductIII\right) = \coproductCap \otimes\coproductTrigII + \coproductXOneXTwo \otimes \coproductBubbleTrigII,
\end{equation}
\begin{equation}
    \Delta'\left(\coproductIV\right) = \coproductCap \otimes\coproductTrigII + \coproductXOneXTwo \otimes \coproductBubbleTrigII,
\end{equation}
\begin{equation}
    \Delta'\left(\coproductV\right) = \coproductCap \otimes \coproductTrigII + \coproductXOneXTwo \otimes \coproductBubbleTrigRedII.
\end{equation}
\end{subequations}
We conclude that only the subtraction
\begin{equation}
    \label{equ:re-routingOOne}
    \coproductIIInox - \coproductIVnox \sim O(\ep^0)\;,
\end{equation}
is regular (its reduced coaction is $0$). The other subtraction
\begin{align}
    \coproductIIInox - \coproductVnox \sim O(\ep^{-1}),
\end{align}
because its reduced coaction is non-zero.

We see that a suitable choice of subtraction can eliminate multiple orders in $\ep$.

Note that we can only use approximations in the first term of \eqref{equ:re-routing} for subgraphs that are not affected by rerouting.

Let us now use the subtraction \eqref{equ:re-routingOOne} to calculate the pole terms of
the first graph.
\begin{equation}
\begin{aligned}
 \coproductIIInox &= \left( \coproductIIInox - \coproductIVnox \right) + \coproductIVnox \\
 &= \coproductIVnox + \mathcal{O}(\ep^0),
 \end{aligned}
\end{equation}
where we have used the fact that the bracket is regular.
Now, we can apply \eqref{equ:twoChain} to remove one internal vertex.
Together with the edge of weight $1$ between $0$ and the other internal vertex, we get
\begin{equation}\label{eqcap}
  \begin{aligned}
  \coproductIIInox &= \frac{1}{\ep(1-2\ep)\Gamma(1-\ep)}\reroutthreestar + \mathcal{O}(\ep^0)\\
  &= \left(\frac1\ep + 2 - \gamma_E\right) \reroutthreestar + \mathcal{O}(\ep^0).
  \end{aligned}
\end{equation}
The graph on the right hand side is constructible.
Here, we use completion to reduce it to a graph that we have already calculated.

We find the completion on the right hand side of Equation~\eqref{equ:munchhausenDiff}, with $a_0 = -3, a_1 = a_2 = -1$. From \eqref{equ:nuxinf} and \eqref{equ:nurest}, we obtain $\nu_{x\infty} = 3\ep/\lambda$ and $\nu_{1\infty} = (1 - 3\ep)/\lambda$.
The double transposition $(0\infty)(1z)$ gives.
Graphically,
\begin{equation}
\begin{aligned}
 \coproductIIInox &= \left(\frac1\ep + 2 - \gamma_E\right) \threestarMunchhausenCompletedVs + \mathcal{O}(\ep^0)\\ &= \left(\frac1\ep + 2 - \gamma_E\right)\coprodDecompletedVs + \mathcal{O}(\ep^0)
 = \frac{\left(\frac1\ep + 2 - \gamma_E\right)}{(z\bar{z})^{1-3\ep}}\coprodDecompletedEvaled + \mathcal{O}(\ep^0).
 \end{aligned}
\end{equation}
The graph on the right hand side is the case $a=3$ of \eqref{equ:IRthreestar}.
With the result $\eqref{equ:IRthreestarSol}$, an expansion in  $\ep$ gives
\begin{equation}\label{eqcap2}
    \coproductIIInox = \frac{\frac{1}{\ep^2} + \frac{1}{\ep}\left(3\lo - \li + 5 - 2\gamma_E\right)}{2z\bar{z}} + \mathcal{O}(1).
\end{equation}
\vspace{5pt}

In general, we want to generalize Theorem \ref{lem:regularity} to determine
the pole order of (sums of) graphical functions.
To do this, we iterate the coaction,
\begin{equation}
    (\Delta')^{n+1} = \left(\underbrace{1 \otimes \dots \otimes \Delta' \otimes \dots \otimes 1}_{n}\right)(\Delta')^n,
\end{equation}
where, by co-associativity, we get the same result no matter on which slot the reduced coaction $\Delta'$ coacts. Here, it is important that the
reduced coproduct in the Hopf-algebra of log-divergent graphs
is $\Delta'=\Delta-1\otimes\mathrm{id}-\mathrm{id}\otimes1$ while
the reduced coaction is $\Delta'=\Delta-1\otimes\mathrm{id}$ (the term
$\mathrm{id}\otimes1$ does not exist because the algebraic
structure on the left of the tensor product differs from the algebraic structure
of the right side).

The case $n=1$,  e.g., is
\begin{equation}
    (\Delta')^{2} = \left(1 \otimes \Delta'\right)\Delta' = \left(\Delta' \otimes 1\right)\Delta'.
\end{equation}
A sum of graphs $X$ has pole order $n$ if and only if
\begin{equation}
(\Delta')^n(X) \neq 0, \qquad (\Delta')^{n+1}(X) = 0.
\end{equation}
The case $n=0$ is Theorem \ref{lem:regularity}.

We recover the pole order two in \eqref{eqcap2} by
\begin{equation}
\begin{aligned}
    (\Delta')^2\coproductIII =&~(\Delta'\otimes1)\Biggl(\coproductCap \otimes\coproductTrigII
    + \coproductXOneXTwo \otimes \coproductBubbleTrigII\Biggr)\\
    =& \coproductXOneXTwo \otimes \coproductBubbleZeroX \otimes \coproductTrigII.
\end{aligned}
\end{equation}
A further application of $\Delta'$ gives zero because all cofactors are primitive.

We conclude the subsection with a more complicated example.
\begin{equation}\label{rerouteex1}
\begin{aligned}
\Delta'\CKExampleII =&~ \coproductXOneXTwo \otimes \CKExampleIIxxcontracted + \coproductBubbleXZ \otimes \CKExampleIIxzcontracted \\
 &\hspace{-2.5cm}+ \coproductCap \otimes \CKExampleIIxxzerocontracted + \coproductDoubleBubble \otimes \CKExampleIIxxxzcontracted \\
 &\hspace{-2.5cm}+ \coproductFull \otimes \CKExampleIIxxzeroxzcontracted.
\end{aligned}
\end{equation}
Two more applications of $\Delta'$ (on any side of the tensor product) give
\begin{equation}
\begin{aligned}
(\Delta')^3\CKExampleII =&~ \coproductXOneXTwo \otimes \coproductBubbleXZ \otimes \coproductBubbleZeroX \otimes \coproductTrigOneZeroZ \\
+& \coproductXOneXTwo \otimes \coproductBubbleZeroX \otimes \coproductBubbleXZ \otimes \coproductTrigOneZeroZ \\
+& \coproductBubbleXZ \otimes \coproductXOneXTwo \otimes \coproductBubbleZeroX \otimes \coproductTrigOneZeroZ.
\end{aligned}
\end{equation}
We obtain all permutations of bubbles in the cofactors where
the bubble $x_1-x_2$ is left of the bubble $0-(x_1=x_2)$.
All cofactors are primitive, which implies that the original graph has pole order three.

The calculation of graph on the left hand side of (\ref{rerouteex1}) to
order $\ep^0$ shows the interplay between rerouting and approximation.
We reroute the graph as follows (compare (\ref{equ:re-routingOOne})).
\begin{equation}
\begin{aligned}
\Delta'\CKExampleIIRR =&~ \coproductXOneXTwo \otimes \CKExampleIIxxcontracted + \coproductBubbleXZ \otimes \CKExampleIIRRxzcontracted \\
 &\hspace{-2.5cm}+ \coproductCap \otimes \CKExampleIIxxzerocontracted + \coproductDoubleBubble \otimes \CKExampleIIxxxzcontracted \\
 &\hspace{-2.5cm}+ \coproductFull \otimes \CKExampleIIxxzeroxzcontracted.
\end{aligned}
\end{equation}
Subtraction of this equation from (\ref{rerouteex1}) gives
\begin{equation}
\begin{aligned}
&\Delta'\left(\CKExampleII - \CKExampleIIRR\right) \\
&\hspace{4cm}=\coproductBubbleXZ \otimes \left(\CKExampleIIxzcontracted - \CKExampleIIRRxzcontracted\right).
\end{aligned}
\end{equation}
The right hand side is in the kernel of $\Delta' \otimes 1$ (or $1\otimes\Delta'$),
so the subtraction has pole order $1$.
We can cure this by also rerouting the horizontal bubble, but in this
example we want to approximate the horizontal bubble that is unaffected
by the rerouting we have done.

We use the approximation \eqref{equ:threestarApprox} simultaneously
in both graphs of the subtraction (approximating individual terms in the subtraction is not possible). Note that we can distribute the external vertices freely to the external vertices in the approximation since the total weight of the approximation equals the total weight
of the original graph.

Because the pole order of the subtraction is 1 and we are only interested in the result
modulo $\mathcal{O}(\ep^0)$, it suffices to keep the lowest order in \eqref{equ:threestarApprox}. We are thus allowed to set $\ep=1/2$ in the
lower left edge of the triangle and drop the right edge ($\ep=0$).
We also only keep the $1/\ep$ term in the prefactor. This yields
\begin{equation}\label{eqrerouteex}
\begin{aligned}
    \CKExampleII = \frac{1}{\ep}\left(\CKExampleIIapprox - \CKExampleIIRRapprox \right) + \CKExampleIIRR + \mathcal{O}(\ep^0).
\end{aligned}
\end{equation}
The graphs in the first term are constructible and contribute to the
Laurent expansion according to \cite{hyperlogprocedures}
\begin{equation}
    \label{equ:reroutParens1}
    \frac{1}{\ep}\left(\CKExampleIIapprox - \CKExampleIIRRapprox \right) = \frac{z\lio - \bar{z}\loi - (z - \bar{z})\lii}{\ep(z-1)(\bar{z}-1)z\bar{z}(z-\bar{z})} + \mathcal{O}(\ep^0).
\end{equation}
In the second term in (\ref{eqrerouteex}), we use \eqref{equ:twoChain} to remove one internal vertex
\begin{equation}\label{eqrerouteexII}
\CKExampleIIRR = \frac{1}{\ep(1-2\ep)\Gamma(1-\ep)}\CKExampleIIRRapproxTwo.
\end{equation}
The new graph has pole order $2$ in $\ep$. We need to calculate it including the
constant term in $\ep$. We use rerouting again and obtain
\begin{equation}
    \label{equ:re-routingTwo}
    \CKExampleIIRRapproxTwo = \left(\CKExampleIIRRapproxTwo - \CKExampleIIRRapproxTwoRR\right) + \CKExampleIIRRapproxTwoRR.
\end{equation}
The subtraction has pole order $1$. Now, we need the full two orders of the approximation \eqref{equ:threestarApprox} to remove the horizontal bubbles in the subtraction.
The edge between $1$ and $z$ is a factor and we get
\begin{align}
    \label{equ:re-routingEx2R2}
    &\CKExampleIIRRapproxTwo - \CKExampleIIRRapproxTwoRR\notag\\
    &= \frac{\frac1\ep + 2 - \gamma_E}{((z-1)(\bar{z} - 1))^{1-2\ep}}\left(\CKExampleIIRRapproxTwoapprox - \CKExampleIIRRapproxTwoRRapprox\right)
    + \mathcal{O}(\ep^1).
\end{align}
Both graphs on the right hand side can be calculated with the ``M\"unchhausen" method
(the last example in Section \ref{extdiffII}).

We are left with the right graph in \eqref{equ:re-routingTwo}.
We drop the edge between $0$ and $1$ and integrate out $x_2$ using \eqref{equ:twoChain} to find
\begin{equation}
    \label{equ:re-routingEx2R3}
    \CKExampleIIRRapproxTwoRR = \frac{1}{2\ep(1-3\ep)\Gamma(1-\ep)} \CKExampleIIRRapproxTwoRRconvolution,
\end{equation}
which is also constructible.

If we substitute the results for \eqref{equ:re-routingEx2R2}, \eqref{equ:re-routingEx2R3}
and \eqref{equ:reroutParens1} into \eqref{eqrerouteexII} and \eqref{eqrerouteex},
we obtain the final result.
\begin{equation}
\begin{aligned}
&\CKExampleII = \frac{z(3+\bar{z})\lio - \bar{z}(3+z)\loi}{\ep z\bar{z}(z-1) (\bar{z} - 1)(z - \bar{z})} \\
&+ \frac{\frac{1}{2\ep^3} + \left(\frac{7}{2} - \frac{3\gamma_E}{2}\right)\frac{1}{\ep^2}+ \left(\frac{33}{2} - \frac{21}{2}\gamma_E + \frac{9}{4}\gamma_E^2-\frac{\pi^2}{8}\right)\frac1\ep + \left(\frac{2}{\ep^2} + \frac{14 - 6\gamma_E}{\ep}\right)\lo + \frac{8}{\ep}\loo - \frac{4}{\ep}\lii}{z\bar{z}(z-1)(\bar{z}-1)} \\
&+ \mathcal{O}(\ep^0).
\end{aligned}
\end{equation}

For a more sophisticated example for the interplay of various calculation techniques in the theory of
graphical functions, see Section 12 of \cite{Schnetz:2022nsc}.

\subsection{The Connes-Kreimer coaction}\label{sectcoact}
We close this section with some more detailed remarks on the Connes-Kreimer coaction
and its connection to the Connes-Kreimer coproduct.

We label all vertices. For internal vertices, the labels
correspond to integration variables that can be assigned arbitrarily.
So, in principle, the internal vertices should be considered unlabelled.
For rerouting, however, we want to apply subtractions in the integrand.
After subtraction, we want to identify the subgraph that is not affected by
subtraction and simplify it with different methods.
In the context of integrands, we factorize the subtraction into a factor where the subtraction is active and an unaffected factor.
In this first factor, the singularity in $\ep$ is reduced by the subtraction,
allowing us to use more powerful techniques on the second factor (an explicit
example is at the end of the previous subsection). This factorization depends on
a consistent choice of integration variables in each term.

Labels for internal vertices cause serious problems in the definition of a product
structure. Like in the integrand, one has to avoid identical labels in
factors of the product. Squaring an integral (e.g.) means to introduce a second copy of
ambient space with a new set of integration variables.
We are not aware of a canonical way to implement this in a combinatorial structure.
This problem does not occur for external vertices (with no integration). The product
of two graphs with identical external vertices is connected at the external vertices
and only the internal vertices are disconnected; see \eqref{eqprod} in Section \ref{sec:prod}.

We only obtain a (Hopf-)algebra structure upon forgetting internal labels.
For rerouting, however, we do not need the algebra structure but only the
coaction and its co-associativity. The coaction lifts naturally to internally labelled
graphs. So, more precisely, we think of algebra structures
on internally unlabelled graphs and lift the coaction to the internally labelled
setup.

To obtain a Hopf-algebra with a coproduct instead of a coaction, it is necessary
to neglect convergent dressings such as external legs. In the Connes-Kreimer
Hopf-algebra of renormalization, all trees are identified with the empty graph.
This is more natural in momentum space, where external legs and trees are purely
rational structures. In our context, it is much more natural to
keep convergent dressings and lift the coproduct to a coaction on actual graphical
functions.

Note that graphs with trivial reduced coproduct are primitive
(they are log-divergent with no subdivergence) but graphs with trivial reduced
coaction are convergent in the limit $\ep\to0$.

Once the consistency of the combinatorial structure with renormalization has been established,
it is easy to see that trivial reduced coaction implies regularity (see Theorem \ref{lem:regularity}).
The BPHZ $R$-operation can be expressed in terms of the reduced coaction \eqref{equ:coaction}
(originally conceived in \cite{Kreimer:1997dp} with a review in \cite{Kreimer:2000ja}).
\begin{equation}\label{equ:convolution}
    R(G) = m\circ(Z\otimes\phi)\circ\Delta(G) = \phi(G)+\sum_{\substack{\varnothing \neq g \subsetneq G \\ g~\text{log div. }, 1\text{PI}}} Z(g)\;\phi( G / g),
\end{equation}
where $Z$ evaluates the UV counterterm of the graph it acts on (denoted by $S_R$ in \cite{Kreimer:2000ja},) while $\phi$ denotes the Feynman integral of the graph. The operation $m$ multiplies both sides of the tensor product.
For any sum of graphs $X$ in the kernel of the reduced coation, we have
$\Delta X = 1\otimes X$.
Substitution into \eqref{equ:convolution} gives
$R(X) = \phi(X)$.
By definition, $R(X)$ is finite for any linear combination of graphs $X$ (as it is the BPHZ $R$-operation). Hence, $\phi(X)= \mathcal{O}(\ep^0)$.

\section{Self-duality and momentum space}
\label{sec:self_duality}
Planar duality in Section \ref{sec:gf_dimreg_sing} is the planar case
of self-duality of massless scalar three-point integrals with total weight
$N_G = \frac{D}{2\lambda}$ \cite{Schnetz:2026mht}.
The three-point integral $A_G(0,z_1, z_2)$ can be expressed in terms of the invariants
$z\bar{z}$ and $(z-1)(\bar{z}-1)$.
\begin{equation}\label{AGdef}
A_G(0,z_1, z_2) = \frac{1}{\norm{z_1}^{2\lambda N_G}}f_G(z) = \frac{1}{\norm{z_1}^{2\lambda N_G}}f_G(z\bar{z}, (z-1)(\bar{z}-1)).
\end{equation}
The momentum space representation of $A_G$ is determined by the Fourier transformation
\begin{equation}
\begin{aligned}
    \tilde{A}_G(p_2, p_1) =& \int \frac{\dd^Dz_1\dd^Dz_2}{(2\pi)^D} e^{iz_1p_2-iz_2p_1} A_G(0, z_1, z_2)\\
    =& \int~\dd^Dz_1 e^{iz_1p_2} \int \frac{\dd^Dz_2}{(2\pi\norm{z_1})^D} e^{-iz_2p_1}f\left(\frac{z_2^2}{z_1^2}, \frac{(z_2-z_1)^2}{z_1^2}\right).
\end{aligned}
\end{equation}
The relative minus sign between the two scalar products in the exponential is related
by the opposite orientation of momenta at the external legs; see (\ref{mompos}).

To make sense of the integrals in a classical sense, we assume that the dimension $D$ is
a positive integer. By Euclidean symmetry, the integral over $z_2$ can only depend on the invariants $\norm{p_1}, \norm{z_1}$, and the scalar product of unit vectors $\hat{p}_1\cdot\hat{z}_1$. In particular, the integral over $z_2$ is invariant under swapping $\hat{p}_1$ and
$\hat{z}_1$.
\begin{equation}
    \tilde{A}_G(p_2, p_1) = \int \dd^Dz_1~e^{iz_1p_2} \int \frac{\dd^Dz_2}{(2\pi\norm{z_1})^D} e^{-iz_2\cdot\hat{z}_1\norm{p_1}}f\left(\frac{z_2^2}{z_1^2}, \frac{z_2^2}{z_1^2} - 2\frac{z_2\cdot\hat{p}_1}{\norm{z_1}} + 1\right)
\end{equation}
Now, we scale $z_2\rightarrow z_2\norm{z_1}$ and obtain
\begin{equation}
    \tilde{A}_G(p_2, p_1) = \int \dd^Dz_1~e^{iz_1p_2} \int \frac{\dd^Dz_2}{(2\pi)^D} e^{-iz_2\cdot z_1\norm{p_1}}f\left(z_2^2, z_2^2 - 2 z_2\cdot\hat{p}_1 + 1\right).
\end{equation}
The function $f$ does not depend on $z_1$. Interchanging the order of integration, we find
\begin{equation}
\begin{aligned}
    \tilde{A}_G(p_2, p_1) &= \int \frac{\dd^Dz_2}{(2\pi)^D} \int~\dd^Dz_1 e^{iz_1(p_2 - \norm{p_1}z_2)} f\left(z_2^2, z_2^2 - 2 z_2\cdot\hat{p}_1 + 1\right) \\
    &= \int\dd^D z_2~\delta^{(D)}(p_2 - \norm{p_1}z_2) f\left(z_2^2, z_2^2 - 2 z_2\cdot\hat{p}_1 + 1\right) \\
    &= \frac{1}{\norm{p_1}^D}f(\frac{p_2^2}{p_1^2}, \frac{p_2^2}{p_1^2} + 1 - 2 \frac{p_2\cdot p_1}{p_1^2}) = A_G(0, p_1, p_2).
\end{aligned}
\end{equation}
For any graph $G$ with $N_G = \frac{D}{2\lambda}$, the position space three-point integral is
invariant under Fourier transformation.

The analogous result holds in momentum space if a momentum space graph $G^p$ with $h_{G^p}$
independent cycles (loops) has the scaling weight 
\begin{equation}
    N_{G^p} = \sum \nu^p - \frac{D}{2}h_{G^p} = \frac{D}{2}.
\end{equation}
We obtain the self-duality identities
\begin{equation}
    \tilde{A}_G=A_G,\quad\tilde{A}_{G^p}=A_{G^p},\quad\text{if}\quad\lambda N_G=N_{G^p} = \frac{D}{2}.
\end{equation}
Self-duality allows us to calculate a momentum space Feynman integral with position
space Feynman rules.
The massless propagator in momentum space is the Fourier transform of the position space propagator
\begin{equation}
\frac{1}{\norm{k}^{2\nu^p}} = \frac{\Gamma(\lambda+1-\nu^p)}{2^{2\nu^p}\Gamma(\nu^p)} \int \frac{\dd x^D}{\pi^{\frac{D}{2}}}\frac{e^{ix\cdot k}}{\norm{x}^{2(\lambda + 1 - \nu^p)}}.
\end{equation}
Upon Fourier transformation, the edge weight in position space $\nu$ and in momentum space
$\nu^p$ are related by
\begin{equation}\label{nunu}
    \nu = 1 - \frac{\nu^p - 1}{\lambda}.
\end{equation}
The normalization of position space weights is such that $\nu^p=1$
implies $\nu=1$.
We define the position space graph $G$ as $G^p$ with $\nu^p\rightarrow \nu$
and external labels opposite to momentum flow.
\begin{equation}\label{mompos}
    \momentumthreepointp \rightarrow \momentumthreepointpPositionized.
\end{equation}
By graph homology, the number of vertices of $G$ and $h_G$ is connected to the number of edges of $G$,
\begin{equation}\label{graphhom}
|\mathcal{V}_G^{\mathrm{int}}|+3+h_G=|\mathcal{E}_G|+1.
\end{equation}
We obtain
\begin{equation}\label{NGNG}
N_{G^p}=\Big(\sum_{e\in G}\frac D2-\lambda\nu_e\Big)-\frac D2(|\mathcal{E}_G|-|\mathcal{V}_G^{\mathrm{int}}|-2)=D-\lambda N_G.
\end{equation}
In particular, $N_{G^p}=D/2$ is equivalent to $N_G=D/2\lambda$.

Using self-duality, the classical link between momentum space and position space Feynman integrals gives
\begin{equation}
    \label{equ:full_duality}
    \left(\prod_e \frac{\Gamma(\nu_e^p)}{\Gamma(\lambda\nu_e)}\right)A_{G^p}(p_1, p_2) =  \tilde{A}_G(0, p_1, p_2) =  A_G(0, p_1, p_2)=\frac{f_G(p)}{\norm{p_1}^D}\quad\text{if}\quad
    N_{G^p}=D/2.
\end{equation}
According to (\ref{mompos}), the graphical function $f_G(p)$ is considered as a function of the complex variable $p$.

For an externally planar graph $G$, the Feynman integral $A_{G^p}$ can be interpreted as the position space integral of the dual graph.
This gives planar duality in Section \ref{sec:gf_integer}.

Although self-duality was effectively proved in 2015 in \cite{Golz:2015rea},
it remained unnoticed until Jiang pointed out the connection
between momentum and position space in the context
of four-dimensional super Yang-Mills theory \cite{Jiang:2025vrr}.
\smallskip

Self-duality has several implications.
Firstly, we can express any massless scalar momentum space three-point function directly in terms of a graphical function.
If the original graph does not have the suitable weight $D/2$, we attach an external leg of suitable weight. In momentum space, the external propagator only contributes by an elementary factor to the Feynman integral.

Secondly, self-duality also provides a new relation between graphical functions.
In momentum space, attaching a leg to an external vertex is a multiplication.
This implies the momentum space relations
\begin{equation}
\momentumthreepointza=||z_1||^{2\nu^p}\momentumthreepointzd=||z_2-z_1||^{2\nu^p}\momentumthreepointzb=||z_2||^{2\nu^p}\momentumthreepointzc.
\end{equation}
If the graph $G^p$ with the attached leg of weight $\nu^p$ has total weight $N_{G^p}+\nu^p= D/2$, we can map it into position space (with an appropriate change of edge weights).
In position space, the scalar factors can be interpreted as edges between the external vertices
opposite to the attached leg.
The factors have the weight $-\nu^p/\lambda=\nu-(\lambda+1)/\lambda$ by (\ref{nunu}).
With (\ref{NGNG}), the condition $N_{G^p}+\nu^p= D/2$ translates into
$N_{G_{\mathrm{tot}}}-(\nu-(\lambda+1)/\lambda)=(\lambda+1)/\lambda$, where $G_{\mathrm{tot}}$
is the full graph with the attached and the external edges. We obtain
\begin{equation}
\label{equ:rotationRel}
\positionthreepointzd=\positionthreepointzb=\positionthreepointzc\text{if }N_{G_{\mathrm{tot}}}=\nu.
\end{equation}
We used the fact that the $\Gamma$ factors in the duality relation \eqref{equ:full_duality}
are identical, so they can be dropped. Note that in (\ref{equ:rotationRel}) only the attached and external edges are rotated, while the $G$ is fixed.

We use the above identity to calculate the following graphical function. The graph
\begin{equation}
    G = \rotationExTwo
\end{equation}
has total weight $N_G = D/2\lambda$, so (\ref{equ:rotationRel}) can be applied
if we attach an edge between $0$ and $1$ with weight $(-1+3\ep)/\lambda$.
This edge does not affect the value of the graphical function $f_G(z)$. We obtain
\begin{equation}\label{selfdualex}
    \begin{aligned}
    \rotationExTwoAttached = \rotationExTwoRotated = \rotationExTwoCombined
    = \frac{1}{(z\bar{z})^{2\ep}} \rotationExTwoDetached,
    \end{aligned}
\end{equation}
where we combined the two edges between $0$ and $z$ in the second step and traded the result for a factor in the third step.
The final graph is constructible (while $G$ was not).

If we are only interested in the first two orders of $f_{G_a}(z)$, we can generalize the graph
$G$ to
\begin{equation}
    G_a = \rotationExA.
\end{equation}
An approximation in analogy to \eqref{equ:edgeApprox} gives
\begin{equation}
    f_{G_a}(z)=\frac{1 + a}{3} \rotationExTwo + \frac{2-a}{3} \rotationExNone + \mathcal{O}(\ep^2).
\end{equation}
With (\ref{selfdualex}), both graphs can be constructed.

A general version of \eqref{equ:rotationRel}, that neither has the edge $0$-$1$ nor is restricted to $N_{G_{\mathrm{tot}}}=\nu$, is
\begin{equation}
    \label{equ:rotationRelFull}
    \graphicalfunctiona = \graphicalfunctionb = \graphicalfunctionc,
\end{equation}
where $G=G_{\mathrm{tot}}$ is the full graph on the left hand side.\\[3pt]

This setup leads to a third consequence of self-duality \eqref{equ:full_duality} which is a new twist identity for Feynman periods (see Section \ref{thetwist} for the classical twist).
If the completed graph $\overline{G}$ of a Feynman period has a four vertex split,
\begin{equation}
  P_{\overline{G}} = \hspace{2pt}\NewTwist,
\end{equation}
then we can decomplete at $\infty$ and use the general identity \eqref{equ:rotationRelFull}
on one side of the split to obtain a twisted graph.
Completion and gluing the two split parts (possibly moving some edges at the split vertices as
for the classical twist) gives a new completed graph with period $P_{\overline{G}}$.

If we consider $\phi^4$ and non-$\phi^4$ graphs in $D=4$ dimensions,
the new twist leads to identities somewhat similar to the classical twist.
However, there exist identities that are unique for the new twist (and others that are unique for
the classical twist).
Up to eight loops, we have 18 new identities that are not classical twists while only four identities are classical twists but not new twists. No new identity gives a new relation inside $\phi^4$
theory (they only relate $\phi^4$ periods to non-$\phi^4$ periods). In particular, the conjectures in
\eqref{conjid} remain open.

\section{The future of graphical functions}
\label{sec:future}
As we have seen throughout this review, many identities for graphical functions are already known, arising from symmetry properties, conformal transformations, differential equations, and their relations to three‑point functions.
Nonetheless, we also know various relations between Feynman periods that we cannot yet explain within the current framework.
This suggests that additional identities remain to be discovered, and that the full algebraic and analytic structure of graphical functions is still not completely understood. 
To that end, contiguous relations between graphical functions of different weights, obtained through IBP identities, are under active investigation.
These relations provide a systematic way to connect graphical functions at successive weights, and may reveal additional hidden structure within the space of such functions.

The discussion in this review has mostly been restricted to scalar quantum field theories.
The recent extension of graphical functions to include numerators, corresponding to particles with spin greater than zero (developed in \cite{Schnetz:2025opm}), represents an important step beyond the scope considered here.

In particular, \cite{Schnetz:2025opm} applies graphical functions with spin to Gross–Neveu–Yukawa theory, demonstrating that the framework can accommodate fermionic and Yukawa interactions. Some results for Feynman periods in this theory were included in Section \ref{sec:periodsEven}, but the full theory is too complex to be included in these
lecture notes.
A further extension to gauge bosons, which would enable applications to gauge theories such as quantum electrodynamics and quantum chromodynamics, is currently under active development.

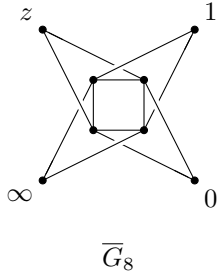
\begin{figure}[t]
\def\scale{1}
\begin{center}
\begin{tikzpicture}[baseline={([yshift=-.7ex]0,0)}]
    \coordinate (vm) at  (0,0);
    \coordinate[label=above left:$z$] (vz) at  (-\scale, \scale);
    \coordinate[label=above right:$1$] (v1) at  ( \scale, \scale);
    \coordinate[label=below left:$\infty$] (voo) at (-\scale,-\scale);
    \coordinate[label=below right:$0$] (v0)  at (\scale, -\scale);
    \coordinate (v2)  at ({ \scale/3}, { \scale/3});
    \coordinate (v3)  at ({-\scale/3}, { \scale/3});
    \coordinate (v4)  at ({-\scale/3}, {-\scale/3});
    \coordinate (v5)  at ({ \scale/3}, {-\scale/3});
    \node (G) at (0,{-2*\scale}) {$\overline{G}_8$};
    \draw (vz) -- (v2);
    \draw (voo) -- (v3);
    \draw (v0) -- (v4);
    \draw (v1) -- (v5);
    \draw[preaction={draw, white, line width=3pt, -}] (vz) -- (v4);
    \draw[preaction={draw, white, line width=3pt, -}] (voo) -- (v5);
    \draw[preaction={draw, white, line width=3pt, -}] (v0) -- (v2);
    \draw[preaction={draw, white, line width=3pt, -}] (v1) -- (v3);
    \draw (vz) -- (v2); 
    \draw (voo) -- (v3); 
    \draw (v0) -- (v4); 
    \draw (v1) -- (v5); 
    \draw (v2) -- (v3) -- (v4) -- (v5) -- (v2);
    \filldraw (v0) circle (1pt);
    \filldraw (v1) circle (1pt);
    \filldraw (vz) circle (1pt);
    \filldraw (voo) circle (1pt);
    \filldraw (v2) circle (1pt);
    \filldraw (v3) circle (1pt);
    \filldraw (v4) circle (1pt);
    \filldraw (v5) circle (1pt);
\end{tikzpicture}
\caption{The irreducible internally completed graphical function $f_{\overline{G}_8}(z)$ in four dimensions with all edges of weight $1$ is conjectured to be elliptic~\cite{Borinsky:2021gkd}.}
\label{fig:elliptic} 
\end{center}
\end{figure}
Regarding the theory of graphical functions itself, future investigations into the types of functions that can appear beyond hyperlogarithms would be of considerable interest.
At higher loop orders, the class of functions associated with Feynman graphs extends beyond hyperlogarithms, with elliptic integrals beginning to appear, and it remains an important open question how to incorporate such functions systematically into the graphical-function framework. 
Figure~\ref{fig:elliptic} illustrates a graphical function that is conjectured to be elliptic, highlighting the need for further development in this direction.

Further generalisations of graphical functions, such as extensions to odd dimensions, to massive theories, or even to cases with more than three external vertices, are, at present, speculative, and it is not yet clear whether they can be achieved.
Since all three of these directions are of significant interest in physical research, they represent important goals for future investigation.

On the side of software development underpinning computations, the \texttt{HyperFORM} package~\cite{Kardos:2025klp} provides an efficient tool for parametric integration, and its functionality is expected to be expanded further in the future.
On the more technical side, the \texttt{HyperlogProcedures} package could be ported to a higher-performance framework---such as \texttt{C++}, \texttt{Rust} or \texttt{FORM}---in order to achieve much greater speed and memory efficiency.

There are many additional directions that can be pursued, and they will undoubtedly be of interest for future research.
\section{Further reading}
\label{sec:further}
We conclude by suggesting a few resources for more in‑depth reading. We first point to standard textbooks covering modern multiloop techniques, followed by references on graphical functions, which provide more detailed background on several of the topics discussed in this review.

Modern and comprehensive treatments of multiloop Feynman‑integral techniques, ranging from introductory concepts to advanced topics such as differential equations, polylogarithms, elliptic integrals, sector decomposition, and asymptotic expansions, are available in \cite{
Weinzierl:2022eaz, Henn:2014qga, Badger:2023eqz, Smirnov:2004ym}. 
A dedicated discussion of Mellin–Barnes techniques, including the construction of Mellin-Barnes representations, analytic continuation, and contour‑integration methods for multi‑scale integrals, can be found in \cite{Dubovyk:2022obc}.

For graphical functions in integer dimensions, including detailed proofs of their structural relations and numerous worked examples, a comprehensive overview is provided in \cite{Borinsky:2021gkd}. Graphical functions defined within dimensional regularisation, together with their analytic properties and connections to multiloop integrals, are treated in \cite{Schnetz:2022nsc}.
This reference includes, for example, detailed discussions of techniques such as appending an edge and various approximation methods, which relate to the material presented in Sections~\ref{sec:leg_dimreg_sing} and~\ref{sec:gf_dimreg_sing}. 
In addition, a complementary framework based on parametric representations, useful for connecting graphical functions to Feynman integrals in Schwinger parameters, is introduced in \cite{Golz:2015rea}, while the specific five‑twist construction discussed in Section \ref{sec:periodsEven} is analysed and proven in detail in \cite{Schnetz:2025mjw}. Self-duality in Section \ref{sec:self_duality} is
treated in detail in \cite{Schnetz:2026mht}.

For readers interested in the broader theory of periods and their significance in quantum field theory, \cite{Schnetz:2008mp} offers an accessible introduction that explains how periods arise naturally in the evaluation of Feynman integrals and how they connect to number‑theoretic structures.
For a more mathematical review on the calculation of periods using graphical functions see \cite{Borinsky:2022lds}.
Additional background on multiple polylogarithms, including their algebraic relations and the shuffle algebra mentioned in the text, can be found in the lecture notes \cite{Duhr:2014woa}, which provide a clear and pedagogical development of their fundamental properties, even though they focus on the non‑single‑valued setup.
For an introduction to Hopf algebras, particularly in the context of renormalization theory, the lecture notes \cite{Manchon:2001bf} give a detailed presentation of the Connes–Kreimer framework, illustrating how Feynman graphs generate connected Hopf algebras and how the associated Birkhoff decomposition encodes the renormalization procedure.

The algorithms for graphical functions are implemented in the {\texttt{HyperlogProcedures}} package~\cite{hyperlogprocedures}, which runs within \texttt{MAPLE}. Additional tools for carrying out direct parametric integrations of Feynman diagrams in terms of hyperlogarithms, as described in Section~\ref{sec:identities}, are also available in \texttt{MAPLE}. The {\texttt{HyperInt}} package \cite{Panzer:2014caa} can be obtained from \cite{hyperint}, and a recent \texttt{FORM}-based implementation, {\texttt{HyperFORM}}, is provided in \cite{Kardos:2025klp,kardos_2025_17706909}.

\subsection*{Acknowledgments}
We acknowledge the ERC Advanced Grant 101095857 {\it Conformal-EIC}.
O.S. has been supported by the DFG through grant SCHN 1240/3-1.

\bibliographystyle{JHEP}
\providecommand{\href}[2]{#2}\begingroup\raggedright\endgroup
\end{document}